\documentclass[article,12pt]{article}

\usepackage{tabu}
\usepackage{amsthm,amsmath}
\usepackage{amssymb}
\usepackage{makecell}
\usepackage{bm}
\usepackage{setspace}
\usepackage{subfigure} 
\usepackage{tabularx}
\usepackage{tutorial}
\usepackage{multirow}
\usepackage{xcolor}
\usepackage{graphicx}
\usepackage{epsfig,wrapfig,colortbl}
\usepackage{verbatim}
\usepackage{enumitem}
\usepackage{framed}

\newtheorem*{theo}{Theorem}

\newcommand{\moment}{\boldsymbol{\rho}} 

\title{A Survey of Monte Carlo Methods \\
 for Parameter Estimation}

\author{D. Luengo$^1$, L. Martino$^2$, M. Bugallo$^3$, V. Elvira$^4$,  S. S\"arkk\"a$^5$  \\
{\footnotesize$^1$  Universidad Polit\'ecnica  de Madrid (UPM), Spain.}   \\
 {\footnotesize$^2$ Universidad Rey Juan Carlos (URJC), Spain.} \\
 {\footnotesize$^3$ Stony Brook University, USA.} \\ 
 {\footnotesize$^4$ IMT Lille Douai, Universit\'e de Lille, France,} \\
{\footnotesize$^5$ Aalto University, Finland.}
}

\date{}

\begin{document}

\noindent
Cite as:  \\
{\it D. Luengo, L. Martino, M. Bugallo, V. Elvira, S. S\"arkk\"a, ``A survey of Monte Carlo methods for parameter estimation''. EURASIP Journal on Advances in Signal Processing, 2020, Article number: 25 (2020). }

\begin{minipage}{\textwidth}
   \maketitle
\end{minipage}

\begin{abstract} 

Statistical signal processing applications usually require the estimation of some parameters of interest given a set of observed data.
These estimates are typically obtained either by solving a multi-variate optimization problem, as in the maximum likelihood (ML) or maximum a posteriori (MAP) estimators, or by performing a multi-dimensional integration, as in the minimum mean squared error (MMSE) estimators.
Unfortunately, analytical expressions for these estimators cannot be found in most real-world applications, and the Monte Carlo (MC) methodology is one feasible approach.
MC methods proceed by drawing random samples, either from the desired distribution or from a simpler one, and using them to compute consistent estimators.
The most important families of MC algorithms are Markov chain MC (MCMC) and importance sampling (IS).
On the one hand, MCMC methods draw samples from a proposal density, building then an ergodic Markov chain whose stationary distribution is the desired distribution by accepting or rejecting those candidate samples as the new state of the chain.
On the other hand, IS techniques draw samples from a simple proposal density, and then assign them suitable weights that measure their quality in some appropriate way.
In this paper, we perform a thorough review of MC methods for the estimation of static parameters in signal processing applications.
A historical note on the development of MC schemes is also provided, followed by the basic MC method and a brief description of the rejection sampling (RS) algorithm, as well as three sections describing many of the most relevant MCMC and IS algorithms, and their combined use.
Finally, five numerical examples (including the estimation of the parameters of a chaotic system, a localization problem in wireless sensor networks and a spectral analysis application) are provided in order to demonstrate the performance of the described approaches.

\end{abstract}

\section{Introduction}
\label{sec:intro}

\vspace*{12pt}

\subsection{Motivation: Parameter estimation in statistical signal processing applications}
\label{sec:motivation}

\vspace*{12pt}

Statistical inference deals with the estimation of a set of unknowns given a collection of observed data contaminated by noise and possibly by some other types of distortions and interferences \cite{casella2002statistical}.
In many signal processing applications, this typically amounts to inferring some static parameters of interest from the noisy observations \cite{Scharf1991,Kay1993,van2004detection}.\footnote{Note that, although we concentrate here on the parameter estimation problem, other related signal processing problems (model selection, prediction, classification, etc.) can be addressed in a similar way.}
For instance, in \emph{denoising} applications the aim is reconstructing the original signal (e.g., an audio recording or an image) from the noisy observations \cite{godsill1997book}.
An extended version of this problem occurs in \emph{blind deconvolution}, where a noisy filtered signal is available and the goal is to recover both the unknown filter and the input \cite{haykin1994blind}.\footnote{Closely related problems are blind equalization (where the input is a digital communications signal), blind identification (where the main goal is to recover the filter) \cite{ding2001blind}, and blind source separation (where multiple input signals have to be separated from a collection of multiple outputs) \cite{comon2010handbook}.}
Finally, as a third application, \emph{target localization/tracking} in wireless sensor networks requires estimating/tracking the location of the target (maybe jointly with some parameters of the system, such as the noise variance, the propagation constant or even the position of the sensors) from measurements recorded by the sensors \cite{zhao2004wireless,swami2007wireless}.

In the \emph{Bayesian framework}, all the aforementioned problems are addressed by formulating a \emph{prior distribution}, which should gather all the available information about the parameters of interest external to the data, and assuming an input-output model (the \emph{likelihood}), that incorporates our knowledge or lack thereof on the way in which the observed data relate to the unknown parameters \cite{bernardo2001bayesian}.
Then, \emph{Bayes theorem} allows us to obtain the \emph{posterior distribution}, which takes into account both the effect of the prior information and the observed data in an optimal way.
Finally, the desired Bayesian point estimators are obtained by minimizing a pre-defined cost function that can typically be expressed either as some integral measure with respect to (w.r.t.) the posterior or as some optimization problem.
For instance, the well-known minimum mean squared error (MMSE) estimator corresponds to the conditional mean of the parameters of interest given the data (i.e., the expected value of the posterior distribution), whereas the maximum a posteriori (MAP) estimator corresponds to the value of the parameters where the posterior attains its highest peak.\footnote{Note that the MAP estimator can also be expressed in an integral form \cite{Scharf1991,Kay1993}, but the maximization approach is much more useful from a practical point of view.}
Note that a similar procedure is also followed by \emph{frequentist methods} (i.e., in the end they also attempt to minimize some cost function which is either expressed as some integral measure or formulated as an optimization problem), even though they are completely different from a conceptual point of view.
Indeed, the frequentist maximum likelihood (ML) estimator simply corresponds to the Bayesian MAP estimator with a uniform prior.
Hence, although we focus on Bayesian approaches in the sequel, all the techniques mentioned here are also applicable in a frequentist context.

Unfortunately, obtaining closed-form expressions for any of these estimators is usually impossible in real-world problems.
This issue can be circumvented by using approximate estimators (e.g., heuristic estimators in the frequentist context or variational Bayesian approximations) or by restricting the class of models that were considered (e.g., in the case of Bayesian inference by using only conjugate priors).
However, with the increase in computational power and the extensive development of Monte Carlo methods, Bayesian inference has been freed from the use of a restricted class of models and much more complicated problems can now be tackled in a realistic way.
In the following section we briefly review the history of Monte Carlo methods, pointing out the key developments and some of the most relevant algorithms that will be described in detail throughout the paper.
Note that, apart from MC methods, there are several alternative techniques for approximating integrals in statistical inference problems \cite{evans1995methods}: asymptotic methods, multiple quadrature approaches, and subregion adaptive integration.
However, these schemes cannot be applied in high-dimensional problems and MC algorithms become the only feasible approach in many practical applications.
{Another related topic which is not covered here due to space constraints is variational Bayesian inference.
However, the interested reader can check some of the existing tutorials (and references therein) for an overview these methods \cite{tzikas2008variational,fox2012tutorial}.}


\subsection{Framework: Monte Carlo methods}
\label{sec:framework}

\vspace*{12pt}

The so-called Monte Carlo (MC) methods encompass a large class of stochastic simulation techniques that can be used to solve many optimization and inference problems in science and engineering.
Essentially, MC methods proceed by obtaining a large pool of potential values of the desired parameters and substituting the integrations by sample averages.
In practice, these parameter values can be obtained either by physically replicating the desired experiment or by characterizing it probabilistically and generating a set of random realizations.

The origin of MC methods can be traced back to Buffon's experiments to compute an empirical value on the \emph{St. Petersburg game},\footnote{The St. Petersburg game consists of tossing a fair coin repeteadly until a head occurs \cite{martin2011stPetersburg}. The payoff then is $2^{k}$, where $k$ is the number of tosses required. Buffon's goal was computing the expected payoff of the game in practice (theoretically it is infinite), which turned out to be $4.9106$ in his experiment.} and the formulation of his famous experiment (nowadays commonly known as \emph{Buffon's needle}) to calculate the value of $\pi$ \cite{buffon1733,buffon1777}.\footnote{Buffon's needle experiment consists of dropping a needle of length $\ell$ on a grid of parallel lines uniformly separated a distance $d > \ell$, and counting the number of times that the needles intersect the lines ($n$) out of the $N$ experiments. This empirical intersection probability, $\hat{p}=\frac{n}{N}$, can be used to obtain an approximate value of $\pi$, since $p = \frac{2\ell}{\pi d}$, and thus $\pi \approx \frac{2\ell}{\hat{p}d}$.}
Buffon's needle experiment became quite well known after it was mentioned by Laplace in 1812 \cite{Laplace1820theorie}, and several scientists attempted to replicate his experiment during the last quarter of the 19th century \cite{deMorgan1872budget,hall1873experimental,lazzarini1902applicazione,riedwyl1990rudolf}.\footnote{Actually, Lazzarini's experimental approximation of $\pi \approx 3.1415929$ (accurate to six decimal places), provided in \cite{lazzarini1902applicazione}, has been disputed and several authors have suggested that he did not perform a fair experiment \cite{gridgeman1960geometric,OBeirne1965puzzles,badger1994lazzarini}.}
Meanwhile, other statisticians were experimenting with different mechanisms to generate random numbers (e.g., using cards, a roulette or dice) to verify empirically, through some kind of primitive stochastic simulation, their complicated statistical procedures \cite{stigler1991stochastic}.
Another example of simulation in statistical computations occurred at the beginning of the 20th century, when William Gosset (``Student'') published his famous papers, where he investigated the distribution of the $t$-statistic and the correlation coefficient \cite{student1908a,student1908b}.\footnote{William S. Gosset published his two famous papers under the pseudonym ``Student'', after attaining permission from his employer \emph{Arthur Guinness \& Sons} of Dublin, to avoid conflicts with other employees who were forbidden from publishing papers in scientific journals \cite{hotelling1930british,zabell2008student}.}
Finally, Leonard H. C. Tippett devised a way to systematically draw random numbers for his experiments on extreme value distributions and published a list of random digits that was used by subsequent researchers \cite{tippett1925extreme,tippett1927random}.
However, all these approaches occurred before the advent of computers and aimed only at solving some particular problem at hand, not at providing some general simulation method (except for Galton's approach \cite{galton1890dice}, which provided a generic way to draw normal random variables (RVs) for all types of applications, but failed to gain widespread acceptance).

In spite of all these early attempts to perform stochastic simulation (a.k.a. statistical sampling), the formulation of the MC method as we know it today did not happen until the construction of the first computers in the 1940s.\footnote{Apparently, Enrico Fermi was the first one to make a systematic use of statistical sampling techniques to compute approximations to all kind of physical quantities of interest while working in Rome (i.e., before 1938). However, he never wrote anything about it and we only have an indirect account of this fact from his student Emilio Segr\`e \cite{segre1980x} (see also \cite{anderson1986,metropolis1987beginning,gass2005model}).}
Stanislaw Ulam, a Polish mathematician working at Los Alamos National Laboratory, devised the MC method while convalescing from an illness in 1946 \cite{metropolis1987beginning,eckhardt1987stan}.
He was playing solitaire and trying to calculate the probability of success (a difficult combinatorial problem) when he realized that an easier way to accomplish that task (at least in an approximate way) would be to play a certain number of hands and compute the empirical success probability.
On his return to Los Alamos, he learnt of the new computers that were being built from his close friend John von Neumann, a consultor both at Los Alamos and the Ballistics Research Laboratory (where the first computer, the ENIAC, was being developed), and discussed the possibility of developing a computer-based implementation of his idea to solve difficult problems in statistical physics.
Von Neumann immediately recognized the relevance of Ulam's idea and sketched an approach to solve neutron diffusion/multiplication problems through computer-based statistical sampling in a 1947 letter to Robert Richtmyer (head of the Theoretical Division at Los Alamos) \cite{eckhardt1987stan}.
The method was then successfully tested on 9 neutron transport problems using ENIAC and Nicholas Metropolis coined the name ``Monte Carlo'', inspired by an uncle of Stan Ulam who borrowed money from relatives because ``he just had to go to Monte Carlo'' \cite{metropolis1987beginning,gass2005model}.
The seminal paper on MC was then published in 1949 \cite{metropolis1949monte}, more powerful computers were developed (like the MANIAC in Los Alamos \cite{anderson1986}), and many physicists started using computer-based MC methods to obtain approximate solutions to their problems \cite{hurd1985note}.
MC methods required an extensive supply of random numbers, and the development of the essential random number generators required by MC methods also started during those years.
For instance, von Neumann described the rejection sampling (RS) method in a 1947 letter to Ulam \cite{eckhardt1987stan} (although it was not published until 1951 \cite{vonNeumann1951various}) and Lehmer introduced linear congruential random number generators in 1951 \cite{lehmer1951mathematical}.

The next milestone in statistical sampling was the development of the Metropolis-Hastings (MH) algorithm.
The MH algorithm was initially devised by Nicholas Metropolis \emph{et al.} in 1953 as a general method to speed up the computation of the properties of substances composed of interacting individual molecules \cite{metropolis1953equation}.
The idea is rather simple: random uniformly distributed moves of particles around their current position were proposed, if the global energy of the system was decreased these moves were always accepted, otherwise they were accepted only with some non-null probability that depended on the energy increase (the larger the increase, the less likely the move to be accepted).
Rejected moves were also used to compute the desired averages.
Metropolis \emph{et al.} proved that the method was ergodic and samples were drawn from the desired distribution.
This approach can be seen as a Markov chain, with an RS sampling step at the core to ensure that the chain has the desired invariant probability density function (PDF), and thus Markov chain Monte Carlo (MCMC) methods were born.
A symmetric proposal density was considered in \cite{metropolis1953equation}.
In 1970, Hastings showed that non-symmetric proposal densities could also be used \cite{hastings1970monte}, thus allowing for much more flexibility in the method, and proposed a generic acceptance probability that guaranteed the ergodicity of the chain.
In the meantime, a different acceptance probability rule had been proposed by Barker in 1965 \cite{Barker1965}, and it remained to be seen which rule was better.
This issue was settled in 1973 by Peskun (a Ph.D. student of Hastings), who proved that Hastings acceptance rule was optimal \cite{Peskun1973}.
The MH algorithm was extensively used by the physics community since the beginning, but few statisticians or engineers were aware of it until the 1990s \cite{hitchcock2003history}.

Another crucial event in the history of MC methods was the introduction, by Stuart Geman and Donald Geman in 1984, of a novel MCMC algorithm, the Gibbs sampler, for the Bayesian restoration of images \cite{geman1984stochastic}.
The Gibbs sampler became very popular soon afterwards, thanks to the classical 1990 paper of Gelfand and Smith \cite{gelfand1990sampling}, who gave examples of how the Gibbs sampler could be applied in Bayesian inference.
Andrew Gelman showed in 1992 that the Gibbs sampler was a particular case of the MH algorithm \cite{gelman1992iterative}, thus causing a renewed interest in the MH algorithm by statisticians.
Then, Tierney wrote an influential paper on the history and theory of the MH algorithm in 1994 \cite{tierney1994markov}, where he showed how it could be used to deal with non-standard distributions in Bayesian inference.
Simple explanations of the Gibbs sampler and the MH algorithm also appeared in the 1990s \cite{casella1992explaining,chib1995understanding}, and those two methods started being applied for all sort of problems during the following years: medicine \cite{gilks1993modelling}, econometrics \cite{geweke1994bayesian}, biostatistics \cite{gelman1996markov}, phylogenetic inference \cite{mau1999bayesian}, etc.
Indeed, the MH algorithm has become so popular since its re-discovery in the early 1990s, that it was named one of the top 10 algorithms in the 20th century by the \emph{IEEE Computing in Science \& Engineering Magazine} \cite{dongarra2000guest}.

The first signal processing applications of MCMC followed soon after Geman and Geman's publication of the Gibbs sampler (indeed, their original application involved a signal processing problem: the denoising of images).
In the 1990s, both the MH algorithm and the Gibbs sampler were applied to several signal processing problems: blind identification, deconvolution and equalization  \cite{li1993blind,chen1995blind,doucet1996fully,clapp1997bayesian,cappe1999simulation}; denoising and restoration of missing samples in digital audio recordings \cite{godsill1997book,ruanaidh1994interpolation,godsill1996robust2,godsill1997bayesian}; reconstruction of the images obtained in computed tomography  \cite{bouman1995tractable,saquib1998ml}; parameter estimation of time-varying autoregressive (AR) models \cite{rajan1995parameter,rajan1997bayesian}; etc.
Then, Fitzgerald published the first tutorial on MCMC methods for signal processing applications in 2001 \cite{fitzgerald2001markov}, and the first special issue on MC methods for statistical signal processing (edited by Petar Djuric and Simon Godsill) appeared in 2002 \cite{djuric2002guest}.
During these years, tutorial papers on the related areas of signal processing for wireless communications and machine learning also appeared \cite{wang2002monte,andrieu2003introduction}, as well as another review paper on MC methods for statistical signal processing \cite{doucet2005monte}.

The second large family of Monte Carlo methods are the so-called importance sampling (IS) and its adaptive versions (AIS).
Unlike MCMC techniques, where candidate samples can be either accepted or discarded, IS methods employ all the generated candidates, assigning them a weight according to their ``quality''.
IS was first used in statistical physics in order to estimate the probability of nuclear particles to penetrate shields \cite{kahn1950random}.
During the following decades, IS was extensively used as a variance reduction technique (especially for rare event simulation) in a large variety of applications: operations research \cite{kahn1953methods}, simulation of stochastic processes \cite{moy1965sampling}, other problems in statistical physics \cite{murthy1986analytical,booth1986monte}, digital communications \cite{davis1986improved,hahn1987developments}, computer reliability \cite{conway1986monte}, inventory systems \cite{hesterberg1987importance}, etc.
In the 1970s and 1980s, several authors also applied the IS principle in Bayesian inference problems when direct sampling from the posterior distribution was either impossible or impractical \cite{kloek1978bayesian,stewart1979multiparameter,hesterberg1991importance}.
The limitations of the IS approach were also recognized at this time: the performance of IS-based estimators critically depends on the choice of the proposal, with good proposals leading to a substantial decrease in variance and bad proposals resulting in a very poor performance (with a potentially infinite variance from a theoretical point of view).
In order to solve these issues, the multiple IS (MIS) approach and alternative weighting schemes (like the so called deterministic mixture (DM)) were proposed in the 1990s \cite{hesterberg1991weighted,hesterberg1995weighted,veach1995optimally,owen2000safe}.
During these years, sequential importance sampling (SIS) methods (a.k.a. particle filters) were also developed as an alternative to the Kalman filter for the estimation of dynamic parameters \cite{gordon1993novel,doucet2009tutorial}.
These methods are also based on the IS methodology, with weights that are sequentially updated as new observations become available.
See the companion tutorial in this special issue for a detailed review of sequential Monte Carlo (SMC) methods, which essentially correspond to SIS with resampling \cite{doucet2009tutorial}.

However, IS techniques did not become widely known to all computational statistics, machine learning and statistical signal processing practitioners until the 2000s.
In 2001, Iba published a cross-disciplinary survey in which he grouped several algorithms where ``a set of ``walkers'' or ``particles'' is used as a representation of a high-dimensional vector'' under the generic name of \emph{population Monte Carlo algorithms} \cite{iba2001population}.
Soon afterwards, Capp\'e et al. published their influential population Monte Carlo (PMC) paper \cite{Cappe04}, where they borrowed the name coined by Iba for their proposed AIS framework.
In short, \cite{Cappe04} showed that previously drawn samples can be used to adapt the proposal in order to reduce the variance of the desired estimators.
The original PMC algorithm considered a set of Gaussian proposals with different variances and means selected from the previous population through a multinomial resampling step, where particles were selected with a probability proportional to their IS weights.
This classical or standard PMC algorithm is numerically unstable and shows a poor performance in many practical applications, but opened the door to other improved PMC algorithms, like the mixture PMC (M-PMC) \cite{Cappe08} or the recent deterministic mixture PMC (DM-PMC) \cite{ElviraPMC15}.
Furthermore, the success of PMC-based approaches renewed the interest in IS techniques for the estimation of static parameters, encouraging authors to develop other AIS methods, like the adaptive multiple importance sampling (AMIS) \cite{CORNUET12} or the adaptive population importance sampling (APIS) \cite{APIS15} algorithms.

Finally, let us remark that many important advances have occurred in the field of Monte Carlo methods during the last 20 years: adaptive MCMC techniques that increase the acceptance rate and decrease the correlation among samples; gradient-based MCMC methods which improve the performance in high-dimensional parameter spaces; multiple candidate MCMC algorithms for a higher efficiency in sample generation; generalized sampling and weighting schemes in MIS algorithms for a reduced variance of the desired estimators; the combination of MCMC and AIS techniques in order to exploit their complementary strong points and minimize their drawbacks; etc.
A detailed description of all these methods is provided in this survey.
Other important topics that are also briefly covered here are the following: the pseudo-marginal MCMC framework \cite{andrieu2009pseudo}; approximate Bayesian computation (ABC) \cite{marin2012approximate,sunnaaker2013approximate}; the application of Monte Carlo algorithms in a big data context \cite{zhu2014big}; noisy MCMC methods; and approximated likelihood algorithms.

\subsection{Related articles, books and software packages}

\vspace*{12pt}

The literature on MC methods is rather vast, with many technical reports, journal papers, conference articles, books and book chapters that cover different aspects of the many existing MC algorithms.
In this section, we provide a brief summary (which intends to be illustrative rather than exhaustive) of the articles and books that provide a tutorial introduction or an overview of several aspects of MC methods and closely related topics.
At the end of the section we also describe some of the most relevant software packages which are freely available to implement several important MC algorithms.
Note that these articles, books and/or software packages often concentrate on some particular class of MC algorithms, and the user has to select the most appropriate family of methods and software for the specific problem.
In particular, note that different MCMC methods have different convergence properties, and therefore we encourage users to be careful and select the most reliable algorithm for their problem.

On the one hand, many excellent books are entirely devoted to the general theory and practice of MC methods \cite{Robert04,gamerman2006markov,Liu04b,liang2011advanced,brooks2011handbook,kroese2013handbook}.
However, none of these books is specifically written with signal processing practitioners in mind and they are 5--14 years old, thus not covering several important recently developed algorithms.
On the other hand, several books are also devoted to specific classes of MC methods.
For instance, \cite{ristic2003beyond} and \cite{ristic2013particle} focus on particle filters for tracking applications and random set models respectively, \cite{candy2016bayesian} details several different state-space processors (including those based on particle filters),
\cite{doucet2001sequential} is entirely devoted to the theoretical and practical aspects of SMC methods, and \cite{sarkka2013bayesian} covers Bayesian filtering and smoothing techniques from Kalman to particle filters.
Finally, several books address the related topic of random variable generation \cite{devroye1986non-uniform,Dagpunar88,Hormann03,Gentle04,MartinoBook18}, which is an essential issue for MC algorithms, and some of these books also contain one or more chapters on MC methods (e.g., Chapter 7 of \cite{Gentle04,MartinoBook18}).

There are also many other journal papers and conference articles that provide tutorial descriptions of MC methods, but they are either more than 10 years old, differ in scope from the present paper or cover only some specific class of MC algorithms.
The first tutorial on MC methods for signal processing practitioners (as far as we know), covering classical MC techniques (e.g., the MH algorithm, the Gibbs sampler and reversible jump MCMC) for parameter estimation and model selection, appeared in 2001 \cite{fitzgerald2001markov}.
Similar tutorials for wireless communications \cite{wang2002monte}, including also SIS and SMC schemes, and machine learning \cite{andrieu2003introduction}, where simulated annealing and the MC-EM algorithm are described, shortly followed.
Then, another tutorial on MC methods for signal processing was published in 2004 and focused on recent advances in MCMC algorithms and particle filters \cite{djuric2004monte}. 
More recently, Green \emph{et al.} published a tutorial on Bayesian computation that partially overlaps with the current survey (e.g., it includes MALA, the HMC algorithm and particle MCMC) \cite{green2015bayesian}.
%
%
A survey specifically focused on different Multiple Try MCMC methods can be found in \cite{martino2018review}, whereas Robert \emph{et al.} \cite{robert2018accelerating} have recently published in arXiv another overview on algorithms to accelerate MCMC that briefly discusses several methods included in this paper (like MTM, HMC or adaptive MCMC).
Several surveys that concentrate exclusively on importance sampling methods have also been published recently \cite{bugallo2015adaptive,SAM16,bugallo2017ais}.

Finally, note that many toolboxes and specific software implementations (in Matlab, Python, R and other programming languages) of the different algorithms described in this survey are freely available on-line.
Due to their importance, let us mention three of the main existing environments for MC computation: BUGS, JAGS and Stan.\footnote{Further information about them can be found in their respective web sites: JAGS (\underline{\texttt{http://mcmc-jags.sourceforge.net}}), BUGS (\underline{\texttt{http://www.openbugs.net/w/FrontPage}}) and Stan (\underline{\texttt{http://mc-stan.org}}).}
On the one hand, BUGS (Bayesian inference Using Gibbs Sampling) is a software package that allows the user to specify a statistical model by simply stating the relationships between related variables \cite{lunn2000winbugs,lunn2009bugs,lunn2012bugs}.
The software includes an ``expert system'' that determines the appropriate MCMC scheme (based on the Gibbs sampler) for analysing the specified model.
On the other hand, JAGS (Just Another Gibbs Sampler) is a program for the analysis of Bayesian hierarchical models using MCMC simulation \cite{plummer2003jags}.
It provides a cross-platform engine for the BUGS language, allowing users to write their own functions, distributions and samplers.
Finally, Stan is a flexible probabilistic programming language that allows users to specify their statistical models and then perform Bayesian inference using MCMC methods (NUTS and HMC), ABC or ML estimation \cite{gelman2015stan,carpenter2017stan}.
Stan has Python and R interfaces, as well as wrapper packages for Matlab, Julia, Stata and Mathematica.

\vspace*{12pt}

\subsection{Acronyms, notation and organization}
\label{sec:notation}

\vspace*{12pt}

Table \ref{tab:acronyms} provides a list of the acronyms used throughout the paper, whereas Table \ref{tab:notation} summarizes the main notation.
Moreover, the following rules will be followed regarding the notation:
\begin{itemize}
	\item Vectors and matrices will be denoted in boldface (e.g., $\datavec$ and $\covmat$), with $\vector{\data_1, \ldots, \data_{\ndata}}$ denoting the vectorization operation,
		i.e. the stacking of a set of vectors ($\data_1, \ldots, \data_{\ndata}$) of dimension $\ddata \times 1$ in order to construct a single vector $\datavec \in \Real^{\ndata\ddata}$. {Capital boldface symbols are used for matrices, whereas lowercase boldface symbols are used for vectors.}
	\item The notation $\parvec_{\neg i}$ will be used to denote a vector with the $i$-th component removed, i.e.,
		$\parvec_{\neg i} = [\param_1, \ldots, \param_{i-1}, \param_{i+1}, \ldots, \param_{\dpar}]^{\top}$.
	\item Capital letters will be used to denote random variables (e.g., $X$), while lowercase letters are used for their realizations (e.g., $x$).
	\item When required, properly normalized PDFs will be indicated by using a bar (e.g., $\normtarget$ and $\normproposal$), whereas their non-negative unnormalized
		versions will be indicated by the same letter without the bar (e.g., $\target$ and $\proposal$).
	\item The notation $x \sim p(X)$ indicates that a realization $x$ of the random variable $X$ is drawn from the PDF $p$.
	\item We use an argument-wise notation for the different normalized and unnormalized densities used throughout the text. For instance, $\target(\parvec)$ denotes the
		$\dpar$-dimensional target, whereas $\target(\param_d|\parvec_{\neg d})$ denotes the one-dimensional full conditional density of the $d$-th parameter.
	\item The notation $\Esp{p}{g}$ will be used to denote the mathematical expectation of the function $g$ w.r.t. the PDF $p$.
\end{itemize}

Regarding the structure of the paper, let us remark that we concentrate on the use of MCMC methods for the estimation of static parameters, although the extension of some of these techniques to a dynamical setting will be occasionally discussed.
This choice is motivated by two facts: the need to keep the length of the tutorial within reasonable bounds, and the existence of two recent review papers on AIS methods \cite{bugallo2017ais,Bugallo15}.
However, two sections detailing the different IS and AIS techniques, as well as the use of IS-within-MCMC, have also been included for the sake of completeness.
Regarding the selection of the methods covered, we have tried to include the most relevant MC algorithms from the different families, following a chronological order from the classical (and usually simpler) MC methods to the more recent and sophisticated ones.
Finally, note that the main focus of the paper is describing the different MC algorithms in a unified way by using a consistent notation which is amenable to signal processing practitioners.
However, some theoretical aspects are also briefly discussed, as well as the main advantages and limitations of each algorithm.

The rest of the paper is organized as follows.
First of all, the mathematical background is provided in Section \ref{sec:formulation}.
The Bayesian framework for statistical inference and the basic MC algorithm are briefly reviewed here (Section \ref{sec:monteCarlo}), altogether with RS, which lies at the heart of MCMC methods (Section \ref{sec:RS}).
Then, Section \ref{sec:mcmc} describes in detail many of the most relevant MCMC algorithms for signal processing applications: the MH algorithm, the Gibbs sampler and their combined use (Section \ref{sec:basicMcmc}); adaptive MCMC methods (Section \ref{sec:adaptMcmc}); gradient-based algorithms (Section \ref{sec:gradMcmc}); and other advanced MCMC schemes (Section \ref{sec:other_advanced_mcmc}).
A short dicussion on MCMC convergence diagnostics (Section \ref{sec:diag}) is also included here.
This is followed by Section \ref{sec:is}, where IS techniques are described: standard IS vs. multiple IS (Section \ref{sec:sisMis}); adaptive IS (Section \ref{sec:ais}); group IS (Section \ref{sec:gis}); and sequential IS (Section \ref{SIS}).
Some convergence results on IS and AIS (Section \ref{sec:convIS}) are also included here, as well as a short discussion on the variance of the IS estimator and the choice of the optimal proposal (Section \ref{sec:varianceIS}), a note on the estimation of the effective sample size (Section \ref{SectionESS_IS}), and a description of proper weighting schemes (Section \ref{LiuSect}).
This is followed by the description of different schemes for the use of IS-within-MCMC in Section \ref{sec:is_within_mcmc}: multiple try approaches for static (Section \ref{sec:advMcmc}) and dynamic parameters (Section \ref{PMH_bigsec}); pseudo-marginal MCMC methods (Section \ref{PSM_MCMC}); noisy MCMC algorithms (Section \ref{Noisy_MCMC}); and approximated likelihood methods (Section \ref{sec:approx_likelihood}).
Finally, the performance of many of the described methods is demonstrated through several numerical simulations in Section \ref{sec:results}: two simple examples for MCMC and IS methods (Sections \ref{sec:toy1} and \ref{sec:toy2}); the estimation of the parameters of a chaotic system (Section \ref{sec:chaos}); a localization problem in wireless sensor networks (Section \ref{sec:localization}); and a spectral estimation application (Section \ref{Simu_SA}).
A discussion of the reviewed methods concludes the paper in Section \ref{sec:discussion}.

\begin{table}[!htb]
\centering
\caption{List of acronyms used.}
{\setstretch{0.9}
\begin{tabularx}{\textwidth}{|l|Z||l|Z|}
	\hline
    ABC & Approximate Bayesian Computation & MC & Monte Carlo \\
    ADS & Adaptive Direction Sampling & MCMC & Markov Chain Monte Carlo \\
    AGM-MH & Adaptive Gaussian Mixture Metropolis-Hastings & MH & Metropolis-Hastings \\
    AIS & Adaptive Importance Sampling & MIS & Multiple Importance Sampling \\
    AISM & Adaptive Independent Sticky Metropolis & ML & Maximum Likelihood \\
    AM & Adaptive Metropolis & MMALA & Riemann Manifold MALA \\
    AMCMC & Adaptive Markov Chain Monte Carlo & MMSE & Minimum Mean Squared Error \\
    AMIS & Adaptive Multiple Importance Sampling & M-PMC & Mixture Population Monte Carlo \\
    APIS & Adaptive Population Importance Sampling & MRF & Markov Random Field \\
    ARS & Adaptive Rejection Sampling & MSE & Mean Squared Error \\
    ARMS & Adaptive Rejection Metropolis Sampling & MTM & Multiple Try Metropolis \\
    CDF & Cumulative Distribution Function & NUTS & No U-Turn Sampler \\
    CLT & Central Limit Theorem & OFDM & Orthogonal Frequency Division Multiplexing \\
    DA & Data Augmentation & PDF & Probability Density Function \\
    DM & Deterministic Mixture & PMC & Population Monte Carlo \\
    DR & Delayed Rejection & PMH & Particle Metropolis-Hastings \\
    FUSS & Fast Universal Self-tuned Sampler & PWC & Piecewise Constant \\
    GMS & Group Metropolis Sampling & PWL & Piecewise Linear \\
    HMC & Hamiltonian Monte Carlo & RMHMC & Riemann Manifold HMC \\
    IA$^2$RMS & Independent Doubly Adaptive Rejection Metropolis Sampling & RS & Rejection Sampling \\
    IID & Independent and Identically Distributed & RV & Random Variable \\
    IS & Importance Sampling & SDE & Stochastic Differential Equation \\
    LAIS & Layered Adaptive Importance Sampling & SIS & Sequential Importance Sampling \\
    MALA & Metropolis Adjusted Langevin Algorithm & SMC & Sequential Monte Carlo \\
    MAP & Maximum A Posteriori & WSN & Wireless Sensor Network \\
    \hline
\end{tabularx}}
\label{tab:acronyms}
\end{table}

\begin{table}[!htb]
\centering
\caption{Summary of the main notation used throughout the paper.}
{\setstretch{1.2}
\begin{tabularx}{\textwidth}{|l|X|}
	\hline
    \textbf{Notation} & \textbf{Description} \\
    \hline
    $\ddata$ & Dimension of the data. \\
    $\ndata$ & Number of data available. \\
    $\datavec \in \mathbb{R}^{\ndata\ddata}$ & $\ndata\ddata$-dimensional observations vector,
    	$\datavec = \vector{\data_1,\ldots,\data_\ndata}$ with $\data_i \in \mathbb{R}^{\ddata}$ for $i = 1, \ldots, L$. \\
    $\dpar$ & Dimension of the parameter space. \\
    $\parspace = \parspace_1 \times \cdots \times \parspace_{\dpar}$ & Feature space for the parameter vector $\parvec$. \\
    $\parvec \in \mathbb{R}^{\dpar}$ & $\dpar$-dimensional parameter vector, $\parvec = [\param_1, \ldots,
    	\param_{\dpar}]$ with $\param_d \in \parspace_d$ for $d = 1, \ldots, \dpar$. \\
	$\parvec^{(m)}$ & $m$-th sample of the parameter vector in MC and RS. \\
	$\parvec^{(t)}$ & Sample of the parameter vector at the $t$-th iteration in MCMC methods. \\
    $\normtarget(\parvec|\datavec) \equiv \normtarget(\parvec)$ & Target (i.e., posterior) PDF. \\
    $\target(\parvec|\datavec) \equiv \target(\parvec)$ & Target function (i.e., non-negative but unnormalized). \\
    $\prior(\parvec)$ & Prior probability density function. \\
    $\likelihood(\datavec|\parvec)$ & Likelihood. \\
    $\partition(\datavec)$ & Normalizing constant of the target (a.k.a. partition function, marginal likelihood or model evidence). \\
    $\normtarget(\param_d|\parvec_{\neg d})$ & Full conditional PDF for the $d$-th parameter given all the other parameters (used in the Gibbs sampler). \\
	$\niter$ & Number of Monte Carlo iterations performed. \\
	$\nburn$ & Number of iterations for the burn-in period in MCMC. \\
    $\nprop$ & Number of proposals used in multiple IS approaches. \\
    $\npart$ & Number of samples drawn in the MC algorithm, RS and IS approaches. Usually $\npart \ge \nprop$ in MIS. \\
    $\normproposal(\parvec)$, $\normproposal_t(\parvec)$, $\normproposal_{m,t}(\parvec)$ & Proposal PDF. \\
    $\proposal(\parvec)$, $\proposal_t(\parvec)$, $\proposal_{m,t}(\parvec)$ & Proposal function (i.e., non-negative but unnormalized) for $t=1,\ldots,T$ and $m=1,\ldots,M$. \\
    $\weight_{m,t}(\parvec)$ & Unnormalized weight of the $m$-th particle ($m=1,\ldots,M$) at the $t$-th iteration ($t=1,\ldots,T$) for AIS approaches. \\
    $\normweight_{m,t}(\parvec)$ & Normalized weight of the $m$-th particle ($m=1,\ldots,M$) at the $t$-th iteration ($t=1,\ldots,T$) for AIS approaches. \\
    $\randmeas(\parvec)$ & Random measure used to approximate the target at the $t$-th iteration. \\
    $\gauss(\meanvec,\covmat)$, $\gauss(\cdot|\meanvec,\covmat)$ & Gaussian PDF with mean $\meanvec$ and covariance $\covmat$. \\
    $\uniform(\interval)$ & Uniform PDF within the interval $\interval$. \\
    \hline
\end{tabularx}}
\label{tab:notation}
\end{table}




\section{Mathematical Formulation}
\label{sec:formulation}

\vspace*{12pt}

\subsection{Bayesian Inference and the Monte Carlo Method}
\label{sec:monteCarlo}

\vspace*{12pt}

Let us assume that we have a dataset, $\datavec=\vector{\data_1, \ldots, \data_{\ndata}} \in \Real^{\ndata\ddata}$ with $\data_i \in \Real^{\ddata}$, which depends on some \emph{static} parameter vector, $\parvec=[\param_1, \ldots, \param_{\dpar}]^{\top} \in \parspace \subseteq \Real^{\dpar}$.
From a Bayesian point of view, all the information required to solve any task related to $\parvec$ (e.g., inference or optimization problems) is contained in the \emph{posterior} or \emph{target} PDF, $\normtarget(\parvec|\data)$.
Using Bayes rule, this posterior can be expressed as
\begin{equation}
	\normtarget(\parvec|\data) = \frac{\likelihood(\datavec|\parvec) \prior(\parvec)}{\partition(\datavec)} = \frac{\target(\parvec|\datavec)}{\partition(\datavec)},
\label{eq:bayesRule}
\end{equation}
where $\likelihood(\datavec|\parvec)$ is the \emph{likelihood}, that depends on the statistical input-output model assumed; $\prior(\parvec)$ is the \emph{prior} PDF, which summarizes all the information available about $\parvec$ {external to the observation of the} data; $\partition(\datavec)$ is the \emph{marginal likelihood} (a.k.a. as model evidence or partition function in some contexts), a normalizing term which does not depend on $\parvec$; and $\target(\parvec|\datavec)$ is the \emph{target function}, a non-negative definite function (i.e., $\target(\parvec|\datavec) \ge 0$ for all $\parvec \in \parspace \subseteq \mathbb{R}^{\dpar}$ and $\datavec \in \Real^{\ndata\ddata}$) such that $\int_{\parspace}{\target(\parvec|\datavec)\ d\parvec} = \partition(\datavec)$ with $\partition(\datavec) \ne 1$ in general.

Now, let us assume that we want to compute the following integral,
\begin{equation}
	I = \BEsp{\normtarget}{g(\parvec)}
		= \int_{\parspace}{g(\parvec) \normtarget(\parvec|\datavec) d\parvec}
		=  \frac{1}{\partition(\datavec)} \int_{\parspace}{g(\parvec) \target(\parvec|\datavec) d\parvec} < \infty,
\label{eq:intBayes}
\end{equation}
where $g(\parvec)$ can be any integrable function w.r.t. $\normtarget(\parvec|\datavec)$.
For instance, when $g(\parvec)=\parvec$ this integral becomes the well-known minimum mean squared error (MMSE) estimator of the parameter $\parvec$ \cite{Scharf1991,Kay1993,van2004detection},
\begin{equation}
	\hat{\parvec}_{\mmse} = \BEsp{\normtarget}{\parvec} = \int_{\parspace}{\parvec \normtarget(\parvec|\datavec) d\parvec},
\label{eq:estMMSE}
\end{equation}
which is widely used in many statistical signal processing applications and corresponds to the conditional expectation of $\parvec$ w.r.t. the posterior PDF.

Unfortunately, obtaining an analytical solution of these integrals is usually unfeasible in many practical problems of interest.
In these cases, an approximate solution of \eqref{eq:intBayes} can be obtained through the Monte Carlo (MC) method shown in Algorithm \ref{alg:mc}.
Essentially, the MC method simply consists of obtaining a set of independent and identically distributed (IID) samples of the parameter vector to be inferred and using them to approximate the desired integral by means of an unweighted sum.
These $M$ samples, $\parvec^{(m)}$, can be obtained either by sampling directly from the target PDF (i.e., the posterior $\normtarget(\parvec|\datavec)$), as shown in Algorithm \ref{alg:mc}, or by replicating the physical procedure where the desired parameters are involved.
Note that the subindex $M$ in $\widehat{I}_M$ denotes the number of samples involved in the estimation.


\begin{alg}{Monte Carlo (MC) approximation of the integral in Eq. \eqref{eq:intBayes}}
{\begin{enumerate}
	\item Draw $\parvec^{(m)} \sim \normtarget(\parvec|\datavec)$ for $m=1,\ldots,\npart$.
    \item Approximate the integral in Eq. \eqref{eq:intBayes} as
    	\begin{equation}
        		\widehat{I}_M = \frac{1}{\npart}\sum_{m=1}^{M}{g(\parvec^{(m)})}.
	\label{eq:estMC}
        \end{equation}
   \end{enumerate}
}
\label{alg:mc}
\end{alg}

The MC estimate of $I$ provided by Eq. \eqref{eq:estMC} is unbiased, i.e., $\Esp{\normtarget}{\widehat{I}_M} = I$.
Moreover, by the \emph{strong law of large numbers}, $\widehat{I}_M \to I$ almost surely (a.s.) as $M \to \infty$ \cite{Robert04}.
Furthermore, if $g(\parvec)$ is square integrable w.r.t. $\normtarget(\parvec|\datavec)$, then we can use the \emph{central limit theorem} (CLT) to state the following result \cite{Robert04}:
\begin{equation}
	\frac{\widehat{I}_M - I}{\sqrt{V_M}} \overset{d}{\to} \gauss(0,1) \qquad \text{as} \quad M \to \infty,
\label{eq:cltMC}
\end{equation}
where $\overset{d}{\to}$ denotes convergence in distribution, and
\begin{equation}
	V_M = \frac{1}{M} \BEsp{\normtarget}{(g(\parvec)-I)^2} = \frac{1}{M} \int_{\parspace}{(g(\parvec)-I)^2 \normtarget(\parvec|\datavec)\ d\parvec}.
\label{eq:varMC}
\end{equation}
Note that \eqref{eq:cltMC} is equivalent to stating that $\widehat{I}_M \overset{d}{\to} \gauss(I,V_M)$ as $M \to \infty$.

Unfortunately, Algorithm \ref{alg:mc} cannot be applied in many practical problems, because we cannot draw samples directly from $\normtarget(\parvec|\datavec)$.
In these cases, if we can perform point-wise evaluations of the target function, $\target(\parvec|\datavec) = \likelihood(\datavec|\parvec) \prior(\parvec)$, we can apply other types of Monte Carlo algorithms: Rejection Sampling (RS) schemes, Markov chain Monte Carlo (MCMC) techniques and importance sampling (IS) methods.
%
These two large classes of algorithms, MCMC and IS, are the core of this paper and will be described in detail in the rest of this work.
Before, we briefly recall the basis of the RS approach, which is one of the key ingredients of MCMC methods, in the following section.

\subsection{Rejection Sampling (RS)}
\label{sec:RS}

\vspace*{12pt}

The RS method is a classical Monte Carlo technique for {\it universal sampling} that can be used to generate samples virtually from any target density $\normtarget(\parvec)$ by drawing from a simpler proposal density $\normproposal(\parvec)$.\footnote{For the sake of simplicity, in the sequel we drop the dependence on the data ($\datavec$) from the target, i.e., we use $\normtarget(\parvec) \equiv \normtarget(\parvec|\datavec)$ and $\target(\parvec) \equiv \target(\parvec|\datavec)$.}
The sample is either accepted or rejected by applying an adequate test to the ratio of the two PDFs, and it can be easily proved that accepted samples are actually distributed according to the target density \cite{devroye1986non-uniform}.
The RS algorithm was originally proposed by John von Neumann in a 1947 letter to Stan Ulam \cite{eckhardt1987stan}, but it was not published until 1951 \cite{vonNeumann1951various}.
In its original formulation, von Neumann considered only a uniform proposal PDF, but the algorithm was later generalized to allow drawing samples from any proposal density from which sampling is straightforward.
In the standard RS algorithm \cite{vonNeumann1951various,devroye1986non-uniform}, we first draw a sample from the proposal PDF, $\parvec'\sim q(\parvec)$, and then accept it with probability
\begin{equation}
	p_A(\parvec') = \frac{\target(\parvec')}{C\proposal(\parvec')} \leq 1,
\end{equation}
where $C$ is a constant such that $C\proposal(\parvec)$ is an envelope function for $\target(\parvec)$, i.e.,  $C\proposal(\parvec) \geq \target(\parvec)$ for all $\parvec \in \parspace$.
%
%
We can summarize this procedure in an equivalent way: at the $t$-th iteration, draw a sample $\parvec^{(t)} \sim \normproposal(\parvec)$ and $u \sim \mathcal{U}([0,1])$; if $u \leq \frac{\target(\parvec^{(t)})}{C\proposal(\parvec^{(t)})}$, accept $\parvec^{(t)}$, otherwise, reject it; when the desired number of samples have been drawn from $\normtarget(\parvec)$, stop. 
Algorithm \ref{alg:rs} summarizes the generation of $\npart$ samples from the target PDF using the standard RS algorithm.

\begin{alg}{Rejection Sampling (RS) method.}
{\begin{enumerate}
	\item \textbf{Initialization:} Choose a proposal function, $\proposal(\parvec)$, and the required number of samples from the target, $\npart$. Find an upper bound,
		$C \geq \frac{\target(\parvec)}{\proposal(\parvec)}$ for all $\parvec \in \parspace$, and let $t=m=1$.
    \item \texttt{WHILE} $m \le \npart$:
    	\begin{enumerate}
        		\item Draw $\parvec^{(t)} \sim \normproposal(\parvec)$ and $u \sim \uniform([0,1))$.
		\item If $u \leq \frac{\target(\parvec^{(t)})}{C\proposal(\parvec^{(t)})}$, accept $\parvec^{(t)}$, setting $\parvec^{(m)} = \parvec^{(t)}$ and letting $m=m+1$.
		\item Set $t=t+1$ regardless of whether $\parvec^{(t)}$ has been accepted or not.
         \end{enumerate}
   \item Approximate the integral in Eq. \eqref{eq:intBayes} using Eq. \eqref{eq:estMC}.
\end{enumerate}
   } 
\label{alg:rs}
\end{alg}

The RS algorithm is a simple MC method for approximating the integral in Eq. \eqref{eq:intBayes} that can be universally applied as long as the upper bound $C$ can be found.
However, it has several important drawbacks that hinder its practical application:
\begin{enumerate}
	\item For complicated targets, finding a bound $C$ such that $C\proposal(\parvec) \geq \target(\parvec)$ for all $\parvec \in \parspace$ can be difficult, especially for
		high-dimensional parameter spaces.
	\item Even if this bound can be found, the RS algorithm can be very inefficient if the ratio $\frac{\target(\parvec)}{C\proposal(\parvec)}$ is small for a large portion
		of the parameter space. Indeed, the acceptance probability of the RS algorithm is given by
		\begin{equation}
			P_A = \int_{\parspace}{\frac{\target(\parvec)}{C\proposal(\parvec)} \normproposal(\parvec)\ d\parvec} = \frac{\partition_{\target}}{C\partition_{\proposal}},
		\label{eq:acceptRS}
		\end{equation}
		where $\partition_{\target} = \int_{\parspace}{\target(\parvec)\ d\parvec}$ and $\partition_{\proposal} = \int_{\parspace}{\proposal(\parvec)\ d\parvec}$.
		Depending on the target and the proposal selected, this $P_A$ can be very low (this happens when $C\partition_{\proposal} \gg \partition_{\target}$), thus rendering
		the RS algorithm useless in practice. For this reason, many RS approaches have been specifically designed for drawing efficiently from a specific target distribution
		\cite{Ahrens74,Damien01}. For example, efficient random number generators based on RS schemes can be found for the Gamma, Beta and Nakagami distributions
		\cite{Cheng77,Beaulieu05,LuengoNak12,Zhu2011,MartinoNak18}. 
	\item The number of iterations required to generate $\npart$ samples, $\niter$, is a random variable with an expected value $\Esp{}{\niter} = \frac{\npart}{P_A}$ and $P_A$
		given by \eqref{eq:acceptRS}. Hence, the exact time required to generate $M$ valid samples cannot be set a priori, and this can be a serious problem in many
		applications.
\end{enumerate}
 One way to tackle some of these difficulties is by constructing the proposal $\proposal(\parvec)$ adaptively, using some of the so called adaptive RS (ARS) methods.
The ARS algorithm was originally proposed by Gilks and Wild in 1992 \cite{Gilks92}, and several generalized ARS algorithms have been proposed since then \cite{Botts11,Martino09,Hoermann95,martino2010generalized,Gorur08rev,martino2011generalization,Evans98,CARS,PARS}.
However, the need to have $C\proposal(\parvec) \ge \target(\parvec)$ for all $\parvec \in \parspace$ and the difficulty of constructing the adaptive proposals in high-dimensional parameter spaces limit the applicability of those generalized ARS algorithms \cite{Hoermann95b,MartinoBook18}, rendering MCMC and IS approaches more efficient in general, and thus preferable for practical applications. 
For further information see Chapters 3 and 4 in \cite{MartinoBook18}.



\section{Markov chain Monte Carlo (MCMC)}
\label{sec:mcmc}

\vspace*{12pt}

According to Definition 7.1 of \cite{Robert04}, an MCMC method is any method producing an ergodic Markov chain whose stationary density is the desired target PDF, $\normtarget({\bm \theta})$.
In the following, we detail some of the most relevant MCMC algorithms, starting from the basic building blocks (the MH algorithm and the Gibbs sampler) in Section \ref{sec:basicMcmc}, and ending up with several advanced adaptive (Section \ref{sec:adaptMcmc}), gradient-based (Section \ref{sec:gradMcmc}), and other advanced MCMC schemes (Section \ref{sec:other_advanced_mcmc}).
Note that we focus on describing the different algorithms rather than on their theoretical properties, although a brief discussion on the validity of the MH algorithm (due to its importance as the basis of most MCMC algorithms) is provided in Section \ref{sec:validity_mh}.

\subsection{MCMC basic building blocks: Metropolis-Hastings and Gibbs samplers}
\label{sec:basicMcmc}

\vspace*{12pt}

\subsubsection{Metropolis-Hastings (MH) algorithm}

\vspace*{12pt}

As mentioned in the introduction, the MH algorithm was initially devised by Nicholas Metropolis \emph{et al.} in 1953 as a general method to speed up the computation of the properties of substances composed of interacting individual molecules \cite{metropolis1953equation}.
In a two-dimensional square with $N$ particles, studying these properties requires computing $2N$-dimensional integrals, an impossible task (both analytically or numerically) for the large values of $N$ required in practice.
A naive MC simulation would consist in drawing particle configurations uniformly at random and assigning them weights proportional to $\exp(-E/K\mathcal{T})$, where $E$ is the energy of the configuration, $K \approx 1.38 \times 10^{-23}$ is Boltzmann's constant and $\mathcal{T}$ is the absolute temperature (in Kelvin degrees).\footnote{Nowadays this would be considered an application of the IS methodology, since the particles are not drawn from the desired target density, $\normtarget(\parvec) \propto \exp(-E(\parvec)/K\mathcal{T})$, but from a uniform random proposal. Thus, they are assigned the corresponding standard IS weights, $w(\parvec) \propto \exp(-E(\parvec)/K\mathcal{T})$, for the subsequent integration (see Section \ref{sec:ais} for a  description of IS methods).}
Nevertheless, Metropolis \emph{et al.} devised the following alternative modified MC scheme:\footnote{Note that we have used $\parvec$ to represent the positions of the particles (which are the parameters to be inferred), as in the rest of the paper, instead of the original notation of \cite{metropolis1953equation}, which used $\x$.}
\begin{enumerate}
	\item Place the $N$ particles in an initial arbitrary configuration (e.g., in a regular lattice) within the square.
	\item At the $t$-th iteration ($t = 1, \ldots, T$) and for $i = 1, \ldots, N$:\footnote{Note that the way in which the samples are drawn
		actually corresponds to the \emph{MH-one-at-a-time} algorithm \cite{doucet2005monte}, which is equivalent to the well-known 
		MH-within-Gibbs algorithm (see Section \ref{sec:MHwithinGibbs}) with one iteration of the internal MH method per iteration of the 
		external systematic scan Gibbs sampler (see Section \ref{sec:gibbs}), since the particles are moved one at a time and the 
		proposed moves are not performed according to the desired target PDF, but using a simpler proposal (and thus they can be 
		accepted or rejected).}
		\begin{enumerate}
			\item Propose a move for the $i$-th particle, located at $\parvec_i^{(t-1)}$, according to the following rule:
				\begin{align}
					\parvec_i' = \parvec_i^{(t-1)} + \kappa \xivec^{(t)},
				\label{eq:metropolis}
				\end{align}
				where $\kappa$ is a user-defined parameter that represents the maximum allowed displacement per iteration, while
				$\xivec^{(t)} \sim \uniform([-1,1) \times [-1,1))$.
			\item Compute the change in energy caused by the move, $\Delta E_i^{(t)}$. If the energy is not increased
				(i.e., $\Delta E_i^{(t)} \le 0$), then accept the move, setting $\parvec_i^{(t)} = \parvec_i'$. Otherwise (i.e., if the energy
				is increased and thus $\Delta E_i^{(t)} > 0$), then accept the move with probability
				$\exp(-\Delta E_i^{(t)}/K\mathcal{T})$, thus setting $\parvec_i^{(t)} = \parvec_i'$, or reject it with probability
				$1-\exp(-\Delta E_i^{(t)}/K\mathcal{T})$, thus letting $\parvec_i^{(t)} = \parvec_i^{(t-1)}$.
		\end{enumerate}
	\item All the different configurations, $\parvec_i^{(t)}$ for $t=1,\ldots,T$, are used to compute the desired averages.
\end{enumerate}
Intuitively, the MH algorithm can be seen as a generalized rejection sampler whose proposal depends on the result of the previous iteration (i.e., on $\parvec^{(t-1)}$).
Furthermore, the acceptance rate also depends on $\parvec^{(t-1)}$ and the value of $\parvec^{(t-1)}$ is re-used whenever a candidate sample $\parvec'$ is rejected.
This creates an undesired effect, since the drawn samples are no longer independent as in the RS algorithm, but allows us to work with proposal densities that may lie below the target.
This is due to the fact that the underlying Markov chain has the desired target as the limiting invariant distribution (e.g., see \cite{Robert04} for a rigorous proof).
Another useful perspective is to view the method as a thinning of a random walk in precisely the right way to ensure convergence to the correct target.
{Loosely speaking, the chain is thinned by discarding those candidates which correspond to moves from the current state that happen too often, and this is done with the right probability to ensure that the invariant distribution of the Markov chain is exactly the desired target.}
{See the excellent tutorial (but rigorous) exposition of the MH algorithm provided by Chib and Greenberg for further information about this issue \cite{chib1995understanding}.}

In this algorithm, the proposal for the $t$-th iteration and the $i$-th particle is $\normproposal(\parvec_i|\parvec_{1:N}^{(t-1)}) = \normproposal(\parvec_i|\parvec_i^{(t-1)}) =
\uniform([\param_{i,1}^{(t-1)}-\kappa,\param_{i,1}^{(t-1)}+\kappa) \times [\param_{i,2}^{(t-1)}-\kappa,\param_{i,2}^{(t-1)}+\kappa))$, whereas the target is $\normtarget(\parvec) \propto \exp(-E(\parvec)/K\mathcal{T})$.
The acceptance probability is then given by
\begin{equation}
	\alpha(\vec{x}_i',\vec{x}_{1:N}^{(t-1)}) = \min\left[1,\exp(-\Delta E_i^{(t)}/K\mathcal{T})\right],
\label{eq:alphaMetropolis}
\end{equation}
with $\Delta E_i^{(t)} = E(\parvec_1^{(t-1)}, \ldots, \parvec_{i-1}^{(t-1)}, \parvec_i', \parvec_{i+1}^{(t-1)}, \ldots, \parvec_{N}^{(t-1)}) - E(\parvec_1^{(t-1)}, \ldots, \parvec_N^{(t-1)})$.
This acceptance probability guarantees the ergodicity of the chain and the convergence of the algorithm to the desired target PDF \cite{metropolis1953equation}, but is not the only valid acceptance rule.
Indeed, in 1965 Barker proposed an alternative acceptance probability for the computation of radial distribution functions in plasmas \cite{Barker1965}:
\begin{equation}
	\alpha(\parvec_i',\parvec_{1:N}^{(t-1)}) = \frac{1}{1+\exp(-\Delta E_i^{(t)}/K\mathcal{T})}.
\label{eq:alphaBarker}
\end{equation}

Soon afterwards, Hastings generalized these two acceptance probabilities, allowing for non-symmetric proposals (unlike the proposals considered both by Metropolis and Barker, which were both symmetric) \cite{hastings1970monte}.
Using our notation, where the parameters to be estimated are denoted as $\parvec$, the two acceptance rules ($\alpha_M$ and $\alpha_B$ denote the generalization of Metropolis' and Barker's acceptance rules, repectively) become:
\begin{subequations}
\begin{align}
	\alpha_M(\parvec',\parvec^{(t-1)}) & =
		\min\left[1,\frac{\target(\parvec')\proposal(\parvec^{(t-1)}|\parvec')}{\target(\parvec^{(t-1)})\proposal(\parvec'|\parvec^{(t-1)})}\right], \label{eq:alphaMH}\\
	\alpha_{B}(\parvec',\parvec^{(t-1)}) & = \frac{\target(\parvec')\proposal(\parvec^{(t-1)}|\parvec')}
		{\target(\parvec')\proposal(\parvec^{(t-1)}|\parvec')+\target(\parvec^{(t-1)})\proposal(\parvec'|\parvec^{(t-1)})}, \label{eq:alphaBH}
\end{align}
\end{subequations}
Finally, in 1973 Peskun proved that the acceptance rule of Eq. \eqref{eq:alphaMH} was optimal \cite{Peskun1973}, and this settled the structure of the algorithm used nowadays \cite{MHWiley17}.

The MH algorithm with the acceptance rule of Eq. \eqref{eq:alphaMH} is summarized in Algorithm \ref{alg:metropolis}.
The burn-in period ($\nburn$) is the number of initial samples removed from the empirical average in Eq. \eqref{EstimatorDavid}, which is used to compute the desired estimator, in order to guarantee that the chain has converged approximately to its stationary distribution.
This period can be estimated automatically (e.g., see Section \ref{sec:diag} for a brief discussion on this issue and \cite{cowles1996markov} for a comparative review of different techniques to assess the convergence of a Markov chain and thus determine the burn-in period) or set to some pre-defined value, and is required by all MCMC algorithms.


\begin{alg}{Metropolis-Hastings (MH) algorithm.}
{\begin{enumerate}
	\item \textbf{Initialization:} Choose a proposal function $\proposal(\parvec|\parvec^{(t-1)})$, an initial state $\parvec^{(0)}$,
    	the total number of iterations ($\niter$), and the burn-in period ($\nburn$).
    \item \texttt{FOR} $t = 1, \ldots, \niter$:
    	\begin{enumerate}
        	\item Draw $\parvec' \sim \proposal(\parvec|\parvec^{(t-1)})$ and $u \sim \uniform([0,1))$.
            \item Compute the acceptance probability:
            	\begin{equation}
                \alpha_t \equiv \alpha(\parvec',\parvec^{(t-1)}) = \min\left[1,\frac{\target(\parvec')\proposal(\parvec^{(t-1)}|\parvec')}
                	{\target(\parvec^{(t-1)})\proposal(\parvec'|\parvec^{(t-1)})}\right].
                \label{eq:alpha}
                \end{equation}
            \item If $u \le \alpha_t$, accept $\parvec'$ and set $\parvec^{(t)} = \parvec'$. Otherwise (i.e., if
            	$u > \alpha_t$), reject $\parvec'$ and set $\parvec^{(t)} = \parvec^{(t-1)}$.
         \end{enumerate}
    \item Approximate the integral in Eq. \eqref{eq:intBayes} as
    	\begin{equation}
        \label{EstimatorDavid}
        	\widehat{I}_{T-T_b} = \frac{1}{\niter-\nburn}\sum_{t=\nburn+1}^{\niter}{g(\parvec^{(t)})}.
        \end{equation}
   \end{enumerate}
}
\label{alg:metropolis}
\end{alg}

One of the main advantages of the MH algorithm is that it is a very generic method that admits the use of almost any proposal and target PDFs.
However, although the algorithm is valid regardless of the shape and parameters of the proposal PDF (see Section~\ref{sec:validity_mh} for a brief review of the specific conditions for the validity of the MH algorithm), the speed of convergence and the quality of the estimators obtained substantially depend on the quality of this proposal.
Many choices are possible, but here we will only consider the two most widely used (see \cite{chib1995understanding} for a brief discussion on five different families of proposals):
\begin{itemize}
	\item \textbf{Independent MH:} The proposal is fixed and does not depend on the current state of the chain, i.e.,
		$\normproposal(\parvec|\parvec^{(t-1)}) = \normproposal(\parvec)$. For instance, a widely used choice in this case is a multi-variate Gaussian PDF with fixed mean
		vector and covariance matrices: $\normproposal(\parvec) = \gauss(\parvec|\meanvec,\covmat)$. An independent proposal can be considered a
		\emph{global proposal}, since it can generate candidate samples in the whole state space regardless of the current state of the chain. This type of proposal fosters
		the exploration of the state space, but its performance can be poor for complicated target PDFs (especially for high-dimensional state spaces, where it can be
		difficult to find a good parameterization).
		
	\item \textbf{Random Walk MH:} The proposal is centered on the current state of the chain, i.e., the proposed candidate at the $t$-th iteration can be expressed as
		$\parvec' = \parvec^{(t-1)} + \varthetavec'$, where $\varthetavec' \sim p(\varthetavec|\vec{0},\covmat_{\vartheta})$ and $p(\varthetavec|\meanvec,\covmat_{\vartheta})$
		is an arbitrary PDF specified using a location parameter $\meanvec$ and a scale parameter $\covmat$. For instance, using a Gaussian PDF for $\varthetavec$
		we have $\varthetavec' \sim \gauss(\varthetavec|\vec{0},\covmat_{\vartheta})$, which implies that $\parvec' \sim \normproposal(\parvec|\parvec^{(t-1)}) =
		\gauss(\varthetavec|\parvec^{(t-1)},\covmat_{\vartheta})$. If the PDF of $\varthetavec$ is symmetric
		(i.e., $q(\varthetavec|\varthetavec^{(t-1)}) = q(\varthetavec^{(t-1)}|\varthetavec)$), then the acceptance rule becomes:
		\begin{equation}
			\alpha(\parvec',\parvec^{(t-1)}) = \min\left[1,\frac{\target(\parvec')}{\target(\parvec^{(t-1)})}\right].
		\label{eq:alphaMHsym}
		\end{equation}
		This is the type of proposal used by Metropolis \emph{et al.} (with a uniform distribution for $\varthetavec$) in \cite{metropolis1953equation}, which led them to the
		simplified acceptance probability shown in Eq. \eqref{eq:alphaMetropolis}. A random walk proposal can be seen as a local proposal, since it is centered on the current
		state of the chain. Hence, the random walk MH algorithm encourages a more local exploration around the current state. 
\end{itemize}

A critical issue for the good performance of the MH algorithm is the acceptance rate (AR), which depends on the variance of the proposal PDF and should be neither too high nor too low.
On the one hand, a high variance typically leads to a low AR, thus implying that the MH algorithm gets stuck because most candidate samples are rejected.
On the other hand, a low variance can easily lead to a high AR, as only local moves around previously accepted samples are proposed, but can result in the MH algorithm failing to explore the target.
The seminal work of Roberts, Gelman and Wilks proved, for the random walk MH algorithm and in a simplified setting, that the proposal's variance should be tuned in such a way that the average acceptance rate is roughly 1/4 \cite{roberts1997weak}.
In \cite{gelman1996efficient}, the same authors delved deeper into this issue, showing that the optimal acceptance rate is approximately 44\% for $\dpar=1$ and declines to 23\% when $\dpar \to \infty$.
These results can be extended to different settings and other methods based on the MH algorithm, like MH-within-Gibbs or Hamiltonian MC (see Section \ref{sec:MHwithinGibbs} and Section \ref{sec:hmc}, respectively), and have lead to the practical rule of thumb of choosing the variance of the proposal in order to ensure and acceptance rate between 25\% and 40\%.
However, let us remark that several authors have proved that the optimal AR can be substantially different for other settings/methods.
For instance, B\'edard and Rosenthal have recently warned that the asymptotically optimal AR can be significantly different from the well-known 0.234 AR when the target's components are not independent \cite{bedard2007optimal}.
Indeed, in \cite{bedard2008optimal,bedard2008efficient} B\'edard showed that 0.234 is the upper limit for the AR in the simplified model considered, but much lower ARs can actually be optimal.
Other authors have also found that higher acceptance rates can be optimal for other algorithms that make use of gradient information, like the simplified Langevin algorithm (SLA) or the modified adaptive Langevin algorithm (MALA) (see Section \ref{sec:mala}) \cite{beskos2009optimal,pillai2012optimal}.

Finally, let us remark that the local and global proposals, used by the independent and random walk MH algorithms respectively, can be combined.
For instance, \cite{guan2006markov} proposes using the following \emph{small world proposal}:
\begin{equation}
	\proposal(\parvec) = (1-p) \cdot \proposal_G(\parvec) + p \cdot \proposal_L(\parvec),
\label{eq:smallWorldProposal}
\end{equation}
where $\proposal_L(\parvec)$ is a local proposal centered around the current state of the chain, $\proposal_G(\parvec)$ is a global proposal that allows for ``wild'' moves far away from the current state, and $p$ is a small probability.
Using this proposal leads to an MH algorithm with improved performance, especially for complicated heterogeneous spaces and multi-modal distributions, and can turn slowly mixing into rapidly mixing chains \cite{guan2006markov,guan2007small}.

\subsubsection{Validity of the Metropolis-Hastings algorithm}
\label{sec:validity_mh}

\vspace*{12pt}

Let us now take at the conditions when the MH algorithm (Alg.~\ref{alg:metropolis}) produces samples from the desired target PDF. In order to analyze its output, let us first notice that the states $\parvec^{(1)},\parvec^{(2)},\ldots$ form a Markov chain with a certain transition density $K(\parvec^{(t)}|\parvec^{(t-1)})$. The key trick of the MH algorithm is that the algorithm has been constructed in such a way that the stationary PDF of the Markov chain is the target PDF:
\begin{equation}
  \normtarget(\parvec') = \int_{\parspace} K(\parvec'|\parvec) \normtarget(\parvec)
   d\parvec.
\end{equation}
One way to ensure the above is the \emph{detailed balance condition}, which demands that
\begin{equation}
\label{DBC_eq}
  K(\parvec'|\parvec) \normtarget(\parvec) = K(\parvec|\parvec') \normtarget(\parvec').
\end{equation}
Integrating both sides over $\parvec$ and recalling $ \int_{\parspace} K(\parvec|\parvec') d\parvec = 1$ now gives
\begin{equation}
  \int_{\parspace} K(\parvec'|\parvec) \normtarget(\parvec) d\parvec
  = \int_{\parspace} K(\parvec|\parvec') \normtarget(\parvec')  d\parvec
  =  \normtarget(\parvec'),
\end{equation}
which shows that $\normtarget(\parvec)$ is the stationary PDF of the Markov chain.
Furthermore, this condition also ensures that the Markov chain is reversible \cite{gamerman2006markov,Robert04,Liu04b,MHWiley17}.
The transition PDF of the MH algorithm consist of two parts -- the PDF of the accepted samples and the PDF of the rejected samples.
It can thus be written in the following form:
\begin{equation}
  K(\parvec'|\parvec)
  = \alpha(\parvec',\parvec) \, \proposal(\parvec'|\parvec)
  + \left(1 - \int_{\parspace} \alpha(\parvec',\parvec)  \proposal(\parvec'|\parvec) d\parvec' \right) \, \delta(\parvec' - \parvec).
\end{equation}
By direct computation, it can be easily verified that the detailed balance condition is satisfied (see also Theorem 7.2 of \cite{Robert04}).

In addition to having the correct stationary PDF we also need to ensure that the Markov chain is ergodic. The ergodicity property ensures that the Markov chain converges to the stationary distribution with a predefined rate so that we can estimate expectations of the state distributions by computing time averages. A sufficient condition for ergodity is to ensure that the Markov chain is also an aperiodic $\normtarget$-irreducible Harris chain, which can be ensured by the following conditions (see Equations 7.4 and 7.5 and Lemma 7.6 in \cite{Robert04}):\footnote{These conditions for the proposal density can be slightly relaxed (e.g., see Lemma 7.6 in \cite{Robert04}).}
\begin{enumerate}
\item The stationary distribution and the proposal PDF satisfy $P[ \normtarget(\parvec) \proposal(\parvec'|\parvec) \le \normtarget(\parvec') \proposal(\parvec|\parvec') ] < 1$.
\item The proposal PDF is strictly positive everywhere in the parameter space, i.e., $\proposal(\parvec'|\parvec) > 0$ for all $\parvec',\parvec \in \parspace$.
\end{enumerate}

Provided that the detailed balance condition and the aforementioned properties are satisfied, then Corollaries 7.5 and 7.7 in \cite{Robert04} ensure the following ergodicity properties for the MH Markov chain: 
\begin{subequations}
\begin{align}
  & \lim_{T \to \infty} \sum_{t=1}^T g(\parvec^{(t)}) = \int_{\parspace} g(\parvec) \normtarget(\parvec) d\parvec', \label{eq:ergodicity}\\
  & \lim_{n \to \infty} \left\| \int_{\parspace} K^n(\cdot|\parvec') \normtarget_0(\parvec')  d\parvec - \normtarget \right\|_{TV} = 0, \label{eq:tv_conv}
\end{align}
\end{subequations}
where $g$ is an arbitrary $L_1$ function, $\|\cdot\|_{TV}$ is the total variation norm, $K^n$ denotes the $n$-step transition kernel, and $\normtarget_0$ is an arbitrary initial PDF.
Eq. \eqref{eq:ergodicity} guarantees that the sample average converges to the true value of the integral, whereas \eqref{eq:tv_conv} ensures that the chain's PDF converges to the target PDF regardless of the initial density.

The aforementioned conditions ensure that the chain converges to the target distribution and that time averages can be used to approximate expectations.
However, the convergence of the algorithm can still be arbitrarily slow.
In order to guarantee that the chain does not get stuck in some region of parameter space for large amounts of time, we need MCMC algorithms which are \emph{geometrically ergodic}.
An MCMC algorithm is geometrically ergodic if
\begin{equation}
	\left\| \int_{\Theta}{K^n(\cdot|\theta') \bar{\pi}_0(\theta')\ d\theta'} - \bar{\pi} \right\|_{\text{TV}} < C_{\bar{\pi}_0} \rho^n,
\label{eq:geometric_ergodicity}
\end{equation}
for some $C_{\bar{\pi}_0}$ and $0 < \rho < 1$ giving the convergence rate.
There are two main reasons why geometric ergodicity is essential.
On the one hand, geometric ergodicity guarantees the existence of a Central Limit Theorem which enables error bounds to be developed.
On the other hand, without geometric ergodicity algorithms are more-or-less guaranteed to give rise to sample paths with ``heavy-tailed excursions'' far away from the centre of the distribution, thus leading to instability and inaccuracy of the subsequent parameter estimation procedures.
See \cite{mengersen1996rates} and \cite{roberts1996geometric} for a more detailed discussion on geometric ergodicity on the one-dimensional and multi-dimensional cases, respectively.

\subsubsection{Gibbs sampler}
\label{sec:gibbs}

\vspace*{12pt}

The Gibbs sampler was introduced by Stuart Geman and Donald Geman in 1984 in order to sample from the Markov Random Field (MRF) induced by the Gibbs distribution \cite{geman1984stochastic}.
The application considered was the Bayesian restoration of images degraded by blurring, nonlinear deformations and multiplicative or additive noise.\footnote{In \cite{geman1984stochastic}, the authors use MAP estimators for the Bayesian restoration task, since they believe that ``the MAP formulation is well-suited to restoration, particularly for handling general forms of spatial degradation.'' However, they also state that ``minimum mean-square error (MMSE) estimation is also feasible by using the (temporal) ergodicity of the relaxation chain to compute \emph{means} w.r.t. the posterior distribution.''}
In order to deal with these distortions, Geman and Geman proposed a stochastic relaxation algorithm that relied on iteratively making local random changes in the image based on current values of the pixels.
A simulated annealing approach, that gradually lowers the system's ``temperature'' \cite{kirkpatrick1983optimization}, was used to avoid local maxima.
More precisely, using our notation the Gibbs sampler proposed in \cite{geman1984stochastic} was the following:
\begin{enumerate}
	\item Select an arbitrary configuration of the pixels, $\parvec^{(0)} = [\param_1^{(0)}, \ldots, \param_{\dpar}^{(0)}]^{\top}$.
	\item Select the sequence of pixels ($n_1, n_2, \ldots$) that will be visited for replacement. The sequence used in \cite{geman1984stochastic} corresponded to a raster
		scan of the image (i.e., repeteadly visiting all the sites in some ``natural'' fixed order), but this sequence does not necessarily have to be periodic.
	\item At the $t$-th ``epoch'' ($t=1, 2, 3, \ldots$), update the $n_t$-th pixel by drawing a sample from the conditional PDF of $\param_{n_t}$ given the current value of the
		remaining pixels, $\param_{n_t}^{(t)} \sim \normtarget(\param_{n_t}|\parvec_{\neg n_t}^{(t-1)})$ with $\parvec_{\neg n_t}^{(t-1)} = [\param_1^{(t-1)}, \ldots,
		\param_{n_t-1}^{(t-1)}, \param_{n_t+1}^{(t-1)},$ $\ldots, \param_{\dpar}^{(t-1)}]^{\top}$.
	\item Repeat step 3 until a pre-specified termination condition (e.g., a fixed number of iterations $T$) is fulfilled.
\end{enumerate}

This approach can be easily generalized and adapted to many practical problems.
Algorithm \ref{alg:gibbs} provides a generic version of the Gibbs sampler with an arbitrary selection of the indices to be sampled.
As already mentioned in the introduction, Gelman showed that the Gibbs sampler is a particular case of the MH algorithm \cite{gelman1992iterative}.
This can be easily seen by considering the MH algorithm (Algorithm \ref{alg:metropolis}) with a proposal at the $t$-th iteration given by $\proposal(\parvec|\parvec^{(t-1)}) = \target(\param_{d_t}|\parvec_{\neg d_t}^{(t-1)}) \delta(\parvec_{\neg d_t}-\parvec_{\neg d_t}^{(t-1)})$, where $\delta(\cdot)$ denotes Dirac's delta.
Then, $\parvec' = [\param_1^{(t-1)}, \ldots, \param_{d_t-1}^{(t-1)}, \param_{d_t}', \param_{d_t+1}^{(t-1)}, \ldots, \param_{\dpar}^{(t-1)}]$ with $\param_{d_t}' \sim \target(\param_{d_t}|\parvec_{\neg d_t}^{(t-1)})$, just like in the $t$-th iteration of the Gibbs sampler of Algorithm \ref{alg:gibbs}.
Now we just need to prove that $\parvec'$ is always accepted, as it happens in the Gibbs sampler.
Noting that $\target(\parvec) = \target(\param_{d_t}|\parvec_{\neg d_t}) \target(\parvec_{\neg d_t})$ by the chain rule of probability, the ratio inside the acceptance probability ($\alpha_t$) of the MH algorithm becomes:
\begin{equation*}
	\frac{\target(\parvec')\proposal(\parvec^{(t-1)}|\parvec')}{\target(\parvec^{(t-1)})\proposal(\parvec'|\parvec^{(t-1)})}
		= \frac{\target(\param_d'|\parvec_{\neg d_t}^{(t-1)}) \target(\parvec_{\neg d_t}^{(t-1)}) \target(\param_d^{(t-1)}|\parvec_{\neg d_t}^{(t-1)})}
			{\target(\param_d^{(t-1)}|\parvec_{\neg d_t}^{(t-1)}) \target(\parvec_{\neg d_t}^{(t-1)}) \target(\param_d'|\parvec_{\neg d_t}^{(t-1)})} = 1.
\end{equation*}
Hence, the proposed sample (drawn from the $d_t$-th full conditional PDF) is always accepted and only the $d_t$-th coordinate is updated at the $t$-th iteration, just like in the Gibbs sampler.


\begin{alg}{Generic Gibbs sampler.}
{\begin{enumerate}
	\item \textbf{Initialization:} Choose an initial state $\parvec^{(0)}$, the total number of iterations ($\niter$), and the burn-in period ($\nburn$).
	\item \texttt{FOR} $t = 1, \ldots, \niter$:
    	\begin{enumerate}
		\item Select the coordinate to be sampled, $d_t \in \{1, \ldots, \dpar\}$, using some of the approaches described below.
        		\item Draw $\param_{d_t}^{(t)} \sim \normtarget(\param_{d_t}|\parvec_{\neg d_t}^{(t-1)})$, with $\parvec_{\neg d_t}^{(t-1)} = [\param_1^{(t-1)}, \ldots,
			\param_{d_t-1}^{(t-1)}, \param_{d_t+1}^{(t-1)},$ $\ldots, \param_{\dpar}^{(t-1)}]^{\top}$.
	\end{enumerate}
	\item Approximate the integral in Eq. \eqref{eq:intBayes} using Eq. \eqref{EstimatorDavid}.
\end{enumerate}
}
\label{alg:gibbs}
\end{alg}

Note that we still have to specify how to select the coordinates to be sampled.
In general it may be difficult to determine the best type of scan for a Gibbs sampler, as shown by Roberts and Rosenthal in \cite{roberts2015surprising}, and many alternative approaches can be devised.
However, the three most widely used schemes are the following \cite{Robert04}:
\begin{itemize}
	\item \textbf{Systematic Scan:} The parameters are updated according to some pre-specified ``canonical'' order. Without loss of generality, let us consider that this order
		is simply $\param_1, \param_2, \ldots, \param_{\dpar}$. Then, we have the following sequence of coordinates to be updated: $d_1=1, d_2=2, \ldots, d_{\dpar}=\dpar,
		d_{\dpar+1}=1, d_{\dpar+2}=2, \ldots, d_{2\dpar}=\dpar, d_{2\dpar+1}=1, \ldots$ This can be expressed more compactly as $d_t = ((t-1))_{\dpar} + 1$, where
		$((t))_{\dpar}$ denotes the modulo operation: $((t))_{\dpar} = m \iff t = k \dpar + m$ for some $k, m \in \mathbb{Z}$ with $m \in \{0, 1, \ldots, \dpar-1\}$ and
		$-\infty < k < \infty$. In this particular case, the Gibbs sampler in Algorithm \ref{alg:gibbs} can be expressed using a double \texttt{FOR} loop, with the inner loop
		running sequentially over the different parameters, as shown in Algorithm \ref{alg:gibbsSystematic}. In this systematic scan Gibbs sampler, which is probably the
		most widely used version of the algorithm in signal processing applications, one iteration of the Gibbs sampler corresponds to one step of the outer loop. Note that
		the total number of samples drawn from the full conditional PDFs in Algorithm \ref{alg:gibbsSystematic} is $\niter \dpar$, whereas in Algorithm \ref{alg:gibbs} only
		$\niter$ samples were drawn. Finally, note that the Markov chain induced by the systematic scan Gibbs sampler is non-reversible \cite{Robert04}.
	\item \textbf{Symmetric Scan:} The coordinates are also explored following a pre-specified deterministic order \cite{Robert04}: first in an ascending order and then in a
		descending order, and this scheme is repeated periodically, i.e.,
		$d_1=1, d_2=2, \ldots, d_{\dpar}=\dpar, d_{\dpar+1}=\dpar-1, d_{\dpar+2}=\dpar-2, \ldots, d_{2\dpar-1}=1, d_{2\dpar}=1, d_{2\dpar+1}=2, \ldots$
		Using the modulo notation, $d_t = \min\{((t-1))_{2\dpar-2},((-t))_{2\dpar-2}\}$.\footnote{Note that the modulo notation is very convenient,
		since it leads to a straightforward computation of the sequence of indexes. For instance, in MATLAB the sequence of indexes for the systematic scan Gibbs sampler
		is obtained as \texttt{dt = mod(t-1,Dpar)}, whereas for the symmetric scan it is given by \texttt{dt = min(mod(t-1,2*Dpar-2),mod(-t,2*Dpar-2))}, with \texttt{Dpar}
		indicating the dimension of the parameter space. Moreover, since these sequences are deterministic, they can be easily pre-computed and stored for further use when
		$T$ is fixed a priori.}
		Unlike the systematic scan, the symmetric scan leads to a reversible Markov chain and can also result in an improved performance. The symmetric Gibbs sampler can
		also be expressed using a double \texttt{FOR} loop, as shown in Algorithm \ref{alg:gibbsSymmetric}, with one iteration of the Gibbs sampler corresponding to one
		step of the outer loop. Now, the total number of samples drawn from the full conditional PDFs is $\niter (2\dpar-1)$.
	\item \textbf{Random Scan:} This method was proposed originally by Liu \emph{et al.} \cite{liu1996metropolized}. In this case, the parameter to be updated is selected
		randomly at each iteration, typically following a uniform distribution, i.e., $d_t \sim \uniform(\{1, 2, \ldots, \dpar\})$. This scheme also produces a reversible Markov
		chain and can lead to an improved performance w.r.t. the symmetric scan Gibbs sampler.\footnote{Note that the sequence of indexes for the random scan Gibbs
		sampler can also be pre-computed when $\niter$ is fixed a priori. In this case, this sequence is obtained by the following Matlab command: \texttt{dt = randi(Dpar,1,T)},
		with \texttt{Dpar} indicating again the dimension of the parameter space.}
\end{itemize}


\begin{alg}{Systematic scan Gibbs sampler.}
{\begin{enumerate}
	\item \textbf{Initialization:} Choose an initial state $\parvec^{(0)}$, the total number of iterations ($\niter$), and the burn-in period ($\nburn$).
	\item \texttt{FOR} $t = 1, \ldots, \niter$:
    	\begin{enumerate}
		\item Draw $\param_{1}^{(t)} \sim \normtarget(\param_{1}|\parvec_{2:\dpar}^{(t-1)})$.
		\item \texttt{FOR} $d = 2, \ldots, \dpar-1$:
		\begin{itemize}
        			\item Draw $\param_{d}^{(t)} \sim \normtarget(\param_{d}|\parvec_{1:d-1}^{(t)},\parvec_{d+1:\dpar}^{(t-1)})$.
		\end{itemize}
		\item Draw $\param_{\dpar}^{(t)} \sim \normtarget(\param_{\dpar}|\parvec_{1:\dpar-1}^{(t)})$.
	\end{enumerate}
	\item Approximate the integral in Eq. \eqref{eq:intBayes} using Eq. \eqref{EstimatorDavid}.
\end{enumerate}
}
\label{alg:gibbsSystematic}
\end{alg}


\begin{alg}{Symmetric scan Gibbs sampler.}
{\begin{enumerate}
	\item \textbf{Initialization:} Choose an initial state $\parvec^{(0)}$, the total number of iterations ($\niter$), and the burn-in period ($\nburn$).
	\item \texttt{FOR} $t = 1, \ldots, \niter$:
    	\begin{enumerate}
		\item Draw $\param_{1}' \sim \normtarget(\param_{1}|\parvec_{2:\dpar}^{(t-1)})$.
		\item \texttt{FOR} $d = 2, \ldots, \dpar-1$:
		\begin{itemize}
        			\item Draw $\param_{d}' \sim \normtarget(\param_{d}|\parvec_{1:d-1}',\parvec_{d+1:\dpar}^{(t-1)})$.
		\end{itemize}
		\item Draw $\param_{\dpar}^{(t)} \sim \normtarget(\param_{\dpar}|\parvec_{1:\dpar-1}')$.
		\item \texttt{FOR} $d = \dpar-1, \ldots, 2$:
		\begin{itemize}
        			\item Draw $\param_{d}^{(t)} \sim \normtarget(\param_{d}|\parvec_{1:d-1}',\parvec_{d+1:\dpar}^{(t)})$.
		\end{itemize}
		\item Draw $\param_{1}^{(t)} \sim \normtarget(\param_{1}|\parvec_{2:\dpar}^{(t)})$.
	\end{enumerate}
	\item Approximate the integral in Eq. \eqref{eq:intBayes} using Eq. \eqref{EstimatorDavid}.
\end{enumerate}
}
\label{alg:gibbsSymmetric}
\end{alg}

Note that only the samples corresponding to the outer loops in Algorithms \ref{alg:gibbsSystematic} and \ref{alg:gibbsSymmetric} (i.e., $\parvec^{(t)} = [\param_1^{(t)}, \ldots, \param_{\dpar}^{(t)}]^{\top}$) are typically used to compute the approximate estimator of Eq. \eqref{EstimatorDavid}.
This entails an inefficient use of the generated samples w.r.t. the generic Gibbs sampler of Algorithm \ref{alg:gibbs}, which uses all the drawn samples to compute the approximate estimator of Eq. \eqref{EstimatorDavid}.
However, ``nothing prevents the use of all the simulations [samples] in integral approximations'', as stated by Robert and Casella \cite{Robert04}.
Indeed, it has been shown very recently that using all the intermediate samples, both in the Gibbs and MH-within-Gibbs (see Section \ref{sec:MHwithinGibbs}) samplers, can result in a substantial improvement in performance in some cases \cite{RG2016}.

Regarding the convergence of the Gibbs sampler, \cite{geman1984stochastic,schervish1992convergence} provide regularity conditions under which the Gibbs sampler is ergodic and the distribution of $\parvec^{(t)}$ converges to the target distribution as $t \to \infty$, whereas \cite{casella1992explaining} provides a simple convergence proof.
In short, the convergence of the Gibbs sampler essentially requires that all the coordinates keep being updated as the algorithm proceeds, implying that every coordinate is visited infinitely often as $t \to \infty$.

Finally, note that there is no need to sample each of the $\dpar$ parameters individually.
Indeed, if a certain subset of parameters can be easily sampled jointly given the rest, then we can group them together inside the loop of Algorithm \ref{alg:gibbs} (and also in Algorithms \ref{alg:gibbsSystematic} and \ref{alg:gibbsSymmetric}).
Let us assume that the $\dpar$ parameters in $\parvec = [\param_1, \ldots, \param_{\dpar}]^{\top}$ can be grouped into $N_g$ disjoint groups in such a way that $\varphivec = [\varphivec_1, \ldots, \varphivec_{N_g}]^{\top}$ contains all the parameters to be inferred.
Then, Algorithm \ref{alg:gibbs} can be applied on $\varphivec$ instead of $\parvec$, drawing $\varphivec_{d_t}^{(t)} \sim \normtarget(\varphivec_{d_t}|\varphivec_{\neg d_t}^{(t-1)})$.
This algorithm is known as the \emph{group} or \emph{block Gibbs sampler}.
Alternatively, if a subset of parameters can be easily sampled given the rest, we can remove them from the loop of the Gibbs sampler.
Without loss of generality, let us assume that we keep the first $\widetilde{D}_{\param}$ parameters and leave the remaining parameters outside of the iterations of the Gibbs sampler, i.e., we only draw samples from the reduced set of parameters $\widetilde{\parvec} = [\param_1, \ldots, \param_{\widetilde{D}_{\param}}]^{\top}$.
Then, Algorithm \ref{alg:gibbs} can be applied on $\widetilde{\parvec}$ instead of $\parvec$, drawing $\param_{d_t}^{(t)} \sim \normtarget(\param_{d_t}|\widetilde{\parvec}_{\neg d_t}^{(t-1)})$ with $d_t \in \{1, \ldots, \widetilde{D}_{\param}\}$.
When the chain has converged, then we can easily sample from the remaining parameters given the samples from the first $\widetilde{D}_{\param}$ parameters obtained using the Gibbs sampler.
This algorithm is known as the \emph{collapsed Gibbs sampler}.
Although the addition of auxiliary variables can speed up the convergence of the Gibbs sampler in some cases (e.g., see the data augmentation algorithm in Section \ref{sec:other}), in general grouping or collapsing down variables leads to improved convergence and decreased sample autocovariances, as shown by Liu in \cite{liu1994collapsed}.
However, let us remark that Liu's proof is highly restrictive and in some cases the uncollapsed sampler can actually converge faster than the collapsed one (e.g., see the counterexample in Appendix A of Terenin et al. \cite{terenin2018polya}).
Finally, note also that finding the optimal variables to group or collapse in order to achieve the optimal performance depends on the problem and can be a very difficult task.

The Gibbs sampler is a fundamental algorithm for parameter estimation in many signal processing and machine learning problems.
Indeed, it may be the only choice for some models, because it is well-defined even on discrete state spaces where gradients are not available and good Metropolis-Hastings proposals are difficult to construct.
Therefore, it has been extensively used in practical applications either as a stand-alone method or combined with the MH algorithm as described in the following section.

\subsubsection{MH-within-Gibbs}
\label{sec:MHwithinGibbs}

\vspace*{12pt}

The Gibbs sampler requires sampling from the full univariate conditional PDFs.
Unfortunately, although this should be a much easier task than sampling from the multi-variate posterior PDF, in many real-world applications these conditional PDFs have non-standard forms and we cannot sample directly from them. 
Initially, some authors tackled this problem by using the RS algorithm (e.g., see \cite{zeger1991generalized}), and the adaptive RS (ARS) algorithm was specifically designed for this task \cite{Gilks92}.
However, as already mentioned before, both the RS and ARS algorithms require finding a bounding constant $C$ such that $C\proposal(\parvec) \ge \target(\parvec)$, a task that  may be difficult for complicated targets and lead to very inefficient sampling if $C$ is large.
In this section, we briefly discuss a widely used technique developed to address this problem, the MH-within-Gibbs algorithm (often also called {\it Component-wise MH} method), as well as two related methods: the griddy Gibbs sampler and the fast universal self-tuned sampler (FUSS).

In order to sample from non-standard full conditional PDFs, Ritter and Tanner proposed the so called \emph{griddy Gibbs sampler} \cite{ritter1991griddy,ritter1992facilitating}.
Their basic idea was using a set of evaluations from the desired full conditional PDF to build a piecewise approximation from which sampling is straightforward.
The $t$-th iteration of the griddy Gibbs sampler for the $d$-th coordinate ($1 \le d \le \dpar$) proceeds as follows:
\begin{enumerate}
	\item Evaluate the target at some pre-specified set of parameters, $\mathcal{S}_d^{(t)}=\{\param_{d,1}^{(t)}, \ldots,$ $\param_{d,K}^{(t)}\}$, obtaining
		$P_{d,1}^{(t)}=\target(\param_{d,1}^{(t)}|\parvec_{\neg d}^{(t-1)}), \ldots, P_{d,K}^{(t)} = \target(\param_{d,K}^{(t)}|\parvec_{\neg d}^{(t-1)})$.
	\item Construct an approximate inverse cumulative distribution function (CDF) of the target, $\hat{\Pi}^{-1}(\param_d|\parvec_{\neg d}^{(t-1)},\mathcal{S}_d^{(t)})$,
		using $P_{d,1}^{(t)}, \ldots, P_{d,K}^{(t)}$ and a piecewise constant (PWC) or piecewise linear (PWL) approximation.
	\item Draw $u \sim \uniform([0,1))$ and apply the \emph{inverse method} \cite{MartinoBook18} to obtain a sample drawn approximately from the target as
		$\theta_d^{(t)} = \hat{\Pi}^{-1}(u|\parvec_{\neg d}^{(t-1)},\mathcal{S}_d^{(t)})$.
\end{enumerate}
The griddy Gibbs sampler can be easily implemented for univariate full conditional PDFs, and its performance can be improved by using an adaptive grid and allowing the grid to grow if necessary (using the so called \emph{grid grower}), as described in \cite{ritter1991griddy,ritter1992facilitating}.
However, the samples obtained are only approximately distributed according to the target, and building an effective approximation of the inverse CDF in the multi-variate case (e.g., for its use within the block Gibbs sampler) is a challenging task.
The first issue can be addressed by using the \emph{Gibbs stopper} \cite{ritter1992facilitating}, where an IS weight is assigned to the drawn samples in order to ensure that they come exactly from the target PDF, but the second one is much more difficult to solve.

In order to sample virtually from any full conditional PDF, the MH algorithm can be used within the Gibbs sampler.
This results in a \emph{hybrid sampler} \cite{Robert04}, where an \emph{internal} Monte Carlo method (the MH algorithm) is used within another \emph{external} Monte Carlo technique (the Gibbs sampler).
Apparently, Geweke and Tanizaki were the first ones to suggest using the MH algorithm within the Gibbs sampler in order to provide a general solution to nonlinear and/or non-Gaussian state space modeling in a Bayesian framework \cite{geweke1999markov,geweke2001bayesian}.
The MH-within-Gibbs sampler is detailed in Algorithm \ref{alg:MHwithinGibbs}.
Note that $T_{MH}$ iterations of the internal MH algorithm are performed per iteration of the external Gibbs sampler and only the last sample drawn from the MH algorithm is typically used for the integral approximation in Eq. \eqref{EstimatorDavid}.
Furthermore, usually $T_{MH}=1$ for the sake of efficiency, but several authors have shown that this is often not the best alternative from the point of view of reducing the variance of the desired estimators for a given computational budget \cite{MartinoIA2RMS15}.
Note also that the internal MH algorithm should be used to sample only those parameters that cannot be sampled directly (Algorithm \ref{alg:MHwithinGibbs} assumes that all the parameters require it), and that it can also be easily applied within the block and collapsed Gibbs samplers.
Finally, note that Neal and Roberts have shown that the optimal scaling rate for the MH algorithm (which leads to an average acceptance rate of 0.234) also holds for the MH-within-Gibbs sampler regardless of the dimensionality of the update rule \cite{neal2006optimal}.


\begin{alg}{MH-within-Gibbs algorithm.}
{\begin{enumerate}
	\item \textbf{Initialization:} Choose a set of proposal PDFs, $\{\normproposal(\param_d|\param_d^{(t-1)},\parvec_{\neg d})\}_{d=1}^{\dpar}$, an initial state $\parvec^{(0)}$,
		the total number of iterations ($\niter$), the number of iterations of the internal MH algorithm ($T_{MH}$), and the burn-in period ($\nburn$).
	\item \texttt{FOR} $t = 1, \ldots, \niter$:
    	\begin{enumerate}
		\item Select the coordinate to be sampled, $d_t \in \{1, \ldots, \dpar\}$, and set $\widetilde{\param}_{d_t}^{(0)} = \param_{d_t}^{(t-1)}$.
		\item \texttt{FOR} $t' = 1, \ldots, T_{MH}$:
		\begin{enumerate}
			\item Draw $\param_{d_t}' \sim \normproposal(\param_{d_t}|\widetilde{\param}_{d_t}^{(t'-1)},\parvec_{\neg d_t}^{(t-1)})$ and $u \sim \uniform([0, 1))$.
			\item Compute the acceptance probability ($\alpha_t$):
				\begin{equation*}
					\alpha(\parvec',\parvec_{\neg d_t}^{(t-1)},\widetilde{\param}_{d_t}^{(t'-1)}) = \min\left[1,
						\frac{\target(\param_{d_t}'|\parvec_{\neg d_t}^{(t-1)}) \proposal(\widetilde{\param}_{d_t}^{(t'-1)}|\param_{d_t}',\parvec_{\neg d_t}^{(t-1)})}
						{\target(\widetilde{\param}_{d_t}^{(t'-1)}|\parvec_{\neg d_t}^{(t-1)}) \proposal(\param_{d_t}'|\widetilde{\param}_{d_t}^{(t'-1)},\parvec_{\neg d_t}^{(t-1)})}\right]
				\end{equation*}
			\item If $u \le \alpha_t$, accept $\param_{d_t}'$ and set $\widetilde{\param}_{d_t}^{(t')} = \param_{d_t}'$. Otherwise (i.e., if $u > \alpha_t$), reject $\param_{d_t}'$ and set
				$\widetilde{\param}_{d_t}^{(t')} = \widetilde{\param}_{d_t}^{(t'-1)}$.
		\end{enumerate}
		\item Set $\param_{d_t}^{(t)} = \widetilde{\param}_{d_t}^{(T_{MH})}$.
	\end{enumerate}
	\item Approximate the integral in Eq. \eqref{eq:intBayes} using Eq. \eqref{EstimatorDavid}.
\end{enumerate}
}
\label{alg:MHwithinGibbs}
\end{alg}

Noting that the piecewise proposal built by the griddy Gibbs sampler could be used to construct very good proposals for the MH-within-Gibbs sampler, Martino \emph{et al.} recently proposed the fast universal self-tuned sampler (FUSS) within Gibbs algorithm \cite{FUSS}.
Essentially, the idea is starting with a very dense grid that roughly covers the whole effective support of the corresponding full conditional PDF and then applying a pruning strategy in order to obtain a sparse grid that contains most of the probability mass of the conditional PDF.
The steps performed by the FUSS algorithm, at the $t$-th step of the Gibbs sampler for the $d$-th parameter, are the following:
\begin{enumerate}
	\item {\bf Initialization:} Choose a large set of support points, $\widetilde{\mathcal{S}}_d^{(t)}=\{\param_{d,1}^{(t)}, \ldots,$ $\param_{d,L}^{(t)}\}$, that densely cover the whole
		effective support of the target.
	\item {\bf Pruning:} Remove support points according to a pre-specified and efficient criterion, attaining a final sparse set of support points,
		$\mathcal{S}_d^{(t)}=\{\param_{d,1}^{(t)}, \ldots,$ $\param_{d,K}^{(t)}\}$ with $K \ll L$.
	\item {\bf Construction:} Build a proposal function $\proposal(\param_d|\parvec_{\neg d}^{(t-1)},\mathcal{S}_d^{(t)})$ using some appropriate pre-defined mechanism, typically
		a PWC or PWL approach.
	\item {\bf MH steps:} Perform $T_{MH}$ steps of the internal MH algorithm, as in Algorithm \ref{alg:MHwithinGibbs}, using
		$\proposal(\param_d|\parvec_{\neg d}^{(t-1)},\mathcal{S}_d^{(t)})$ as the proposal PDF.
\end{enumerate}
Since the FUSS algorithm builds a proposal tailored to the target, the acceptance rate of the internal MH algorithm is usually very high and the correlation among the drawn samples very small.
This leads to estimators with a reduced variance, especially for very peaky proposals, where other Monte Carlo methods fail (see \cite{FUSS} for further details).
Finally, note that it is again possible to employ all the $T_{MH}$ samples generated by the internal MH algorithm in the final estimators, as shown in \cite{RG2016}.

\subsubsection{Other Classical MCMC Techniques}
\label{sec:other}

\vspace*{12pt}

In this section, we describe other classical approaches for sampling from non-standard multi-variate densities: data augmentation, slice sampling, the hit-and-run algorithm, and adaptive direction sampling.
We also discuss briefly the issue of thinning or subsampling the Markov chain, which is often used in signal processing applications to reduce the computational cost and the correlation among the generated samples.

\vspace*{12pt}
\noindent
\emph{Data Augmentation (DA)}
\vspace*{12pt}

The \emph{data augmentation} method was originally devised by Tanner and Wong in order to compute posterior distributions for Bayesian inference \cite{tanner1987calculation}.
The basic idea of data augmentation (DA) is the same one that underlies the well-known and widely used expectation-maximization (E-M) algorithm \cite{dempster1977maximum}: in many practical problems, augmenting the observed dataset ($\datavec$) with a set of latent data ($\z$) leads to an easier analysis of the problem.
In the Bayesian inference case, the DA algorithm is based on the assumption that $\target(\parvec|\datavec,\z)$ is straightforward to analyze, whereas $\target(\parvec|\datavec)$ is intractable.
Another important assumption regards the generation of the latent data ($\z$): they should be easy to draw given the parameters and the observed data.
Under these two assumptions, drawing samples from the desired target can be easily accomplished following the iterative approach shown in Algorithm \ref{alg:DA}.
Note that the DA procedure shown in Algorithm \ref{alg:DA} is equivalent to the application of the Gibbs sampler of Algorithm \ref{alg:gibbs} on the augmented parameter vector $\thetavec_a = [\parvec, \z_1, \ldots, \z_K]$ \cite{liu1994collapsed}.\footnote{In fact, Algorithm \ref{alg:DA} corresponds to the block Gibbs sampler with $K+1$ groups: $\varphivec_1 = \parvec$, $\varphivec_2=\z_1$, \ldots, $\varphivec_{K+1} = \z_K$. However, it can be easily converted into a component-wise Gibbs algorithm (with $K\ddata+\dpar$ components) by decomposing steps 2(a) and 2(b) into $\dpar$ and $K\ddata$ draws from the corresponding univariate full conditional PDFs, respectively.}
Note also that data augmentation is the opposite of integrating out parameters from a model in closed form, as done in the collapsed Gibbs sampler described in Section \ref{sec:gibbs}.
{Finally, let us remark that, just like it happens with the collapsed Gibbs sampler (cf. the previously mentioned discussion of Liu et al. in Section \ref{sec:gibbs}), DA can either increase or reduce the mixing efficiency.}

\newpage

\begin{alg}{Basic Data Augmentation (DA) algorithm.}
{\begin{enumerate}
	\item \textbf{Initialization:} Select the number of latent data generated per iteration ($K$), the total number of iterations ($\niter$) and the burn-in period ($\nburn$).
		Obtain an initial set of latent data ($\z_1^{(0)}, \ldots, \z_K^{(0)}$) and construct an initial approximation of the target,
		$\hat{\target}^{(0)}(\parvec|\datavec,\z_1^{(0)},\ldots,\z_K^{(0)}) = \frac{1}{K}\sum_{k=1}^{K}{\target(\parvec|\datavec,\z_k^{(0)})}$.
	\item \texttt{FOR} $t = 1, \ldots, \niter$:
    	\begin{enumerate}
        		\item Draw $\parvec^{(t)} \sim \hat{\target}^{(t-1)}(\parvec|\datavec,\z_1^{(0)},\ldots,\z_K^{(0)})$.
		\item Draw $\z_k^{(t)} \sim p(\z|\parvec^{(t)},\datavec)$ for $k=1, \ldots, K$.
		\item Update the approximation of the target:
			\begin{equation}
				\hat{\target}^{(t)}(\parvec|\datavec,\z_1^{(0)},\ldots,\z_K^{(0)}) = \frac{1}{K}\sum_{k=1}^{K}{\target(\parvec|\datavec,\z_k^{(t)})}.
			\label{eq:targetDA}
			\end{equation}
	\end{enumerate}
	\item Approximate the integral in Eq. \eqref{eq:intBayes} using Eq. \eqref{EstimatorDavid}.
\end{enumerate}
}
\label{alg:DA}
\end{alg}

\vspace*{12pt}
\noindent
\emph{Slice sampling}
\vspace*{12pt}

Several Monte Carlo techniques, like direct methods (e.g., the inverse-of-density method) \cite{MartinoBook18}, the rejection sampler (see Section \ref{sec:RS}), and some MCMC algorithms (e.g., the so-called {\it slice sampler}) rely on a simple result, known as {\it the fundamental theorem of simulation}.
\begin{theo} 
 Drawing samples from a random variable ${\bm \Theta}$ with density $\bar{\pi}({\bm \theta})\propto \pi({\bm \theta})$ is equivalent to sampling uniformly on the region defined by
	        \begin{equation}
	        \label{Azero}
                        \mathcal{A}_\pi=\{({\bm \theta},z)\in \mathbb{R}^2: \  0\leq z \leq \pi({\bm \theta})\}.
                  \end{equation}                             
Namely,  considering a realization $({\bm \theta}',z')$, if it is distributed uniformly on $\mathcal{A}_\pi$, then ${\bm \theta}'$ is a sample from $\bar{\pi}({\bm \theta})$  \cite{Robert04,MartinoBook18}.  
\end{theo} 

Therefore, if we are able to draw a vector  $({\bm \theta}',z')$ uniformly on $\mathcal{A}_\pi$ (i.e., the area below the unnormalized target function $\pi({\bm \theta})$), then the coordinate ${\bm \theta}'$ is marginally distributed according to $\bar{\pi}({\bm \theta})$.
The variable $z$ plays the role of an auxiliary variable which is introduced in order to ease the sampling procedure, just like the latent data in the data augmentation algorithm.

The slice sampler is precisely a Gibbs sampling method that can be applied for drawing samples uniformly from $\mathcal{A}_\pi$. Let us define the set  
\begin{equation}
\mathcal{O}(z)=\{\mbox{all } {\bm \theta}\in \mathbb{R}^{\mathcal{D}_\theta} \mbox{ such that } \pi({\bm \theta})\geq z \}.
\end{equation}
The slice sampler is given in Algorithm \ref{alg:SliceSampling}.

\newpage

\begin{alg}{The slice sampler.}
{\begin{enumerate}
	\item \textbf{Initialization:} Choose an initial state $\parvec^{(0)}$, the total number of iterations ($\niter$), and the burn-in period ($T_b$).
	\item \texttt{FOR} $t = 1, \ldots, \niter$:
    	\begin{enumerate}
		\item Draw $z^{(t)}$ uniformly in the interval $[0,\pi({\bm \theta}^{(t-1)})]$ ,
\item\label{EsteStep} Draw $\parvec^{(t)}$ uniformly in the set $\mathcal{O}(z^{(t)})$. 
\end{enumerate}
\item Approximate the integral in Eq. \eqref{eq:intBayes} using Eq. \eqref{EstimatorDavid}.
\end{enumerate}
}
\label{alg:SliceSampling}
\end{alg}
The slice sampling algorithm generates a Markov chain over $\mathcal{A}_\pi$, producing samples uniformly distributed in $\mathcal{A}_\pi$ after the burn-in period.
However, performing step \ref{EsteStep} is often virtually impossible (even for unidimensional target PDFs), since it requires the inversion of $\pi({\bm \theta})$ in order to determine the set $\mathcal{O}(z)$.
The difficulty of this inversion is due to the fact that $\pi({\bm \theta})$ is usually a non-monotonic function, implying that the set $\mathcal{O}(z)$ is typically formed by the union of disjoint sets which are difficult to determine.
Fortunately, several practical procedures have been suggested for this purpose.
See \cite{neal2003} for further information on this issue.

\vspace*{12pt}
\noindent
\emph{Hit-and-Run}
\vspace*{12pt}

Another important class of methods that can be used both for global optimization and Bayesian inference are the so called \emph{hit-and-run algorithms}, which are a collection of efficient sampling techniques that use random walks to explore the parameter space.
Sampling through random walks was independently proposed by Boneh and Golan \cite{boneh1979constraints} and Smith \cite{smith1980monte,smith1984efficient}, and this class of methods were later renamed as hit-and-run algorithms \cite{berbee1987hit}.
The generic hit-and-run algorithm is shown in Algorithm \ref{alg:hit-and-run}.
The basic idea is determining a random direction in the $\dpar$-dimensional parameter space using the proposal $\proposal(\parvec)$, and then selecting a random point along that direction with a probability proportional to the target PDF evaluated along the chosen direction.

\begin{alg}{Hit-and-run algorithm.}
{\begin{enumerate}
	\item \textbf{Initialization:} Choose a proposal function over the unit hypersphere,
		$\proposal(\parvec) \ne 0 \iff \parvec \in \mathcal{B}=\{\parvec \in \parspace \subseteq \mathbb{R}^{\dpar}: \|\parvec\| = 1\}$, the initial state, $\parvec^{(0)}$, the total
		number of iterations ($\niter$) and the burn-in period ($\nburn$).
	\item \texttt{FOR} $t = 1, \ldots, \niter$:
    	\begin{enumerate}
        		\item Draw $\parvec' \sim \normproposal(\parvec)$.
		\item Find the set $\Lambda_t=\{\lambda \in \mathbb{R}:\parvec^{(t-1)}+\lambda\parvec' \in \parspace \subseteq \mathbb{R}^{\dpar}\}$.
		\item Draw $\lambda_t \sim p^{(t)}(\lambda) \propto \target(\parvec^{(t-1)}+\lambda\parvec')$ with $\lambda \in \Lambda_t$.
		\item Set $\parvec^{(t)} = \parvec^{(t-1)}+\lambda_t\parvec'$.
	\end{enumerate}
	\item Approximate the integral in Eq. \eqref{eq:intBayes} using Eq. \eqref{EstimatorDavid}.
\end{enumerate}
}
\label{alg:hit-and-run}
\end{alg}

Different hit-and-run algorithms are obtained depending on the proposal function $\proposal(\parvec)$.
For instance, the original \emph{hypersphere directions} (HD) hit-and-run algorithm considered a uniform proposal $\proposal(\parvec)$ and a uniform target on some bounded region $\Theta \subset \mathbb{R}^{\dpar}$ \cite{boneh1979constraints,smith1980monte,smith1984efficient}, whereas the \emph{coordinate directions} (CD) hit-and-run randomly chooses one of the $\dpar$ coordinates of the parameter space \cite{kaufman1998direction}.
Regarding the connections with other methods, the hit-and-run algorithm has similarities both with the MH algorithm and the Gibbs sampler.
On the one hand, the hit-and-run algorithm resembles the random walk MH algorithm, but the generated samples are always accepted, since they are drawn from the target.\footnote{Note that drawing samples directly from the target for an arbitrary direction of the parameter space may be unfeasible in many problems. In these cases, the Metropolised hit-and-run sampler proposed in \cite{chen1993performance} can be used.}
On the other hand, the CD hit-and-run algorithm is equivalent to the random scan Gibbs sampler.
However, note that the generic hit-and-run algorithm is more flexible than the Gibbs sampler, since it can choose any arbitrary direction, not only one of the directions corresponding to the different parameters.

\vspace*{12pt}
\noindent
\emph{Adaptive Direction Sampling (ADS)}
\vspace*{12pt}

A third important family of methods that attempt to improve the convergence speed of the Gibbs sampler is \emph{adaptive direction sampling} (ADS) \cite{gilks1994adaptive,roberts1994convergence}.
The basic idea of ADS is maintaining a set of support points that are constantly updated, with the current support set being used to determine the sampling direction.
The general adaptive direction sampler is shown in Algorithm \ref{alg:ADS}.

\begin{alg}{Adaptive direction sampling (ADS).}
{\begin{enumerate}
	\item \textbf{Initialization:} Choose an initial support set, $\mathcal{S}^{(0)} = \{\parvec_1^{(0)}, \ldots, \parvec_K^{(0)}\}$ with $K > \dpar$, the total
		number of iterations ($\niter$), and the burn-in period ($\nburn$).
	\item \texttt{FOR} $t = 1, \ldots, \niter$:
    	\begin{enumerate}
        		\item Draw $i \sim \uniform(\{1, \ldots, K\})$ and set $\parvec_c^{(t)} = \parvec_i^{(t-1)}$.
		\item Select a second vector $\varthetavec^{(t)}$ according to some pre-specified scheme, depending on the specific ADS algorithm to be implemented (see below).
		\item Draw $r' \sim p(r) \propto \target(\parvec_c^{(t)} + r(\varthetavec^{(t)}+\lambda_t \parvec_c^{(t)})) \times |1+r\lambda_t|^{\dpar-1}$.
		\item Set $\parvec_i^{(t)} = \parvec_c^{(t)} + r' (\varthetavec^{(t)}+\lambda_t \parvec_c^{(t)})$.
	\end{enumerate}
	\item Approximate the integral in Eq. \eqref{eq:intBayes} using Eq. \eqref{EstimatorDavid}.
\end{enumerate}
}
\label{alg:ADS}
\end{alg}

The procedure shown in Algorithm \ref{alg:ADS} is very general and many different algorithms can be obtained by considering different choices for $\varthetavec^{(t)}$ and $\lambda_t$ \cite{gilks1994adaptive}:
\begin{itemize}
	\item \textbf{Snooker algorithm:} Important special case of the general ADS algorithm obtained by setting
		$\varthetavec^{(t)} = \parvec_a^{(t)} \sim \uniform(\mathcal{S}^{(t-1)} \setminus \{\parvec_c^{(t)}\})$ and $\lambda_t=-1$. In this specific algorithm, $\parvec_a^{(t)}$ sets
		the direction along which $\parvec_c^{(t)}$ is moved in order to obtain the new sample.
	\item \textbf{Parallel ADS:} Obtained by setting $\varthetavec^{(t)}=\parvec_a^{(t)}-\parvec_b^{(t)}$, with
		$\parvec_a^{(t)} \sim \uniform(\mathcal{S}^{(t-1)} \setminus \{\parvec_c^{(t)}\})$ and
		$\parvec_b^{(t)} \sim \uniform(\mathcal{S}^{(t-1)} \setminus \{\parvec_a^{(t)},\parvec_c^{(t)}\})$, and $\lambda_t=0$.
		In this case, the direction for the movement of $\parvec_c^{(t)}$ is set by the two auxiliary points drawn from $\mathcal{S}^{(t-1)}$, $\parvec_a^{(t)}$ and $\parvec_b^{(t)}$.
	\item \textbf{Hit-and-run:} Obtained as a particular case of the general ADS algorithm by setting $\lambda_t=0$ and $\varthetavec^{(t)}$ to some random direction uniformly
		selected in the parameter space.
	\item \textbf{Random scan Gibbs sampler:} Obtained by setting $\lambda_t=0$ and $\varthetavec^{(t)}$ to be some randomly chosen parameter out of the $\dpar$ available,
		i.e., $\varthetavec^{(t)}=[0, \ldots, 0, \param_d^{(t)}, 0, \ldots, 0]$).
\end{itemize}

\vspace*{12pt}
\noindent
\emph{Subsampling or Thinning of Markov Chains}
\vspace*{12pt}

Finally, let us briefly discuss the issue of thinning or subsampling a Markov chain, which is often used to reduce the correlation among the generated samples of MCMC methods and also serves to decrease the computational/storage burden.
Thinning consists of discarding $K-1$ out of every $K$ outputs of the obtained Markov chain (i.e., downsampling or decimating by a factor $K$ in signal processing terminology), thus resulting in the following estimator:
\begin{equation}
	\widehat{I}_{M_{\textrm{thin}}} = \frac{1}{M_{\textrm{thin}}}\sum_{m=0}^{M_{\textrm{thin}}-1}{g(\parvec^{(\nburn+1+mK)})},
\label{eq:estimatorThinning}
\end{equation}
with $M_{\textrm{thin}} = \left\lfloor\frac{\niter-(\nburn+1)}{K}\right\rfloor$ and $\lfloor \cdot \rfloor$ denoting the integer part approximated from below.
It is well-known that the estimator in Eq. \eqref{eq:estimatorThinning} has a larger variance than the estimator of Eq. \eqref{EstimatorDavid}, see \cite{geyer1992practical} for a formal proof for reversible Markov chains or \cite{maceachern1994subsampling} for a simpler justification which does not rely on reversibility.
Hence, many authors have warned practitioners against subsampling, which should be used only when strictly required due to computation/storage constraints \cite{link2012thinning}.
However, Owen has shown very recently that thinning can actually be advantageous in some cases \cite{owen2015statistically}.
Assuming that it costs one unit of time to advance a Markov chain and $\Delta t > 0$ units of time to compute a sampled quantity of interest, he shows that thinning will improve the statistical efficiency (as quantified by the variance of the resulting estimator) when $\Delta t$ is large and the autocorrelations decay slowly enough.
Hence, even when practical restrictions do not apply, signal processing practitioners should check the correlation structure of their problems in order to determine whether thinning can be advantageous or not, and which is the optimal thinning factor (see \cite{owen2015statistically} for further details).

\subsection{Adaptive MCMC}
\label{sec:adaptMcmc}

\vspace*{12pt}

The MH algorithm produces samples distributed according to any desired target distribution after the burn-in period by using an arbitrary proposal density that fulfills some mild regularity conditions.
However, the choice of the proposal PDF is crucial for the practical operation of the algorithm.
In order to sample from a given target distribution efficiently, we must carefully choose the proposal distribution in such a way that we obtain independent-enough samples with high-enough rate so that we achieve a good approximation of the distribution in a reasonable time (e.g., in minutes or hours instead of hundreds of years).
The manual tuning of the proposal distribution can be a tedious task.
For that reason, researchers have developed adaptive MCMC methods (e.g., see \cite{Andrieu+Thoms:2008}), which aim at tuning the algorithm's performance automatically, typically by adapting the proposal distribution based on the already seen samples.
In the following, we review some of the most relevant adaptive MCMC approaches.

\subsubsection{Parametric approaches}
\label{sec:parametricAMCMC}

\vspace*{12pt}

The history of adaptive MCMC methods can probably be considered to start with the adaptive Metropolis (AM) algorithm \cite{Haario+Saksman+Tamminen:2001}, where the idea is to adapt the covariance of a random-walk Metropolis algorithm using previous samples drawn by the same algorithm.
Let us recall that the random-walk MH algorithm typically uses a Gaussian proposal PDF of the form
\begin{equation}
  \proposal(\parvec | \parvec^{(t-1)}) = \gauss(\parvec | \parvec^{(t-1)}, \covmat),
\end{equation}
where $\covmat$ is some suitable covariance matrix.
The tuning of this algorithm is then reduced to the selection of the covariance matrix $\covmat$. 

One way to approach the tuning problem is to consider an idealized case where we actually know the true covariance $\boldsymbol{\Sigma}$ of the target distribution.
It turns out that the optimal covariance matrix is $\covmat^* = \lambda \boldsymbol{\Sigma}$ under certain idealized conditions.
Furthermore, in the Gaussian case we can compute the optimal $\lambda^* = 2.38^2 / \dpar$ \cite{Gelman_et_al:1996}.
This result can now be used to adapt the proposal's covariance by replacing the target distribution covariance with its empirical counterpart.

The adaptive Metropolis (AM) \cite{Haario+Saksman+Tamminen:2001} uses an adaptation rule where the covariance of the target distribution is estimated via
\begin{equation}
  \boldsymbol{\Sigma}_{t} = \mathrm{Cov}[\parvec^{(0)}, \ldots, \parvec^{(t-1)}, \parvec^{(t)}] + \epsilon \mathbf{I},
  \label{eq:amcov}
\end{equation}
where $\epsilon$ is a small positive constant which is used to ensure that $\boldsymbol{\Sigma}_{t}$ is not ill-conditioned.
In practice, this amounts to adding an extra step to the MH Algorithm~\ref{alg:metropolis}, just after the acceptance step, to update the current estimate of the covariance.
Furthermore, the adaptation rule can be easily implemented recursively \cite{Haario+Laine:2006}.
Numerous modifications and improvements to this rule have been proposed.
For example, Gaussian mixture approximations are considered in \cite{Giordani+Kohn:2010,Luengo+Martino:2013}, the combination with early rejection is proposed in \cite{Haario+Laine:2006}, and adaptive Kalman filter based covariance estimation is considered in \cite{Mbalawata_et_al:2015}.

Because the identity for the optimal $\lambda^*$ only applies to Gaussian target PDFs, it is often also desirable to adapt the $\lambda$-parameter as well \cite{Andrieu+Thoms:2008,Atchade+Fort:2010,Vihola:2011,Vihola:2012}.
In this case, the optimization criterion consists typically in trying to reach ``an optimal" acceptance rate of $\alpha^* = 0.234$.
Note that this optimal acceptance rate also corresponds to certain idealized conditions, but still provides a good rule of thumb.
A typical rule for the adaptation then has the form
\begin{equation}
  \log \lambda_t = \log \lambda_{t-1} + \gamma_t \, (\alpha_t - \bar{\alpha}),
  \label{eq:amlambda}
\end{equation}
where $\gamma_t$ is a suitable gain sequence, $\alpha_t$ is the acceptance probability at the current step, and $\bar{\alpha}$ is the target acceptance rate (e.g., $\bar{\alpha} = \alpha^*$).

The general AM algorithm, including both covariance and acceptance rate adaptation, is shown in Algorithm \ref{alg:am}.
Comparing Algorithms \ref{alg:metropolis} and \ref{alg:am}, we note that the difference simply lies in the introduction of two additional steps (steps 2(d) and 2(e)), where the empirical covariance matrix $\boldsymbol{\Sigma}_{t}$ and the scale factor $\lambda_t$ are computed.
However, these two simple steps can lead to a substantially improved proposal w.r.t. the initial one and thus to a much better performance of the resulting MH algorithm.


\begin{alg}{General Adaptive Metropolis (AM) Algorithm.}
{\begin{enumerate}
	\item \textbf{Initialization:} Choose an initial covariance $\boldsymbol{\Sigma}_0$, an initial adaptation parameter $\lambda_0$,
	a target acceptance rate $\bar{\alpha}$, an initial state $\parvec^{(0)}$, the total number of iterations ($\niter$) and the burn-in period ($\nburn$).
    \item \texttt{FOR} $t = 1, \ldots, \niter$:
    	\begin{enumerate}
        	\item Set $\covmat_t = \lambda_{t-1} \boldsymbol{\Sigma}_{t-1}$. Draw $\parvec' \sim \gauss(\parvec^{(t-1)},\covmat_t)$ and $u \sim \uniform([0,1))$.
            \item Compute the acceptance probability:
            	\begin{equation}
                \alpha_t \equiv \alpha(\parvec',\parvec^{(t-1)}) = \min\left[1,\frac{\target(\parvec')} {\target(\parvec^{(t-1)})}\right].
                \label{eq:alphaAM}
                \end{equation}
            \item If $u \le \alpha_t$, accept $\parvec'$ and set $\parvec^{(t)} = \parvec'$. Otherwise (i.e., if $u > \alpha_t$), reject $\parvec'$ and set $\parvec^{(t)}= \parvec^{(t-1)}$.
         \item Obtain a new estimate of the target's covariance $\boldsymbol{\Sigma}_{t}$, e.g., by using the estimator of Eq. \eqref{eq:amcov}.
         \item Compute a new $\lambda_t$, e.g., by applying the rule of Eq. \eqref{eq:amlambda}.
         \end{enumerate}
    \item Approximate the integral in Eq. \eqref{eq:intBayes} using Eq. \eqref{EstimatorDavid}.
   \end{enumerate}
}
\label{alg:am}
\end{alg}

\subsubsection{Non-parametric approaches}
\label{sec:nonParametricAMCMC}

\vspace*{12pt}

Designing parametric adaptive MCMC approaches that attain a good performance directly in high-dimensional parameter spaces can be a very challenging task.
For this reason, some authors have concentrated on developing very efficient non-parametric adaptive MH algorithms in low-dimensional parameter spaces (typically one-dimensional spaces).
These adaptive schemes can then be used within the Gibbs sampler (see Section \ref{sec:gibbs}) in order to perform estimations in higher dimensional spaces.
In this section we review the most widely known approach, Adaptive Rejection Metropolis Sampling (ARMS), as well as some very recent extensions.

\vspace*{12pt}
\noindent
\emph{Adaptive Rejection Metropolis Sampling (ARMS)}
\vspace*{12pt}

The adaptive rejection Metropolis sampling (ARMS) technique combines the ARS method and the MH algorithm in order to sample virtually from any univariate PDF \cite{Gilks95}.
ARMS is summarized in Algorithm \ref{alg:arms} for a one-dimensional parameter $\param$.
For multi-dimensional parameter spaces, ARMS can simply be embedded within a Gibbs sampler in a similar way as done in the MH-within-Gibbs algorithm (see Algorithm \ref{alg:MHwithinGibbs}).
Essentially, ARMS performs first an RS test, and the rejected samples are used to improve the proposal PDF with the aim of constructing a proposal that becomes closer and closer to the target.
Then, the accepted samples from the RS test go through an MH test, where they can still be rejected.
This MH step removes the main limitation of rejection sampling approaches: requiring that $C \proposal_t(\param) \ge \target(\param)$ for some constant $C$ and all the possible values of $\param \in \parspace$.
This allows ARMS to generate samples from a wide variety of target densities, becoming virtually a universal sampler from a theoretical point of view.


\begin{alg}{Adaptive Rejection Metropolis Sampling (ARMS).}
{\begin{enumerate}
	\item \textbf{Initialization:} Set $t=0$ (chain's iteration) and $m=0$ (algorithm's iteration). Choose an initial state ($\parvec^{(0)}$), an initial number of support points ($K_0$),
		an initial support set $\mathcal{S}^{(0)} = \{\param_1^{(0)}, \ldots, \param_{K_0}^{(0)}\}$, the total number of iterations ($\niter$) and the burn-in period ($\nburn$).
    \item \texttt{WHILE} $t \le \niter$:
    	\begin{enumerate}
		\item Build a proposal function, $\proposal_m(\param)$, given a set of support points $\mathcal{S}^{(m)} = \{\param_1^{(m)}, \param_2^{(m)}, \ldots, \param_{K_m}^{(m)}\}$,
			according to Eq. \eqref{EqWtARMS}.
		\item Draw $\param' \sim \normproposal_m(\param)$ and $u \sim \uniform([0,1))$.
		\item If $u > \frac{\target(\param')}{\proposal_m(\param')}$, then reject  $\param'$, update $\mathcal{S}^{(m+1)}=\mathcal{S}^{(m)} \cup \{\param'\}$, $K_{m+1} = K_m+1$,
			and set $m=m+1$. Go back to step 2(a).
		\item Otherwise (i.e., if $u \le \target(\param')/\proposal_m(\param')$), draw $u' \sim \uniform([0,1))$ and compute the acceptance probability:
            	\begin{equation}
			\alpha_t \equiv \alpha(\parvec',\parvec^{(t-1)}) = \min\left[1,\frac{\target(\parvec')\min\{\target(\parvec^{(t-1)}),\proposal_m(\parvec^{(t-1)})\}}
				{\target(\parvec^{(t-1)})\min\{\target(\parvec'),\proposal_m(\parvec')\}}\right].
		\label{eq:alphaARMS}
                \end{equation}
		\item If $u' \le \alpha_t$, accept $\param'$ and set $\param^{(t)}=\param'$. Otherwise (i.e., if $u' > \alpha_t$), reject $\param'$ and set $\param^{(t)}=\param^{(t-1)}$.
		\item Set $\mathcal{S}^{(m+1)}=\mathcal{S}^{(m)}$, $K_{m+1}=K_m$, $m=m+1$, $t=t+1$ and return to step 2(a).
         \end{enumerate}
    \item Approximate the integral in Eq. \eqref{eq:intBayes} using Eq. \eqref{EstimatorDavid}.
   \end{enumerate}
}
\label{alg:arms}
\end{alg}

The mechanism used to construct the proposal is critical for the good performance of ARMS \cite{Gilks95}.
Let us consider the set of  support points at the $m$-th iteration of the algorithm, $\mathcal{S}^{(m)} = \{\param_1^{(m)}, \param_2^{(m)}, \ldots, \param_{K_m}^{(m)}\}$, and define the intervals $\mathcal{I}_0^{(m)}=(-\infty,\param_1^{(m)}]$, $\mathcal{I}_j^{(m)}=(\param_j^{(m)},\param_{j+1}^{(m)}]$ for $j=1,...,K_m-1$, and $\mathcal{I}_{K_m}^{(m)}=(\param_{K_m}^{(m)},+\infty)$.
Moreover, let us denote as $L_{j,j+1}^{(m)}(\param)$ the line passing through $(\param_{j}^{(m)},V(\param_{j}^{(m)}))$ and $(\param_{j+1}^{(m)},V(\param_{j+1}^{(m)}))$, with $V=\log(\target(\param))$ and $j=1,...,K_m-1$.
Then, a PWL potential function $W_m(x)$ is constructed in ARMS as
\begin{equation}
\label{EqWtARMS}
\small
W_m(\param) = \left\{ 
\begin{array}{ll}
 L_{1,2}^{(m)}(\param),  &  \param \in \mathcal{I}_0^{(m)},	\\
  \max\big\{L_{1,2}^{(m)}(\param), L_{2,3}^{(m)}(\param) \big\},  & \param \in \mathcal{I}_1^{(m)},	\\
  \varphi_j^{(m)}(\param), & \param \in \mathcal{I}_j,\\
  \max\big\{L_{K_m-2,K_m-1}^{(m)}(\param), L_{K_m-1,K_m}^{(m)}(\param)\big\}, & \param \in \mathcal{I}_{K_m-1}^{(m)},  \\ 
 L_{K_m-1,K_m}^{(m)}(\param), & \param \in \mathcal{I}_{K_m}^{(m)},	\\ 
\end{array}
\right. 
\end{equation}
where
\begin{equation*}
	\varphi_j^{(m)}(\param) = \max\left\{L_{j,j+1}^{(m)}(\param), \min\left\{L_{j-1,j}^{(m)}(\param),L_{j+1,j+2}^{(m)}(\param)\right\} \right\},
\end{equation*}
and $j=2,\ldots,K_m-1$.
Hence, the proposal PDF, $\proposal_m(\param) \propto \exp(W_m(\param))$, is formed by exponential pieces.
However, note that the number of linear pieces that form the proposal with this construction is larger than $K_m$ in general, since the proposal can be formed by two segments rather than one in some intervals.
Hence, the computation of intersection points among these two segments is also required to implement the algorithm.         
More sophisticated approaches to build $W_m(\param)$ (e.g., using quadratic segments when possible \cite{Meyer08}) have been proposed, but none of them solves the structural problem of ARMS that is briefly described next.

\vspace*{12pt}
\noindent
\emph{Independent Doubly Adaptive Rejection Metropolis Sampling (IA$^2$RMS)}
\vspace*{12pt}

Unfortunately, ARMS cannot always guarantee the convergence of the sequence of proposals to the target, since the proposal PDF is only updated when a sample $\param'$ is rejected by the RS test, something that can only happen when $\proposal_m(\param') > \target(\param')$.
When a sample is initially accepted by the RS test, as it always happens when $\proposal_m(\param') > \target(\param')$, the proposal is never updated.
Thus, the satisfactory performance of ARMS depends on two issues:
\begin{enumerate}
	\item $W_{m}(\param)$ should be constructed in such a way that $W_m(\param) \geq V(\param)$ almost everywhere (a.e.), i.e.,
		$\proposal_m(\param) \ge \target(\param)$ a.e., so that support points can be added a.e.
	\item The addition of a support point inside an interval must entail a change of the proposal PDF inside other neighbor intervals when building $W_{m+1}(\param)$.
		This allows the proposal to improve inside regions where $\proposal_m(\param) < \target(\param)$.
\end{enumerate}
These two conditions can be relaxed by allowing support points to be added inside regions where $\proposal_m(\param) < \target(\param)$ in a controlled way.
The independent doubly adaptive rejection Metropolis sampling (IA$^2$RMS) algorithm achieves this task by introducing a second control test that allows rejected samples from the MH stage to be incorporated to the support set with a certain non-null probability \cite{MartinoIA2RMS15,martino2014independent}.
The IA$^2$RMS algorithm is shown in Algorithm \ref{alg:ia2rms}.
Note that the only difference w.r.t. ARMS lies in step 2(f).
However, this crucial step allows samples to be added to the support set everywhere regardless of whether the proposal is below or above the target, and guarantees that $\proposal_m(\param) \to \target(\param)$ as $t \to \infty$.
Moreover, this allows IA$^2$RMS to effectively decouple the proposal construction from the algorithm's evolution, thus allowing the use of simpler proposals than the one used in ARMS (see \cite{MartinoIA2RMS15} for further details).

\newpage

\begin{alg}{Independent Doubly Adaptive Rejection Metropolis Sampling (IA$^2$RMS).}
{\begin{enumerate}
	\item \textbf{Initialization:} Set $t=0$ (chain's iteration) and $m=0$ (algorithm's iteration). Choose an initial state ($\parvec^{(0)}$), an initial number of support points ($K_0$),
		an initial support set $\mathcal{S}^{(0)} = \{\param_1^{(0)}, \ldots, \param_{K_0}^{(0)}\}$, the total number of iterations ($\niter$) and the burn-in period ($\nburn$).
    \item \texttt{WHILE} $t \le \niter$:
    	\begin{enumerate}
		\item Build a proposal function, $\proposal_m(\param)$, given a set of support points $\mathcal{S}^{(m)} = \{\param_1^{(m)}, \param_2^{(m)}, \ldots, \param_{K_m}^{(m)}\}$,
			using some simple non-parametric approach (see \cite{MartinoIA2RMS15} for several possibilities).
		\item Draw $\param' \sim \normproposal_m(\param)$ and $u \sim \uniform([0,1))$.
		\item If $u > \frac{\target(\param')}{\proposal_m(\param')}$, then reject  $\param'$, update $\mathcal{S}^{(m+1)}=\mathcal{S}^{(m)} \cup \{\param'\}$, $K_{m+1} = K_m+1$,
			and set $m=m+1$. Go back to step 2(a).
		\item Otherwise (i.e., if $u \le \target(\param')/\proposal_m(\param')$), draw $u' \sim \uniform([0,1))$ and compute the acceptance probability:
            	\begin{equation}
			\alpha_t \equiv \alpha(\parvec',\parvec^{(t-1)}) = \min\left[1,\frac{\target(\parvec')\min\{\target(\parvec^{(t-1)}),\proposal_m(\parvec^{(t-1)})\}}
				{\target(\parvec^{(t-1)})\min\{\target(\parvec'),\proposal_m(\parvec')\}}\right].
		\label{eq:alphaIA2RMS}
                \end{equation}
		\item If $u' \le \alpha_t$, accept $\param'$, setting $\param^{(t)}=\param'$ and $\vartheta = \param^{(t-1)}$. Otherwise (i.e., if $u' > \alpha_t$), reject $\param'$, setting
			$\param^{(t)}=\param^{(t-1)}$ and $\vartheta = \param'$.
		\item Draw $u'' \sim \uniform([0,1))$. If $u'' > \proposal_m(\vartheta)/\target(\vartheta)$, set $\mathcal{S}^{(m+1)}=\mathcal{S}^{(m)} \cup \{\vartheta\}$ and $K_{m+1}=K_m+1$.
			Otherwise (i.e., if $u'' \le \proposal_m(\vartheta)/\target(\vartheta)$), set $\mathcal{S}^{(m+1)}=\mathcal{S}^{(m)}$ and $K_{m+1}=K_m$.
		\item Set $m=m+1$, $t=t+1$ and return to step 2(a).
         \end{enumerate}
    \item Approximate the integral in Eq. \eqref{eq:intBayes} using Eq. \eqref{EstimatorDavid}.
   \end{enumerate}
}
\label{alg:ia2rms}
\end{alg}

Finally, let us remark that the mechanism used to accept/reject samples and to build the support set can be generalized further, as shown in \cite{Sticky13}, where the adaptive procedure is also extended to the framework of multiple try Metropolis (MTM) algorithms (see Section \ref{sec:mtm}).

\subsubsection{Convergence of adaptive MCMC algorithms}

Note that, since the adaptation can depend on all past samples in adaptive MCMC algorithms, the {Markovian} nature of classical MCMC methods is lost.
Therefore, ensuring the convergence of these techniques is much more complicated, as it cannot rely on standard tools and usually it has to be analyzed on a case by case basis depending on the method.
Furthermore, note that adaptive MCMC methods do not always converge, even if the adaptation itself converges.
A cautionary example about this issue is provided by Roberts and Rosenthal in \cite{roberts2007coupling}.
For these reasons, it is advisable either to follow a finite adaptation policy (i.e., adapting only during a finite number of initial iterations) or to adapt increasingly less often.\footnote{In some methods, the adaptation rate is automatically controlled by the algorithm. For instance, this happens in the parametric methods (ARMS, IA$^2$RMS and its variants) described in Section \ref{sec:nonParametricAMCMC}: as new samples are incorporated to the support set the probability of adding new samples to this set decreases and so the adaptation is performed more and more rarely.}
Indeed, Chimisov et al. \cite{chimisov2018air} have recently analyzed this second scenario, showing that a Central Limit Theorem can be proven for such chains.

\subsection{Gradient-based techniques}
\label{sec:gradMcmc}



\vspace*{12pt}

In this section, we consider MCMC methods which use the gradient of the log-posterior, $\nabla \log \target(\parvec)$, to enhance the efficiency of the sampling procedure.
The intuition is that, by using the gradient, we can form proposal distributions that allow for longer jumps without pushing the acceptance ratio of the method too low. 

\subsubsection{Metropolis adjusted Langevin algorithm}
\label{sec:mala}

\vspace*{12pt}

{\em The Metropolis adjusted Langevin algorithm (MALA)} is an MH algorithm \cite{Roberts:2002,Girolami+Calderhead:2011} which uses a stochastic differential equation (SDE) to form the proposal distribution.
Let us consider the following Langevin diffusion type of SDE:
\begin{equation}
  d\parvec(\tau) = \mathbf{f}(\parvec(\tau)) \, d\tau + d\mathbf{b}(\tau),
\end{equation}
where $\mathbf{b}(\tau)$ is a $\dpar$-dimensional Brownian motion. The Fokker--Planck equation giving the probability density $p(\parvec,\tau)$ of the diffusion state is
\begin{equation}
  \frac{\partial p(\parvec,\tau)}{\partial \tau}
  = -\nabla \cdot [ \mathbf{f}(\parvec) \, p(\parvec,\tau)]
  + \frac{1}{2} \nabla^2 p(\parvec,\tau).
\end{equation}
If we now select the drift of the diffusion as
\begin{equation}
  \mathbf{f}(\parvec) = \frac{1}{2} \nabla \log \target(\parvec),
\end{equation}
then the stationary solution $\partial p(\parvec,\tau) / \partial \tau = 0$ is given by
\begin{equation}
  p(\parvec,\tau) = \normtarget(\parvec).
\end{equation}
By starting from a value $\parvec^{(0)} \sim \normtarget(\parvec)$ and solving the SDE for $\tau > 0$ we can ``generate" more samples from $\normtarget(\parvec)$, because the marginal distribution of the SDE solution $\parvec^{(t)}$ is $\normtarget(\parvec)$ for all $t \ge 0$.

MALA uses an SDE of the kind described above as the proposal distribution in an MH algorithm.
Unfortunately, we cannot solve or simulate the SDE exactly.
Hence, we typically approximate its solution using the Euler--Maruama method \cite{Kloeden+Platen:1999}:
\begin{equation}
  \parvec^{(\tau_{n+1})} \approx \parvec^{(\tau_{n})}
  + \frac{\Delta \tau}{2} \nabla \log \target(\parvec^{(\tau_{n})})
  + \sqrt{\Delta \tau} \, \mathbf{z}_n,
\end{equation}
where $\mathbf{z}_n \sim \gauss(0,\mathbf{I})$ and $\Delta \tau = \tau_{n+1} - \tau_{n}$.
Nevertheless, let us remark that it would also be possible to use other numerical solution methods for SDEs in MALA as well \cite{Kloeden+Platen:1999}.
Algorithm \ref{alg:mala} summarizes the resulting MALA algorithm with one step of the Euler--Maruyama method for the numerical integration.

\newpage

\begin{alg}{Metropolis adjusted Langevin algorithm (MALA).}
{\begin{enumerate}
	\item \textbf{Initialization:} Choose an initial state $\parvec^{(0)}$,
	the discretization step $\Delta \tau$, 
    	the total number of iterations ($\niter$), and the burn-in period ($\nburn$).
    \item \texttt{FOR} $t = 1, \ldots, \niter$:
    	\begin{enumerate}
        	\item Draw $\mathbf{z} \sim \gauss(0,\mathbf{I})$, $u \sim \uniform([0,1))$
	and simulate a new sample from the Langevin diffusion:
\begin{equation}
  \parvec' = \parvec^{(t-1)}
  + \frac{\Delta \tau}{2} \nabla \log \target(\parvec^{(t-1)})
  + \sqrt{\Delta \tau} \, \mathbf{z},
\end{equation}
            \item Compute the acceptance probability ($\alpha_t$):
            	\begin{equation}
                	\alpha(\parvec',\parvec^{(t-1)})
                    	= \min\left[1,
	\frac{\target(\parvec')
	\gauss(\parvec^{(t-1)}|\parvec' + \frac{\Delta \tau}{2} \nabla \log \target(\parvec'),\Delta \tau \mathbf{I})}	
                {\target(\parvec^{(t-1)})
	\gauss(\parvec'|\parvec^{(t-1)} + \frac{\Delta \tau}{2} \nabla \log \target(\parvec^{(t-1)}),\Delta \tau \mathbf{I})}
		\right]
                \end{equation}
            \item If $u \le \alpha_t$, accept $\parvec'$ and set $\parvec^{(t)} = \parvec'$. Otherwise (i.e., if
            	$u > \alpha_t$), reject $\parvec'$ and set $\parvec^{(t)} = \parvec^{(t-1)}$.
         \end{enumerate}
    \item Approximate the integral in Eq. \eqref{eq:intBayes} using Eq. \eqref{EstimatorDavid}.
   \end{enumerate}
}
\label{alg:mala}
\end{alg}

It would be tempting to use more than one step of the Euler--Maruyama (or other SDE simulation method) in order to improve the proposal distribution.
Unfortunately this is not possible, because with multiple steps the evaluation of the transition density becomes intractable.
This limits the proposal to very local moves, and hence the algorithm only provides a small improvement w.r.t. the random walk MH algorithm.
A related algorithm which allows for larger jumps (and thus a greater improvement) is the Hamiltonian Monte Carlo (HMC) algorithm which we discuss next.

However, before we describe the HMC algorithm, let us make a short remark regarding the scaling of MALA.
The seminar work of Roberts and Rosenthal \cite{roberts1998optimal} concluded that the optimal asymptotic AR (i.e., as $\dpar \to \infty$) for MALA is approximately $0.574$.
Furthermore, they showed that the proposal variance should scale as $\dpar^{-1/3}$, and thus $\mathcal{O}(\dpar^{1/3})$ are required to ensure the algorithm's convergence.
Pillai \emph{et al.} also studied the efficiency of MALA on a natural class of target measures supported on an infinite dimensional Hilbert space, confirming that the optimal scaling AR is $0.574$ \cite{pillai2012optimal}.
Intuitively, this increased optimal AR w.r.t. the random walk MH algorithm (whose optimal AR is around $0.234$) is due to the incorporation of additional information about the target into the sampling through the use of the SDE.

\subsubsection{Hamiltonian Monte Carlo}
\label{sec:hmc}

\vspace*{12pt}

The {\em Hamiltonian Monte Carlo (HMC)} or the {\em hybrid Monte Carlo (HMC)} method \cite{Duane+Kennedy+Pendleton+Roweth:1987,Brooks_et_al:2011}, uses a statistical physical simulation of a physical system to form the proposal distribution for the MH algorithm.
It is based on considering a particle system with the following Hamiltonian:
\begin{equation}
  H(\parvec,\moment) = -\log \target(\parvec) + \frac{1}{2} \moment^T \moment,
\end{equation}
where $\parvec$ can be interpreted as the generalized coordinate and $\moment$ is the corresponding momentum.
Assuming a suitable temperature, the distribution of the particles is then given by
\begin{equation}
  p(\parvec,\moment) = \frac{1}{\partition} \exp(-H(\parvec,\moment)) = \normtarget(\parvec) \gauss(\moment | \vec{0},\mathbf{I}),
\end{equation}
which has the target density, $\normtarget(\parvec)$, as its marginal PDF.

The Hamiltonian equations for the dynamics of the particles in fictious time $\tau$ are now given by
\begin{subequations}
\begin{align}  
  \frac{d\parvec}{d\tau} &= \nabla_{\moment} H = \moment, \label{eq:hameq1} \\
   \frac{d\moment}{d\tau} &= -\nabla_{\parvec} H = \nabla \log \target(\parvec). \label{eq:hameq2}
\end{align}
\end{subequations}
The HMC algorithm constructs the proposal distribution by simulating trajectories from the Hamiltonian equations.
Because an exact simulation is not possible, we need to use again numerical methods to simulate the trajectories.
In order to construct a valid MH algorithm, a symplectic integrator such as the Leapfrog method (e.g., see \cite{Neal:2011}) needs to be used.
Then we can ensure both the preservation of the volume element as well as the time-reversibility of the simulation, which enables us to correct for the numerical solution inaccuracy by using a single MH acceptance step.

One step of the Leapfrog method for the Hamiltonian equations starting from $\tau$ with step size $\Delta \tau$ is given as
\begin{subequations}
\begin{align}  
  \widetilde{\moment}^{(\tau + \Delta \tau/2)} &=
  \widetilde{\moment}^{(\tau)}
  + \frac{\Delta \tau}{2} \, \nabla \log \target(\widetilde{\parvec}^{(\tau)}), \\
  \widetilde{\parvec}^{(\tau + \Delta \tau)} &=
  \widetilde{\parvec}^{(\tau)} + \Delta \tau \, \widetilde{\moment}^{(\tau + \Delta \tau/2)}, \\
  \widetilde{\moment}^{(\tau + \Delta \tau)} &=
  \widetilde{\moment}^{(\tau + \Delta \tau/2)}
  + \frac{\Delta \tau}{2} \, \nabla \log \target(\widetilde{\parvec}^{(\tau + \Delta \tau))}).
\end{align}
\label{eq:leapfrog}
\end{subequations}

The resulting HMC method is shown in Algorithm \ref{alg:hmc}.

\newpage

\begin{alg}{Hamiltonian Monte Carlo (HMC) Algorithm.}
{\begin{enumerate}
	\item \textbf{Initialization:} Choose an initial state $\parvec^{(0)}$,
	the discretization step $\Delta \tau$, 
	the number of integration steps $L$, 
    	the total number of iterations ($\niter$), and the burn-in period ($\nburn$).
    \item \texttt{FOR} $t = 1, \ldots, \niter$:
    	\begin{enumerate}
        	\item Draw $u \sim \uniform([0,1))$, numerically solve the Hamiltonian equations, \eqref{eq:hameq1} and \eqref{eq:hameq2}, using $L$ steps of a Leapfrog method \eqref{eq:leapfrog} starting from
		$\widetilde{\parvec}^{(0)} = \parvec^{(t-1)}$ and $\widetilde{\moment}^{(0)} \sim \gauss(0,\mathbf{I})$, setting $\parvec' = \widetilde{\parvec}^{(L \Delta \tau)}$ and
		$\moment' = -\widetilde{\moment}^{(L \Delta \tau)}$.
            \item Compute the acceptance probability:
            	\begin{align}
                	\alpha_t & = \alpha(\parvec',\moment';\parvec^{(t-1)},\moment_{t-1}) \nonumber \\
                    	& = \min\left[1, \exp\left( - H(\parvec',\moment') + H(\parvec^{(t-1)},\moment_{t-1}) \right)
		\right]
                \end{align}
            \item If $u \le \alpha_t$, accept $\parvec'$ and set $\parvec^{(t)} = \parvec'$. Otherwise (i.e., if
            	$u > \alpha_t$), reject $\parvec'$ and set $\parvec^{(t)} = \parvec^{(t-1)}$.
         \end{enumerate}
    \item Approximate the integral in Eq. \eqref{eq:intBayes} using Eq. \eqref{EstimatorDavid}.
    \end{enumerate}
    }
\label{alg:hmc}
\end{alg}

As discussed, for example, in \cite{Girolami+Calderhead:2011}, a single-step of the HMC algorithm is equivalent to MALA, and hence the two methods are closely related.
There are numerous improvements and modifications to this basic algorithm -- for example, we do not need to randomize the momenta fully at each time step, we can use preconditioning to improve the numerical stability {and the mixing rate}, and we can adapt the step sizes as well as the number of steps.

Finally, the optimal scaling of the HMC algorithm has been analyzed by Beskos \emph{et al.}
In \cite{beskos2013optimal}, they prove that it requires $\mathcal{O}(d^{1/4})$ steps to traverse the state space and that the asymptotically optimal AR for the HMC is $0.651$, even higher than the optimal AR for MALA ($0.574$).
This shows that HMC is more efficient than MALA in the incorporation of information about the target into the sampling approach.

\subsubsection{Riemann manifold MALA and HMC}
\label{sec:riemannMalaHmc}

\vspace*{12pt}

A practical challenge in both MALA and HMC methods is that their performance heavily depends on the particular parameterization chosen for $\parvec$.
Although this problem can be diminished by using a preconditioning matrix to compensate for the uneven scaling of the parameters, its manual selection can be hard. Fortunately, a more general automatic selection procedure is provided by the Riemann manifold Monte Carlo methods that we briefly discuss here.

The idea of the \emph{Riemann manifold Langevian and Hamiltonian Monte Carlo} methods is to perform the Langevin or Hamiltonian simulations in a suitable Riemann manifold instead of the Euclidean space \cite{Girolami+Calderhead:2011,Zlochin:2001}.
Although the idea was already proposed by \cite{Zlochin:2001}, the introduction of proper symplectic integrators by \cite{Girolami+Calderhead:2011} led to an exact MCMC algorithm.
Let us recall that the squared distance between two locations $\parvec$ and $\parvec + d\parvec$ in Euclidean space is given by $d^2 = d\parvec^T d\parvec$.
In the Riemann manifold the distance is generalized to be $d^2 = d\parvec^T \mathbf{G}^{-1}(\parvec) d\parvec$, where $\mathbf{G}(\parvec)$ is the metric tensor, which is a positive definite matrix for any given $\parvec$.
A particularly useful metric tensor in the context of probability distributions is the one arising from information geometry, which is given as
\begin{equation}
  \mathbf{G}(\parvec) = \E_{\datavec|\parvec}[\nabla \log \target(\parvec|\datavec) \nabla \log \target(\parvec|\datavec)^T].
\end{equation}
We can now modify the MALA method such that the SDE evolves along the Riemann manifold instead of the Euclidean space as follows:
\begin{equation}
  d\parvec(\tau) = \tilde{\mathbf{f}}(\parvec(t)) d\tau + d\tilde{\mathbf{b}}(\tau),
\end{equation}
where
\begin{subequations}
\begin{align}
  \tilde{\mathbf{f}}(\parvec) &= \frac{1}{2} \mathbf{G}^{-1}(\parvec) \nabla \log \target(\parvec), \label{eq:mmala1} \\
  d\tilde{b}_i &= | \mathbf{G}(\parvec) |^{-1/2} \sum_j \frac{\partial}{\partial \param_j}
    [ \mathbf{G}^{-1}(\parvec) ]_{ij} | \mathbf{G}(\parvec) |^{-1/2} dt
    + [ \mathbf{G}^{-1/2}(\parvec) d\mathbf{b}]_i. \label{eq:mmala2}
\end{align}
\end{subequations}
The {\em Riemann manifold Langevian Monte Carlo (MMALA)} algorithm can now be constructed by replacing the SDE in the basic MALA (see Algorithm \ref{alg:mala}) with the SDE defined above in Eqs. \eqref{eq:mmala1} and \eqref{eq:mmala2}.
For further details, the reader is referred to \cite{Girolami+Calderhead:2011}.

In the {\em Riemann manifold Hamiltonian Monte Carlo (RMHMC)} we construct the particle system dynamics in the Riemann manifold.
This results in the following Hamiltonian:
\begin{equation}
  H(\parvec,\moment) = -\log \target(\parvec)
  + \frac{1}{2} \log | 2 \pi \mathbf{G}(\parvec) |
  + \frac{1}{2} \moment^T \mathbf{G}^{-1}(\parvec) \moment,
\end{equation}
and the Hamiltonian equations are now given as
\begin{subequations}
\begin{align}  
  \frac{d\parvec}{d\tau} &= \nabla_{\moment} H = \mathbf{G}^{-1}(\parvec) \moment, \label{eq:hameq3} \\
   \frac{d\moment}{d\tau} &= -\nabla_{\parvec} H = \nabla \log \target(\parvec) + \mathbf{h}(\parvec), \label{eq:hameq4}
\end{align}
\end{subequations}
where the additional term in Eq. \eqref{eq:hameq4} is given by
\begin{equation}
\begin{split}  
  h_i(\parvec) = -\frac{1}{2} \mathrm{tr}
    \left\{ \mathbf{G}^{-1}(\parvec) \frac{\partial \mathbf{G}(\parvec)}{\partial \param_i} \right\}
  + \frac{1}{2} \moment^T \mathbf{G}^{-1}(\parvec) \frac{\partial \mathbf{G}(\parvec)}{\partial \param_i}
  \mathbf{G}^{-1}(\parvec)
   \moment.
\end{split}  
\end{equation}
Although the construction of the RMHMC is analogous to that of the HMC, the selection of integration method requires much more care.
The simple Leapfrog method is no longer enough now, and we need to use a more general symplectic integrator.
For further details on this issue, see again \cite{Girolami+Calderhead:2011}.

\subsubsection{Step-size and trajectory-length adaptation methods}
\label{sec:stepSize}

\vspace*{12pt}

Selecting the step size of the Leapfrog method is important for the performance of both MALA and HMC based methods.
As discussed in \cite{Neal:2011}, fortunately, the step size selection does not have a huge impact on the error in the Hamiltonian provided that it is small enough to make the discrete dynamics stable.
For analysis on practical selection of step sizes as well as the lengths of trajectories see \cite{Neal:2011}.
In \cite{Hoffman+Gelman:2014} the authors propose No-U-Turn Sampler (NUTS) method which approaches the problem by limiting the trajectory to a length where it would change the direction.
In NUTS, the step length is adapted using a stochastic optimization method.
Some step size estimation methods for Leapfrog and HMC have also been provided in \cite{Chen_et_al:2000,Holder:2001}.
Finally, the optimal step size scaling with the number of dimensions as well as the optimal acceptance rate of HMC has also been analyzed (e.g., see \cite{Beskos_et_al:2013,Betancourt_et_al:2014}).

\subsubsection{The geometric foundations of HMC and further considerations}
\label{sec:geometricHmc}

\vspace*{12pt}

Before concluding this section, let us remark that gradient-based algorithms have obtained a wide success in many applications.
For instance, in high-dimensional problems, where the probability mass is typically concentrated in a very small portion of the space, they should probably be the first choice for practitioners.
Recently, several researchers have developed a rigorous understanding of the reasons for their good performance on difficult problems, as well as suggestions about their practical application.
Intuitively, in high dimensional applications, exploiting the gradient information is crucial to make large jumps away from the current state and, at the same time, ensure that they are highly likely to be accepted.
In \cite{Betancourt_et_al:2014b}, the authors have suggested that, in order to build a good proposal mechanism in an MH-type method, it is required to define a family of transformations preserving the target measure which form a Lie group on composition, as this ensures that proposals generated are both far away from the previous point and highly likely to be accepted.
They also show that HMC emerges naturally when attempting to construct these transformations using ideas from differential geometry \cite{Betancourt_et_al:2014b,Betancourt_et_al:2017}.
However, it is also important to emphasize that gradient-based methods suffer from some important problems, {as noted by Nishimura and Dunson \cite{Nishimura16}: one is the efficient tuning and/or adaptation of the parameters of the algorithm (which is not an easy task, in general), and another one is their application in multimodal scenarios; see \cite{Nishimura16,LAN14} and the discussion in \cite{Girolami+Calderhead:2011} for further information about this issue.}

More generally, note that \emph{deterministic procedures} have been included within the sampling algorithms in order to reduce the computational demand of the MC methods and the variance of the resulting estimators.
In this section we have explored one widely used possibility: exploiting the gradient information.
Another idea, employed in Quasi Monte Carlo methods, is using deterministic sequences of points based on the concept of {\it low-discrepancy} \cite{Niederreiter92}.

\subsection{Other advanced schemes}
\label{sec:other_advanced_mcmc}

\vspace*{12pt}

In this section, we briefly discuss other important topics and relevant methodologies which are related to the algorithms described in the rest of this work. 

\subsubsection{Parallel schemes and implementations}

\vspace*{12pt}

A long run of a single chain can remain trapped in a local mode (e.g., when the parameters of the proposal PDF are not well-tuned) or the convergence can be very slow.
In some cases (e.g., when the posterior density is multimodal), the use of shorter parallel MCMC chains can be advantageous.
Thus, in order to speed up the exploration of the state space, and especially in order to deal with high-dimensional applications, several schemes employing  parallel chains have been proposed \cite{Craiu09,cascralei:2012,OMCMC}, as well as multiple try and interacting schemes (see Section \ref{sec:mtm}).
Several other population-based techniques can also be found in the literature \cite{altekar2004parallel,Geyer91,Smelly,APIM,Geyer95,Jasra07}.
Finally, the use of non-reversible parallel MH algorithms has been also proposed \cite{Corander06,Corander08}.

In addition, the interest in the parallel computation can also be due to other motivations.
For instance, several authors have studied the parallelization of MCMC algorithms, that have traditionally been implemented in an iterative non-parallel fashion, in order to reduce their computation time \cite{Calderhead14}.
Furthermore, parallel MCMC schemes are required in \emph{big data} problems, where one possible approach consists of splitting the complete posterior distribution into several partial sub-posteriors \cite{bardenet2015markov,EmbaraMCMC,Wang2014,Wang2015}.
Moreover, in the literature there is a great interest in the parallel implementation of MCMC algorithms so that the computation is distributed across a bunch of parallel processors \cite{brockwell2006parallel,huelsenbeck2001mrbayes,rosenthal2000parallel,strid2010efficient}.

Finally, note that the optimal parallelization strategy for an MCMC algorithm may differ depending on the properties of the parallel system under consideration.
For instance, on GPUs parallelizing the individual MCMC steps can yield large performance improvement \cite{terenin2019gpu}.
On the other hand, on distributed systems a better approach can consist in parallelizing the MCMC algorithm itself \cite{terenin2018asynchronous}.

\subsubsection{Delayed Rejection Metropolis Sampling}
\label{DRMS}
\vspace*{12pt}

In Section \ref{sec:advMcmc}, we will describe several multiple try schemes which extend the classical MH method, using (and comparing) different candidates at each iteration. 
Here, we show an alternative use of different candidates in one iteration of an MH-type method \cite{Tierney1999}. 
The idea behind the so called {\it Delayed Rejection Metropolis (DRM)} algorithm is the following.
In the standard MH algorithm, one sample is proposed at each iteration, $\widetilde{{\bm \theta}}^{(1)} \sim \proposal_1({\bm \theta}|{\bm \theta}^{(t-1)})$,\footnote{Note that the $\widetilde{{\bm \theta}}^{(1)}$ and $\proposal_1({\bm \theta}|{\bm \theta}^{(t-1)})$ used here play the role of ${\bm \theta}'$ and $\proposal({\bm \theta}|{\bm \theta}^{(t-1)})$, respectively, in Algorithm \ref{alg:metropolis}.}
and accepted with probability
\begin{equation}
	\alpha_1(\widetilde{{\bm \theta}}^{(1)},{\bm \theta}^{(t-1)})
		= \min\left[1,\frac{\target(\widetilde{{\bm \theta}}^{(1)})\proposal_1({\bm \theta}^{(t-1)}|\widetilde{{\bm \theta}}^{(1)})}
			{\target(\widetilde{{\bm \theta}}^{(t-1)}) \proposal_1(\widetilde{{\bm \theta}}^{(1)}|{\bm \theta}^{(t-1)})}\right].
\end{equation}
If $\widetilde{{\bm \theta}}^{(1)}$ is accepted, then ${\bm \theta}^{(t)} = \widetilde{{\bm \theta}}^{(1)}$ and the chain is moved forward, as in the standard MH algorithm.
However, if $\widetilde{{\bm \theta}}^{(1)}$ is rejected, the DRM method suggests drawing another sample, $\widetilde{{\bm \theta}}^{(2)}\sim \proposal_2({\bm \theta}|{\bm \theta}^{(t-1)}, \widetilde{{\bm \theta}}^{(1)})$, from a different proposal PDF, $\proposal_2$, which takes into account the previous candidate, $\widetilde{{\bm \theta}}^{(1)}$, and accepting it with a suitable acceptance probability:
\begin{equation}
	\alpha_2^{(t)} \equiv \alpha_2(\widetilde{{\bm \theta}}^{(2)},{\bm \theta}^{(t-1)}) = \min\left[1,\frac{\rho(\widetilde{{\bm \theta}}^{(2)},{\bm \theta}^{(t-1)}|\widetilde{{\bm \theta}}^{(1)})}
		{\rho({\bm \theta}^{(t-1)},\widetilde{{\bm \theta}}^{(2)}|\widetilde{{\bm \theta}}^{(1)}) }\right],
\label{eq:alphaDRM}
\end{equation}
where
\begin{align}
	& \rho({\bm \theta}^{(t-1)},\widetilde{{\bm \theta}}^{(2)}|\widetilde{{\bm \theta}}^{(1)}) \nonumber \\
	& \quad = \target({\bm \theta}^{(t-1)}) \proposal_1(\widetilde{{\bm \theta}}^{(1)}|{\bm \theta}^{(t-1)})(1-\alpha_1(\widetilde{{\bm \theta}}^{(1)},{\bm \theta}^{(t-1)}))
		\target(\widetilde{{\bm \theta}}^{(2)}) \proposal_2(\widetilde{{\bm \theta}}^{(2)}|{\bm \theta}^{(t-1)},\widetilde{{\bm \theta}}^{(1)}).
\label{eq:rhoDRM}
\end{align}
The acceptance function of Eq. \eqref{eq:alphaDRM}, $\alpha_2(\widetilde{{\bm \theta}}^{(2)},{\bm \theta}^{(t-1)})$, is designed in order to ensure the ergodicity of the chain.
Indeed, the kernel of the DRM algorithm satisfies the detailed balance condition \cite{Tierney1999}.
If $\widetilde{{\bm \theta}}^{(2)}$ is rejected, then we can either set ${\bm \theta}^{(t)} = {\bm \theta}^{(t-1)}$ and perform another iteration of the algorithm, or continue with the sequential strategy, drawing $\widetilde{{\bm \theta}}^{(3)} \sim \proposal_3({\bm \theta}|{\bm \theta}^{(t-1)},\widetilde{{\bm \theta}}^{(1)},\widetilde{{\bm \theta}}^{(2)})$ and testing it with a proper acceptance probability $\alpha_3(\widetilde{{\bm \theta}}^{(3)},{\bm \theta}^{(t-1)})$.
The DRM method with two sequential steps is outlined in Algorithm \ref{alg:DelayedRejection}.
Unlike the multiple try schemes described in Section \ref{sec:advMcmc}, in the DRM sampler the candidates are not compared together at the same time, but each candidate is drawn from a proposal PDF and then tested with an MH-type acceptance probability.

\begin{alg}{Delayed Rejection Metropolis with two acceptance steps.}
{
\begin{enumerate}
	\item \textbf{Initialization:} Choose two proposal functions, $\proposal_1({\bm \theta})$ and $\proposal_2({\bm \theta})$, an initial state ${\bm \theta}^{(0)}$, and the total number of iterations ($\niter$).
	\item \texttt{FOR} $t = 1, \ldots, \niter$:
    		\begin{enumerate}
        			\item Draw $\widetilde{{\bm \theta}}^{(1)} \sim \proposal_1({\bm \theta}|{\bm \theta}^{(t-1)})$ and $u_1\sim \mathcal{U}([0,1])$.
      			\item Compute the acceptance probability of $\widetilde{{\bm \theta}}^{(1)}$:
       				\begin{equation}
          				\alpha_1^{(t)} \equiv \alpha_1(\widetilde{{\bm \theta}}^{(1)},{\bm \theta}^{(t-1)}) = \min\left[1,\frac{\target(\widetilde{{\bm \theta}}^{(1)})
						\proposal_1({\bm \theta}^{(t-1)}|\widetilde{{\bm \theta}}^{(1)})} {\target({\bm \theta}^{(t-1)}) \proposal_1(\widetilde{{\bm \theta}}^{(1)}|{\bm \theta}^{(t-1)})}\right], 
                			\label{eq:alphaDelay1} 
      				\end{equation}
    			\item  If $u_1 \leq \alpha_1^{(t)}$, accept $\widetilde{{\bm \theta}}^{(1)}$, set ${\bm \theta}^{(t)} = \widetilde{{\bm \theta}}^{(1)}$ and go back to step 2(a).
    			\item Otherwise (i.e., if $ u_1 > \alpha_1^{(t)}$):
    				\begin{enumerate}
					\item Draw $\widetilde{{\bm \theta}}^{(2)} \sim \proposal_2({\bm \theta}|{\bm \theta}^{(t-1)},\widetilde{{\bm \theta}}^{(1)})$ and  $u_2 \sim \uniform([0,1))$.
					\item Compute the acceptance probability of $\widetilde{{\bm \theta}}^{(2)}$:
						\begin{equation*}
							\alpha_2^{(t)} \equiv \alpha_2(\widetilde{{\bm \theta}}^{(2)},{\bm \theta}^{(t-1)})
								= \min\left[1,\frac{\rho(\widetilde{{\bm \theta}}^{(2)},{\bm \theta}^{(t-1)}|\widetilde{{\bm \theta}}^{(1)})}
									{\rho({\bm \theta}^{(t-1)},\widetilde{{\bm \theta}}^{(2)}|\widetilde{{\bm \theta}}^{(1)}) }\right],
						\end{equation*}
						with $\rho({\bm \theta}^{(t-1)},\widetilde{{\bm \theta}}^{(2)}|\widetilde{{\bm \theta}}^{(1)})$ given by Eq. \eqref{eq:rhoDRM}.
					\item If $u_2 \leq \alpha_2^{(t)}$, accept $\widetilde{{\bm \theta}}^{(2)}$ and set ${\bm \theta}^{(t)} = \widetilde{{\bm \theta}}^{(2)}$.
						Otherwise (i.e., if $u_2 > \alpha_2^{(t)}$), reject $\widetilde{{\bm \theta}}^{(2)}$ and set ${\bm \theta}^{(t)} = {\bm \theta}^{(t-1)}$.
				\end{enumerate}
      		\end{enumerate}
	\item Approximate the integral in Eq. \eqref{eq:intBayes} using Eq. \eqref{EstimatorDavid}.
\end{enumerate}
}
\label{alg:DelayedRejection}
\end{alg}

\subsubsection{Non-reversible chains}

\vspace*{12pt}

The MCMC techniques described so far fulfill the so-called {\it detailed balance condition}. In this case, the generated chain is {\it reversible}. Denoting as $K({\bm \theta}'|{\bm \theta})$ the transition density of a specific MCMC method, the detailed balance condition is given by the Eq. \ref{DBC_eq} in Section \ref{sec:validity_mh}, i.e.,
\begin{equation}
{\bar \pi}({\bm \theta})K({\bm \theta}'|{\bm \theta})={\bar \pi}({\bm \theta}')K({\bm \theta}|{\bm \theta}').
\end{equation}
However, recent studies have shown that \emph{non-reversible} chains can provide better performance \cite{Bierkens2016,NealnonRev2,FERNANDES20111856,TURITSYN2011410,SCHRAM201588,Vucelja14}. Non-reversible chains can be generated using different procedures. For instance, defining a  {\it vorticity density} $\gamma({\bm \theta},{\bm \theta}')$, such that 
$$
\gamma({\bm \theta},{\bm \theta}')=-\gamma({\bm \theta}',{\bm \theta}),
$$
$$
\int_{{\bf A}\times {\bm \Theta}} \gamma({\bm \theta},{\bm \theta}') d{\bm \theta}'d{\bm \theta} =0, \qquad  {\bf A} \in \mathcal{B}({\bm \Theta}),
$$
where $\mathcal{B}({\bm \Theta})$ is a Borel $\sigma$-algebra generated by open sets in ${\bm \Theta}$ and, additionally,
$$
\gamma({\bm \theta},{\bm \theta}')+\pi({\bm \theta}')q({\bm \theta}|{\bm \theta}')\geq 0,
$$
for all ${\bm \theta},{\bm \theta}'\in {\bm \Theta}$ for which $\pi({\bm \theta}')q({\bm \theta}|{\bm \theta}') \neq 0$. Moreover, in the case that $\pi({\bm \theta}')q({\bm \theta}|{\bm \theta}') = 0$, we need $\gamma({\bm \theta},{\bm \theta}')=0$.
In this scenario, we can run an MH-type algorithm with acceptance probability 
$$
\alpha=\min\left[1,\frac{\gamma({\bm \theta},{\bm \theta}')+\pi({\bm \theta}')q({\bm \theta}|{\bm \theta}')}{\pi({\bm \theta})q({\bm \theta}'|{\bm \theta})}\right].
$$
It is possible to show that the generated chain by the MH method with the $\alpha$ above still has ${\bar \pi}$ as invariant density, although is non-reversible \cite{Bierkens2016,NealnonRev2}.

Many non-reversible MCMC methods have recently been proposed, and some authors have also developed general frameworks to construct different irreversible MC algorithms \cite{turitsyn2011irreversible,schram2015monte}.
However, the amount of improvement provided by these schemes in complex practical problems still remains to be seen.



\subsection{MCMC Convergence Diagnostics}
\label{sec:diag}

\vspace*{12pt}

Properly designed MCMC algorithms automatically produce samples from the target distribution after an initial transitory period.
However, theoretically determining the length of this transitory period may be very difficult.
Hence, during the finite running time of a simulation the Markov chain could fail to converge to its stationary distribution.
In this case, the  generated samples might not well represent the target PDF and any inference performed using them is bound to produce erroneous results.
For this reason it is important to study the output of the algorithm to determine if the MCMC simulation has properly converged. Methods and good practices for convergence diagnostics can be found, for example, in Chapter 11 of Gelman et al. (2013) \cite{Gelman_et_al:2013}, or Chapter 6 of Brooks et al. (2011) \cite{Brooks_et_al:2011}. 

\subsubsection{General principles of convergence diagnostics}

\vspace*{12pt}

Gelman and Shirley, in Chapter 6 of Brooks et al. (2011) \cite{Brooks_et_al:2011}, summarize the recommended convergence diagnostics as follows:
\begin{enumerate}
\item Run three or more chains in parallel with varying starting points. The starting points can be selected randomly around or within a simpler approximation.
\item The chains should be then split to two halves, where only the second half is retained. Diagnosis methods for between-chain and within-chain analysis can be then used to monitor the mixing of the chain in the second half.
\item After approximate convergence is obtained, the second halves of the chains should be then mixed together to summarize the target distribution. The autocorrelations should not matter at this stage any more.
\item Adaptive Markov chain Monte Carlo (AMCMC) methods can be used for tuning the proposal densities and other properties of the MCMC method. Provided that we always restart the MCMC method after adaptation, any adaptation method can be applied.
\end{enumerate}
Additionally, the algorithm can be debugged by running it on a model with known parameters and checking that the posterior distributions are consistent with the true values.

\subsubsection{Between-chain and within-chain diagnostics}

\vspace*{12pt}

Given a set of simulated MCMC samples -- such as the second halves of the chains from step 2 in the previous section -- it is then possible to investigate whether the samples have converged to the target distribution.
Although there are a number of possible approaches for the convergence diagnostics (see, e.g., \cite{brooks1998general,brooks2003nonparametric}), the potential scale reduction factor (PSRF) \cite{gelman1992inference,Gelman_et_al:2013} is the diagnostic tool that is often recommended for practical use \cite{Brooks_et_al:2011,Gelman_et_al:2013}.

Gelman et al. (2013) \cite{Gelman_et_al:2013} define the PSRF ($\hat{R}$) as follows.
Let us denote the chain consisting of samples from a scalar variable $\theta$ as $\theta^{(i,j)}$, where $i=1,\ldots,M$ is the sample index and $j=1,\ldots,S$ is the chain index.
Compute then $B$ and $W$, which correspond to the between-chain and within-chain variances:
\begin{subequations}
\begin{align}
  B &= \frac{M}{S -1} \sum_{j=1}^S (\bar{\theta}^{(\cdot,j)}-\bar{\theta}^{(\cdot,\cdot)})^2, \\
  W &= \frac{1}{S} \sum_{j=1}^S s_j^2 = \frac{1}{S (M-1)} \sum_{j=1}^S  \sum_{i=1}^M (\theta^{(i,j)} - \bar{\theta}^{(\cdot,j)})^2,
\end{align}
\end{subequations}
where
\begin{subequations}
\begin{align}
	\bar{\theta}^{(\cdot,j)} & = \frac{1}{M} \sum_{i=1}^M \theta^{(i,j)}, \\
	\bar{\theta}^{(\cdot,\cdot)} & = \frac{1}{S} \sum_{j=1}^S \bar{\theta}^{(\cdot,j)}, \\
	s_j^2 & = \frac{1}{M-1} \sum_{i=1}^M (\theta^{(i,j)} - \bar{\theta}^{(\cdot,j)})^2.
\end{align}
\end{subequations}
The PSRF can then be defined as \cite{Gelman_et_al:2013}
\begin{equation}
  \hat{R} = \sqrt{  \frac{\widehat{\mathrm{var}}^+(\theta)}{W} },
\end{equation}
where
\begin{equation}
  \widehat{\mathrm{var}}^+(\theta)
  = \frac{M-1}{M} \, W + \frac{1}{M} \, B
\end{equation}
is an estimator for the posterior variance.
In the multivariate case the PSRF values can be computed for each dimension separately.

The PSRF value should approach 1 when the convergence occurs.
If the value is significantly higher than 1, then convergence has probably not occurred.
Although in PSRF we should use multiple independent chains, a single chain can also be analyzed by splitting it into two or more parts and computing the PSRF as if the splits where independent chains. 

\subsubsection{Effective number of samples}

\vspace*{12pt}

Generally, the MCMC algorithms present a positive correlation among the generated samples.
This clearly implies a loss of efficiency with respect to the case of independent samples, i.e., with respect to the classical Monte Carlo approach.
Namely, positively correlated samples provide less statistical information than independent samples, meaning that the corresponding estimators will be less efficient in general, i.e., with a higher variance for a given sample size.
The concept of Effective Sample Size (ESS) has been introduced to measure  this loss of efficiency \cite{gamerman2006markov}, \cite[Section 9.1]{Sticky13}.
If $T$ is the length of the chain (without removing any burn-in period) and denoting as $\rho(\tau)$ the autocorrelation at lag $\tau$, the effective sample size for an MCMC algorithm is defined as
\begin{equation}
	\textrm{ESS} = \frac{T}{1+2\sum_{\tau=1}^{\infty} \rho(\tau)}.
\end{equation}
Clearly, positive correlations $\rho(\tau)$ decrease the ESS value, and hence $ESS<T$.
If the correlation is zero (i.e., $\rho(\tau)=0$ for all $\tau$), then $ESS=T$, as in the classical Monte Carlo scheme using independent samples.
In a similar fashion, it is also possible to define ESS measures for other Monte Carlo approaches, as shown in Section  \ref{SectionESS_IS}.

\subsubsection{Other Recent Approaches}

\vspace*{12pt}

More recently, alternative approaches to the method described in the previous sections have been proposed to measure the convergence of MCMC algorithms.
On the one hand, Gorham \emph{et al.} have introduced a novel family of discrepancy measures based on Stein's method \cite{gorham2015measuring}.
These measures bound the discrepancy between sample and target expectations over a large class of test functions, and some specific members of this family can be computed by solving a linear program \cite{gorham2018measuring}.
Finally, using zero mean reproducing kernel theory, several authors have shown that other members of the Stein discrepancy family have a closed-form solution involving the sum of kernel evaluations over pairs of sample points \cite{chwialkowski2016kernel,liu2016kernelized,oates2017control,gorham2017measuring}.
On the other hand, Johndrow \emph{et al.} have applied computational complexity theory to analyze how the computational efficiency of MCMC algorithms degrades with the problem's size \cite{johndrow2018mcmc}.
Their goal is determining whether an MCMC algorithm will perform well in a given context/problem, rather than providing performance bounds which are often very loose to be of any practical use.
Note that the aforementioned techniques are still not mature enough for their widespread use in practical applications (especially in high-dimensional problems), but this is a very active research field where we can expect new contributions in coming years.

\vspace*{12pt}

\section{Importance Sampling}
\label{sec:is}

\vspace*{12pt}

Importance sampling (IS) is a Monte Carlo methodology that can be used to characterize virtually any target PDF and to approximate its moments.
Like any other MC method, IS techniques proceed by drawing samples from one or several proposal PDFs.
However, unlike MCMC methods, that accept or discard the drawn samples according to some appropriate test, IS approaches accept all the samples and assign them a weight according to their quality in approximating the desired target distribution.
Regarding the number of proposals, IS methods can be divided into classical or standard IS, where a single proposal is used to draw all the samples, and multiple IS (MIS), where a collection of proposals are used to draw the samples.
With respect to the temporal evolution of the proposals, IS methods can be classified as non-adaptive or ``static", where the proposals are fixed (i.e., their parameters are selected a priori and remain fixed for the whole simulation), and adaptive, where the parameters of the proposals are adapted iteratively in order to better approximate the desired target density.
In the following section we briefly review ``static'' IS (both using a single and multiple proposals), whereas in Section \ref{sec:ais} we consider adaptive IS (AIS).
Then, some remarks on the convergence of IS and AIS are provided in Section \ref{sec:convIS}.
This is followed by a discussion on the optimal proposal choice and the variance of the IS estimators in Section \ref{sec:varianceIS}, and the definition of the effective sample size in Section \ref{SectionESS_IS}.
Afterwards, the concept of proper weights is introduced in Section \ref{LiuSect} and this leads naturally to the group importance sampling (GIS) approach described in Section \ref{sec:gis}.
Finally, a short introduction to sequential importance sampling (SIS) for the estimation of dynamic parameters is also provided in Section \ref{SIS}.

\subsection{Standard and Multiple Importance Sampling}
\label{sec:sisMis}

\vspace*{12pt}

\subsubsection{Importance sampling with a single proposal}

\vspace*{12pt}

Let us consider a single proposal PDF, $\normproposal(\parvec)$, with heavier tails than the target, $\normtarget(\parvec)$.{\footnote{{Note that one of the main disadvantages of IS is the fact that the variance of the IS estimator becomes infinite when the tails of the proposal, $\normproposal(\parvec)$, decay faster than $\normtarget(\parvec)^2$ due to the appearance of $\normproposal(\parvec)$ in the denominator of the weights in \eqref{is_weights_static} \cite{owen2000safe}. Therefore, in order prevent this situation, which leads to the complete failure of the IS sampler, a common restriction is selecting a proposal with heavier tails than the target.}}}
The proposal is used to draw a set of $\npart$ IID samples, $\{\parvec^{(m)}\}_{m=1}^{\npart}$ with $\parvec^{(m)} \sim \normproposal(\parvec)$.
An importance weight is then associated to each sample according to
\begin{equation} 
\weight_m=w(\parvec_m) = \frac{\target(\parvec_m)}{{\normproposal(\parvec_m)}}, \quad m=1,\ldots,\npart.
\label{is_weights_static}
\end{equation}
If the normalizing constant of the target, $\partition$, is known, then we can approximate the targeted integral of Eq. \eqref{eq:intBayes} by the so-called \emph{unnormalized estimator}:
\begin{equation}
	\widehat{I}_{\npart} = \frac{1}{\npart\partition}  \sum_{m=1}^{\npart} \weight_m g(\parvec^{(m)}).
\label{eq_unnorm_est}
\end{equation}
Unfortunately, $\partition$ is unknown in many practical problems, and we have to resort to the alternative \emph{self-normalized estimator}, which is given by
\begin{equation}
	\widetilde{I}_{\npart} = \frac{1}{\npart\widehat{\partition}}  \sum_{m=1}^{\npart} \weight_m g(\parvec^{(m)}),
\label{eq_norm_est}
\end{equation}
where $\widehat{\partition} = \frac{1}{\npart}\sum_{m=1}^{\npart} \weight_m$ is an unbiased estimator of $\partition$ \cite{Robert04}.
Note that both $\widetilde{I}_{\npart}$ and $\widehat{I}_{\npart}$ are consistent estimators of $I$, i.e., both $\widetilde{I}_{\npart} \to I$ and $\widehat{I}_{\npart} \to I$ as $\npart \to \infty$ (see Section \ref{sec:convIS} for further details).
However, their performance (as measured by their variance) is highly related to the discrepancy between $\normtarget(\parvec)|g(\parvec)|$ and the proposal $\normproposal(\parvec)$  \cite{Robert04}.
Indeed, with a properly selected proposal IS methods can provide a lower variance than the direct application of the MC method, whereas a poor selection can easily lead to infinite variance estimators.
In practice, since IS approaches are often used to simultaneously approximate several functions ($g_1(\parvec), g_2(\parvec), \ldots$), a common approach is simply minimizing the mismatch between the proposal $\normproposal$ and the target $\normtarget$ \cite[Section 3.2]{doucet2009tutorial}.

\subsubsection{Importance sampling with multiple proposals}

\vspace*{12pt}

Several reasons justify the use of more than one proposal.
On the one hand, the target can be multimodal and therefore it can be better fitted with a mixture of proposals.
On the other hand, choosing a single good proposal a priori is usually very difficult, and adaptive processes must be performed (as described in Section \ref{sec:ais}) in order to tailor the proposal to the target.
In this situation, the exploration of the parameter space, $\parspace$, is more efficient when multiple proposals, $\{q_n(\parvec)\}_{n=1}^{\nprop}$, are available \cite{Veach95,owen2000safe}.
The use of several proposals is usually known as \emph{multiple importance sampling (MIS)} in the IS literature, and it is a key feature of most state-of-the-art adaptive {IS} algorithms (e.g., \cite{Cappe04, CORNUET12, APIS15,martino2016layered,elvira2017improving}).
A generic MIS framework has been recently proposed in \cite{elvira2015generalized}, where it is shown that several sampling and weighting schemes can be used.
For the sake of conciseness, here we present a single sampling scheme with two different common weighting schemes that evidence the flexibility of MIS.
Let us consider a sampling scheme where exactly one sample per proposal (i.e., $\npart = \nprop$) is drawn,
\begin{equation}
	\parvec^{(m)} \sim \normproposal_m(\parvec), \qquad m=1,...,\npart.
\label{eq_sampling_mis}
\end{equation}
Several proper weighting strategies for this sampling approach have been proposed in the literature (e.g., see \cite{elvira2015generalized} and the references therein for a review of different valid weighting schemes), but the two most common ones are the following:
\begin{itemize}
\item \emph{standard MIS (s-MIS)} \cite{Cappe04}:
\begin{equation} 
\weight_m = \frac{\target(\parvec^{(m)})}{{\normproposal_m(\parvec^{(m)})}}, \quad m=1,\ldots, \npart; \label{eq_w_smis}
\end{equation}
\item \emph{deterministic mixture MIS (DM)} \cite{owen2000safe}:
\begin{equation} 
	\weight_m = \frac{\target(\parvec^{(m)})}{\psi(\parvec^{(m)})} = \frac{\target(\parvec^{(m)})}{\frac{1}{M}\sum_{j=1}^{M}\normproposal_j(\parvec^{(m)})}, \quad m=1,\ldots, \npart,
\label{eq_w_dmmis}
\end{equation}
where $\psi(\parvec)=\frac{1}{M}\sum_{j=1}^{M}\normproposal_j(\parvec)$ is the mixture PDF, composed of all the proposal PDFs with equal weights in the mixture.
\end{itemize}
On the one hand, the unnormalized estimator using any of those two sets of weights (i.e., Eq. \eqref{eq_w_smis} or Eq. \eqref{eq_w_dmmis}) is consistent and unbiased.
On the other hand, the self-normalized estimator is consistent and asymptotically unbiased using both sets of weights.
However, the performance of the estimators may differ substantially depending on which set of weights is used.
For instance, the performance of the unnormalized estimator $\widehat{I}_M$ with the DM approach is superior (in terms of attaining a reduced variance) w.r.t. the s-MIS approach \cite{elvira2015generalized}. 
Finally, note that both weighting alternatives require the same number of target evaluations, but the DM estimator is computationally more expensive w.r.t. the number of proposal evaluations ($\nprop^2$ evaluations for the DM weights vs. $\nprop$ evaluations for the s-MIS ones).
Several efficient approaches to reduce the variance of the estimators, while limiting the computational complexity, have been proposed \cite{elvira2015efficient,elvira2016heretical}.

\subsection{Adaptive importance sampling}
\label{sec:ais}

\vspace*{12pt}

An adaptive importance sampler is an iterative algorithm where the proposals are iteratively improved using either the previously drawn samples or some other independent mechanism.
In particular, here the focus is on the more general case of adaptive MIS, where the set of $\nprop$ proposals $\{\normproposal_n(\parvec)\}_{n=1}^{\nprop}$ are adapted over the subsequent iterations of the algorithm.
Introducing the time-step into the notation, the set of available proposals becomes $\{\normproposal_{n,t}(\parvec)\}_{n=1}^{\nprop}$, where $t$ indicates the $t$-th iteration.
Therefore, not only the parameters of the proposals can be updated, but even the family of distributions can be changed.
For the  sake of simplicity, in this review we only consider location-scale densities, such that each proposal, $\normproposal_{n,t}$, is completely characterized by a mean vector, $\meanvec_{n,t}$, and a covariance matrix, $\covmat_{n,t}$.

Algorithm \ref{alg:apis} describes a generic AIS algorithm, where only the location parameters are adapted (i.e., $\covmat_{n,t}=\covmat_n$ for all $t$).\footnote{Adapting the scale parameters is dangerous and must be done with a lot of care, since it can lead to ill conditioned proposals that result in estimators with huge variances (potentially infinite). For this reason, many AIS algorithms use multiple proposals with different scales and only adapt their locations. However, several algorithms that adapt the scale parameter have also been proposed \cite{Cappe08,CORNUET12,elvira2015gradient}.} 
At the $t$-th iteration, $\npart$ independent samples are drawn from each proposal (step 1 of Algorithm \ref{alg:apis}), i.e.,
\begin{equation}
\parvec_{n,t}^{(m)}\sim \proposal_{n,t}(\parvec|\bmu_{n,t},\bC_n).
\end{equation}
Each sample is then associated with an importance weight of the form 
\begin{equation}
\weight_{n,t}^{(m)}=\frac{\target(\parvec_{n,t}^{(m)})}{\Phi_{n,t}(\parvec_{n,t}^{(m)})},
\label{Gen_W_EqEsto}
\end{equation}
for $m=1,\ldots,M$ and $n=1,\ldots,N$ (step 2).
Note that we use a generic function $\Phi_{n,t}$ at the denominator of the weight.
Similarly to the static MIS methodology, different weighting schemes are possible (in the adaptive setup there is an even larger selection of valid weighting schemes).
In particular, a basic requirement is choosing the set of functions $\{ \Phi_{n,t} \}_{n=1,t=1}^{N,T}$ in such a way that the estimator of Eq. \eqref{eq_unnorm_est} is unbiased \cite[Section 2.5.4]{Liu04b}, \cite{martino2016layered,elvira2015generalized}.
Different choices of $\Phi_{n,t}$ available in the literature can be found in Table \ref{table_weight_denominator}.
Finally, the location parameters of the multiple proposals are adapted in step 3.
Regarding the adaptation of the location parameter, here we divide the methods that can be found in the literature into two main groups:

\begin{enumerate}

\item Algorithms that employ the previous samples for the adaptation. This  approach is summarized in Table \ref{adaptatation_family_1}, and  includes the possible application of resampling steps over a subset of these weighted samples \cite{Cappe04,Koblents14,elvira2017improving} or fitting the moments of the proposals \cite{CORNUET12,APIS14}.

\item Algorithms with independent adaptive procedures (detached from the sampling procedure), e.g. the gradient of $\normtarget$ used in \cite{elvira2015gradient,Schuster15}, or the MCMC techniques applied to adapt the location parameters in \cite{martino2016layered,EBEB14,Nguyen15,Botev13,Yuan13}. While these approaches usually present a superior performance, they tend to be computationally more expensive. Table \ref{adaptatation_family_2} summarizes them.  

\end{enumerate}

\begin{table}[!htb]
\caption{\textbf{Examples of $\Phi_{n,t}$ (function used in the denominator for the estimation) and $\Omega_{n,t}$ (function used in the denominator for the adaptation) that can be found in the literature.}}
\vspace{-0.2cm}
	\begin{tabular}{|c|c|c|c|}
    \hline
\footnotesize
 {\bf Weight den.} & {\bf PMC} \cite{Cappe04} &{\bf AMIS ($N=1$)} \cite{CORNUET12} &{\bf APIS} \cite{APIS15}  \\
 \hline
  \hline 
 $\Phi_{n,t}(\parvec)$ & $q_{n,t}(\parvec)$ & $\frac{1}{t}\sum_{\tau=1}^t q_{\tau}(\parvec)$ & $\frac{1}{N}\sum_{n=1}^N q_{n,t}(\parvec)$ \\
 \hline 
  \hline 
 $\Omega_{n,t}(\parvec)$ & $q_{n,t}(\parvec)$ & $\frac{1}{t}\sum_{\tau=1}^t q_{\tau}(\parvec)$ & $q_{n,t}(\parvec)$ \\
\hline
\end{tabular}
\label{table_weight_denominator}
\end{table}

\newpage

\begin{alg}{Generic Adaptive Population Importance Sampler.}
{
\begin{enumerate}
	\item {\bf Initialization:} Choose the number of proposals ($\nprop$), the number of samples drawn per proposal and iteration ($\npart$), the initial proposals,
		$q_{n,0}(\parvec|\meanvec_{n,0},\covmat_n)$ for $n=1,\ldots,N$ with appropriate values of $\meanvec_{n,0}$ and $\covmat_n$, and the total number of iterations ($\niter$).		\item \texttt{FOR} $t=1,\ldots,T$:
		\begin{enumerate}
			\item {\bf Sampling:} Draw $M$ samples $\parvec_{n,t}^{(1)},\ldots,\parvec_{n,t}^{(M)}$ from each of the $N$ proposal PDFs in the population
				$\left\{\normproposal_1,\ldots,\normproposal_N\right\}$, i.e.,
				\begin{equation}
					\parvec_{n,t}^{(m)} \sim \proposal_{n,t}(\parvec|\bmu_{n,t},\bC_n),
				\end{equation}
				for $m=1,\ldots,M$.
			\item {\bf Weighting:} Weight the samples, $\{\parvec_{n,t}^{(m)}\}_{n=1}^N$, using
				\begin{equation}
					\weight_{n,t}^{(m)}=\frac{\target(\parvec_{n,t}^{(m)})}{\Phi_{n,t}(\parvec_{n,t}^{(m)})}
					\label{Gen_W_Eq}
				\end{equation}
			\item {\bf Adaptation of the means:} Apply some suitable procedure to update the mean vectors,
				\begin{equation}
					\{\bmu_{n,t-1}\}_{n=1}^N \longrightarrow \{\bmu_{n,t}\}_{n=1}^N,
				\end{equation}
				without jeopardizing the consistency of the IS estimators.
		\end{enumerate}
	\item {\bf Output:} Approximate the integral in Eq. \eqref{eq:intBayes} using either the unnormalized estimator when the normalizing constant is known,
		\begin{equation}
			\widehat{I}_{\npart\nprop\niter}
				= \frac{1}{\npart\nprop\niter\partition} \sum_{t=1}^{\niter}{\sum_{n=1}^{\nprop}{\sum_{m=1}^{\npart}{\weight_{n,t}^{(m)} g(\parvec_{n,t}^{(m)})}}}, 
					\label{eq:aisUnNormEst}
		\end{equation}
		or the self-normalizing estimator when the normalizing constant is unknown,
		\begin{subequations}
		\begin{align}
			\widetilde{I}_{\npart\nprop\niter} & = \frac{1}{\npart\nprop\niter\hat{\partition}}
				\sum_{t=1}^{\niter}{\sum_{n=1}^{\nprop}{\sum_{m=1}^{\npart}{\weight_{n,t}^{(m)} g(\parvec_{n,t}^{(m)})}}}, \label{eq:aisSelfNormEst} \\
			\widehat{\partition} & = \frac{1}{\npart\nprop\niter}
				\sum_{t=1}^{\niter}{\sum_{n=1}^{\nprop}{\sum_{m=1}^{\npart}{\weight_{n,t}^{(m)}}}}. \label{eq:aisPartitionEst}
		\end{align}
		\end{subequations}
\end{enumerate}
}
\label{alg:apis}
\end{alg}

\begin{table}[!htb]
\caption{\textbf{Adaptation based on weighted samples.}}
	\begin{tabular}{|p{0.95\columnwidth}|}
    \hline
Let us compute the set of IS weights (potentially different from those of Eq. \eqref{Gen_W_Eq}),
$$
\rho_{n,t}^{(m)} = \frac{\target(\parvec_{n,t}^{(m)})}{\Omega_{n,t}(\parvec_{n,t}^{(m)})}, 
$$
where $\Omega_{n,t}$ are chosen in such a way that they do not jeopardize the consistency of the IS estimators, and they can be equal to $w_{n,t}^{(m)}$ in Eq. \eqref{Gen_W_Eq} or not (see Table \ref{table_weight_denominator}). Two different procedures are used in literature: 
\begin{enumerate}
\item[{\bf P1}] Apply some resampling strategy to $\{\bmu_{n,t-1}\}_{n=1}^N$, with probabilities according to the weights $\rho_{n,t}^{(m)}$, to obtain $\{\bmu_{n,t}\}_{n=1}^N$ \cite{Cappe04,elvira2017improving,Moral06}. Nonlinear transformations of $\rho_{n,t}^{(m)}$ can also be applied \cite{Koblents14}.
\item[{\bf P2}] Build estimators of some moments of $\target$ employing $\rho_{n,t}^{(m)}$, and use this information to obtain $\{\bmu_{n,t}\}_{n=1}^N$ \cite{Cappe08,CORNUET12,oh1992adaptive}.
\end{enumerate}  \\
\hline 
\end{tabular}
\label{adaptatation_family_1}
\end{table}

\begin{table}[!htb]
\caption{\textbf{More sophisticated adaptation procedures.}}
	\begin{tabular}{|p{0.95\columnwidth}|}
    \hline
\begin{enumerate}
\item[{\bf P3}]  Adaptation by using MCMC transitions to obtain $\{\bmu_{n,t}\}_{n=1}^N$ given $\{\bmu_{n,t-1}\}_{n=1}^N$, as in \cite{martino2016layered,Moral06,Nguyen15,Botev13,Yuan13}. 
\item[{\bf P4}]  Adaptation by using stochastic gradient search of $\target$ for moving $\{\bmu_{n,t-1}\}_{n=1}^N$ to $\{\bmu_{n,t}\}_{n=1}^N$ \cite{elvira2015gradient,Schuster15}.
\end{enumerate} \\
\hline 
\end{tabular}
\label{adaptatation_family_2}
\end{table}

\subsection{Convergence of IS and AIS methods}
\label{sec:convIS}

\vspace*{12pt}

In this section, we briefly discuss the convergence of IS and AIS estimators.
We consider both the unnormalized estimator of Eq. \eqref{eq_unnorm_est} and the self-normalized estimator of Eq. \eqref{eq_norm_est}.

First of all, let us consider a fixed set of proposals, $\proposal_n(\parvec)$ for $n=1,\ldots,\nprop$.
Then, it can be easily shown that the unnormalized estimator of Eq. \eqref{eq_unnorm_est} is an unbiased estimator of the desired integral, i.e., $\Esp{}{\widehat{I}_{\npart}} = I$ for any value of $\npart$.
On the other hand, the strong law of large numbers guarantees that the self-normalized estimator is asymptotically unbiased, i.e., $\widetilde{I}_{\npart} \to I$ a.s. as $\npart \to \infty$.
These two results hold, regardless of whether the s-MIS or the DM-MIS weights are used, as long as $\proposal(\parvec) > 0$ whenever $\target(\parvec) > 0$ \cite{geweke1989bayesian}.
Furthermore, under some additional mild regularity conditions (see \cite{geweke1989bayesian}), the following CLTs can be established:
\begin{subequations}
\begin{align}
	\frac{\widehat{I}_{\npart}-I}{\sqrt{\widehat{V}_{\npart}}} & \overset{d}{\to} \gauss(0,1), \\
	\frac{\widetilde{I}_{\npart}-I}{\sqrt{\widetilde{V}_{\npart}}} & \overset{d}{\to} \gauss(0,1),
\end{align}
\end{subequations}
where
\begin{subequations}
\begin{align}
	\widehat{V}_{\npart} & = \frac{1}{M} \BEsp{\normtarget}{(g(\parvec)-I)^2 \weight(\parvec)}, \\
	\widetilde{V}_{\npart} & = \frac{1}{M} \BEsp{\normtarget}{(g(\parvec)-I)^2 \bar{\weight}(\parvec)},
\end{align}
\end{subequations}
$\weight(\parvec)$ is the weighting function used to construct $\widehat{I}_{\npart}$, and $\bar{\weight}(\parvec)$ is its normalized counterpart, which is used in the formulation of $\widetilde{I}_{\npart}$.
Hence, $\widehat{I}_{\npart} \overset{d}{\to} \gauss(I,\widehat{V}_{\npart})$ and $\widetilde{I}_{\npart} \overset{d}{\to} \gauss(I,\widetilde{V}_{\npart})$ as $\npart \to \infty$.
Note that, even though the convergence of both estimators to the desired integral for any proper weighting scheme is ensured, the differences in convergence rate can be quite large (e.g., see \cite{elvira2015generalized} for variance proofs and a discussion on this issue).

Now, let us briefly consider adaptive IS schemes, where the proposals are iteratively updated.
First of all, note that the previous results also hold for AIS methods.
However, a second question that arises in this case is the convergence of the estimators as the proposals are adapted.
This issue is tackled by Oh and Berger in \cite{oh1992adaptive}, where they analyze the estimator obtained by aggregating weighted samples produced through several consecutive iterations using different proposal PDFs.
More precisely, they consider the estimator in Eq. \eqref{eq:aisSelfNormEst} and prove, under fairly general conditions, that $\widetilde{I}_{\npart\nprop\niter} \to I$ a.s. and $\widetilde{I}_{\npart\nprop\niter} \overset{d}{\to} \gauss(I,\widetilde{V}_{\npart\nprop\niter})$ (with the decay of $\widetilde{V}_{\npart\nprop\niter}$ proportional to $\frac{1}{\npart\nprop\niter}$, the optimal Monte Carlo approximation rate) as $\npart\nprop\niter \to \infty$.
See \cite{oh1992adaptive} for further details.

\subsection{Variance of the IS estimators and optimal proposal}
\label{sec:varianceIS}

\vspace*{12pt}

In this section, we analyze the variance of the IS estimators, briefly discussing which is the optimal proposal in terms of variance minimization.
Let assume first that $Z$ is known.
Recalling that $I=\int_{{\bm \Theta}} g({\bm \theta}) \bar{\pi}({\bm \theta}) d{\bm \theta}$, the variance of the IS estimator ${\widehat I}_M$ in Eq. \eqref{eq_unnorm_est} is $\widehat{V}_M=\mbox{var}_q[{\widehat I}_M]= \frac{\sigma_q^2}{M}$, where
\begin{align}
	\sigma_q^2 & = \int_{{\bm \Theta}} \left(\frac{g({\bm \theta}) \bar{\pi}({\bm \theta})}{\normproposal({\bm \theta})}\right)^2 \normproposal({\bm \theta}) d{\bm \theta}- I^2, 
		\nonumber \\
		& = \int_{{\bm \Theta}} \frac{\left(g({\bm \theta}) \bar{\pi}({\bm \theta})\right)^2}{\normproposal({\bm \theta})} d{\bm \theta}- I^2
			= \int_{{\bm \Theta}} \frac{\left(g({\bm \theta}) \bar{\pi}({\bm \theta})-I \normproposal({\bm \theta}) \right)^2}{\normproposal({\bm \theta})} d{\bm \theta}, 
\end{align}
where we have used that 
\begin{align}
	\int_{{\bm \Theta}} \frac{\left(g({\bm \theta}) \bar{\pi}({\bm \theta})-I \normproposal({\bm \theta}) \right)^2}{\normproposal({\bm \theta})} d{\bm \theta} & = 
			\int_{{\bm \Theta}} \frac{\left(g({\bm \theta}) \bar{\pi}({\bm \theta})\right)^2+I^2 q^2({\bm \theta})-2 I g({\bm \theta}) \bar{\pi}({\bm \theta})\normproposal({\bm \theta})}
				{\normproposal({\bm \theta})} d{\bm \theta} \nonumber \\
		& = \int_{{\bm \Theta}} \frac{\left(g({\bm \theta}) \bar{\pi}({\bm \theta})\right)^2}{\normproposal({\bm \theta})} d{\bm \theta}+I^2-2 I^2,  \nonumber \\
		& = \int_{{\bm \Theta}} \frac{\left(g({\bm \theta}) \bar{\pi}({\bm \theta})\right)^2}{\normproposal({\bm \theta})} d{\bm \theta}- I^2.
\end{align}
For a specific function $g({\bm \theta})$, the optimal proposal PDF is 
\begin{equation}
q_{opt}({\bm \theta})=\frac{|g({\bm \theta})| \bar{\pi}({\bm \theta})}{I}= \frac{g({\bm \theta}) \bar{\pi}({\bm \theta})}{\int_{{\bm \Theta}} g({\bm \theta}) \bar{\pi}({\bm \theta}) d{\bm \theta}}\propto |g({\bm \theta})| \bar{\pi}({\bm \theta}).
\end{equation}
However, in many applications practitioners are not interested in estimating a specific integral $I$, but in approximating the measure of ${\bar \pi}$.
In  this case, an appropriate choice for the proposal is $\normproposal({\bm \theta}) = {\bar \pi}({\bm \theta}) \propto \pi({\bm \theta})$, which leads to $w_n=\frac{1}{Z}$ and ${\bar w}_n=\frac{1}{N}$ for all $n=1,\ldots,N$, i.e., we come back to the original Monte Carlo scheme described in Section \ref{sec:monteCarlo}.
Furthermore, the variance of the random variable $w({\bm \theta})=\frac{\pi({\bm \theta})}{\normproposal({\bm \theta})}$, with ${\bm \theta}\sim \normproposal({\bm \theta})$, is given by
\begin{align}
	\mbox{var}_q\left[w({\bm \theta})\right] & = \int_{{\bm \Theta}} \left(\frac{\pi({\bm \theta})}{\normproposal({\bm \theta})}\right)^2 \normproposal({\bm \theta}) d{\bm \theta}
			- \left(\int_{{\bm \Theta}} \left(\frac{\pi({\bm \theta})}{\normproposal({\bm \theta})}\right) \normproposal({\bm \theta}) d{\bm \theta} \right)^2, \nonumber \\
		& = \int_{{\bm \Theta}}\frac{\pi^2({\bm \theta})}{\normproposal({\bm \theta})}  d{\bm \theta}-  Z^2, \nonumber \\
		& = \int_{{\bm \Theta}}\frac{\left(\pi({\bm \theta})-Z\normproposal({\bm \theta})\right)^2}{\normproposal({\bm \theta})}  d{\bm \theta} \nonumber \\
		& = Z^2\int_{{\bm \Theta}}\frac{\left( {\bar \pi}({\bm \theta})-\normproposal({\bm \theta})\right)^2}{\normproposal({\bm \theta})}  d{\bm \theta}
			= Z^2 \chi^2({\bar \pi},\normproposal), \label{DSPJoaquin} 
\end{align}
where we have used ${\bar \pi}({\bm \theta}) =\frac{1}{Z}  \pi({\bm \theta})$ in the last step of the derivation, and $\chi^2({\bar \pi},\normproposal)=\int_{{\bm \Theta}}\frac{\left({\bar \pi}({\bm \theta})-\normproposal({\bm \theta})\right)^2}{\normproposal({\bm \theta})}  d{\bm \theta}$ is the Pearson divergence between ${\bar \pi}$ and $\normproposal$ \cite{Chen05}. Finally, the variance of $\widehat{Z}=\frac{1}{M}\sum_{m=1}^M  w({\bm \theta}^{(m)})$ is 
$\mbox{var}_q[\widehat{Z}]=\frac{Z^2}{M^2}\chi^2({\bar \pi},\normproposal)$.

\subsection{Effective Sample Size}
\label{SectionESS_IS}

\vspace*{12pt}

Let us denote in this section the standard Monte Carlo estimator as 
\begin{equation}
\label{St_MC}
{\overline I}_M=\frac{1}{M}\sum_{m=1}^M g({\bm \theta}^{(m)}),
\end{equation}
where the samples ${\bm \theta}^{(1)},\ldots, {\bm \theta}^{(M)}$ are directly drawn from $\bar{\pi}({\bm \theta})$.
Moreover, let us  define the normalized IS weights,
\begin{equation}
\label{NormWeights}
{\bar w}_m=\frac{1}{M \widehat{Z}} w_m =\frac{ w_m }{\sum_{i=1}^M w_i},  \quad m=1,\ldots,M,
\end{equation}
then the self-normalized IS estimator can be written as ${\widetilde I}_M=\sum_{m=1}^M  {\bar w}_m g({\bm \theta}^{(m)})$.
In general, the estimator ${\widetilde I}_M$ is less efficient than ${\overline I}_M$ in Eq. \eqref{St_MC}.
In several applications of importance sampling, it is required to measure  this loss in efficiency, when ${\widetilde I}_M$ is used instead of ${\overline I}_M$.
The idea is to define the ESS as the ratio of the variances of the estimators \cite{Kong92},
\begin{equation}
\label{DEF_ESS_1}
\textrm{ESS}=M\frac{\mbox{var}_{\normtarget}[{\widehat I}]}{\mbox{var}_{\normproposal}[{\widetilde I}]}.
\end{equation}
The ESS value represents the number of samples from $\bar{\pi}$ required  to obtain a Monte Carlo estimator ${\widehat I}$  with the same efficiency of the IS estimator ${\widetilde I}$, considering $\normproposal$ as the proposal density. 
Finding a useful expression of ESS derived analytically from the theoretical definition above is not straightforward.
Different derivations proceed by using several approximations and assumptions to yield an expression which is useful from a practical point of view \cite{Kong92,Kong94}, \cite[Chapter 11]{SMC01}, \cite[Chapter 4]{Robert10b}.
A well-known ESS approximation, widely used in the literature \cite{SMC01,Liu04b,Robert10b}, is 
\begin{equation}
	\widehat{\textrm{ESS}}=\frac{1}{\sum_{m=1}^M {\bar w}_m^2}.
\label{FirstDef_P}
\end{equation} 
An interesting property of the $\widehat{\textrm{ESS}}$ in \eqref{FirstDef_P} is that $1 \leq \widehat{\textrm{ESS}}\leq M$.
Although Eq. \eqref{FirstDef_P} is often considered a suitable approximation of the theoretical ESS definition, its derivation \cite{Kong92,RobertBlog,Robert10b},\cite[Section 3]{Chen05} contains several approximations and strong assumptions \cite{ESSmartino}.
As a consequence, $\widehat{\textrm{ESS}}$ can differ substantially from the original definition of the ESS in many scenarios. 
In \cite{ESSmartino}, different alternative approximations are discussed.
For instance,  
\begin{equation}
\widehat{\textrm{ESS}}=\frac{1}{\max {\bar w}_m}
\label{Max_ESS}
\end{equation} 
results again in $1 \leq \widehat{\textrm{ESS}}\leq M$: the minimum is obtained when all the samples have zero weight except only one, whereas the maximum is reached when all the weights are equal to ${\bar w}_m=\frac{1}{M}$ \cite{ESSmartino}. Other related discussions and results can be found in \cite{Agapiou15,Whiteley16,Huggins15}. 

\subsection{Proper Weighting}
\label{LiuSect}

\vspace*{12pt}

Although widely adopted, the standard IS weights in Eq. \eqref{is_weights_static} are not the unique possibility.
The definition of a \emph{properly weighted sample} can be extended as suggested in \cite[Section 14.2]{Robert04}, \cite[Section 2.5.4]{Liu04b} and in \cite{elvira2015generalized}.
More specifically, given a set of samples, they are properly weighted with respect to the target ${\bar \pi}$ if, for any integrable function $g$, 
\begin{equation}
E_Q[w({\bm \theta}^{(m)}) g({\bm \theta}^{(m)})]=c E_{\bar \pi}[g({\bm \theta}^{(m)})], \quad \forall m\in\{1,\dots,M\},
\label{eq_liu_1}
\end{equation} 
where $c>0$ is a constant value, independent from the index $m$, and the expectation of the left hand side is performed w.r.t. to the joint PDF of $w({\bm \theta})$ and ${\bm \theta}$, i.e., $Q({\bm \theta},w)$.
%
%
Thus, in order to obtain consistent estimators, one has to design a joint PDF $Q({\bm \theta},w)$ which guarantees that the restriction of Eq. \eqref{eq_liu_1} is fulfilled.  An example is provided below.

\subsubsection{Proper weighting of a resampled particle} 
\label{PropW_RES}

\vspace*{12pt}

Let us consider the particle approximation of the measure of $\bar{\pi}$ obtained by the IS approach drawing $M$ IID particles ${\bm \theta}^{(m)} \sim q({\bm \theta})$,  
\label{ProperGISWeight}
\begin{equation}
\label{PIEQ}
\widehat{\pi}({\bm \theta}|{\bm \theta}_{1:M})= \frac{1}{M \widehat{Z}} \sum_{m=1}^M w({\bm \theta}^{(m)}) \delta({\bm \theta}-{\bm \theta}^{(m)}),
\end{equation}
where $w({\bm \theta}^{(m)})=w_m=\frac{\pi({\bm \theta}^{(m)})}{q({\bm \theta}^{(m)})}$ and  $\delta({\bm \theta})$ is the Dirac delta function.
Therefore,  given the set of weighted samples $\{{\bm \theta}^{(m)},w_m\}_{m=1}^M$, a resampled particle $\widetilde{{\bm \theta}}$ is a sample drawn from $\{\bm \theta^{(m)}\}_{m=1}^M$ according to the probability mass $ \frac{w_m}{M \widehat{Z}}$, i.e.,
\begin{equation}
\widetilde{{\bm \theta}}\sim \widehat{\pi}({\bm \theta}|{\bm \theta}_{1:M}).
\end{equation}
 Let us denote the joint PDF $\widetilde{Q}({\bm \theta},{\bm \theta}_{1:M})= \widehat{\pi}({\bm \theta}|{\bm \theta}_{1:M}) \left[\prod_{i=1}^M q({\bm \theta}^{(i)})\right]$. The marginal PDF $\widetilde{q}({\bm \theta})$ of a resampled particle $\widetilde{{\bm \theta}}$, integrating out ${\bm \theta}_{1:M}$, i.e., $\widetilde{{\bm \theta}}'\sim \widetilde{q}({\bm \theta})$, is
\begin{align}
	\widetilde{q}({\bm \theta}) = \int_{\mathcal{X}^M}  \widetilde{Q}({\bm \theta},{\bm \theta}_{1:M}) d{\bm \theta}_{1:M}
		& = \int_{\mathcal{X}^M}\widehat{\pi}({\bm \theta}|{\bm \theta}_{1:M}) \left[\prod_{i=1}^M q({\bm \theta}^{(i)})\right]  d{\bm \theta}_{1:M},  \nonumber\\
		& = \pi({\bm \theta}) \sum_{j=1}^{M} \left({\int_{\mathcal{X}^{M-1}} \frac{1}{M\widehat{Z}} \left[\prod_{\substack{i=1 \\ i \neq j}}^M q({\bm \theta}^{(i)})\right]
			d{\bm \theta}_{\neg j}} \right), \nonumber \\
		& = \pi({\bm \theta})  {\int_{\mathcal{X}^{M-1}} \frac{1}{\widehat{Z}} \left[\prod_{\substack{i=1 \\ i \neq j}}^M q({\bm \theta}^{(i)})\right]  d{\bm \theta}_{\neg j}}, \label{ESTA_q_dif}
\end{align}
and the standard IS weight of a resampled particle $\widetilde{{\bm \theta}}$ is
\begin{equation}
	w(\widetilde{{\bm \theta}})=\frac{\pi(\widetilde{{\bm \theta}})}{\widetilde{q}(\widetilde{{\bm \theta}})}.
\end{equation}
However, usually $\widetilde{q}({\bm \theta})$ in Eq. \eqref{ESTA_q_dif} cannot be evaluated, and thus the standard IS weight cannot be computed \cite{Lamberti16,lamberti2017independent,GISssp16,NSMC}, \cite[App. C1]{martino2016layered}, \cite[App. B3]{OMCMC}.
An alternative is to use Liu's definition of proper weighting in Eq. \eqref{eq_liu_1} and look for a weight function $\rho({\widetilde {\bm   \theta}})=\rho({\widetilde {\bm   \theta}},{\bm \theta}_{1:M})$ such that 
\begin{equation}
E_{\widetilde{Q}({\bm \theta},{\bm \theta}_{1:M})}[\rho({\bm \theta},{\bm \theta}_{1:M}) h({\bm \theta})]=c E_{\bar \pi}[h({\bm \theta})],
\label{eq_liu_2}
\end{equation} 
where $\widetilde{Q}({\bm \theta},{\bm \theta}_{1:M})= \widehat{\pi}({\bm \theta}|{\bm \theta}_{1:M}) \left[\prod_{i=1}^M q({\bm \theta}^{(i)})\right]$. A suitable choice is 
\begin{equation}
\label{SUPER_IMP_DEFINITION}
\rho({\widetilde {\bm   \theta}},{\bm \theta}_{1:M})=\widehat{Z}=\frac{1}{M}\sum_{i=1}^M w({\bm \theta}^{(i)}),
 \end{equation}
since it holds in Eq.  \eqref{eq_liu_2}. For the proof and further discussions see \cite{GISssp16,GIS17,Martino15PF}.
The proper weighting of a resampled particle is used in several Monte Carlo approaches, like the group IS described in Section \ref{sec:gis}.

\subsection{Group Importance Sampling}
\label{sec:gis}

\vspace*{12pt}

Here, we use the results of the previous section to assign one single weighted sample to a set of weighted samples to summarize all the statistical information.
Let us consider $L$ sets of weighted samples,
$\mathcal{S}_1=\{{\bm \theta}_1^{(m)},w_{1,m}\}_{m=1}^{M_1}$, $\mathcal{S}_2=\{{\bm \theta}_2^{(m)},w_{2,m}\}_{m=1}^{M_2}$, ...., $\mathcal{S}_L=\{{\bm \theta}_L^{(m)},w_{L,m}\}_{m=1}^{M_L}$,
where ${\bm \theta}_\ell^{(m)} \sim q_\ell({\bm \theta})$, i.e., a different proposal PDF can be used to generate each set $\mathcal{S}_\ell$ and in general $M_i\neq M_j$ for all $i \neq j$, $i,j\in\{1,...,L\}$.

In different Monte Carlo applications, it is convenient (and often required) to compress the statistical information contained in all these sets by using a summary sample, $\widetilde{{\bm \theta}}_\ell$, and summary weight, $W_\ell$, $\ell=1,\ldots,L$, in such a way that
\begin{equation}
	\widetilde{I}_L=\frac{1}{\sum_{j=1}^L  W_j} \sum_{\ell=1}^L  W_\ell g(\widetilde{{\bm \theta}}_\ell)
\label{GroupIS}
\end{equation}
is still a consistent estimator of $I$, for a generic integrable function $g({\bm \theta})$ \cite{GIS17}.
Thus, although the compression is lossy, we still have a suitable particle approximation by the set of weighted samples $\{\widetilde{{\bm \theta}}_\ell,W_\ell \}_{\ell=1}^L$ of the target ${\bar \pi}$, as shown in the following.
Let us denote the IS of the $m$-th sample in the $\ell$-th group as  $w_{\ell,m}=\frac{\pi({\bf \theta}_{\ell,m})}{q_m({\bf \theta}_{\ell,m})}$, and $\widehat{Z}_\ell=\frac{1}{M_\ell}\sum_{m=1}^{M_\ell}w_{\ell,m}$.
Then, it is possible to show that, with the choice
\begin{equation}
\widetilde{{\bm \theta}}_\ell \sim  \frac{1}{\sum_{i=1}^{M_\ell}  w_{i,\ell} }   \sum_{m=1}^{M_\ell} w_{\ell,m}  \delta({\bm \theta}-{\bm \theta}_\ell^{(m)}) 
\end{equation}
and 
\begin{equation}
W_\ell=M_\ell \widehat{Z}_\ell,
\end{equation}
then $\widetilde{I}_L$ in Eq. \eqref{GroupIS} is a consistent estimator of $I$.
Note that $\widetilde{{\bm \theta}}_\ell$ is a resampled particle within the $\ell$-th group and $W_\ell$ takes into account the proper weight of a resampled particle, $ \widehat{Z}_\ell$, and the number of samples in the $\ell$-th set, $M_\ell$.
Let us consider the normalized weights $\bar{w}_{\ell,m}=\frac{w_{m,\ell}}{\sum_{i=1}^{M_\ell}  w_{i,\ell} }$.
Since $\bar{w}_{\ell,m}=\frac{w_{\ell,m}}{M_\ell \widehat{Z}_\ell}=\frac{w_{\ell,m}}{W_{\ell}}$, the unnormalized weight of the particle ${\bm \theta}_\ell^{(m)}$ can be expressed as $w_{\ell,m}=W_{\ell} \bar{w}_{\ell,m}$.
This confirms that after a particle is resampled according to $\bar{w}_{\ell,m}$, in order to represent the $\ell$-th group of $M_\ell$ weighted samples, it must be weighted as $W_{\ell}$.
The idea of a summary sample/weight has been implicitly used in different SMC schemes proposed in literature, for instance, for the communication among parallel particle filters \cite{Bolic05,Miguez16,Read2014}, and in the particle island methods \cite{Whiteley16,Pisland15,Pisland15_2}.
GIS also appears indirectly in particle filtering for model selection \cite{Martino15PF,Drovandi14,PetarCopy}, and in the so-called Nested Sequential Monte Carlo techniques  \cite{NSMC,NosCitan,RafaelThesis}.
For further observations and applications of GIS see \cite{GISssp16,GIS17}.

\subsection{Sequential Importance Sampling (SIS)}
\label{SIS}

\vspace*{12pt}

In this section, we describe the Sequential Importance Sampling (SIS) scheme.
In some applications, the parameters of interest ${\bm \theta}$ can be split in two disjoint groups, ${\bm \theta}=[{\bf x},{\bm \lambda}]$, where the first one, ${\bf x}$, is related to a dynamical system (for instance, $\bf x$ can be the hidden state in a state-space model) and the other, ${\bm \lambda}$, is a static parameter (for instance, an unknown parameter of the model).
The strategies for making inference about ${\bf x}$ and ${\bm \lambda}$ should take into account the different nature of the two parameters.
In the previous sections we have considered ${\bm \theta}={\bm \lambda}$.
In Section \ref{PMMH}, we tackle the general case ${\bm \theta}=[{\bf x},{\bm \lambda}]$, whereas here we address the case ${\bm \theta}={\bf x}$.
Namely, we assume that the variable of interest is a dynamical variable, i.e., ${\bm \theta}={\bf x}=x_{1:D}=[x_1\dots,x_D]^{\top}$ with $x_d\in\mathbb{R}$ for the sake of simplicity, and the target can be factorized as
\begin{equation}
\bar{\pi}({\bf x})\propto \pi({\bf x}) = \gamma_1(x_1) \prod_{d=2}^D  \gamma_d(x_d|x_{d-1}).
\end{equation}
Given a proposal $q({\bf x})=q_1(x_1) \prod_{d=2}^D  q_d(x_d|x_{d-1})$, and a sample ${\bf x}^{(m)}=x_{1:D}^{(m)}\sim q({\bf x})$ with $x_{d}^{(m)}\sim q_d(x_d|x_{d-1}^{(m)})$, we assign the importance weight  
\begin{equation}
w({\bf x}^{(m)})=w_D^{(m)}=\frac{\pi({\bf x}^{(m)})}{q({\bf x}^{(m)})}=\frac{ \gamma_1(x_1^{(m)}) \gamma_2(x_2^{(m)}|x_1^{(m)}) \cdots  \gamma_D(x_D^{(m)}|x_{D-1}^{(m)})}{q_1(x_1^{(m)}) q_2(x_2^{(m)}|x_1^{(m)}) \cdots  q_D(x_D^{(m)}|x_{D-1}^{(m)})}.
\end{equation}
 The weight above can be computed efficiently by following a recursive procedure to compute the importance weights: starting with $w_1^{(m)}=\frac{\pi(x_1^{(m)})}{q(x_1^{(m)})}$ and then obtaining
 \begin{equation}
 \label{RecWeights}
 \begin{split}
  w_d^{(m)}=w_{d-1}^{(m)} \beta_d^{(m)}=\prod_{j=1}^{d} \beta_j^{(m)},   \quad \quad d=1,\ldots,D,   
 \end{split}
 \end{equation}
 where
\begin{subequations}
\begin{align}
	\beta_1^{(m)} & = w_1^{(m)} \label{AumWeights1} \\
	\beta_d^{(m)} & = \frac{\gamma_d(x_d^{(m)}|x_{d-1}^{(m)})}{q_d(x_d^{(m)}|x_{d-1}^{(m)})}, \label{AumWeights}
\end{align}
\end{subequations}

\subsubsection{Sequential Importance Resampling (SIR) }
\label{SIR}

\vspace*{12pt}

Sequential Importance Resampling, a.k.a., standard particle filtering, is a SIS scheme where resampling steps are incorporated during the recursion, as shown in Table \ref{alg:SIRpartialRes} \cite{Martino15PF,Djuric03,Doucet08tut,MartinoCCA}.
Resampling consists in drawing particles from the current cloud according to the normalized importance weights.
In general, the resampling steps are applied only in certain iterations in order to avoid the path degeneration, taking into account an ESS approximation, such as $\widehat{\textrm{ESS}}=\frac{1}{\sum_{m=1}^M (\bar{w}_d^{(m)})^2}$ or $\widehat{\textrm{ESS}}=\frac{1}{\max \bar{w}_d^{(m)}}$ with $\bar{w}_d^{(m)}=\frac{w_d^{(m)}}{\sum_{i=1}^M w_d^{(i)}}$  \cite{ESSmartino}.
If $\frac{1}{M}\widehat{\textrm{ESS}}$ is smaller than a pre-established threshold $\eta\in[0,1]$, the particles are resampled.
Thus, the condition for the adaptive resampling can be expressed as $\widehat{\textrm{ESS}} < \eta M$.
When $\eta=1$, the resampling is applied at each iteration and in this case SIR is often called bootstrap particle filter \cite{Djuric03,Doucet08tut}.
If $\eta= 0$, no resampling steps are applied, and we have the SIS method described above.
Consider the Algorithm \ref{alg:SIRpartialRes}. Let us define 
{
\begin{equation}
\widehat{Z}_{d}=\frac{1}{M} \sum\limits_{m=1}^Mw_{d}^{(m)}=\frac{1}{M} \sum\limits_{m=1}^M \prod_{j=1}^{d} \beta_{j}^{(m)},
\end{equation} 
where we have used the recursion for the weights in Alg. \ref{alg:SIRpartialRes}.}
Note that in Algorithm \ref{alg:SIRpartialRes} we have employed a proper weighting for resampling particles (see Section \ref{PropW_RES} and \cite{GISssp16}),
\begin{equation}
w_{d}^{(1)}=w_{d}^{(2)}=\ldots= w_{d}^{(M)}=\widehat{Z}_{d}.
\end{equation}
In many works regarding particle filtering it is noted that { the {\it unnormalized} weights of the resampled particles},  $w_{d}^{(1)}=w_{d}^{(2)}=\ldots= w_{d}^{(M)}$, but a specific value is not given. if a different value $c\neq \widehat{Z}_{d}$ is employed, i.e., $w_{d}^{(1)}=\ldots= w_{d}^{(M)}=c$, the algorithm is still valid (if the resampling is applied considering {\it all} the particles), {\it but} the weight recursion loses some statistical meaning.
{ In the case of standard SIR scheme, i.e., when the resampling is performed considering {\it all} the $M$ particles, the {\it normalized} weights of the resampled particles are   
$$
\bar{w}_{d}^{(1)}=\bar{w}_{d}^{(2)}=\ldots= \bar{w}_{d}^{(M)}=\frac{1}{M},
$$
for any possible choice of $c$. Moreover, people usually employs a different marginal likelihood estimator
\begin{equation}
\widehat{Z}_{d}^{\text{{\bf alt}}}=\prod_{j=1}^{d}\left[\sum_{m=1}^M\bar{w}_{j-1}^{(m)}\beta_j^{(m)}\right],
\end{equation}
which involves only the normalized weights, $\bar{w}_{d}^{(m)}$, instead of the unnormalized ones, $w_{d}^{(m)}$. Hence, this is a suitable and consistent estimator, in this scenario. However, the standard marginal likelihood estimator 
\begin{equation}
\widehat{Z}=\widehat{Z}_{D}=\frac{1}{M} \sum\limits_{m=1}^M w_{D}^{(M)}=\sum\limits_{m=1}^M \prod_{j=1}^{D} \beta_{j}^{(m)},
\end{equation}}
is consistent only if a proper weighting after resampling is used \cite{GISssp16,GIS17,Martino15PF}. { Moreover, if the resampling is performed considering {\it only} a subset of the particles of cardinality $R<M$  (instead over all the $M$ particles), the proper weighting is strictly needed.}

\newpage

\begin{alg}{Sequential importance resampling (SIR).}
{\begin{enumerate}
	\item \textbf{Initialization:} Choose the number of particles ($M$), the initial particles $x_0^{(m)}$ for $m=1,\ldots,M$, an ESS approximation, and a constant value
		$\eta\in[0,1]$.
	\item \texttt{FOR} $d = 1, \ldots, D$:
    		\begin{enumerate}
        			\item {\bf Propagation:}\label{PropStep} Draw  $x_d^{(m)} \sim q_d(x_d|x_{d-1}^{(m)})$, for $m=1,\ldots,M$.
        			\item  {\bf Weighting:}\label{WStep} Compute the weights       
                 		\begin{equation}
			                   w_d^{(m)}=w_{d-1}^{(m)} \beta_d^{(m)} =\prod_{j=1}^{d} \beta_j^{(m)},   \quad \quad m=1,\ldots,M,   
                 		\label{RecWeights_2}
		                \end{equation}       
             			where  $\beta_d^{(m)}=\frac{\gamma_d(x_d^{(m)}|x_{d-1}^{(m)})}{q_d(x_d^{(m)}|x_{d-1}^{(m)})}$. 
	             \item\label{StepResampling}  \texttt{IF} $\widehat{ESS} < \eta M$ \texttt{THEN}: 
             			\begin{enumerate}
					\item {\bf Resampling:}\label{StepResampling2} Resample $M$ times within the set $\{x_{d-1}^{(m)}\}_{m=1}^M$ according to the probabilities
						$\bar{w}_d^{(m)}=\frac{w_{d}^{(m)}}{ \sum_{j=1}^M w_{d}^{(j)}}$, obtaining $M$ resampled particles $\{\bar{x}_d^{(m)}\}_{m=1}^M$.
						Then, set $x_d^{(m)}=\bar{x}_d^{(m)}$, for $m=1,\ldots,M$.
					\item\label{StepProperGIS}  {\bf Proper weighting:} Compute $\widehat{Z}_{d}=\frac{1}{M} \sum\limits_{m=1}^Mw_{d}^{(m)}$ and set
						$w_{d}^{(m)}=\widehat{Z}_{d}$ for $m=1,\ldots,M$.
				\end{enumerate}             
         	\end{enumerate}
         \item Return $\{\x^{(m)}=x_{1:D}^{(m)}, w^{(m)}=w_D^{(m)}\}_{m=1}^M$.
\end{enumerate}}
\label{alg:SIRpartialRes}
\end{alg}

\subsubsection{Conditional particle filter}
\label{CPF}

\vspace*{12pt}

The conditional particle filter (CPF) is a modification of the particle filter algorithm which takes a reference state sequence ${\bf x}^*=x_{1:D}^*$ as input \cite{PMCMC,Lindsten14}.
Namely, the CPF is a standard particle filter (e.g., the SIR in Table \ref{alg:SIRpartialRes}) setting as the first particle $x_{1:D}^{(1)}=x_{1:D}^*$, the reference path.
Hence, the implementation of the CPF algorithm is exactly like a standard particle filter, except for the following two points:
\begin{enumerate}
\item The first path $x_{1:D}^{(1)}$ is not sampled, i.e., it is not randomly generated but fixed in advance. Indeed, each component of the first path $x_{1:D}^{(1)}$ is copied from the reference path $x_{1:D}^*$. 
\item In the resampling step, the first particle is guaranteed to survive.
\end{enumerate}
Considering a resampling step at each iteration (just for the sake of simplicity), the CPF is outlined in Algorithm \ref{CPF_table}.
It is important to remark that the procedure (a) picking a reference path ${\bf x}^*$, (b) running the CPF, (c) picking a path ${\bf x}'=x_{1:D}'$ by resampling once with probabilities proportional to the final weights $w_m$ for $m=1,\ldots,M$, and (d) repeating from (a) considering ${\bf x}^*={\bf x}'$, leaves invariant the target density ${\bar \pi}(\x)$.
Indeed, this procedure virtually coincides with the Ensemble MCMC method that will be described in Section \ref{EnMCMC} (see \cite[Appendix C]{OMCMC} for a proof).
For this reason, the CPF method is often applied within sophisticated MCMC techniques, called Particle Gibbs algorithms (see Section \ref{PG_sect} for further details).
The CPF provides a particle approximation of the target measure given the reference path ${\bf x}'$, i.e.,  
\begin{equation}
\widehat{\pi}(\x|{\bf x}')=\sum_{m=1}^M {\bar w}_D^{(m)}\delta(\x-\x^{(m)}).
\end{equation}
Finally, note that we have considered a CPF method that is slightly different from the technique proposed in \cite{PMCMC}.
Indeed, here we have described the CPF version given in \cite{Lindsten14,Kokkala16}.

\begin{alg}{Conditional particle filter (CPF).}
{\begin{enumerate}
\item \textbf{Initialization:} Determine the reference path $x_{1:D}^*$. Choose the number of particles ($M$) and other $M-1$ initial particles $x_0^{(m)}$ for $m=2,\ldots,M$.
    \item \texttt{FOR} $d = 1, \ldots, D$:
    	\begin{enumerate}
        \item {\bf Propagation:} Set $x_{d}^{(1)}=x_d^*$ and draw  $x_d^{(m)} \sim q_d(x_d|x_{d-1}^{(m)})$, for $m=2,\ldots,M$.
        \item  {\bf Weighting:} Compute the weights       
                 \begin{equation}
                   w_d^{(m)}=w_{d-1}^{(m)} \beta_d^{(m)} =\prod_{j=1}^{d} \beta_j^{(m)},   \quad \quad m=1,\ldots,M,   
                \end{equation}       
                   where  $\beta_d^{(m)}=\frac{\gamma_d(x_d^{(m)}|x_{d-1}^{(m)})}{q_d(x_d^{(m)}|x_{d-1}^{(m)})}$. 
             \item   {\bf Conditional Resampling:}
     \begin{enumerate}
    \item  Set $\bar{x}_d^{(1)}=x_{d}^{(1)}$.
    \item  Resample $M-1$ times within the set of $M$ samples $\{x_{d-1}^{(m)}\}_{m=1}^M$ according to the probabilities $\bar{w}_d^{(m)}=\frac{w_{d}^{(m)}}{ \sum_{j=1}^M w_{d}^{(j)}}$, $m=1,\ldots,M$, obtaining $M-1$ resampled particles $\{\bar{x}_d^{(m)}\}_{m=2}^M$. 
    \item Set $x_d^{(m)}=\bar{x}_d^{(m)}$, for $m=1,\ldots,M$.
\end{enumerate}
\item  {\bf Proper weighting:} Compute $\widehat{Z}_{d}=\frac{1}{M} \sum\limits_{m=1}^Mw_{d}^{(m)}$ and set $w_{d}^{(m)}=\widehat{Z}_{d}$ for all $m=1,\ldots,M$ \cite{GISssp16, GIS17}.            
  \end{enumerate}
         \item Return $\{\x^{(m)}=x_{1:D}^{(m)},  w^{(m)}=w_D^{(m)}\}_{m=1}^M$.
   \end{enumerate}
}
\label{CPF_table}
\end{alg}

\section{MC-within-MCMC methods}
\label{sec:is_within_mcmc}

\vspace*{12pt}

In this section, we describe several MCMC techniques that use other inner MC estimators at each iteration.\footnote{Note that we have already described an MC-within-MCMC method in Section \ref{sec:MHwithinGibbs}: the MH-within-Gibbs algorithm.}
The resulting hybrid methods are still MCMC algorithms, since they rely on a Markov chain to sample from the target PDF, but they require these inner MC techniques for different reasons.
They can be divided into two classes.
The methods in the first class (see Sections \ref{sec:advMcmc} and \ref{PMH_bigsec}) use IS and a resampling step to generate better candidates for the MH acceptance test.
The methods in the second class need some inner MC algorithm (either IS or MCMC) to obtain unbiased estimators of the likelihood function (see Sections \ref{PSM_MCMC} and \ref{Noisy_MCMC}).
There is a connection between these two classes of algorithms which is apparent between the methods in Section \ref{PMMH} and Section \ref{PSM_MCMC}.
We also split the first class in two sub-families.
In the first one (Section \ref{sec:advMcmc}), we describe the MCMC techniques that propose multiple candidates at each iteration and work in a batch way (i.e., directly in the entire space of ${\bm \theta}$).
The methods contained in the second sub-family (Section \ref{PMH_bigsec}) also generate several candidates at each iteration, but they assume that a factorization of the target density is available.
This assumption allows the sequential generation of the candidates (via particle filtering, for instance).

\vspace*{12pt}



\subsection{MCMC with multiple candidates for the estimation of a static parameter}
\label{sec:advMcmc}

\vspace*{12pt}

In the MH algorithm, at each iteration one new sample, ${\bm \theta}'$, is generated and tested w.r.t. the previous state, ${\bm \theta}^{(t-1)}$, by using the acceptance probability $\alpha_t = \alpha({\bm \theta}',{\bm \theta}^{(t-1)})$.
Other generalized MH schemes generate several candidates at each iteration to be tested as the new possible state with the aim of increasing the acceptance rate of candidate samples.
In all these schemes, an extended acceptance probability, $\alpha_t$, has to be properly designed in order to guarantee the ergodicity of the chain. 
Below we describe the most important examples of this kind of generalized MH algorithms \cite{martino2018review}.
%
%
Furthermore, most of these techniques use an AIS approximation of the target density (see Section \ref{sec:ais}) in order to  improve the proposal procedure within an MH-type algorithm.
Namely, they build an IS approximation adaptively and then draw one sample from this approximation (resampling step).
Finally, the selected sample is compared with the previous state of the chain, ${\bm \theta}^{(t-1)}$, according to a suitable generalized acceptance probability $\alpha_t$.

\subsubsection{Multiple Try Metropolis (MTM)}
\label{sec:mtm}

\vspace*{12pt}

The \emph{Multiple Try Metropolis} (MTM) algorithms are examples of this class of methods \cite{martino2018review,Liu00,LucaJesse2,LucaJesse1,Bedard12,Craiu07}.
In this case, $N$ samples (a.k.a. ``tries'' or ``candidates'') are drawn from the proposal PDF, one of them is selected according to some suitable weights, and the selected candidate is finally accepted or rejected according to a generalized probability function $\alpha_t$. 

The standard MTM scheme is shown in Algorithm \ref{alg:MTM}.
For the sake of simplicity, we have considered the use of the standard importance weights $w({\bm \theta})=\frac{\pi({\bm \theta})}{q({\bm \theta})}$ (see Section \ref{sec:ais}), but other more sophisticated alternatives are also possible \cite{Liu00,LucaJesse2,MTMissue17}.
In its general form, when the proposal depends on the previous state of the chain, $q({\bm \theta}|{\bm \theta}^{(t-1)})$, MTM requires the generation of $N-1$ auxiliary samples, ${\bf v}^{(1)},\ldots, {\bf v}^{(N-1)}$, which are employed in the computation of the acceptance probability $\alpha_t$.
These samples are required in order to guarantee the ergodicity of the underlying Markov chain.
Indeed, it can be proved the resulting MTM kernel satisfies the detailed balance condition, implying that the chain is reversible.

\newpage

\begin{alg}{Multiple Try Metropolis (MTM) method.}
{\begin{enumerate}
	\item \textbf{Initialization:} Choose a proposal function $\proposal({\bm \theta}|{\bm \theta}^{(t-1)})$, an initial state ${\bm \theta}^{(0)}$,
    		the total number of iterations ($\niter$), and the number of tries ($N$).
	\item \texttt{FOR} $t = 1, \ldots, \niter$:
    		\begin{enumerate}
        			\item Draw $\widetilde{{\bm \theta}}^{(1)},\widetilde{{\bm \theta}}^{(2)},\ldots, \widetilde{{\bm \theta}}^{(N)}  \sim \proposal({\bm \theta}|{\bm \theta}^{(t-1)})$.
       			\item Compute the importance weights:
       				\begin{equation}
					w(\widetilde{{\bm \theta}}^{(n)}) = \frac{\target(\widetilde{{\bm \theta}}^{(n)})}{\proposal(\widetilde{{\bm \theta}}^{(n)}|{\bm \theta}^{(t-1)})}, \quad \text{for} \quad  n=1,\ldots,N.
        				\end{equation}
     			\item Select one sample $\widetilde{{\bm \theta}}^{(j)} \in \{\widetilde{{\bm \theta}}^{(1)},\ldots,\widetilde{{\bm \theta}}^{(N)}\}$, according to the probability mass function
				\begin{equation}
					\bar{w}_n = \frac{w(\widetilde{{\bm \theta}}^{(n)})}{\sum_{i=1}^N w(\widetilde{{\bm \theta}}^{(i)})}.
				\end{equation}
			\item Draw $N-1$ auxiliary samples ${\bf v}^{(1)},\ldots, {\bf v}^{(j-1)}, {\bf v}^{(j+1)}, \ldots, {\bf v}^{(N)} \sim q({\bm \theta}|\widetilde{{\bm \theta}}^{(j)})$, and set
				${\bf v}^{(j)}={\bm \theta}^{(t-1)}$.
			\item Compute the weights of the auxiliary samples,
				\begin{equation}
					w({\bf v}^{(n)}) = \frac{\target({\bf v}^{(n)})}{\proposal({\bf v}^{(n)}|\widetilde{{\bm \theta}}^{(j)})}, \quad \text{for} \quad  n=1,\ldots,N,
				\end{equation}
				and the acceptance probability of $\widetilde{{\bm \theta}}^{(j)}$:
				\begin{equation}
					\alpha_t \equiv \alpha(\widetilde{{\bm \theta}}^{(j)},{\bm \theta}^{(t-1)}) = \min\left[1,\frac{\sum_{n=1}^N w(\widetilde{{\bm \theta}}^{(n)})}{\sum_{n=1}^N w({\bf v}^{(n)})}\right],
				\label{eq:alphaMTM}
				\end{equation}
			\item Draw $u \sim \uniform([0,1))$. If $u \le \alpha_t$, accept $\widetilde{{\bm \theta}}^{(j)}$ and set ${\bm \theta}^{(t)} = \widetilde{{\bm \theta}}^{(j)}$.
				Otherwise (i.e., if $u > \alpha_t$), reject $\widetilde{{\bm \theta}}^{(j)}$ and set ${\bm \theta}^{(t)} = {\bm \theta}^{(t-1)}$.
		\end{enumerate}
    \item Approximate the integral in Eq. \eqref{eq:intBayes} using Eq. \eqref{EstimatorDavid}.
\end{enumerate}
}
\label{alg:MTM}
\end{alg}

Note that, for $N=1$, we have $\widetilde{{\bm \theta}}^{(j)} = \widetilde{{\bm \theta}}^{(1)}$, ${\bf v}^{(1)} = {\bm \theta}^{(t-1)}$, and the acceptance probability of the MTM method becomes
\begin{align}
   \alpha_t = \alpha(\widetilde{{\bm \theta}}^{(1)},{\bm \theta}^{(t-1)}) &= \min\left[1,\frac{w(\widetilde{{\bm \theta}}^{(1)})}{w({\bf v}^{(1)})}\right], \nonumber \\
   	&= \min\left[1,\frac{w(\widetilde{{\bm \theta}}^{(1)})}{w({\bm \theta}^{(t-1)})}\right],  \nonumber \\
	&= \min\left[1,\frac{\pi(\widetilde{{\bm \theta}}^{(1)}) q({\bm \theta}^{(t-1)}|\widetilde{{\bm \theta}}^{(1)})}{\target({\bm \theta}^{(t-1)}) \proposal(\widetilde{{\bm \theta}}^{(1)}|{\bm \theta}^{(t-1)})}\right],  
\end{align}
which is the  acceptance probability of the classical MH technique shown in Algorithm \ref{alg:metropolis} with $\widetilde{{\bm \theta}}^{(1)}$ playing the role of ${\bm \theta}'$.
Several variants of the standard MTM method shown in Algorithm \ref{alg:MTM} have been studied.
For instance, some authors have considered the use of correlated tries or different proposal PDFs \cite{cascralei:2012,LucaJesse2}.

\subsubsection{Independent Multiple Try Metropolis (I-MTM) schemes}

\vspace*{12pt}

The MTM method described in Algorithm \ref{alg:MTM} requires drawing $2N-1$ samples at each iteration ($N$ candidates and $N-1$ auxiliary samples) and only $N-1$ of those samples are used in the acceptance probability function.
The generation of the auxiliary points,
\begin{equation*}
	{\bf v}^{(1)},\ldots, {\bf v}^{(j-1)}, {\bf v}^{(j+1)}, \ldots, {\bf v}^{(N)} \sim \proposal({\bm \theta}|\widetilde{{\bm \theta}}^{(j)}),
\end{equation*}
can be avoided if the proposal PDF is independent from the previous state, i.e., $\proposal({\bm \theta}|{\bm \theta}^{(t-1)}) = q({\bm \theta})$.
In this case, we should draw $N-1$ samples again from $\proposal({\bm \theta})$ at step 2(d) of Algorithm \ref{alg:MTM}.
However, since we have already drawn $N$ samples from $\proposal({\bm \theta})$ at step 2(a) of Algorithm \ref{alg:MTM}, we can set 
\begin{equation}
\label{AuxP_2}
{\bf v}^{(1)}=\widetilde{{\bm \theta}}^{(1)},\ldots,{\bf v}^{(j-1)}=\widetilde{{\bm \theta}}^{(j-1)}, {\bf v}^{(j)}={\bm \theta}^{(t-1)}, {\bf v}^{(j+1)}=\widetilde{{\bm \theta}}^{(j+1)}\dots {\bf v}^{(N-1)}=\widetilde{{\bm \theta}}^{(N)},
\end{equation}
without jeopardizing the ergodicity of the chain.
Hence, we can avoid step 2(d) in Algorithm \ref{alg:MTM}, and  the acceptance probability becomes  
\begin{equation}
	\alpha_t \equiv \alpha(\widetilde{{\bm \theta}}^{(j)},{\bm \theta}^{(t-1)})
		= \min\left[1,\frac{w(\widetilde{{\bm \theta}}^{(j)})+\sum_{n=1,n\neq j}^N w(\widetilde{{\bm \theta}}^{(n)})}{w({\bm \theta}^{(t-1)})+\sum_{n=1,n\neq j}^N w(\widetilde{{\bm \theta}}^{(n)})}\right].
\label{eq:alphaMTM_fora}
\end{equation}
The I-MTM technique is shown in Algorithm \ref{alg:I-MTM}.
Note that Eq. \eqref{eq:alphaMTM_fora} can be expressed alternatively as 
\begin{equation}
   \alpha(\widetilde{{\bm \theta}}^{(j)},{\bm \theta}^{(t-1)}) = \min\left[1,\frac{\widehat{\partition}_1}{\widehat{\partition}_2}\right],   
\label{eq:alphaMTM_fora2}
\end{equation}
where we have denoted 
\begin{subequations}
\begin{align} 
	\widehat{\partition}_1 & = \frac{1}{N}\sum_{n=1}^N w(\widetilde{{\bm \theta}}^{(n)}), \label{eq:Z1imtm}\\
	\widehat{\partition}_2 & = \frac{1}{N} \left[w({\bm \theta}^{(t-1)})+\sum_{\substack{n=1\\n\neq j}}^N w(\widetilde{{\bm \theta}}^{(n)})\right]. \label{eq:Z2imtm}
\end{align}
\end{subequations}

\newpage

\begin{alg}{Independent Multiple Try Metropolis (I-MTM).}
{\begin{enumerate}
	\item \textbf{Initialization:} Choose a proposal function $\proposal({\bm \theta})$, an initial state ${\bm \theta}^{(0)}$,
    		the total number of iterations ($\niter$), and the number of tries ($N$).
	\item \texttt{FOR} $t = 1, \ldots, \niter$:
    	\begin{enumerate}
         	\item Draw $\widetilde{{\bm \theta}}^{(1)},\widetilde{{\bm \theta}}^{(2)},\ldots, \widetilde{{\bm \theta}}^{(N)}  \sim \proposal({\bm \theta})$.
        		\item Compute the importance weights:
			\begin{equation}
				w(\widetilde{{\bm \theta}}^{(n)}) = \frac{\target(\widetilde{{\bm \theta}}^{(n)})}{\proposal(\widetilde{{\bm \theta}}^{(n)})}, \quad \text{for} \quad  n=1,\ldots,N.
			\end{equation}
		\item Select one sample $\widetilde{{\bm \theta}}^{(j)} \in \{\widetilde{{\bm \theta}}^{(1)},\ldots,\widetilde{{\bm \theta}}^{(N)}\}$, according to the following probability mass function:
			\begin{equation*}
				\bar{w}_n = \frac{w(\widetilde{{\bm \theta}}^{(n)})}{\sum_{i=1}^N w(\widetilde{{\bm \theta}}^{(i)})}.
			\end{equation*}
		\item Compute the acceptance probability of $\widetilde{{\bm \theta}}^{(j)}$:
			\begin{subequations}
			\begin{align}
				\alpha_t \equiv \alpha(\widetilde{{\bm \theta}}^{(j)},{\bm \theta}^{(t-1)})
                    			&= \min\left[1,\frac{w(\widetilde{{\bm \theta}}^{(j)})+\sum_{n=1,n\neq j}^N w(\widetilde{{\bm \theta}}^{(n)})}{w({\bm \theta}^{(t-1)})
						+\sum_{n=1,n\neq j}^N w(\widetilde{{\bm \theta}}^{(n)})}\right],\\
                 			&= \min\left[1,\frac{\widehat{\partition}_1}{\widehat{\partition}_2}\right],   
                		\label{eq:alphaIMTM}
			\end{align}
			\end{subequations}
			where $\widehat{\partition}_1$ and $\widehat{\partition}_2$ are given by \eqref{eq:Z1imtm} and \eqref{eq:Z2imtm}, respectively.
		\item Draw $u \sim \uniform([0,1))$. If $u \le \alpha_t$, accept $\widetilde{{\bm \theta}}^{(j)}$ and set ${\bm \theta}^{(t)} = \widetilde{{\bm \theta}}^{(j)}$.
				Otherwise (i.e., if $u > \alpha_t$), reject $\widetilde{{\bm \theta}}^{(j)}$ and set ${\bm \theta}^{(t)} = {\bm \theta}^{(t-1)}$.
	\end{enumerate}
    \item Approximate the integral in Eq. \eqref{eq:intBayes} using Eq. \eqref{EstimatorDavid}.
\end{enumerate}
}
\label{alg:I-MTM}
\end{alg}

From the IS theory (see Section \ref{sec:is}), we know that both $\widehat{\partition}_1$ and $\widehat{\partition}_2$ are unbiased estimators of the normalizing constant (a.k.a, partition function or marginal likelihood) of the target, $\partition$.
Moreover, Eq. \eqref{eq:Z2imtm} suggests that other more sophisticated unbiased estimators of $\partition$ could be used without jeopardizing the ergodicity of the I-MTM algorithm.
For instance, instead of recycling the samples generated in the same iteration as the auxiliary points in Eq. \eqref{AuxP_2},  we could reuse samples generated in the previous iteration. %
This alternative version of the I-MTM method (I-MTM2) is described in Algorithm \ref{alg:I-MTM2}.
The I-MTM2 method is related to the well-known Particle Metropolis-Hastings (PMH) algorithm \cite{PMCMC} (see \cite{GIS17,MTM_PMH14} for further considerations).
The ergodicity of I-MTM2 is thus ensured, since it can be interpreted as a PMH algorithm where no resampling is applied (implying that the resulting candidates are independent from each other).


\newpage

\begin{alg}{Alternative version of I-MTM method (I-MTM2)}
{\begin{enumerate}
	\item \textbf{Initialization:} Choose a proposal function $\proposal({\bm \theta})$, an initial state ${\bm \theta}^{(0)}$, an initial estimate of the normalizing constant of the target
		$\widehat{Z}_0$, the total number of iterations ($\niter$), and the number of tries ($N$).
    \item \texttt{FOR} $t = 1, \ldots, \niter$:
    	\begin{enumerate}
         	\item Draw $\widetilde{{\bm \theta}}^{(1)},\widetilde{{\bm \theta}}^{(2)},\ldots, \widetilde{{\bm \theta}}^{(N)}  \sim \proposal({\bm \theta})$.
       		\item Compute the importance weights:
       			\begin{equation}
				w(\widetilde{{\bm \theta}}^{(n)}) = \frac{\target(\widetilde{{\bm \theta}}^{(n)})}{\proposal(\widetilde{{\bm \theta}}^{(n)})}, \quad \mbox{for} \quad  n=1,\ldots,N.
			\end{equation}
		\item Select one sample $\widetilde{{\bm \theta}}^{(j)} \in \{\widetilde{{\bm \theta}}^{(1)},\ldots,\widetilde{{\bm \theta}}^{(N)}\}$, according to the following probability mass function:
			\begin{subequations}
			\begin{align}
				\bar{w}_n & = \frac{1}{N\widehat{\partition}'} w(\widetilde{{\bm \theta}}^{(n)}), \\
				\widehat{\partition}' & = \frac{1}{N}\sum_{i=1}^N w(\widetilde{{\bm \theta}}^{(i)}).
			\end{align}
			\end{subequations}
		\item Compute the acceptance probability of $\widetilde{{\bm \theta}}^{(j)}$:
			  \begin{equation}
			  	\alpha_t \equiv \alpha(\widetilde{{\bm \theta}}^{(j)},{\bm \theta}^{(t-1)}) = \min\left[1,\frac{\widehat{Z}'}{\widehat{Z}_{t-1}}\right].
                		  \label{eq:alphaIMTM2}
		  	  \end{equation}
		\item Draw $u \sim \uniform([0,1))$. If $u \le \alpha_t$, accept $\widetilde{{\bm \theta}}^{(j)}$, setting ${\bm \theta}^{(t)} = \widetilde{{\bm \theta}}^{(j)}$ and $\widehat{Z}_t=\widehat{Z}'$.
				Otherwise (i.e., if $u > \alpha_t$), reject $\widetilde{{\bm \theta}}^{(j)}$, setting ${\bm \theta}^{(t)} = {\bm \theta}^{(t-1)}$ and $\widehat{Z}_t=\widehat{Z}_{t-1}$.
	\end{enumerate}
    \item Approximate the integral in Eq. \eqref{eq:intBayes} using Eq. \eqref{EstimatorDavid}.
\end{enumerate}
}
\label{alg:I-MTM2}
\end{alg}

\subsubsection{Group Metropolis Sampling}

\vspace*{12pt}

The  auxiliary weighted samples in the previous I-MTM schemes (i.e., the $N-1$ samples drawn at each iteration which are not selected for comparison with the previous state ${\bm \theta}^{(t-1)}$) can be recycled in order to provide a final Monte Carlo estimator \cite{GISssp16,GIS17}.
This leads to Algorithm \ref{alg:GMS}, known as Group Metropolis Sampling (GMS).
GMS can be considered an extension (for $N>1$ candidates) of the algorithm described in \cite{Casella96}, where the authors show how to recycle and include the samples rejected in one run of a standard MH method (i.e., $N=1$ in this case) into a unique consistent estimator. 
GMS yields a sequence of sets of weighted samples, $\mathcal{S}_t=\{{\bm \theta}^{(t,n)},\rho^{(t,n)}\}_{n=1}^N$ for $t=1,\ldots,T$, where we have denoted as $\rho^{(t,n)}$ the importance weights assigned to the samples ${\bm \theta}^{(t,n)}$.
All the samples are then employed to obtain a joint particle approximation of the target.
This approximation can then be used to compute any desired moment of the target PDF as
\begin{align}
	\widehat{I}_{N(\niter-\nburn)} & = \frac{1}{\niter-\nburn}\sum_{t=\nburn+1}^{\niter} \sum_{n=1}^N \frac{\rho^{(t,n)}}{\sum_{i=1}^N \rho^{(t,i)}} g({\bm \theta}^{(t,n)})
			\nonumber \\
		& = \frac{1}{\niter-\nburn}\sum_{t=\nburn+1}^{\niter} \widehat{I}_N^{(t)},
\label{Echecazzo}
\end{align}
where $\widehat{I}_N^{(t)} = \sum_{n=1}^N \frac{\rho^{(t,n)}}{\sum_{i=1}^N \rho^{(t,i)}} g({\bm \theta}^{(t,n)})$ and $\nburn$ is the burn-in period, as usual.

\begin{alg}{Group Metropolis Sampling (GMS).}
{\begin{enumerate}
	\item \textbf{Initialization:} Choose a proposal function $\proposal({\bm \theta})$, an initial state ${\bm \theta}^{(0)}$, an initial estimate of the normalizing constant of the target
		$\widehat{Z}_0$, the total number of iterations ($\niter$), and the number of tries ($N$).
	\item \texttt{FOR} $t = 1, \ldots, \niter$:
    		\begin{enumerate}
        			\item Draw $\widetilde{{\bm \theta}}^{(1)},\widetilde{{\bm \theta}}^{(2)},\ldots, \widetilde{{\bm \theta}}^{(N)}  \sim \proposal({\bm \theta})$.
       			\item Compute the importance weights:
				\begin{equation}
					w(\widetilde{{\bm \theta}}^{(n)}) = \frac{\target(\widetilde{{\bm \theta}}^{(n)})} {\proposal(\widetilde{{\bm \theta}}^{(n)})}, \quad \text{for} \quad  n=1,\ldots,N.
				\end{equation}
				Define $\mathcal{S}'=\{\widetilde{{\bm \theta}}^{(n)},w(\widetilde{{\bm \theta}}^{(n)})\}_{n=1}^N$ and compute
				$\widehat{Z}'=\frac{1}{N}\sum_{n=1}^N w(\widetilde{{\bm \theta}}^{(n)})$.
			\item Compute the acceptance probability:
				\begin{equation}
					\alpha_t \equiv \alpha(\mathcal{S}',\mathcal{S}_{t-1}) = \min\left[1, \frac{\widehat{Z}'}{\widehat{Z}_{t-1}}\right].
				\label{AlfaGMS}
				\end{equation}
		\item Draw $u \sim \uniform([0,1))$. If $u \le \alpha_t$, accept $\mathcal{S}'$, setting $\widehat{Z}_{t}=\widehat{Z}'$ and
			\begin{equation}
				\mathcal{S}_t = \left\{{\bm \theta}^{(t,n)} = \widetilde{{\bm \theta}}^{(n)}, \rho^{(t,n)} = w(\widetilde{{\bm \theta}}^{(n)})\right\}_{n=1}^N.
			\end{equation}
			Otherwise (i.e., if $u > \alpha_t$), reject $\mathcal{S}'$, setting $\widehat{Z}_{t}=\widehat{Z}_{t-1}$ and $\mathcal{S}_t=\mathcal{S}_{t-1}$.
	\end{enumerate}
	\item Approximate the integral in Eq. \eqref{eq:intBayes} using Eq. \eqref{Echecazzo}.
\end{enumerate}
}
\label{alg:GMS}
\end{alg}

GMS is related to the MTM schemes previously described \cite{LucaJesse2,MTM_PMH14}, even though no resampling steps are applied at each iteration in GMS.
Nevertheless, we can recover an MTM chain from the GMS output by applying one resampling step when $\mathcal{S}_t\neq  \mathcal{S}_{t-1}$, i.e.,       
\begin{equation}
	{\bm \theta}^{(t)} = 
		\begin{cases}
			{\bm \theta}' \sim \sum_{n=1}^N \frac{\rho^{(t,n)}}{\sum_{i=1}^N \rho^{(t,i)}}  \delta({\bm \theta}-{\bm \theta}^{(t,n)}), & \text{if}  \quad \mathcal{S}_t \neq  \mathcal{S}_{t-1}, \\
			{\bm \theta}^{(t-1)}, & \text{if} \quad \mathcal{S}_t =  \mathcal{S}_{t-1},
		\end{cases}
\label{RecChain}
\end{equation}
for $t=1,\ldots,\niter$.
More specifically, $\{{\bm \theta}^{(t)}\}_{t=1}^T$ is equivalent to the Markov chain obtained in one run of the I-MTM2 technique shown in Algorithm \ref{alg:I-MTM2}.
GMS can also be interpreted as an iterative importance sampling scheme, where an IS approximation using $N$ samples is built at each iteration and compared with the previous IS approximation.
This procedure is iterated $\niter$ times, and all the accepted IS estimators, $\widehat{I}_N^{(t)}$, are finally combined to provide a unique global approximation using $N(\niter-\nburn)$ samples.
Note that the temporal combination of the IS estimators is obtained dynamically by the random repetitions due to the rejections in the MH test.
Therefore, the complete procedure for weighting the samples generated by GMS can be interpreted as the composition of two weighting schemes: (a) by an importance sampling approach building $\{\rho^{(t,n)}\}_{n=1}^N$ and (b) by the possible random repetitions due to the rejections in the MH test.

\subsubsection{Ensemble MCMC}
\label{EnMCMC}
\vspace*{12pt}

Another alternative procedure, called Ensemble MCMC and involving several tries at each iteration, has been proposed in \cite{OMCMC,Calderhead14,Neal11}.
In this section, we present the simplest version, which employs a proposal PDF, $\proposal({\bm \theta}),$ independent of the previous state of the chain.
At each iteration, the ensemble MCMC method (summarized in Algorithm \ref{alg:En-MCMC}) generates $N$ new samples, $\widetilde{{\bm \theta}}^{(1)},\widetilde{{\bm \theta}}^{(2)},\ldots, \widetilde{{\bm \theta}}^{(N)}$ and then draws the new state ${\bm \theta}^{(t)}$ from a set of $N+1$ samples that includes the previous state, $\{\widetilde{{\bm \theta}}^{(1)},\ldots,\widetilde{{\bm \theta}}^{(N)},\widetilde{{\bm \theta}}^{(N+1)}={\bm \theta}^{(t-1)}\}$, according to the following probabilities:
\begin{equation}
     \bar{w}_j = \frac{w(\widetilde{{\bm \theta}}^{(j)})}{\sum_{i=1}^N w(\widetilde{{\bm \theta}}^{(i)})+ w({\bm \theta}^{(t-1)})}, \qquad j=1,\ldots,N+1,
\label{A_enMCMC}
\end{equation}
where $w({\bm \theta}) = \frac{\pi({\bm \theta})}{q({\bm \theta})}$ denotes again the standard IS weight.
Note that, for $N=1$, Eq. \eqref{A_enMCMC} becomes 
\begin{align}
     \bar{w}_j & = \frac{w(\widetilde{{\bm \theta}}^{(j)})}{ w(\widetilde{{\bm \theta}}^{(i)})+ w({\bm \theta}^{(t-1)})} \nonumber \\
     & = \frac{\frac{\target(\widetilde{{\bm \theta}}^{(j)})}{\proposal(\widetilde{{\bm \theta}}^{(j)})}}
     	{\frac{\target(\widetilde{{\bm \theta}}^{(j)})}{\proposal({\bm \theta}^{(j)})}+ \frac{\target({\bm \theta}^{(t-1)})}{q({\bm \theta}^{(t-1)})}} \nonumber \\
     & = \frac{\pi(\widetilde{{\bm \theta}}^{(j)})\proposal({\bm \theta}^{(t-1)})}{\target(\widetilde{{\bm \theta}}^{(j)})\proposal({\bm \theta}^{(t-1)})+ \target({\bm \theta}^{(t-1)})\proposal(\widetilde{{\bm \theta}}^{(j)})},  
 \label{A_enMCMC_2}
\end{align}
which is Barker's acceptance function, as given by Eq. \eqref{eq:alphaBH}, with an independent proposal density and $\widetilde{{\bm \theta}}^{(j)}$ playing the role of ${\bm \theta}'$ in \eqref{eq:alphaBH}. See \cite[Appendix C]{OMCMC} for a proof of the ergodicity.

\newpage

\begin{alg}{Ensemble MCMC with an independent proposal PDF.}
{\begin{enumerate}
	\item \textbf{Initialization:} Choose a proposal function $\proposal({\bm \theta})$, an initial state ${\bm \theta}^{(0)}$,
    		the total number of iterations ($\niter$), and the number of tries ($N$).
    	\item \texttt{FOR} $t = 1, \ldots, \niter$:
    		\begin{enumerate}
        			\item Draw $\widetilde{{\bm \theta}}^{(1)},\widetilde{{\bm \theta}}^{(2)},\ldots, \widetilde{{\bm \theta}}^{(N)}  \sim \proposal({\bm \theta})$.
       			\item Compute the importance weights:
       				\begin{equation}           
       					w(\widetilde{{\bm \theta}}^{(n)}) = \frac{\target(\widetilde{{\bm \theta}}^{(n)})}{\proposal(\widetilde{{\bm \theta}}^{(n)})}, \quad \mbox{for} \quad  n=1,\ldots,N.
        				\end{equation}
     			\item Select one sample $\widetilde{{\bm \theta}}^{(j)} \in \{\widetilde{{\bm \theta}}^{(1)},\ldots,\widetilde{{\bm \theta}}^{(N)},\widetilde{{\bm \theta}}^{(N+1)}={\bm \theta}^{(t-1)}\}$,
				according to the probability mass function
				\begin{equation}
					\bar{w}_j = \frac{w(\widetilde{{\bm \theta}}^{(j)})}{\sum_{i=1}^N w(\widetilde{{\bm \theta}}^{(i)})+ w({\bm \theta}^{(t-1)})}.
				\end{equation}
				Set ${\bm \theta}^{(t)} = \widetilde{{\bm \theta}}^{(j)}$.
		\end{enumerate}
  \item Approximate the integral in Eq. \eqref{eq:intBayes} using Eq. \eqref{EstimatorDavid}.
\end{enumerate}
}
\label{alg:En-MCMC}
\end{alg}
\subsection{MCMC with multiple candidates for the estimation of a dynamic parameter}
\label{PMH_bigsec}

\vspace*{12pt}

In this section, we consider that the parameter of interest to be estimated (or at least part of it) is a dynamical variable, such as the state in a state-space model.
In Section \ref{PMHsection}, the parameter of interest consists of a dynamical variable ${\bf x}$, i.e.,  ${\bm \theta}={\bf x}$.
In Section \ref{PMMH}, we consider the more general scenario where the parameter of interest is formed by both  a dynamical variable ${\bf x}$  and static variable ${\bm \lambda}$, i.e., ${\bm \theta}=[{\bf x}, {\bm \lambda}]^{\top}$.

\subsubsection{Particle Metropolis-Hastings (PMH) algorithms}
\label{PMHsection}

\vspace*{12pt}

Let us assume that the variable of interest is a dynamical variable, i.e., ${\bm \theta}={\bf x}=x_{1:D}=[x_1\dots,x_D]^{\top}$.
This is the case of inferring a hidden state in state-space model, for instance. More generally, let assume that we are able to factorize the target density as
\begin{equation}
\bar{\pi}({\bf x})\propto \pi({\bf x}) = \gamma_1(x_1) \prod_{d=2}^D  \gamma_d(x_d|x_{d-1}).
\end{equation}
The Particle Metropolis Hastings (PMH) method \cite{PMCMC,KOKKALA201584,Kokkala16,Lindsten14} is an efficient MCMC technique, proposed independently from the MTM algorithm, specifically designed for being applied in this framework.
Indeed, we can take advantage of the factorization of the target PDF and consider a proposal PDF decomposed in the same fashion
$$
q({\bf x})=q_1(x_1) \prod_{d=2}^D  q_d(x_d|x_{d-1}).
$$
Then, as in a batch IS scheme, given an $n$-th sample ${\bf x}^{(n)}=x_{1:D}^{(n)}\sim q({\bf x})$ with $x_{d}^{(n)}\sim q_d(x_d|x_{d-1}^{(n)})$, we assign the importance weight  
\begin{equation}
\label{EqFinRecW}
w({\bf x}^{(n)})=w_D^{(n)}=\frac{\pi({\bf x}^{(n)})}{q({\bf x}^{(n)})}=\frac{ \gamma_1(x_1^{(n)}) \gamma_2(x_2^{(n)}|x_1^{(n)}) \cdots  \gamma_D(x_D^{(n)}|x_{D-1}^{(n)})}{q_1(x_1^{(n)}) q_2(x_2^{(n)}|x_1^{(n)}) \cdots  q_D(x_D^{(n)}|x_{D-1}^{(n)})}.
\end{equation}
The structure above suggests the use of a sequential approach.
Thus, PMH uses an SIR approach (see Section \ref{SIS}) to provide the particle approximation $
 \widehat{\pi}({\bf x}|{\bf x}^{(1:N)})= \sum_{i=1}^N \bar{w}_D^{(i)}\delta({\bf x}-{\bf x}^{(i)})$, where $ \bar{w}_D^{(i)}=\frac{w_D^{(i)}}{\sum_{n=1}^Nw_D^{(n)}}$ and $w_D^{(i)}=w({\bf x}^{(i)})$ is given by Eq. \eqref{EqFinRecW}.
Then, one particle is drawn from this approximation, i.e., with a probability proportional to the corresponding normalized weight.

\vspace*{12pt}
 
\noindent
{\bf Estimation of the marginal likelihood $Z$.} SIR combines the SIS approach with the application of resampling procedures. In SIR, a consistent estimator of $Z$ is given by 
\begin{equation}
\label{EstZ2}
\widetilde{Z}=\prod_{d=1}^D\left[\sum_{n=1}^N{\bar w}_{d-1}^{(n)}\beta_d^{(n)}\right],
\end{equation}
where
\begin{equation*}
	\bar{w}_{d-1}^{(i)}=\frac{w_{d-1}^{(i)}}{\sum_{n=1}^Nw_{d-1}^{(n)}}.
\end{equation*}
Due to the application of the resampling, in SIR the standard estimator 
\begin{equation}
\widehat{Z}=\frac{1}{N}\sum_{n=1}^N w_{D}^{(n)}=\frac{1}{N}\sum_{n=1}^N w({\bf x}^{(n)}),
\end{equation}
is a possible alternative {\it only if} a proper weighting of the resampled particles is applied \cite{GISssp16,GIS17}.
If a proper weighting of a resampled particle is employed, both $\widetilde{Z}$ and $\widehat{Z}$ are equivalent estimators of $Z$ \cite{GISssp16,GIS17,MTM_PMH14}.
Without the use of resampling steps (i.e., in SIS),  $\widetilde{Z}$ and $\widehat{Z}$ are also equivalent estimators \cite{GIS17}. 

The complete description of PMH is provided in Algorithm \ref{alg:PMH} considering the use of $\widetilde{Z}$.
At each iteration, a particle filter is run to obtain an approximation of the measure of the target with $N$ weighted samples.
Then, a sample among the $N$ weighted particles is chosen by applying a single resampling step.
This selected sample is then accepted or rejected as the next state of the chain according to an MH-type acceptance probability, which involves two estimators of the marginal likelihood $Z$. 

\newpage

\begin{alg}{Particle Metropolis-Hastings (PMH).}
{\begin{enumerate}
\item \textbf{Initialization:} Choose a initial state ${\bf x}_0$ and obtain an initial estimation $\widetilde{Z}_{0}\approx Z$.
\item \texttt{For} $t=1,\ldots,T$:
\begin{enumerate}
\item\label{PMHahora} Employ an SIR approach to draw $N$ particles and weight them, $\{{\bf x}^{(i)},w_D^{(i)}\}_{i=1}^N$, i.e., sequentially obtain a particle approximation
$\widehat{\pi}({\bf x})= \sum_{i=1}^N \bar{w}_D^{(i)}\delta({\bf x}-{\bf x}^{(i)})$ where ${\bf x}^{(i)}=[x_{1}^{(i)},\ldots,x_{D}^{(i)}]^{\top}$. Furthermore, also obtain $\widetilde{Z}^*$ as in Eq. \eqref{EstZ2}.
\item Draw ${\bf x}^* \sim  \widehat{\pi}({\bf x}|{\bf x}^{(1:N)})$, i.e., choose a particle ${\bf x}^*=\{{\bf x}^{(1)},\ldots,{\bf x}^{(N)}\}$ with probability  $\bar{w}_D^{(i)}$, $i=1,...,N$.	 
\item Set  ${\bf x}_t={\bf x}^*$ and $\widetilde{Z}_{t}=\widetilde{Z}^*$ with probability
\begin{equation}
\label{A1pmh}
\alpha=\min\left[1, \frac{{\widetilde Z}^*}{{\widetilde Z}_{t-1}}\right],
\end{equation}
otherwise set ${\bf x}_t={\bf x}_{t-1}$ and $\widetilde{Z}_{t}=\widetilde{Z}_{t-1}$.
\end{enumerate}
\item \textbf{Return:} $\{{\bf x}_t\}_{t=1}^{T}$ with ${\bf x}_t=[x_{1,t},\ldots,x_{D,t}]^{\top}$.
\end{enumerate}
}
\label{alg:PMH}
\end{alg}

\vspace*{12pt}
 
\noindent
{\bf Relationship between MTM and PMH schemes.} A simple look at I-MTM2 and PMH shows that they are closely related \cite{MTM_PMH14}.
Indeed, the structure of the two algorithms coincides.
The main difference lies in the fact that the candidates in PMH are generated sequentially using an SIR scheme.
If no resampling steps are applied, then I-MTM2 and PMH are {\it exactly} the same algorithm, with candidates being drawn either in a \emph{batch} setting or in a \emph{sequential} way.
Indeed, both PMH and I-MTM2 can be interpreted as a standard  MH method with an independent proposal PDF and a proper weighting of a resampled particle \cite{GISssp16,GIS17}.
See \cite{martino2018review,GIS17,MTM_PMH14} for further discussions on this issue. 

\subsubsection{Particle Marginal Metropolis-Hastings (PMMH) method}
\label{PMMH}

\vspace*{12pt}

Assume now that the variable of interest is formed by both dynamical and static variables, i.e., ${\bm \theta}=[{\bf x}, {\bm \lambda}]^{\top}$.
For instance, this is the case of inferring both the hidden state ${\bf x}$ in state-space model and the static parameters ${\bm \lambda}$ of the model.
The Particle Marginal Metropolis-Hastings (PMMH) technique is an extension of PMH which addresses this problem \cite{PMCMC,Kokkala16,KOKKALA201584}.

Let us consider ${\bf x}=x_{1:D}=[x_1,x_2,\ldots,x_D]\in\mathbb{R}^{d_x}$, and an additional model parameter ${\bm \lambda} \in \mathbb{R}^{d_\lambda}$ to be inferred as well (${\bm \theta}=[{\bf x}, {\bm \lambda}]^{\top}\in \mathbb{R}^D$, with $D=d_x+d_\lambda$). Assuming a prior PDF $g_\lambda({\bm \lambda})$ over ${\bm \lambda}$, and a factorized complete posterior PDF $\bar{\pi}({\bm \theta})=\bar{\pi}({\bf x},{\bm \lambda})$, 
\begin{equation}
\label{target_PMMH}
\bar{\pi}({\bf x},{\bm  \lambda})\propto \pi({\bf x},{\bm  \lambda}) = g_\lambda({\bm \lambda}) \pi({\bf x}|{\bm \lambda}),
\end{equation}
where $\pi({\bf x}|{\bm \lambda})=\gamma_1(x_1|{\bm \lambda}) \prod_{d=2}^D\gamma_d(x_d|x_{1:d-1},{\bm \lambda})$.
For a specific value of ${\bm \lambda}$, we can use a particle filter approach, obtaining the approximation $\widehat{\pi}(\x|{\bm \lambda})=\sum_{n=1}^N {\bar w}_D^{(n)}\delta(\x-\x^{(n)})$ and the estimator $\widetilde{Z}({\bm \lambda})$, as described above.
The PMMH technique is then summarized in Algorithm \ref{alg:PMMH}.
The PDF $q_\lambda({\bm \lambda}|{\bm \lambda}_{t-1})$ denotes the proposal density for generating possible values of ${\bm \lambda}$.
Observe that, with the specific choice $q_\lambda({\bm \lambda}|{\bm \lambda}_{t-1})=g_\lambda({\bm \lambda})$, the acceptance function becomes
\begin{equation}
\alpha = \min\left[1, \frac{\widetilde{Z}({\bm \lambda}^*)}{\widetilde{Z}({\bm \lambda}_{t-1})}\right].
\end{equation}
Note also that PMMH  w.r.t. to ${\bm \lambda}$ can be interpreted as MH method where the posterior cannot be evaluated point-wise. Indeed, $\widetilde{Z}({\bm \lambda})$ approximates the marginal likelihood $p({\bf y}|{\bm \lambda})$, i.e., it can also be interpreted as a special case of the pseudo-marginal approach described below \cite{andrieu2009}.

\begin{alg}{Particle Marginal MH (PMMH).}
{\begin{enumerate}
\item  \textbf{Initialization:} Choose the initial states $\x_0$, ${\bm \lambda}_0$, and an initial approximation $\widetilde{Z}_0({\bm \lambda})\approx Z({\bm \lambda}) \approx p({\bf y}|{\bm \lambda})$.
\item \texttt{FOR} $t=1,\ldots,T$:
\begin{enumerate}
\item Draw ${\bm \lambda}^*\sim q_\lambda({\bm \lambda}|{\bm \lambda}_{t-1})$.
\item Given ${\bm \lambda}^*$, run a particle filter obtaining $\widehat{\pi}(\x|{\bm \lambda}^*)=\sum_{n=1}^N {\bar w}_D^{(n)}\delta(\x-\x^{(n)})$ and $\widetilde{Z}({\bm \lambda}^*)$, as in Eq. \eqref{EstZ2}.
\item Draw ${\bf x}^* \sim  \widehat{\pi}({\bf x}|{\bm \lambda}^*, {\bf x}^{(1:N)})$, i.e., choose a particle ${\bf x}^*=\{{\bf x}^{(1)},\ldots,{\bf x}^{(N)}\}$ with probability  $\bar{w}_D^{(i)}$, $i=1,...,N$.	 
\item Set  ${\bm \lambda}_t={\bm \lambda}^*$ and $\x_t={\bf x}^*$ with probability 
\begin{equation}
\label{AlfaPMMH}
\alpha = \min\left[1, \frac{\widetilde{Z}({\bm \lambda}^*) g_\lambda({\bm \lambda}^*) q_\lambda({\bm \lambda}_{t-1}|{\bm \lambda}^*)}{\widetilde{Z}({\bm \lambda}_{t-1})  g_\lambda({\bm \lambda}_{t-1}) q_\lambda({\bm \lambda}^*|{\bm \lambda}_{t-1})}\right].
\end{equation}
Otherwise, set ${\bm \lambda}_t={\bm \lambda}^*$ and $\x_t=\x_{t-1}$.
\end{enumerate}
\item \textbf{Return:} $\{\x_t\}_{t=1}^T$ and $\{{\bm \lambda}_t\}_{t=1}^T$.
\end{enumerate}
 }
\label{alg:PMMH}
\end{alg}

\subsubsection{Particle Gibbs algorithm}
\label{PG_sect}

\vspace*{12pt}

Note that, in order to draw from $\bar{\pi}({\bf x},{\bm  \lambda})\propto \pi({\bf x},{\bm  \lambda})=\pi(x_{1:D},{\bm  \lambda})$ in Eq. \eqref{target_PMMH}, we could use a simple Gibbs sampling approach: draw first from the conditional PDF ${\bm  \lambda}' \sim \bar{\pi}({\bm  \lambda}|{\bf x}')$ given a reference path ${\bf x}'$, and then sample a new path from the other conditional ${\bf x}''\sim\bar{\pi}({\bf x}|{\bm  \lambda}')$.
This procedure continues iteratively, drawing ${\bm  \lambda}'' \sim \bar{\pi}({\bm  \lambda}|{\bf x}'')$ and ${\bf x}'''\sim\bar{\pi}({\bf x}|{\bm  \lambda}'')$, in a Gibbs sampling fashion.
We can draw approximately the paths ${\bf x}=x_{1:D}$ from the conditional PDF $\bar{\pi}({\bf x}|{\bm  \lambda})$ by running a particle filter and then resampling once within the cloud of paths, as described in the sections above (exactly as in PMMH).
However, note that is  this procedure does not take into account the previous path ${\bf x}_{t-1}$ in order to generate the next sample ${\bf x}_{t}$, but only the ${\bm  \lambda}_{t-1}$.
Namely, ${\bf x}''$ is drawn from $\bar{\pi}({\bf x}|{\bm  \lambda}')$ that does not depend on ${\bf x}'$.
The {\it Particle Gibbs} (PG) technique is an extension of the simple Gibbs approach previously described that also considers the last sample generated, ${\bf x}_{t-1}$, to draw the next path ${\bf x}_{t}$ \cite{PMCMC,Lindsten14,Kokkala16,KOKKALA201584}.\footnote{Related ideas about taking into account the previous path have been also discussed in \cite{MTM_PMH14}.}
Algorithm \ref{alg:PG} summarizes the PG algorithm, which is guaranteed to generate a Markov chain with $\bar{\pi}({\bf x},{\bm  \lambda})$ as invariant density \cite{PMCMC,Lindsten14,Kokkala16}.

\begin{alg}{Particle Gibbs (PG).}
{\begin{enumerate}
\item  \textbf{Initialization:} Choose the initial states $\x_0$, ${\bm \lambda}_0$, the number of particles $N$ and the total number of Gibbs iterations $T$.
\item \texttt{FOR} $t=1,\ldots,T$:
\begin{enumerate}
\item Run the Conditional Particle Filter (CPF) described in Section \ref{CPF} (only $N-1$ particles are randomly generated), given ${\bm \lambda}_{t-1}$ and the reference path ${\bf x}_{t-1}$. Thus, we obtain $\widehat{\pi}(\x|{\bm \lambda}_{t-1},{\bf x}_{t-1})=\sum_{n=1}^N {\bar w}_D^{(n)}\delta(\x-\x^{(n)})$.
\item Draw ${\bf x}_{t} \sim  \widehat{\pi}(\x|{\bm \lambda}_{t-1},{\bf x}_{t-1})$, i.e., choose a particle ${\bf x}^*=\{{\bf x}^{(1)},\ldots,{\bf x}^{(N)}\}$ with probability  $\bar{w}_D^{(i)}$, $i=1,...,N$.
\item Draw ${\bm \lambda}_t\sim \bar{\pi}({\bm  \lambda}|{\bf x}_t)$, as in a standard Block Gibbs sampling \cite{geman1984stochastic,gelfand1990sampling}.
\end{enumerate}
\item \textbf{ Return:} $\{\x_t\}_{t=1}^T$ and $\{{\bm \lambda}_t\}_{t=1}^T$.
\end{enumerate}
}
\label{alg:PG}
\end{alg}

\subsection{Pseudo-marginal MCMC methods}
\label{PSM_MCMC}

\vspace*{12pt}

There are numerous applications where the target density ${\bar \pi}$ is not available in closed form and cannot be evaluated pointwise exactly but only approximately.
For instance, in some situations we can evaluate the joint target ${\bar \pi}({\bm \lambda},\x)$, but we are actually only interested on the marginal target PDF, ${\bar \pi}({\bm \lambda}) =\int_{\mathcal{X}} {\bar \pi}({\bm \lambda},\x)d\x$.
If we cannot compute this integral, we cannot evaluate ${\bar \pi}({\bm \lambda})$.
One simple possibility is to run an MCMC algorithm in the extended space $[{\bm \lambda},\x]$ and then consider only the first component.
However, this approach can be very inefficient in many cases.
An alternative is to run an MCMC algorithm in the subspace of ${\bm \lambda}$, addressing ${\bar \pi}({\bm \lambda})$ but using an unbiased estimator $\widehat{\pi}({\bm \lambda})$ of ${\bar \pi}({\bm \lambda})$.
This unbiased estimator can be provided by another Monte Carlo method.
This is exactly the case of the PMMH algorithm described in Algorithm \ref{alg:PMMH}.
Note that, if we are interested only in making inference about ${\bm \lambda}$, then the variable $\x$ can be considered integrated out using a Monte Carlo approximation \cite{PMCMC}.  

In other related scenarios, the likelihood function $\ell(\y|{\bm \theta})$ cannot be evaluated and, fixing a generic value ${\bm \theta}$, an \emph{unbiased} estimator $\widehat{\ell}(\y|{\bm \theta})$  of the probability $\ell(\y|{\bm \theta})$ is available, i.e.,
\begin{equation}
E\left[\widehat{\ell}(\y|{\bm \theta})\right]=\ell(\y|{\bm \theta}), \qquad  \forall {\bm \theta}\in {\bm \Theta}.
\label{eq:unbiased_ell}
\end{equation}
Note that this estimator must be unbiased and valid for all possible values of $ {\bm \theta}\in {\bm \Theta}$.
If $\widehat{\ell}(\y|{\bm \theta})$ is available, then different Monte Carlo algorithms, such as MCMC techniques, can be applied considering the approximated posterior density \cite{andrieu2009}
 \begin{equation}
\widehat{ \pi}({\bm \theta})=\widehat{ \pi}({\bm \theta}|\y)\propto \widehat{\ell}(\y|{\bm \theta})p_0({\bm \theta}),
\label{eq:posterior_approx}
\end{equation}
where $p_0({\bm \theta})$ represents the prior PDF.
Since the IS method is often used to provide the unbiased estimator $\widehat{\ell}(\y|{\bm \theta})$ \cite{andrieu2009}, usually we have IS-within-MCMC algorithms in the pseudo-marginal setup.
The generic Pseudo-marginal MH method is summarized in Algorithm \ref{PSM_MCMC_Alg}.
This method is also known in the literature as {\it Group Independence MH} (GIMH) and a variant of this method is called {\it Monte Carlo within Metropolis} (MCWM) \cite{andrieu2009}.
They differ in the estimator, ${\widehat \pi}({\bm \theta}^{(t-1)})$, used in the denominator of the acceptance probability $\alpha$: in MCWM, ${\widehat \pi}({\bm \theta}^{(t-1)})$ is recomputed at each iteration, whereas in GIMH the value estimated in the previous iteration is recycled (as in Algorithm \ref{PSM_MCMC_Alg}).  

\begin{alg}{Generic Pseudo-Marginal MH method}
{\begin{enumerate}
	\item \textbf{Initialization:} Choose a proposal function $\proposal({\bm \theta}|{\bm \theta}^{(t-1)})$, an initial state ${\bm \theta}^{(0)}$,
    	the total number of iterations ($\niter$), and the burn-in period ($\nburn$).
    \item \texttt{FOR} $t = 1, \ldots, \niter$:
    	\begin{enumerate}
        	\item Draw ${\bm \theta}' \sim \proposal({\bm \theta}|{\bm \theta}^{(t-1)})$. 
	\item Build an unbiased estimator $\widehat{\ell}(\y|{\bm \theta}')$ of the likelihood function $\ell(\y|{\bm \theta}')$ and ${\widehat \pi}({\bm \theta}') \propto \widehat{\ell}(\y|{\bm \theta}')p_0({\bm \theta}')$.
            \item Set ${\bm \theta}^{(t)} = {\bm \theta}'$  with probability,
            \begin{eqnarray}
             \alpha &=& \min\left[1,\frac{{\widehat \pi}({\bm \theta}')\proposal({\bm \theta}^{(t-1)}|{\bm \theta}')}{{\widehat \pi}({\bm \theta}^{(t-1)})\proposal({\bm \theta}'|{\bm \theta}^{(t-1)})}\right], 
             \end{eqnarray}
              otherwise, with probability $1-\alpha$, set ${\bm \theta}^{(t)} = {\bm \theta}^{(t-1)}$.
         \end{enumerate}
         \item \textbf{Return:} ${\bm \theta}^{(t)}$ for $t = 1, \ldots, \niter$.
   \end{enumerate}
}
\label{PSM_MCMC_Alg}
\end{alg}

In the following subsections we describe four different frameworks where the pseudo-marginal approach is either required or indirectly used.
However, before describing potential applications of the pseudo-marginal approach, {let us} remark that the variance of the unbiased estimator used needs to be small in order to obtain a useful output.
Otherwise, pseudo-marginal methods can result in very slowly-mixing chains even if they converge asymptotically in the limit.
This emphasizes the importance of ensuring the geometric convergence of any MCMC algorithm to guarantee that it converges with a non-arbitrarily-slow rate.

\vspace*{12pt}

\noindent
{\bf Latent variable models.} In latent variable models, the likelihood  is often only available as an intractable integral
$$
\ell(\y|{\bm \theta})=\int_{\mathcal{Z}} \psi(\y,{\bf z}|{\bm \theta}) d{\bf z},
$$
and hence ${\pi}({\bm \theta}|\y) \propto p_0({\bm \theta})\left[\int_{\mathcal{Z}} \psi(\y,{\bf z}|{\bm \theta}) d{\bf z}\right]$, which is also intractable.
The simplest solution is to apply an MCMC algorithm for generating vectors $[{\bm \theta}',{\bf z}']$ from the joint target PDF, ${\pi}({\bm \theta},{\bf z}|\y)$, and then considering only the first component of the drawn vectors \cite{Robert04}.
More generally, an approximation of the integral $\int_{\mathcal{Z}} \psi(\y,{\bf z}|{\bm \theta}) d{\bf z}$ is required.
In some cases, this can be obtained using another Monte Carlo technique such as the IS technique.

\vspace*{12pt}

\noindent
{\bf Doubly-intractable likelihoods.} Another scenario where the posterior PDF cannot be completely evaluated is the case of the so-called ``doubly-intractable'' likelihood functions. In this situation, a ``portion'' of the likelihood is unknown or cannot be evaluated, e.g.,
\begin{equation}
\ell(\y|{\bm \theta})=\frac{1}{C({\bm \theta})} \phi(\y|{\bm \theta}),
\end{equation}
where $\phi(\y|{\bm \theta})$ can be evaluated, but 
\begin{equation}
C({\bm \theta})=\int_{{\bm \Theta}} \phi(\y|{\bm \theta}) d{\bf y}
\end{equation}
is unknown.
Hence, the value $C({\bm \theta})$ must be approximated \cite{Murray06}. A first algorithm for handling this kind of distributions was proposed in \cite{Moller06}. As an example, The Single Variable Exchange (SVE) algorithm is described in Algorithm \ref{SVE_Alg} (see \cite{Murray06}). If we denote as  $\y_{\texttt{true}}$ the actual observed data, the posterior PDF is 
\begin{align}
{\bar \pi}({\bm \theta}|\y_{\texttt{true}}) \propto \pi({\bm \theta}|\y_{\texttt{true}}) & =  \ell(\y_{\texttt{true}}|{\bm \theta}) p_0({\bm \theta}), \nonumber \\
& = \frac{1}{C({\bm \theta})} \phi(\y_{\texttt{true}}|{\bm \theta})p_0({\bm \theta}).
\end{align}
Note that, if we are able to draw samples $\y^{(k)} \sim \ell(\y|{\bm \theta}) \propto \phi(\y|{\bm \theta})$ for $k=1,...,L$, then we can approximate the constant $C({\bm \theta})$ via Monte Carlo approximation, i.e., 
\begin{equation}
C({\bm \theta})=\int_{{\bm \Theta}} \phi(\y|{\bm \theta}) d{\bf y} \approx \frac{1}{L}\sum_{k=1}^L \phi(\y^{(k)}|{\bm \theta}).
\end{equation}
If we are able to draw from $\ell(\y|{\bm \theta})$, we can use the IS method, i.e., $\y^{(k)} \sim q_y(\y) $ and then we have
\begin{equation}
C({\bm \theta})=\int_{{\bm \Theta}} \phi(\y|{\bm \theta}) d{\bf y} \approx \frac{1}{L}\sum_{k=1}^L \frac{\phi(\y^{(k)}|{\bm \theta})}{q_y(\y^{(k)})}.
\end{equation}

For the sake of simplicity, let us assume that we are able to draw from $\ell(\y|{\bm \theta})$.
Moreover, we set $L=1$ and denote $\y'=\y^{(1)} \sim \ell(\y|{\bm \theta})$. Hence, we have $C({\bm \theta})\approx \phi(\y'|{\bm \theta})$. Then, we can write the approximate posterior function as
\begin{equation}
 \pi({\bm \theta}|\y_{\texttt{true}})\approx {\widehat \pi}({\bm \theta}|\y_{\texttt{true}}, \y')=\frac{1}{\phi(\y'|{\bm \theta})} \phi(\y_{\texttt{true}}|{\bm \theta})p_0({\bm \theta}), \qquad \y'\sim \ell(\y|{\bm \theta}). 
\end{equation}
The SVE algorithm is an MH method with the target function ${\widehat \pi}({\bm \theta}|\y_{\texttt{true}}, \y')$. Note that 
$$
\frac{C({\bm \theta}^{(t-1)})}{C({\bm \theta}')}\approx \frac{\phi(\y^{(t-1)}|{\bm \theta}^{(t-1)})}{
          \phi(\y'|{\bm \theta}')},
$$
where $\y^{(t-1)} \sim \ell(\y|{\bm \theta}^{(t-1)})$ and $\y'\sim \ell(\y|{\bm \theta}')$.

\newpage

\begin{alg}{Single Variable Exchange (SVE) algorithm.}
{\begin{enumerate}
	\item \textbf{Initialization:} Choose a proposal function $\proposal({\bm \theta}|{\bm \theta}^{(t-1)})$, an initial state ${\bm \theta}^{(0)}$,
   	the total number of iterations ($\niter$), and the burn-in period ($\nburn$).
    \item \texttt{FOR} $t = 1, \ldots, \niter$:
    	\begin{enumerate}
        	\item Draw ${\bm \theta}' \sim \proposal({\bm \theta}|{\bm \theta}^{(t-1)})$. 
	\item Draw $\y' \sim \ell(\y|{\bm \theta}') \propto \phi(\y|{\bm \theta}')$.
            \item Set ${\bm \theta}^{(t)} = {\bm \theta}'$ and $\y^{(t)}=\y'$ with probability,
            \begin{eqnarray}
             \alpha &=& \min\left[1,\frac{{\widehat \pi}({\bm \theta}'|\y_{\texttt{true}}, \y')\proposal({\bm \theta}^{(t-1)}|{\bm \theta}')}{{\widehat \pi}({\bm \theta}^{(t-1)}|\y_{\texttt{true}}, \y^{(t-1)})\proposal({\bm \theta}'|{\bm \theta}^{(t-1)})}\right], \nonumber\\  
           &=&   \min\left[1, \frac{\phi(\y_{\texttt{true}}|{\bm \theta}')}{\phi(\y_{\texttt{true}}|{\bm \theta}^{(t-1)})}
           \frac{p_0({\bm \theta}')}{p_0({\bm \theta}^{(t-1)})}
           \frac{\proposal({\bm \theta}^{(t-1)}|{\bm \theta}')}{ \proposal({\bm \theta}'|{\bm \theta}^{(t-1)})} \frac{\phi(\y^{(t-1)}|{\bm \theta}^{(t-1)})}{
          \phi(\y'|{\bm \theta}')}\right] \nonumber
             \end{eqnarray}
              otherwise, with probability $1-\alpha$, set ${\bm \theta}^{(t)} = {\bm \theta}^{(t-1)}$  and $\y^{(t)}=\y^{(t-1)}$.
         \end{enumerate}
 	\item \textbf{Return:} ${\bm \theta}^{(t)}$ for $t = 1, \ldots, \niter$.
   \end{enumerate}
}
\label{SVE_Alg}
\end{alg}

\vspace*{12pt}

\noindent
 {\bf Approximate Bayesian Computation (ABC).} In many applications, the likelihood cannot be evaluated for several different reasons: (a) it is to costly and/or (b) it is unknown analytically.
 However, in some of these scenarios it is possible to generate artificial data according to the likelihood, i.e., we can simulate synthetic data from the observation model \cite{Marjoram03,Beaumont03,Marin06}.  Namely, in the  Approximate Bayesian Computation (ABC) framework, we can draw samples $[{\bm \theta}',\y']$ from the joint target density,
$$
p({\bm \theta},\y)=\ell(\y|{\bm \theta})p_0({\bm \theta}),
$$ 
with the following procedure:
\begin{enumerate}
\item Draw ${\bm \theta}'\sim p_0({\bm \theta})$ (i.e., draw ${\bm \theta}'$ from the prior).
\item Draw $\y' \sim \ell(\y|{\bm \theta}')$ (i.e., draw $\y'$ from the observation model given ${\bm \theta}'$). 
\end{enumerate}
However, we are interested in having samples from the posterior density,
\begin{equation}
{\bar \pi}({\bm \theta}|\y_{\texttt{true}}) \propto p({\bm \theta},\y_{\texttt{true}}),
\end{equation}
where $\y_{\texttt{true}}$ represents the actual observed data. To solve this issue, the underlying idea in ABC is to apply Monte Carlo techniques considering the generalized posterior PDF,
\begin{equation}
\label{ABC_target}
{\bar \pi}_\epsilon({\bm \theta},\y |\y_{\texttt{true}})  \propto \pi_\epsilon({\bm \theta},\y |\y_{\texttt{true}})= h_{\epsilon}(||\y-\y_{\texttt{true}}||)\ell(\y|{\bm \theta})p_0({\bm \theta}),
\end{equation}
where $||\cdot ||$ denotes a norm, and $h_{\epsilon}(\xi) \in [0,1]$ is a weighting function defined for $\xi\geq 0$ (with a parameter $\epsilon$) which satisfies the following conditions: the maximum value is reached at $0$ (i.e., $h_{\epsilon}(0)>h_{\epsilon}(\xi)$ for any $\xi > 0$), and the two following limits must be  fulfilled, $\lim\limits_{\epsilon \rightarrow 0}h_{\epsilon}(\xi)=\delta(\xi)$ and $\lim\limits_{\epsilon \rightarrow \infty}h_{\epsilon}(\xi)=0$.
For instance, one possible choice is 
\begin{equation}
h_{\epsilon}(||\y-\y_{\texttt{true}}||)=\exp\left[-\frac{||\y-\y_{\texttt{true}}||^2}{2\epsilon^2}\right],
\end{equation}
whereas another common alternative is 
\begin{gather}
\label{RS_h}
  h_\epsilon(||\y-\y_{\texttt{true}}||)=\left\{
   \begin{split}
         &1 \quad \mbox{ if } \quad || \y - \y_{\texttt{true}}|| \leq \epsilon, \\
         &0 \quad \mbox{ if } \quad || \y - \y_{\texttt{true}}|| > \epsilon. 
   \end{split}
   \right.
\end{gather}
Considering the weighting function of Eq. \eqref{RS_h} it is straightforward to see that, as $\epsilon \rightarrow 0$, then the generalized target ${\bar \pi}_\epsilon({\bm \theta},\y |\y_{\texttt{true}})$ becomes more and more similar to ${\bar \pi}({\bm \theta} |\y_{\texttt{true}})$, and indeed
\begin{equation}
\lim_{\epsilon \rightarrow 0} {\bar \pi}_\epsilon({\bm \theta},\y |\y_{\texttt{true}})={\bar \pi}({\bm \theta} |\y_{\texttt{true}}).
\end{equation}
An Metropolis-Hastings ABC (MH-ABC) algorithm addressing  the target density  ${\bar \pi}_\epsilon({\bm \theta},\y |\y_{\texttt{true}})$ defined in Eq. \eqref{ABC_target} with weighting function defined in Eq. \eqref{RS_h}, can be described as in Table \ref{MH-ABC}.  
Note that the extended proposal PDF in this case is
\begin{equation}
q_e({\bm \theta}, \y|{\bm \theta}^{(t-1)})=\ell(\y|{\bm \theta})\proposal({\bm \theta}|{\bm \theta}^{(t-1)}).
\end{equation}
Drawing a $[{\bm \theta}', \y'] \sim q_e({\bm \theta}, \y|{\bm \theta}^{(t-1)})$, the acceptance probability of the MH method in this case is 
\begin{equation}
             \alpha = \min\left[1,\frac{ \pi_\epsilon({\bm \theta}',\y' |\y_{\texttt{true}})\proposal_e({\bm \theta}^{(t-1)},\y^{(t-1)}|{\bm \theta}')}{\pi_\epsilon({\bm \theta}^{(t-1)},\y^{(t-1)} |\y_{\texttt{true}})\proposal_e({\bm \theta}',\y'|{\bm \theta}^{(t-1)})}\right].
\end{equation}
Then, replacing the expressions of $\pi_e$ an $q_e$, we have
      \begin{eqnarray}
             \alpha &=& \min\left[1,\frac{h_{\epsilon}(||\y'-\y_{\texttt{true}}||)\ell(\y'|{\bm \theta}')p_0({\bm \theta}')
             \ell(\y^{(t-1)}|{\bm \theta}^{(t-1)})\proposal({\bm \theta}^{(t-1)}|{\bm \theta}')}{h_{\epsilon}(||\y^{(t-1)}-\y_{\texttt{true}}||)\ell(\y^{(t-1)}|{\bm \theta}^{(t-1)})p_0({\bm \theta}^{(t-1)})\ell(\y'|{\bm \theta}')\proposal({\bm \theta}'|{\bm \theta}^{(t-1)})}\right], \nonumber \\
             &=& \min\left[1,\frac{h_{\epsilon}(||\y'-\y_{\texttt{true}}||)p_0({\bm \theta}')
             \proposal({\bm \theta}^{(t-1)}|{\bm \theta}')}{h_{\epsilon}(||\y^{(t-1)}-\y_{\texttt{true}}||)p_0({\bm \theta}^{(t-1)})\proposal({\bm \theta}'|{\bm \theta}^{(t-1)})}\right]. \nonumber 
             \end{eqnarray}
             It is important to remark that in the previous expression we do not need to evaluate the likelihood function. Finally, note that, if $h_\epsilon$ is given by Eq. \eqref{RS_h}, in order to avoid zeros in the denominator, the acceptance test involving the probability $\alpha$ can be split in two parts \cite{Beaumont03,Marin06},
\begin{gather}
\label{MH_ABC_alfa}
\alpha=\left\{
\begin{split}
& \min\left[1,\frac{p_0({\bm \theta}')
             \proposal({\bm \theta}^{(t-1)}|{\bm \theta}')}{p_0({\bm \theta}^{(t-1)})\proposal({\bm \theta}'|{\bm \theta}^{(t-1)})}\right]  \quad \mbox{ if } \quad || \y - \y_{\texttt{true}}|| \leq \epsilon, \\
&0 \qquad  \qquad  \qquad  \qquad \qquad  \qquad  \qquad   \mbox{ if } \quad || \y - \y_{\texttt{true}}|| > \epsilon. 
\end{split}
\right.
\end{gather}
Algorithm \ref{MH-ABC} uses the acceptance probability in Eq. \eqref{MH_ABC_alfa} \cite{Marjoram03,Beaumont03,Marin06}.

\newpage

\begin{alg}{Metropolis-Hastings ABC (MH-ABC) algorithm}
{\begin{enumerate}
	\item \textbf{Initialization:} Choose a proposal function $\proposal({\bm \theta}|{\bm \theta}^{(t-1)})$, an initial state ${\bm \theta}^{(0)}$,
   	the total number of iterations ($\niter$), and the burn-in period ($\nburn$).
    \item \texttt{FOR} $t = 1, \ldots, \niter$:
    	\begin{enumerate}
        	\item Draw ${\bm \theta}' \sim \proposal({\bm \theta}|{\bm \theta}^{(t-1)})$. 
	\item Draw $\y' \sim \ell(\y|{\bm \theta}')$.
         \item If  $|| \y - \y_{\texttt{true}}|| > \epsilon$, set  ${\bm \theta}^{(t)} = {\bm \theta}^{(t-1)}$  and $\y^{(t)}=\y^{(t-1)}$.	
            \item If  $|| \y - \y_{\texttt{true}}|| \leq \epsilon$, then:
            \begin{itemize}
                 \item   Set ${\bm \theta}^{(t)} = {\bm \theta}'$ and $\y^{(t)}=\y'$ with probability,
            \begin{equation}
             \alpha =  \min\left[1,\frac{p_0({\bm \theta}')
             \proposal({\bm \theta}^{(t-1)}|{\bm \theta}')}{p_0({\bm \theta}^{(t-1)})\proposal({\bm \theta}'|{\bm \theta}^{(t-1)})}\right].
             \end{equation}
              \item Otherwise, with probability $1-\alpha$, set ${\bm \theta}^{(t)} = {\bm \theta}^{(t-1)}$  and $\y^{(t)}=\y^{(t-1)}$.
            \end{itemize}
           \item \textbf{Return:} ${\bm \theta}^{(t)}$ for $t = 1, \ldots, \niter$.
         \end{enumerate}
 
   \end{enumerate}
}
\label{MH-ABC}
\end{alg}


\vspace*{12pt}

\noindent
{\bf Big Data context}. The ABC method completely avoids the evaluation of the likelihood.
As a counterpart, ABC requires the ability of drawing artificial data from the observation model.
Clearly, ABC fits very well in applications where evaluating the likelihood is expensive.
The likelihood function can be costly due to the complexity of the model, or because the size of the full dataset prohibits many evaluations of the likelihood.
Specific methodologies have been designed for this second scenario, i.e.,  when a big number of data is available.
All these techniques consider a cheaper likelihood function including only a subset of data at each iteration.
One possible strategy, often known as {\it adaptive subsampling}, consists in computing the approximate acceptance probability $\alpha$ of the standard MH method, obtained considering only a random subset  of data.
Namely, an approximate implementation of the MH test is performed in some suitable way, in order to guarantee that the performance of the resulting technique is not jeopardized (e.g., the total variation distance between the  perturbed invariant distribution and the desired target distribution is controlled) \cite{Bardenet2014, Korattikara2014,Christen2005,Sherlock2017}.
Other methods are based on the so-called {\it delayed acceptance} approach: divide the acceptance MH test into several parts involving likelihood functions with an increasing number of data \cite{Bardenet2014, Korattikara2014,Sherlock2017}.
A related strategy, called {\it early rejection}, was proposed in \cite{solonen2012}.
However, in the early rejection MCMC technique the acceptance of a new proposed state still requires the evaluation of the full-likelihood, whereas the rejection may require only the evaluation of a partial likelihood based on a subset of data.\footnote{Despite their denominations, this kind of methods (called ``delayed acceptance'' or ``early rejection'') are not directly related to the delayed rejection MH algorithm described in Section \ref{DRMS}, which always considers the complete likelihood.}
Another simple approach consists in dividing the full dataset into mini-batches, running different parallel Monte Carlo algorithms and combining all the partial estimators to obtain the global one \cite{bardenet2015markov,EmbaraMCMC,luengo2015bias,FIREFLY, LuengoBigData,Scott2013}.

\subsection{Noisy and approximate likelihood methods}
\label{Noisy_MCMC}

\vspace*{12pt}

\subsubsection{Noisy MCMC methods}
\label{sec:noisyMCMC}

\vspace*{12pt}

The aforementioned methods can be grouped in the unique framework shown in Algorithm \ref{Noisy_MCMC_Alg} (e.g., see \cite{Nicholls12,Alquier14NoisyMCMC,Medina-Aguayo2016}).
Let us assume that we are not able to evaluate the standard acceptance probability function  of the MH method,
$$
\alpha=\min\left[1, \rho({\bm \theta}^{(t-1)},{\bm \theta}')\right],
$$
where $\rho({\bm \theta}^{(t-1)},{\bm \theta}')=\frac{\pi({\bm \theta}')\proposal({\bm \theta}^{(t-1)}|{\bm \theta}')}{\pi({\bm \theta}^{(t-1)}) \proposal({\bm \theta}'|{\bm \theta}^{(t-1)}) }$.
Then, an approximation 
 $$
 \widehat{\alpha}=\min\left[1,  \widehat{\rho}({\bm \theta}^{(t-1)},{\bm \theta}',\y')\right],
 $$
where $\y' \sim \ell(\y|{\bm \theta}')$, can be used \cite{Alquier14NoisyMCMC}.
It is possible to show that, if $\widehat{\rho}({\bm \theta}^{(t-1)},{\bm \theta}',\y)$ fulfills the following condition,
$$
\int_{\mathcal{Y}} \big|\widehat{\rho}({\bm \theta}^{(t-1)},{\bm \theta}',\y)- \rho({\bm \theta}^{(t-1)},{\bm \theta}')\big| \ell(\y|{\bm \theta}') d\y \leq \delta({\bm \theta}^{(t-1)},{\bm \theta}'), \quad \forall {\bm \theta}^{(t-1)},{\bm \theta}' \in {\bm \Theta}
$$ 
the stationary density of the generated chain will approximate the desired posterior PDF \cite{Alquier14NoisyMCMC}.
Namely, $\widehat{\rho}({\bm \theta}^{(t-1)},{\bm \theta}',\y')$ with $\y' \sim \ell(\y|{\bm \theta}')$ is a randomized version of $\rho({\bm \theta}^{(t-1)},{\bm \theta}')$, and it is reasonable to require $ \big|\widehat{\rho}({\bm \theta}^{(t-1)},{\bm \theta}',\y)- \rho({\bm \theta}^{(t-1)},{\bm \theta}')\big|$ be small in order to obtain a useful approximation  \cite{Alquier14NoisyMCMC,Medina-Aguayo2016}.

\begin{alg}{Noisy MH method}
{\begin{enumerate}
	\item \textbf{Initialization:} Choose a proposal function $\proposal({\bm \theta}|{\bm \theta}^{(t-1)})$, an initial state ${\bm \theta}^{(0)}$,
   	the total number of iterations ($\niter$), and the burn-in period ($\nburn$).
    \item \texttt{FOR} $t = 1, \ldots, \niter$:
    	\begin{enumerate}
        	\item Draw ${\bm \theta}' \sim \proposal({\bm \theta}|{\bm \theta}^{(t-1)})$. 
	\item Draw $\y' \sim \ell(\y|{\bm \theta}')$.
	\item Build an estimator $\widehat{\rho}({\bm \theta}^{(t-1)},{\bm \theta}',\y')$ of the standard acceptance probability of the MH method, $\rho({\bm \theta}^{(t-1)},{\bm \theta}')$.
            \item Set ${\bm \theta}^{(t)} = {\bm \theta}'$  with probability 
            $$
             \widehat{\alpha}=\min\left[1,\widehat{\rho}({\bm \theta}^{(t-1)},{\bm \theta}',\y')\right],
           $$
              Otherwise, with probability $1-\widehat{\alpha}$, set ${\bm \theta}^{(t)} = {\bm \theta}^{(t-1)}$.
         \end{enumerate}
         \item \textbf{Return:} ${\bm \theta}^{(t)}$ for $t = 1, \ldots, \niter$.
   \end{enumerate}
}
\label{Noisy_MCMC_Alg}
\end{alg}

\subsubsection{Approximate likelihood methods}
\label{sec:approx_likelihood}

\vspace*{12pt}

In state-space models, and especially in models involving non-linear stochastic differential equations, the (marginal) likelihood $\ell(\y|{\bm \theta})$ often cannot be evaluated exactly, but we may have a simple approximation $\widehat{\ell}(\y|{\bm \theta})$ available.
For example, in non-linear state-space models we might have a non-linear Kalman filter-based Gaussian approximation of the system \cite{Sarkka:2013}, which also provides us with an approximation of the likelihood.

As discussed above, if $\widehat{\ell}(\y|{\bm \theta})$ is unbiased in the sense of Eq.~\eqref{eq:unbiased_ell}, then using the corresponding posterior distribution  \eqref{eq:posterior_approx} in an MH algorithm leads to a valid algorithm for sampling the parameters.
In the case of non-linear Kalman filter approximations (like extended, unscented or cubature Kalman filters) the estimate is not unbiased{, but this has not prevented researchers from using them.}
{Indeed, several researchers have shown that Kalman filters can provide good approximations of the true posterior distribution in non-linear discrete-time state-space models \cite{Sarkka:2013}, as well as in non-linear models involving stochastic differential equations \cite{Mbalawata2013,Sarkka_et_al:2015}.}

\subsubsection{Analysis of noisy/approximate likelihood methods}
\label{sec:analysis_approx_likelihood}

\vspace*{12pt}

Note that, if we have a Markov chain $\mathcal{M}$, and another Markov chain $\mathcal{M}'$ close to $\mathcal{M}$ in some sense, the stationary distribution $\pi'$ of $\mathcal{M}'$ need not exist, and if it does, it need not be close to the stationary distribution of $\mathcal{M}$.
Consequently, studying noisy/approximate MCMC methods is a rather delicate task.
In this sense, it is worth mentioning the work of Johndrow et al. \cite{johndrow2015optimal}, which includes a general perturbation bound for uniformly ergodic chains, as well as Negrea and Rosenthal's work \cite{negrea2017error}, which presents a more complicated bound for geometrically ergodic chains.


\section{Numerical simulations}
\label{sec:results}

\vspace*{12pt}

In this section, we present several examples where the performance of many of the previously described algorithms is evaluated.
We start with two simple examples (univariate and bivariate Gaussians), where the true estimators can be computed analytically, and thus we can gauge exactly the performance of the different methods.
Then we address a challenging problem that appears in several scientific fields: the estimation of the parameters of a chaotic system.
Finally, we also tackle two classical signal processing problems: localization in a wireless sensor network and a spectral analysis example.

\subsection{Illustrative Example for Adaptive MCMC algorithms}
\label{sec:toy1}

\vspace*{12pt}

For the sake of simplicity, in this first example we consider a univariate target density which is a mixture of Gaussian PDFs.
More specifically, the target PDF is formed by $M$ Gaussians, i.e.,
\begin{equation}
	\bar{\pi}(\theta) = \frac{1}{M} \sum_{i=1}^{M} \mathcal{N}(\theta |  \eta_i, \rho_i^2),
\end{equation}
with variances $\rho_i^2=4$ for all $i=1,...,M$.
We consider three different cases, with $M\in\{2,3,6\}$.
The means are: $\eta_1=-10$ and $\eta_2=10$ for $M=2$; $\eta_1=-10$, $\eta_2=0$ and $\eta_3=10$ for $M=3$; $\eta_1=-15$, $\eta_2=-10$, $\eta_3=-5$, $\eta_4=5$, $\eta_5=10$ and $\eta_6=15$ for $M=6$.

We test the adaptive Metropolis (AM) scheme, that uses an adaptive random walk Gaussian proposal \cite{Haario+Saksman+Tamminen:2001}, $q_t(\param|\param^{(t-1)},\sigma_t^2)=\mathcal{N}(\param|\param^{(t-1)},\sigma_t^2)$, and the Adaptive Gaussian Mixture Metropolis-Hastings (AGM-MH) algorithm of \cite{Luengo+Martino:2013}, that uses the following proposal PDF:
\begin{equation}
	q_t(\theta|\mu_{1,t}, \ldots, \mu_{N,t},\sigma_{t,1}^2, \ldots, \sigma_{t,\nprop}^2) = \sum_{n=1}^{\nprop}{\weight_n^{(t)}\varphi_i(\theta|\mu_i^{(t)},\sigma_{i,t}^2)},
\end{equation}
formed by $\nprop$ Gaussian which are independent from the previous state of the chain, i.e., $\varphi_i(\theta|\mu_i^{(t)},\sigma_{i,t}^2)=\mathcal{N}(\theta|\mu_i^{(t)},\sigma_{i,t}^2)$. 
Moreover, we also compare the correlations obtained by the adaptive MH schemes with those obtained using a non-adaptive standard MH algorithm with a random walk proposal PDF. 
In AGM-MH, we set $N=M$ Gaussians and each initial mean is chosen uniformly in $[-20,20]$.
The initial variances and weights are set as $\sigma_{i,0}^2=10$ and $w_{i,0}=1/N$ for all $i=1,...,N$.
The same initialization of the variance is employed for the single component of AM: $\sigma_0^2=10$.
The goal of this example is to show that performing Monte Carlo estimation on multi-modal targets without specialised algorithms (like adiabatic MC \cite{betancourt2014adiabatic}) is challenging, but can still be tackled by properly designed adaptive algorithms with mixture proposals.

\begin{figure}[htb]
\centering 
\centerline{
 \includegraphics[width=9cm]{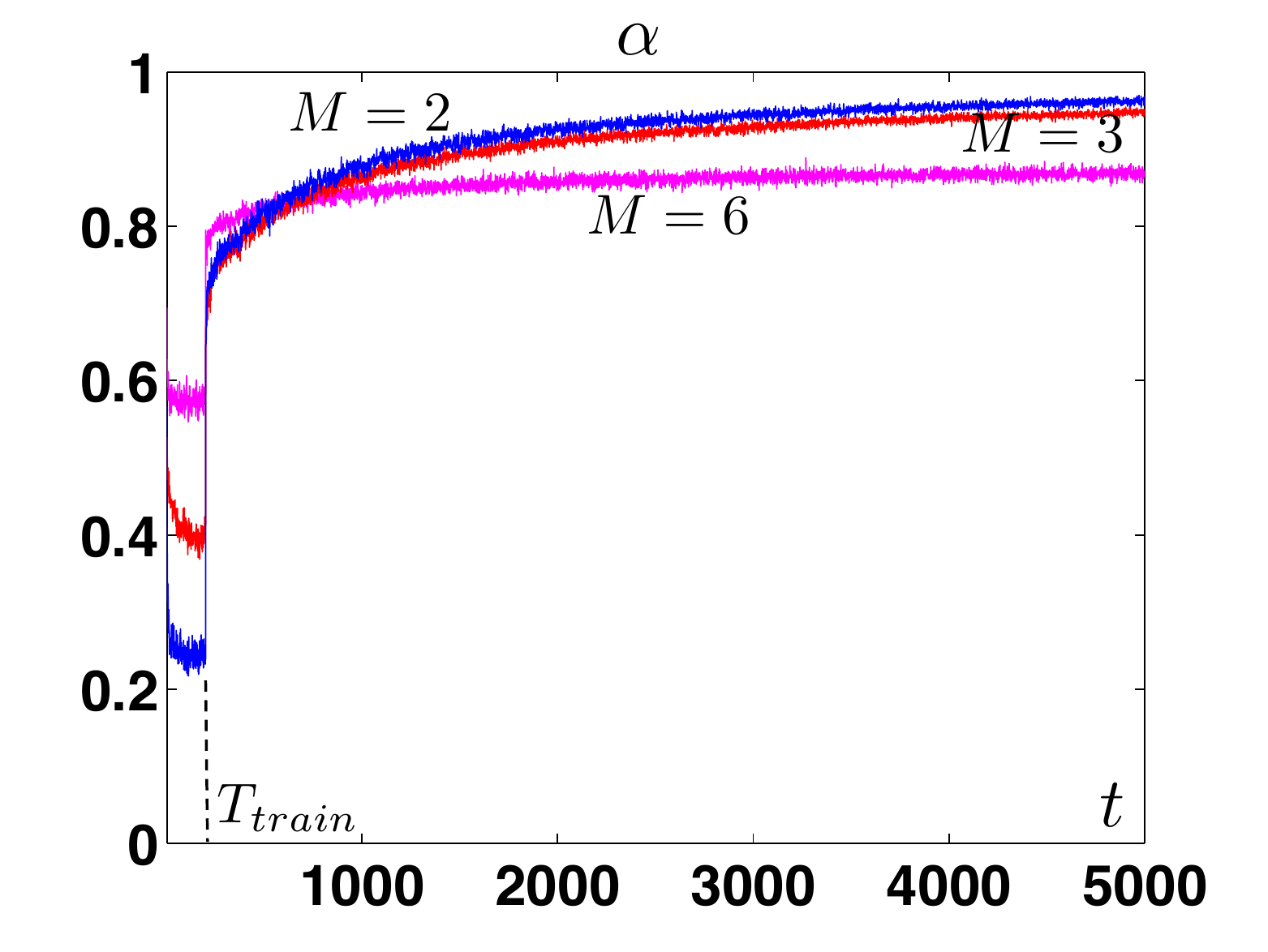}
 }
  \caption{Averaged values of the acceptance probability, $\alpha_t$, as function of the iteration index $t$ and different values of $M \in \{2,3,6\}$, for the univariate Gaussian target in Section \ref{sec:toy1}, using the AGM-MH algorithm. For $t > T_{train}$, $\alpha_t$ grows as a result of the adaptation of the proposal, which becomes closer to the target.}
\label{figAdptiveMCMC}
\end{figure}

We perform $\niter=T_{tot}=5000$ iterations of the chain, setting $T_{train}=200$ (the number of iterations for the initial training period) and $T_{stop}=T_{tot}$ (i.e., the adaptation is never stopped) for the AGM-MH algorithm (see \cite{Luengo+Martino:2013} for a detailed description of these two parameters).
The initial state of the chain is randomly chosen as $\theta^{(0)} \sim \gauss(\theta|0,1)$ in all cases.
Then we use all the generated samples (i.e., $\nburn=0$) to estimate the normalizing constant of the target.  
Table \ref{tab:mseEx1} shows the mean squared error (MSE), averaged over $1000$ independent runs, for the AM and AGM-MH algorithms in the estimation of the expected value of the target PDF, whereas Table \ref{tab:corrEx1} shows the auto-correlation (at lag one) of AM, AGM-MH and a standard MH algorithm without adaptation.
Note the improvement, both in terms or MSE and auto-correlation, attained by both of the adaptive MH algorithms (especially by AGM-MH) even in this simple example.
Finally, Figure \ref{figAdptiveMCMC} depicts the averaged values of the acceptance probability, $\alpha_t$, of the AGM-MH algorithm as function of $t$ and for different values of $M$.
Note the increase in the averaged values of $\alpha_t$ for $t > T_{train}$ as a result of the adaptation.

\begin{table}[!htb]
\caption{\textbf{Mean squared error (MSE) for the univariate Gaussian target in Section \ref{sec:toy1}.}}
\begin{tabular}{|l|c|c|c|}
	\hline
	Algorithm & $M=2$ & $M=3$ & $M=6$ \\
	\hline
	AM & $2 \times 10^{-2}$ & $2 \times 10^{-2}$ & $6 \times 10^{-3}$ \\
	AGM-MH & $1.6 \times 10^{-4}$ & $1.1 \times 10^{-4}$ & $2 \times 10^{-5}$ \\
	\hline
\end{tabular}
\label{tab:mseEx1}
\end{table}

\begin{table}[!htb]
\caption{\textbf{Normalized auto-correlation (at lag one) for the univariate Gaussian target in Section \ref{sec:toy1}.}}
\begin{tabular}{|l|c|c|c|}
	\hline
	Algorithm & $M=2$ & $M=3$ & $M=6$ \\
	\hline
	Standard MH & $0.81$ & $0.72$ & $0.46$ \\
	AM & $0.33$ & $0.26$ & $0.20$ \\
	AGM-MH & $0.13$ & $0.14$ & $0.16$ \\
	\hline
\end{tabular}
\label{tab:corrEx1}
\end{table}

This example shows that a classical adaptive algorithm (AM) fails, as clearly shown by the large MSE and auto-correlation values, whereas a properly designed adaptive MC method (AGM-MH) with an adequate proposal can attain very good results: an MSE two orders of magnitude lower and an auto-correlation up to $2.5$ times smaller.
The performance of the random walk MH largely depends on the variance of the proposal, which should be optimized in order to attain a 25\%--40\% average acceptance rate, as discussed earlier.
Note that this can be easily achieved in this simple example, but becomes a much more challenging task for more complex problems.
Therefore, properly designed adaptive algorithms should be preferred when applicable.

\begin{figure}[htb]
\centering 
  \centerline{
\subfigure[]{\includegraphics[width=6.5cm]{ 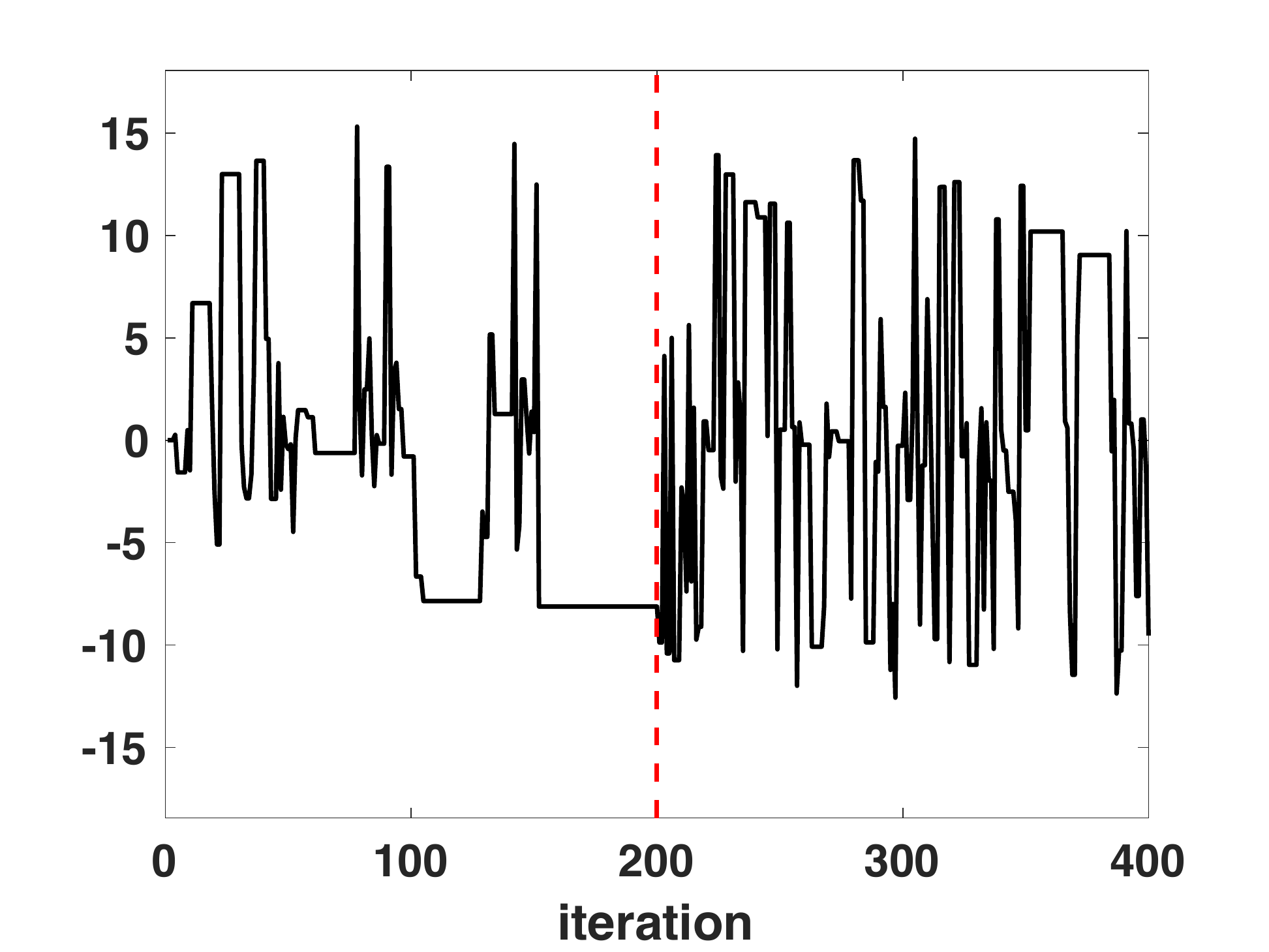}}
\subfigure[]{\includegraphics[width=6.5cm]{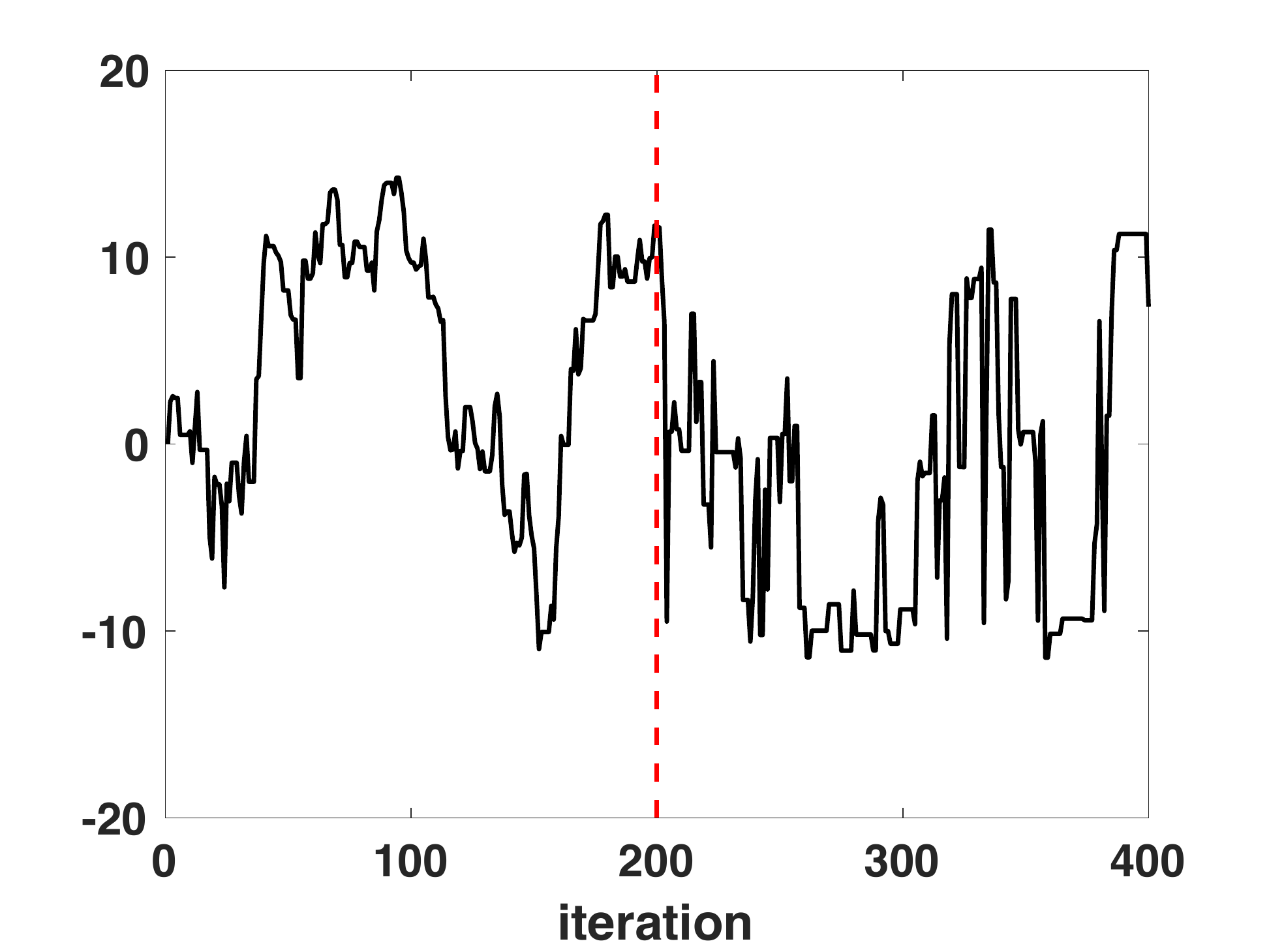}}
  }
  \centerline{  
   \subfigure[]{\includegraphics[width=6.5cm]{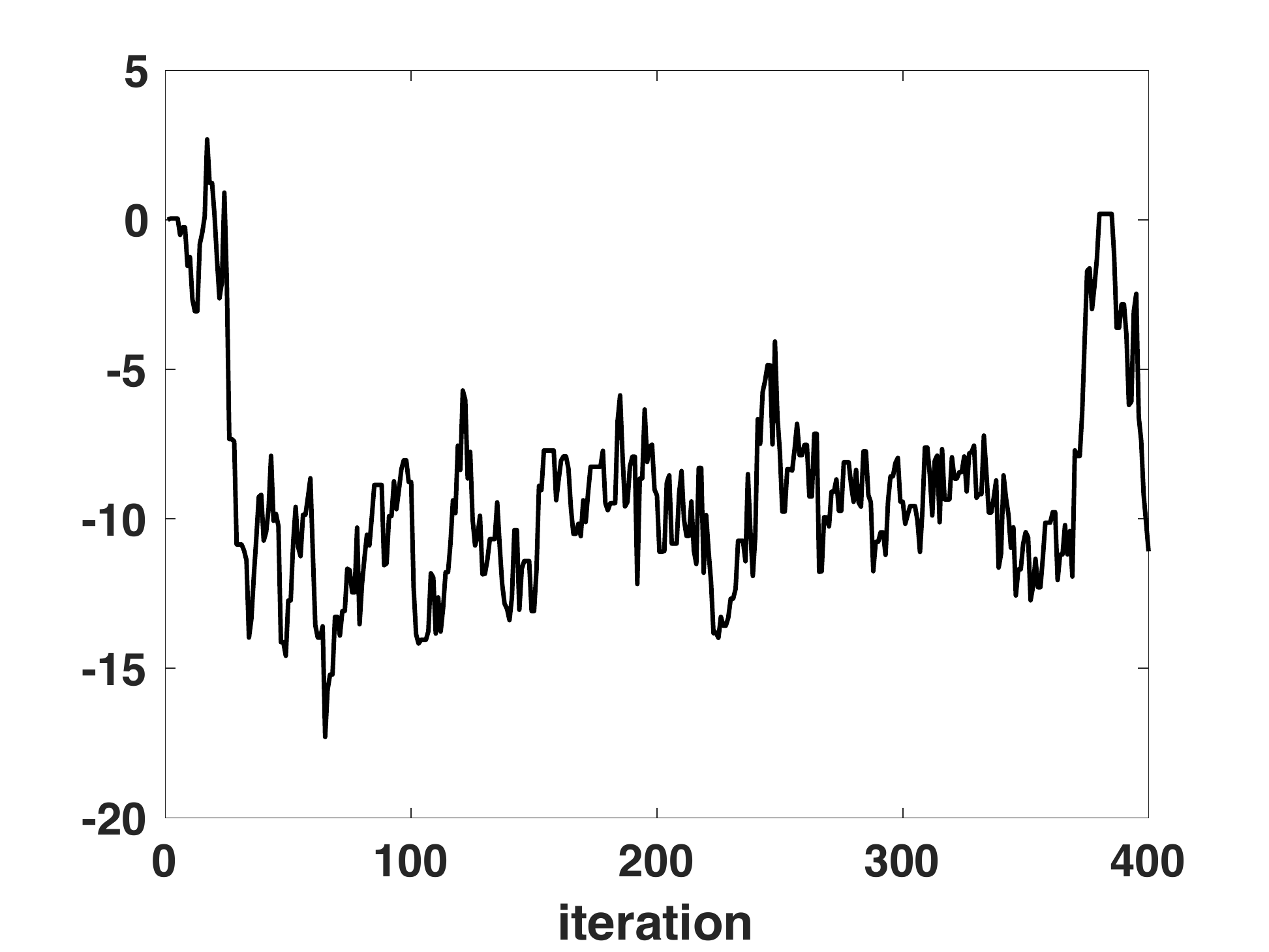}}
    \subfigure[]{\includegraphics[width=6.5cm]{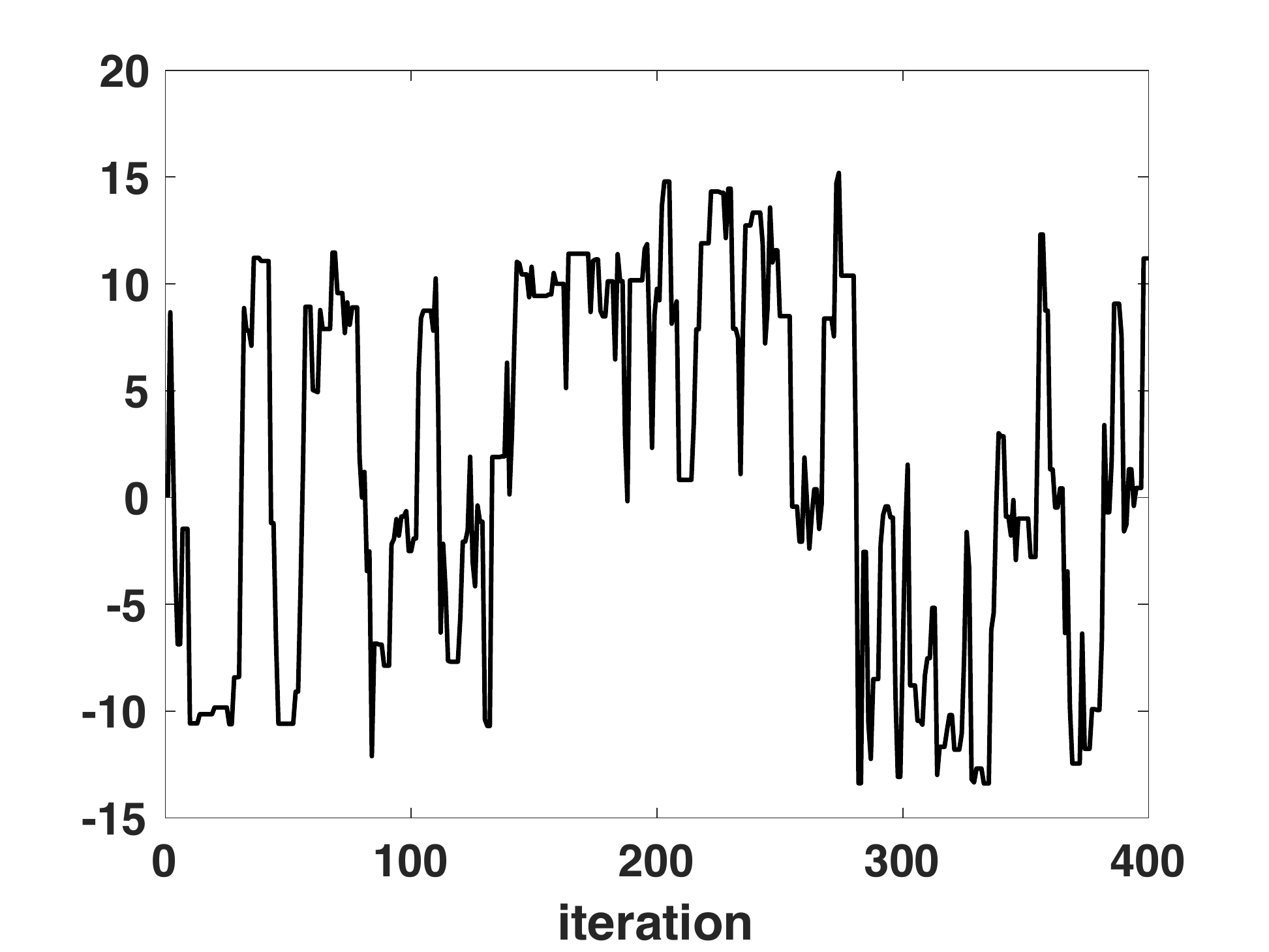}}
 }
  \caption{Trace plots for the univariate Gaussian target in Section \ref{sec:toy1} with $M=3$. (a) AGM-MH algorithm. (b) AM algorithm. The red dashed line in (a) and (b) marks the beginning of the adaptation period. (c) Random walk MH with $\sigma=2$. (d) Random walk MH with $\sigma=5$.}
\label{fig:traceEx1}
\end{figure}

In order to further illustrate the behaviour of the three considered algorithms (RWMH, AM and AGM-MH), Figure \ref{fig:traceEx1} shows typical trace plots for $M=3$ of the two adaptive techniques (AGM-MH and AM), as well as the RWMH algorithm with two different values of $\sigma$.
From Figure \ref{fig:traceEx1}(a) we see that the chain's state for AGM-MH constantly switches to locations around the three modes of the target (placed at $\eta_1=-10$, $\eta_2=0$ and $\eta_3=10$ for $M=3$), showing that the chain is frequently exploring all the modes.
Then, Figure \ref{fig:traceEx1}(b) shows that the chain attained by AM also explores the three modes, but the jumps from one mode to another occur less frequently.
Finally, Figures \ref{fig:traceEx1}(c)--(d) show that the performance of the RWMH algorithm critically depends on the variance: for $\sigma=2$ the resulting chain in the example completely fails to explore one of the modes, whereas for $\sigma=5$ all the three modes are properly covered.

\subsection{Illustrative example for Adaptive Importance Sampling}
\label{sec:toy2}

\vspace*{12pt}

As a second simple example, let us consider a multimodal target PDF consisting of a mixture of five bivariate Gaussians, i.e., 
\begin{equation}
\normtarget(\parvec)=\frac{1}{5}\sum_{i=1}^5 \mathcal{N}(\parvec;{\bm \nu}_i,{\bf \Sigma}_i),
\label{Target1b}
\end{equation}
with $\parvec \in \mathbb{R}^2$, means ${\bm \nu}_1=[-10, -10]^{\top}$, ${\bm \nu}_2=[0, 16]^{\top}$, ${\bm \nu}_3=[13, 8]^{\top}$, ${\bm \nu}_4=[-9, 7]^{\top}$ and ${\bm \nu}_5=[14, -14]^{\top}$, and covariances ${\bf \Sigma}_1=[2, \ 0.6; 0.6, \ 1]$, ${\bf \Sigma}_2=[2, \ -0.4; -0.4, \ 2]$, ${\bf \Sigma}_3=[2, \ 0.8; 0.8, \ 2]$, ${\bf \Sigma}_4=[3, \ 0; 0, \ 0.5]$ and ${\bf \Sigma}_5=[2, \ -0.1; -0.1, \ 2]$.
Since we can analytically compute the moments of the target, this example is very useful to validate the performance of different Monte Carlo techniques.
In particular, we consider the computation of the mean of the target, $E(\parvec)=[1.6, 1.4]^{\top}$, and the normalizing constant, $Z=1$.
We use the MSE (averaged over both components in the computation of $E(\parvec)$) as the figure of merit of the different estimators.

For simplicity, we use again Gaussian proposal densities for all the MC methods.
The proposals are ``poorly'' initialized on purpose in order to test the robustness and the adaptation capabilities of the methods.
More specifically, the location parameters of the proposals are initialized uniformly within the $[-4,4]\times[-4,4]$ square, i.e., ${\bm \mu}_i^{(1)}\sim \mathcal{U}([-4,4]\times[-4,4])$ for $i=1,\ldots,N$.
Note that none of the modes of the target falls within the initialization square.
We test all the alternative methods using the same isotropic covariance matrices for all the Gaussian proposals, $\bC_i = \sigma^2\bI_2$ with $\sigma \in \{1,2,5,10,20,70\}$.
All the results have been averaged over $500$ independent experiments, where the computational cost of the different techniques (in terms of the total number of evaluations of the target distribution, which is usually the most costly step in practice) is fixed to $L=KNT$.\footnote{Note that $L=KNT$ also corresponds to the total number of samples generated in all the schemes.}
We compare the following schemes:

\begin{itemize}

\item {\bf Standard PMC \cite{Cappe04}:} The standard PMC algorithm proposed in \cite{Cappe04} with $N=100$ proposals and $T=2000$ iterations. The total number of samples drawn is $L = NT = 2 \cdot 10^5$.

\item {\bf M-PMC \cite{Cappe08}:} The M-PMC algorithm proposed in \cite{Cappe08} with $\nprop=100$ proposals, $\npart=100$ samples per iteration, and $\niter=2000$ iterations. The total number of samples drawn is $L = \npart\niter = 2 \cdot 10^5$.

\item {\bf SMC \cite{Moral06}:}  A Sequential Monte Carlo (SMC) scheme combining resampling and MCMC steps. More precisely, we consider MH steps as forward reversible kernels. In this example, we do not employ a sequence of tempered target PDFs, i.e., we consider always the true target density. The proposal PDFs for the MH kernels coincide with the Gaussian proposals employed in the propagation resampling steps, with the scale parameters ${\bf C}_i$ of the other tested methods. Due to the application of the MH steps, in this case, $L>2\cdot 10^5$.

\item {\bf K-PMC \cite{elvira2017improving}:} The standard PMC scheme using $\nprop=100$ proposals, but drawing $K>1$ samples per proposal at each iteration and performing global resampling (GR). In order to keep the total number of samples constant, the number of iterations of the algorithm is now $T = 2 \cdot 10^5 /(KN)$. 

\item {\bf DM-PMC \cite{elvira2017improving}:} The standard PMC using the weights of Eq. \eqref{eq_w_dmmis} (i.e., the mixture of all proposals at each iteration), $N=100$ proposals, $T=2000$ iterations, and drawing $K=1$ samples per proposal (i.e., $\npart=\nprop=100$ samples per iteration). The total number of samples drawn is again $L = \npart\niter = 2 \cdot 10^5$.

\item {\bf GR-PMC \cite{elvira2017improving}:} The standard PMC scheme with multiple samples per proposal ($K$), weights computed as in DM-PMC, and global resampling (GR). We use $N=100$ proposals and $T=L/(KN)$ iterations with $L=2 \cdot 10^5$ again. In particular, we test the values $K \in \{2,5,20,100,500\}$, and thus $T \in \{1000,400,100,20,4\}$.

\item {\bf LR-PMC \cite{elvira2017improving}:} The standard PMC scheme with multiple samples per proposal ($K$) and local resampling (LR). All the parameters are selected as in the GR-PMC scheme.

\item {\textbf{Improved  SMC \cite{Moral06,elvira2017improving}:} The SMC scheme with the improvements proposed in those two papers. In all cases, we use the importance weights as in DM-PMC (deterministic mixture of the proposals at each iteration), and we try the GR-SMC and LR-SMC variants. We test $K \in \{5,20\}$ }

\item {\bf APIS \cite{APIS15}:} The adaptive population importance sampling (APIS) scheme with $N=100$ proposals and $T=2000$ iterations. The IS weights are again the spatial deterministic mixture weights.

\item {\bf AMIS \cite{CORNUET12}:} The adaptive multiple importance sampling (AMIS) algorithm, which uses a single proposal, drawing $K$ samples per iteration and running for $T$ iterations. We  use values of $K$ and $T$ such that $ L=KT=2\cdot 10^5$, for a fair comparison. Specifically, we have run different simulations using $K\in\{500,1000,2000,5000\}$ and, as a consequence,  $T\in\{40,20,10,4\}$. Since the samples are reweighted using the whole set of past temporal proposals in the denominator (i.e., a sort of temporal deterministic mixture), AMIS becomes more costly when $T$ increases. In Table \ref{table_mean} we show the best and worst performance for each value of $\sigma$.
\end{itemize}

\begin{figure}[htb]
  \centering
 \centerline{
  \subfigure[]{\includegraphics[width=5.4cm]{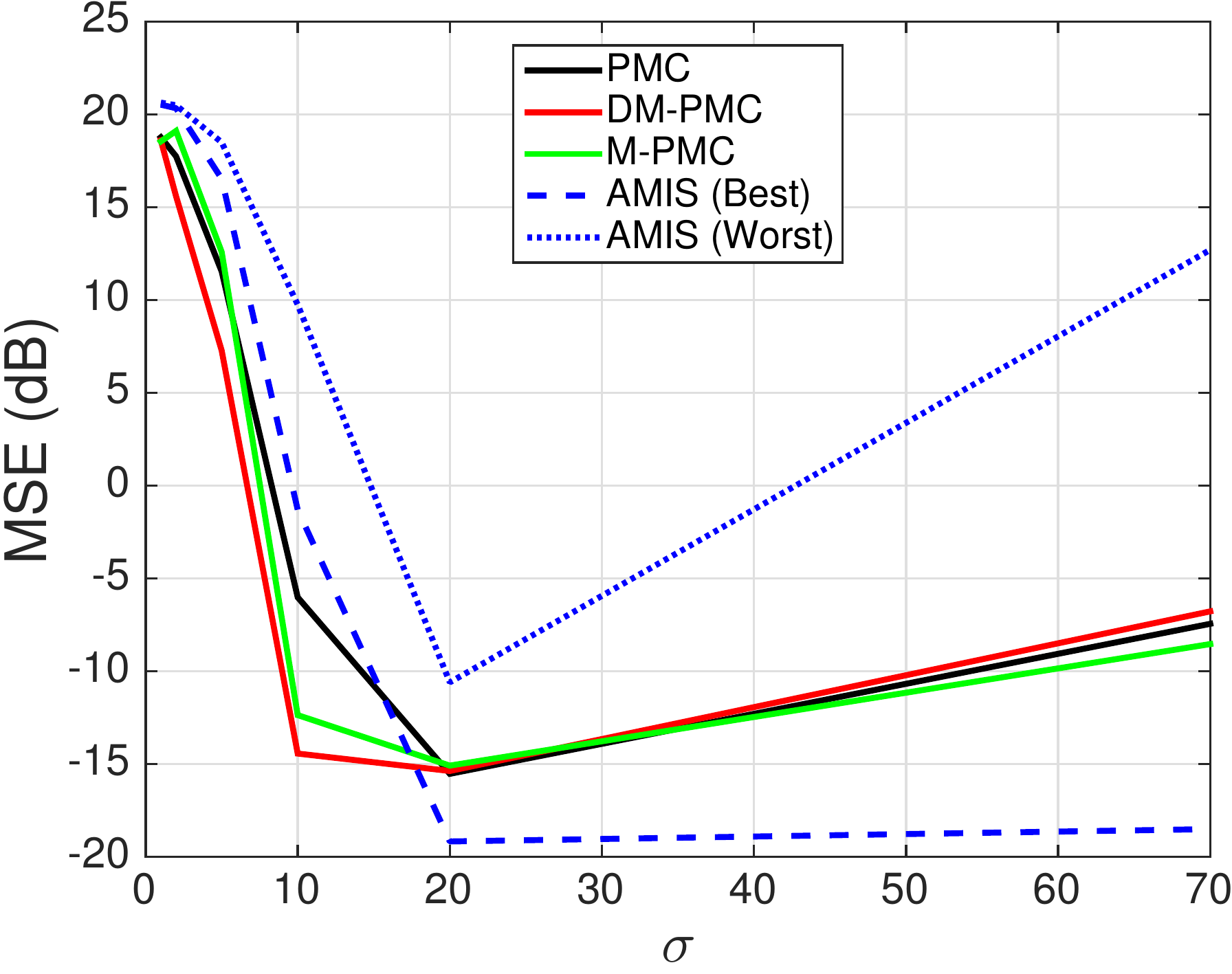}} 
  \subfigure[]{\includegraphics[width=5.4cm]{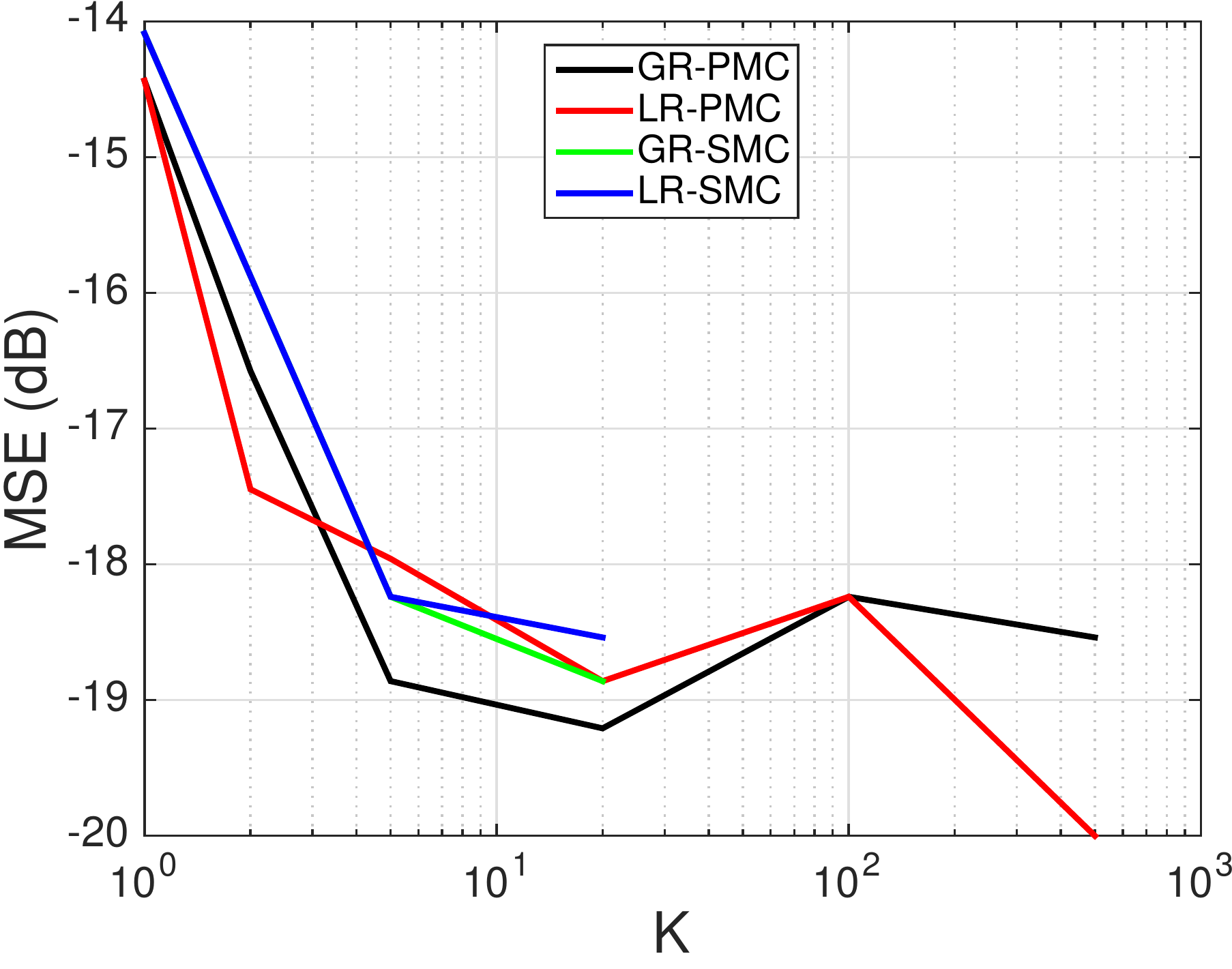}} 
  \subfigure[]{\includegraphics[width=5.4cm]{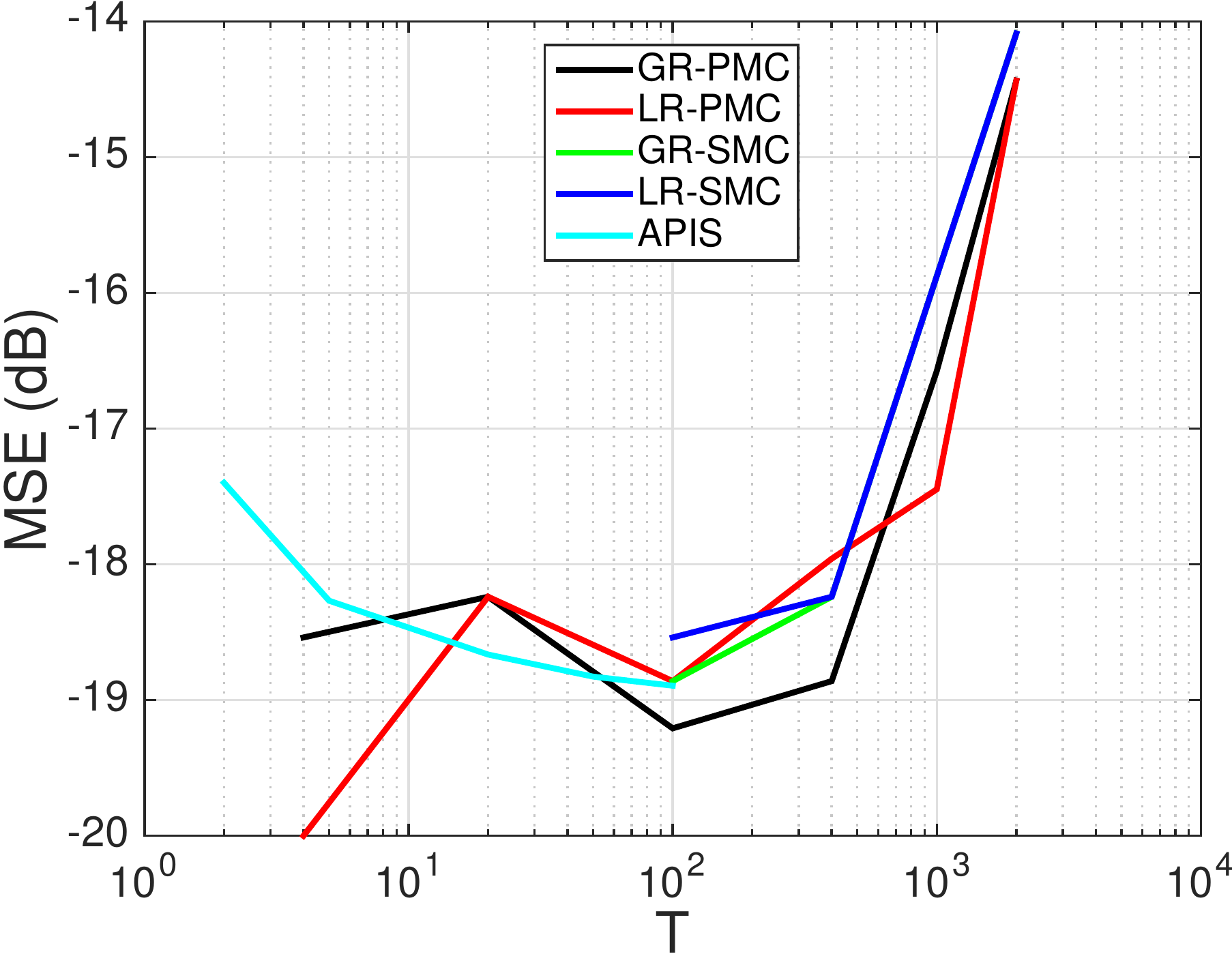}}
  }
\caption{Example of Section \ref{sec:toy2}. {\bf (a)} MSE of several algorithms as a function of $\sigma$ for $N=100$. {\bf (b)} MSE as a function of $K$ for $\sigma=10$. {\bf (c)} MSE as a function of $T$ for $\sigma=10$.}
\label{fig:ais}
\end{figure}

Table \ref{table_mean} shows the full results for the MSE in the estimation of $E(\parvec)$ averaged over both components, whereas Figure \ref{fig:ais} graphically displays some selected cases.
We can see that the compared schemes outperform the standard PMC for any value of $\sigma$.
In general, the local resampling (LR-PMC) works better than the global resampling (GR-PMC).
APIS obtains a good performance for several intermediate values of $\sigma$, while AMIS behaves well with large values of $\sigma$.
Moreover, we note that the optimum value of $K$ in GR-PMC and LR-PMC depends on the value of $\sigma$, the scale parameter of the proposals: for small values of $\sigma$ (e.g., $\sigma=1$ or $\sigma=2$) small values of $K$ lead to better performance, whereas a larger value of $K$ (and thus less iterations $T$) can be used for larger values of $\sigma$ (e.g., $\sigma=10$ or $\sigma=20$).

\begin{table} 
\setlength{\tabcolsep}{2pt}
\def\marginwidth{1.5mm}
\begin{center}
\caption{MSE in the estimation of $E(\parvec)$, keeping the total number of evaluations of the target fixed to $L=KNT=2 \cdot 10^5$ in all algorithms, for the bivariate target in Section \ref{sec:toy2}. The best results for each value of $\sigma$ are highlighted in bold-face.}
{\small
\begin{tabular}{|c@{\hspace{\marginwidth}}|l@{\hspace{\marginwidth}}|c@{\hspace{\marginwidth}}|c@{\hspace{\marginwidth}}|c@{\hspace{\marginwidth}}|c@{\hspace{\marginwidth}}|c@{\hspace{\marginwidth}}|c@{\hspace{\marginwidth}}|}
\hline
  \multicolumn{8}{|c|}{ $\bm L= \bm N\bm K\bm T = 2 \cdot 10^5$} \\
\hline
$\bm N$ &{\bf Algorithm} &  ${\bm \sigma{\bf=1}}$ &  ${\bm \sigma{\bf=2}}$ &  ${\bm \sigma{\bf=5}}$ &  ${\bm \sigma{\bf=10}}$ & ${\bm \sigma{\bf=20}}$ & ${\bm \sigma{\bf=70}}$  \\
\hline
\hline
5 & \multirow{ 3}{*}{Standard PMC \cite{Cappe04}} &92.80 &38.71  &12.65  &0.38  &0.047 &37.44 \\
\cline{1-1}
\cline{3-8}
100 & &75.17  &59.42 &14.24 &0.25 &0.028 &0.18\\
\cline{1-1}
\cline{3-8}
$5 \cdot 10^4$ & &68.29 &37.44 &7.01 &0.25 &0.033 &0.17 \\
\hline
\hline
\multirow{ 15}{*}{100} &DM-PMC ($K=1$) & 72.48  &36.21  &5.34 &0.036 &0.029  &0.21 \\
\cline{2-8}
&GR-PMC ($K=2$) &69.41  &26.23 &3.09 &0.022  &0.028  &0.17 \\
&LR-PMC ($K=2$) &{2.68} & {0.007} &0.010 &0.018 &0.102 &32.88 \\

\cline{2-8}
&GR-PMC ($K=5$) &67.04 &17.44  &0.11 &0.013  &\textbf{0.023}  &0.15 \\
&LR-PMC ($K=5$) &8.04 &0.012  &{0.008} &0.016 &0.027  &2.00 \\
\cline{2-8}
&GR-PMC ($K=20$) &61.58 &15.13 &0.42 &0.012 &0.024 &{0.14} \\
&LR-PMC ($K=20$) &9.51  &1.16 &0.011  &0.013  &0.023  &0.22\\

\cline{2-8}
&GR-PMC ($K=100$) &64.94 &12.50 &0.08 &0.015 &0.026 &0.18 \\
&LR-PMC ($K=100$) &9.60 &1.21 &0.022 &0.015 &0.026 &0.20 \\
\cline{2-8}
&GR-PMC ($K=500$) &58.49  &9.63 &0.08&0.014  &0.024  &0.16 \\
&LR-PMC ($K=500$) &14.79  &6.72  &0.10  &\textbf{0.010}  &0.024  &0.20 \\
\hline 
 100 & {M-PMC \cite{Cappe08}}& 71.39 &81.33  &18.14  &0.058  &0.031  &{0.14} \\
\hline
 10 &   &84.14  &81.68 &6.49 &0.76  &{0.024}  &4.60 \\ 
 \cline{1-1}
 \cline{3-8}
  100 & {SMC \cite{Moral06}}&77.00 & 76.5 &15.98 &0.79  &0.068&0.86\\ 
  \cline{1-1}
 \cline{3-8}
$5 \cdot 10^4$  & &69.08 &51.29 &20.48&0.22 &0.038  &0.68  \\
\hline
   &  {DM-SMC ($K=1$)}  &70.95 &42.40  &1.91  &0.039  &0.027 &0.19 \\
    \cline{2-8}
   &  {GR-SMC ($K=5$)}  &66.64 &41.54 &0.16 &0.015  &{0.024} &0.19\\
       \cline{3-8}
    100 &  {LR-SMC ($K=5$)}  &8.16  &2.32  &0.007  &0.015  &0.027  &2.19 \\
        \cline{2-8}
       &  {GR-SMC ($K=20$)} &65.48 &37.91  &0.10&0.013  &0.025  &0.19 \\
           \cline{3-8}
     &  {LR-SMC ($K=20$)} &8.88  &4.15 &0.010  &0.014  &0.026  &0.20 \\
\hline
\multirow{5}{*}{100}  &{ APIS} ($T=100$)  &  0.0318 &0.0011 & 0.0054 & {\bf 0.0129} & \textbf{0.0211} &0.1794 \\   
\cline{2-8}
&{ APIS} ($T=50$) & 0.0144 &0.0007 & 0.0051 &0.0131 & \textbf{0.0221} &0.1772 \\   
\cline{2-8}
&{\ APIS} ($T=20$) & 0.0401 &0.0006 & {\bf 0.0047} & 0.0136 & \textbf{0.0245} & 0.1732 \\
\cline{2-8}
& { APIS} ($T=5$)  & {\bf 0.0008} &{\bf 0.0005}  &0.0064 &0.0149 & 0.0270 &0.2076 \\
\cline{2-8}
& { APIS} ($T=2$)  &0.0017 &0.0116 & 0.0103 &0.0182 & 0.0387 &0.1844 \\
\hline
\hline
\multirow{2}{*}{1} & { AMIS}(best) & 112.70 &107.85 & 44.93 & 0.7404 & {\bf 0.0121} & {\bf 0.0141}  \\
\cline{2-8}
&{ AMIS} (worst) & 115.62 &111.83 & 70.62 & 9.43 &0.0871 &18.62 \\
\hline 
\end{tabular}
}
\end{center}
\label{table_mean}
\end{table}

\subsection{Parameter estimation in a chaotic system}
\label{sec:chaos}

\vspace*{12pt}

In this numerical experiment, we address the estimation of the parameters of a chaotic system, which is considered a very challenging problem in the literature \cite{Perretti13,Perretti13b}, since the resulting PDFs typically present very sharp full-conditionals.
This type of systems are often utilized for modelling the evolution of population sizes, for instance in ecology \cite{Perretti13}.
Let us consider a logistic map \cite{Boyarsky97} perturbed by multiplicative noise,
\begin{equation}
	z_{t+1}=R\left[\ z_{t}\left(1-\frac{z_t}{\Omega}\right)\right]\exp(\epsilon_t), 
\label{LogNoisy}
\end{equation}
with $\epsilon_t \sim \mathcal{N}(0,\lambda^2)$, $z_1 \sim \mathcal{U}([0,1])$, and unknown parameters $R>0$ and $\Omega>0$.
Let us assume that a sequence $\z_{1:T}=[z_1,\ldots, z_T]$ is observed and, for the sake of simplicity, that $\lambda$ is known.
Under these circumstances, the likelihood function is given by
\begin{equation*}
p(\z_{1:T}|R,\Omega)=\prod_{t=1}^{T-1} p(z_{t+1}|z_t,R,\Omega),
\end{equation*}
where, defining $g(z_t,R,\Omega)=R\left[\ z_{t}\left(1-\frac{z_t}{\Omega}\right)\right]$, we have 
\begin{equation*}
	p(z_{t+1}|z_t,R,\Omega)\propto \left|\frac{g(z_t,R,\Omega)}{z_{t+1}}\right| \exp\left(-\frac{\log\left(\frac{z_{t+1}}{g(z_t,R,\Omega)}\right)^2}{2\lambda^2}\right), 
\end{equation*}
if $g(z_t,R,\Omega)>0$, and $p(z_{t+1}|z_t,R,\Omega)=0$, if $g(z_t,R,\Omega)\leq 0$.
Considering uniform priors, $R \sim \mathcal{U}([0,10^4])$ and $\Omega \sim \mathcal{U}([0,10^4])$, our goal is computing the mean of the bivariate posterior PDF, 
\begin{equation}
	{\bar\pi}(\parvec| \z_{1:T}) = {\bar\pi}(R,\Omega| \z_{1:T}) \propto p(\z_{1:T}|R,\Omega),
\end{equation}
which corresponds to the minimum mean squared error (MMSE) estimate of the parameters.
Note that the parameter vector to be inferred in this example is $\parvec=[\theta_1=R,\theta_2=\Omega]$.

In the experiments, we set $R=3.7$, $\Omega=0.4$ and $T=20$.
Furthermore, we take into account different values of $\lambda$ of the same order of magnitude as considered in \cite{Perretti13}.
Then we apply FUSS-within-Gibbs \cite{FUSS} (with $\delta=10^{-3}$, $K=10$ and an initial grid $\widetilde{\mathcal{S}}_M=\{10^{-4},2 \cdot 10^{-4},\ldots,20\}$), using only $N_G=50$ iterations of the Gibbs sampler.
We also consider an MH-within-Gibbs approach with a random walk proposal, $\normproposal(\theta_{i}^{(t)}|\theta_{i}^{(t-1)}) \propto \exp\left(\frac{-(\theta_{i}^{(t)}-\theta_{i}^{(t-1)})^2}{2\sigma_p^2}\right)$, with $i \in \{1,2\}$, and two different values of $\sigma_p \in \{1,2\}$.
The initial states of the chains are chosen randomly from $\mathcal{U}([1,5])$ and $\mathcal{U}([0.38,1.5])$, respectively.
In order to compare the performance of both approaches, we also perform an approximate computation of the true value of the mean via an expensive deterministic numerical integration procedure.

\begin{figure}[!htb]
  \centering
\centerline{
 \subfigure[]{\includegraphics[width=5.4cm]{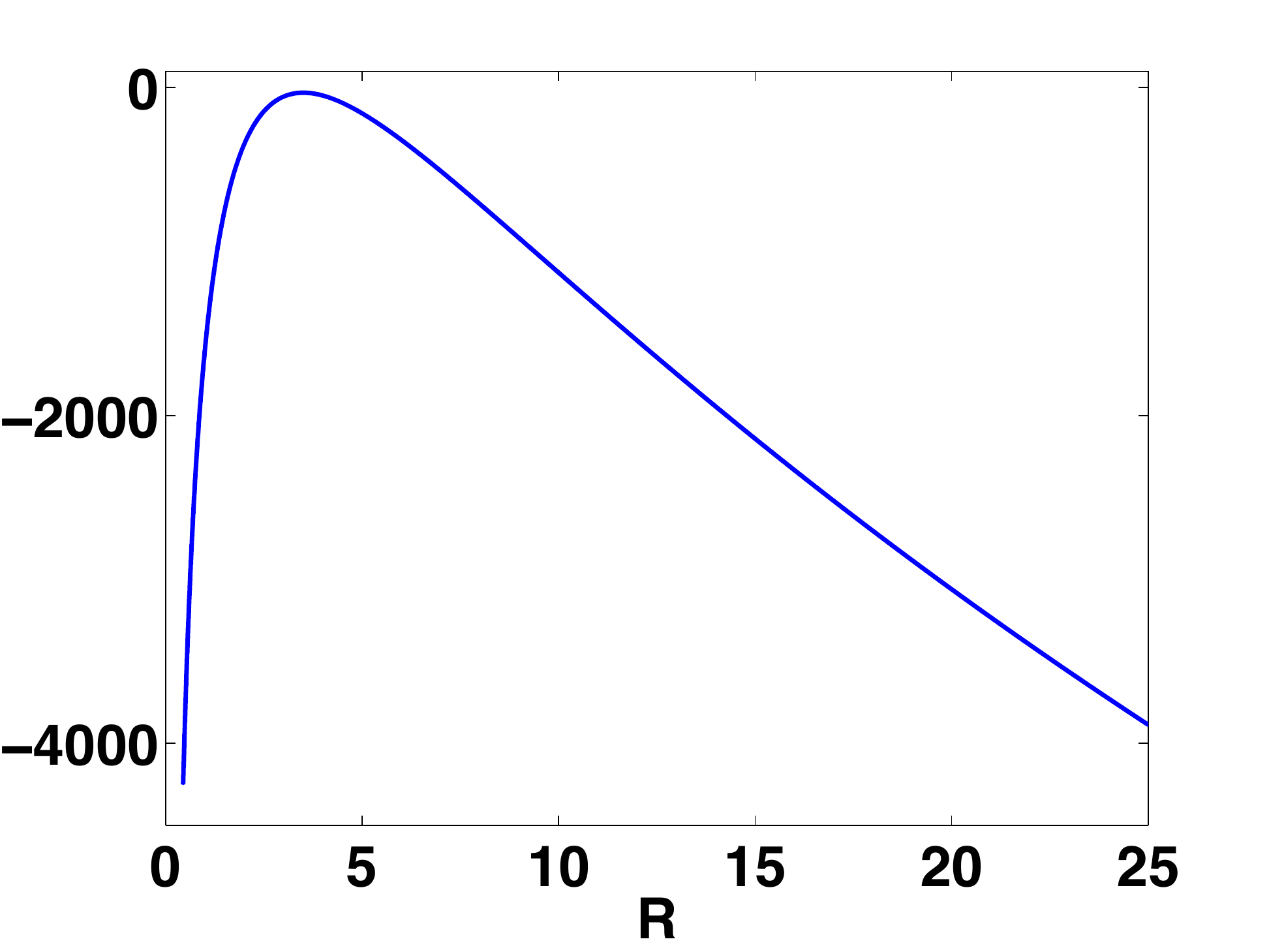}}
  \subfigure[]{\includegraphics[width=5.4cm]{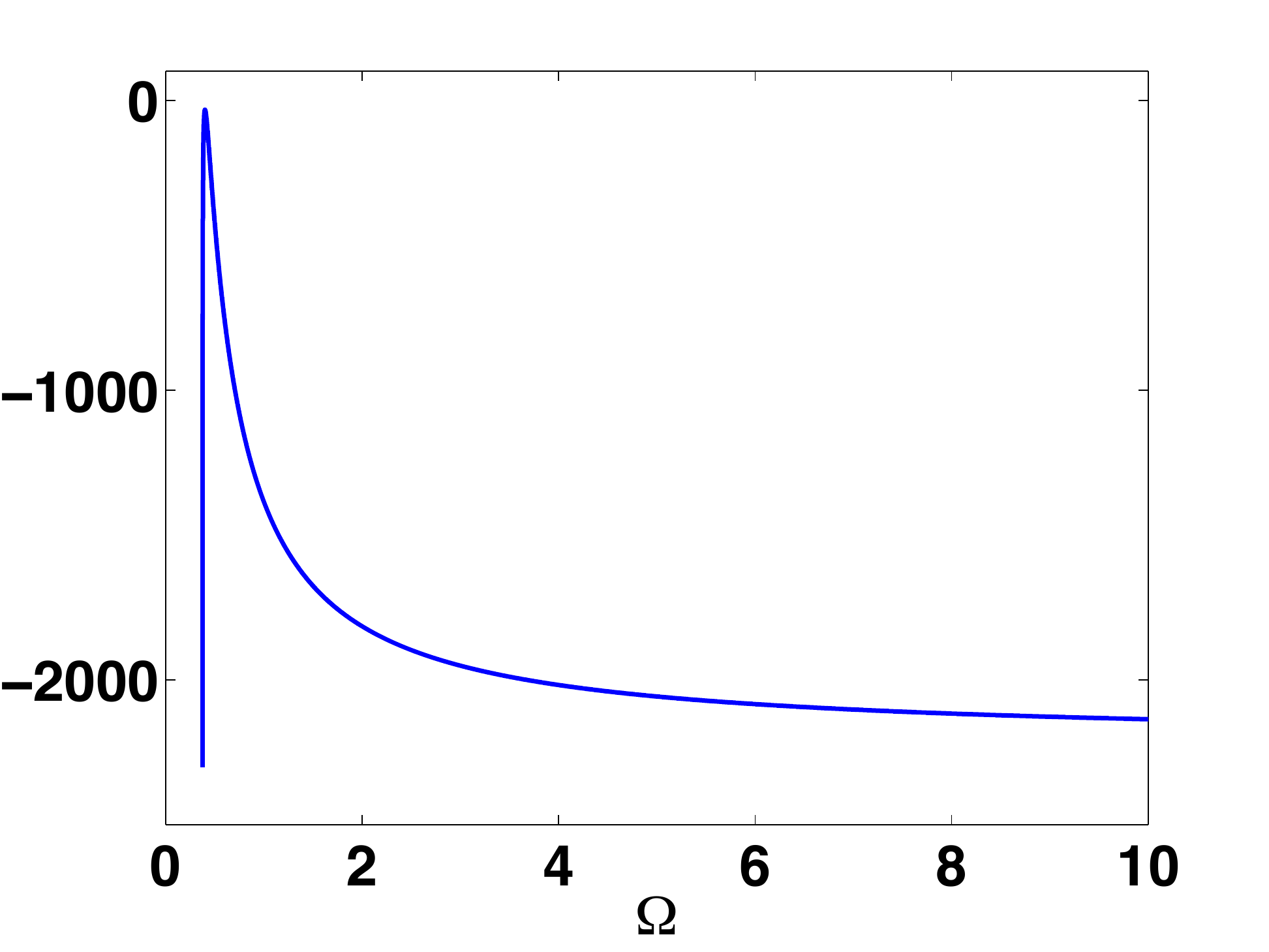}} 
  \subfigure[]{\includegraphics[width=5.4cm]{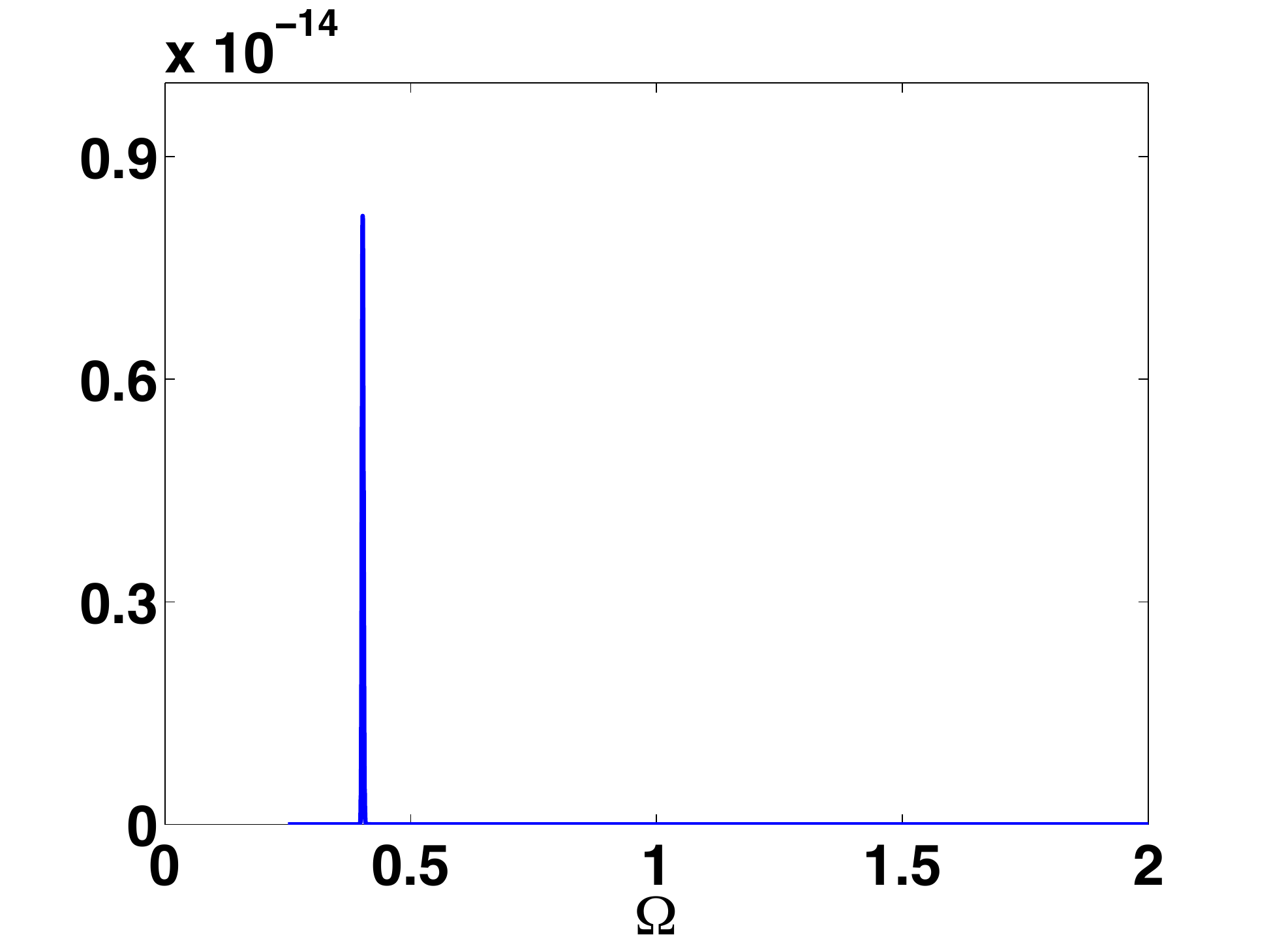}}
  }
\caption{Example of Section \ref{sec:chaos}. {\bf (a)}-{\bf (b)} Examples of (unnormalized) conditional log-PDFs with $\lambda=0.1$ and considering $N=20$ observations. {\bf (a)} Fixing $\Omega=4$. {\bf (b)} Fixing $R=0.7$. {\bf (c)} The (unormalized) conditional PDF corresponding to Figure (b).}
\label{fig:Perretti}
\end{figure}

The results, averaged over $1000$ independent runs, are shown in Table \ref{TablaResPeretti}.
It can be clearly seen that FUSS-within-Gibbs achieves a very small MSE in the estimation of the two desired parameters (especially in the case of $\Omega$) for any value of $\lambda$.
Comparing with the MSE obtained by the MH algorithm, the benefit of building a proposal tailored to the full-conditionals (as done by FUSS) becomes apparent.
Figures \ref{fig:Perretti}(a)--(b) provide two examples of conditional log-PDFs, whereas Figure \ref{fig:Perretti}(c) shows the ``sharp'' conditional density corresponding to Figure \ref{fig:Perretti}(b).
This PDF resembles a delta function: even using sophisticated adaptive techniques it is difficult to recognize the mode of this kind of target PDF.
However, by constructing a proposal which is adapted to the full conditionals using the FUSS algorithm, very good results can be obtained even in this extreme case.

\begin{table*}[!htb]
\setlength{\tabcolsep}{2pt}
\def\marginwidth{1.5mm}
\caption{MSEs in estimation of $R$ and $\Omega$ using FUSS and MH inside a Gibbs sampler, with $\delta=10^{-3}$, $K=10$ and $N_G=50$, for the example of Section \ref{sec:chaos}. The observed sequence, $\z_{1:T}$, is generated with $R=3.7$, $\Omega=0.4$, and $N=20$ and different values of $\lambda$.}
\label{TablaResPeretti}
\begin{center}
\footnotesize
\begin{tabu}{|[1pt]l@{\hspace{\marginwidth}}|c@{\hspace{\marginwidth}}|[1pt]c@{\hspace{\marginwidth}}|c@{\hspace{\marginwidth}}|c@{\hspace{\marginwidth}}|c@{\hspace{\marginwidth}}|c@{\hspace{\marginwidth}}|c@{\hspace{\marginwidth}}|[1pt]}
\cline{3-8}
 \multicolumn{2}{c|[1pt]}{} & $\lambda=0.001$ & $\lambda=0.005$  & $\lambda=0.01$ & $\lambda=0.05$  & $\lambda=0.08$ & $\lambda=0.10$      \\
\Xhline{2\arrayrulewidth}
\multirow{2}{*}{\textbf{FUSS}} &MSE($R$) & 0.0071 & 0.0089  & 0.0093  & 0.0138 & 0.0150 & 0.0778   \\
\cline{2-8}
  &MSE($\Omega$)  & 5.01  $10^{-5}$ &  6.15  $10^{-5}$ &  6.15  $10^{-5}$ & 5.26  $10^{-5}$ & 7.33  $10^{-5}$ & 1.78  $10^{-4}$  \\
\Xhline{2\arrayrulewidth}
 \multirow{2}{*}{\textbf{MH} (${\bm \sigma_p}{\bf =1}$)} &MSE($R$) &  0.6830& 0.7264 & 0.7067  & 1.1631  &  1.3298 &  1.3293 \\
\cline{2-8}
  &MSE($\Omega$)  & 0.0373 & 0.0402 & 0.0423 &0.0399  & 0.0471 &  0.0440   \\
\Xhline{2\arrayrulewidth}
 \multirow{2}{*}{\textbf{MH} (${\bm \sigma_p}{\bf =2}$)} &MSE($R$) & 1.3566 & 1.4906 & 1.4247 & 2.0015 & 2.3042  & 2.2401  \\
\cline{2-8}
  &MSE($\Omega$)  & 0.0897 &  0.1117 & 0.1041  & 0.0989 & 0.1089 &  0.1125  \\
 \cline{2-8}
\Xhline{2\arrayrulewidth}
\hline
\end{tabu}
\end{center}
\end{table*}

\begin{figure}[!htb]
  \centering
	\centerline{
	  \subfigure[]{\includegraphics[width=5.4cm]{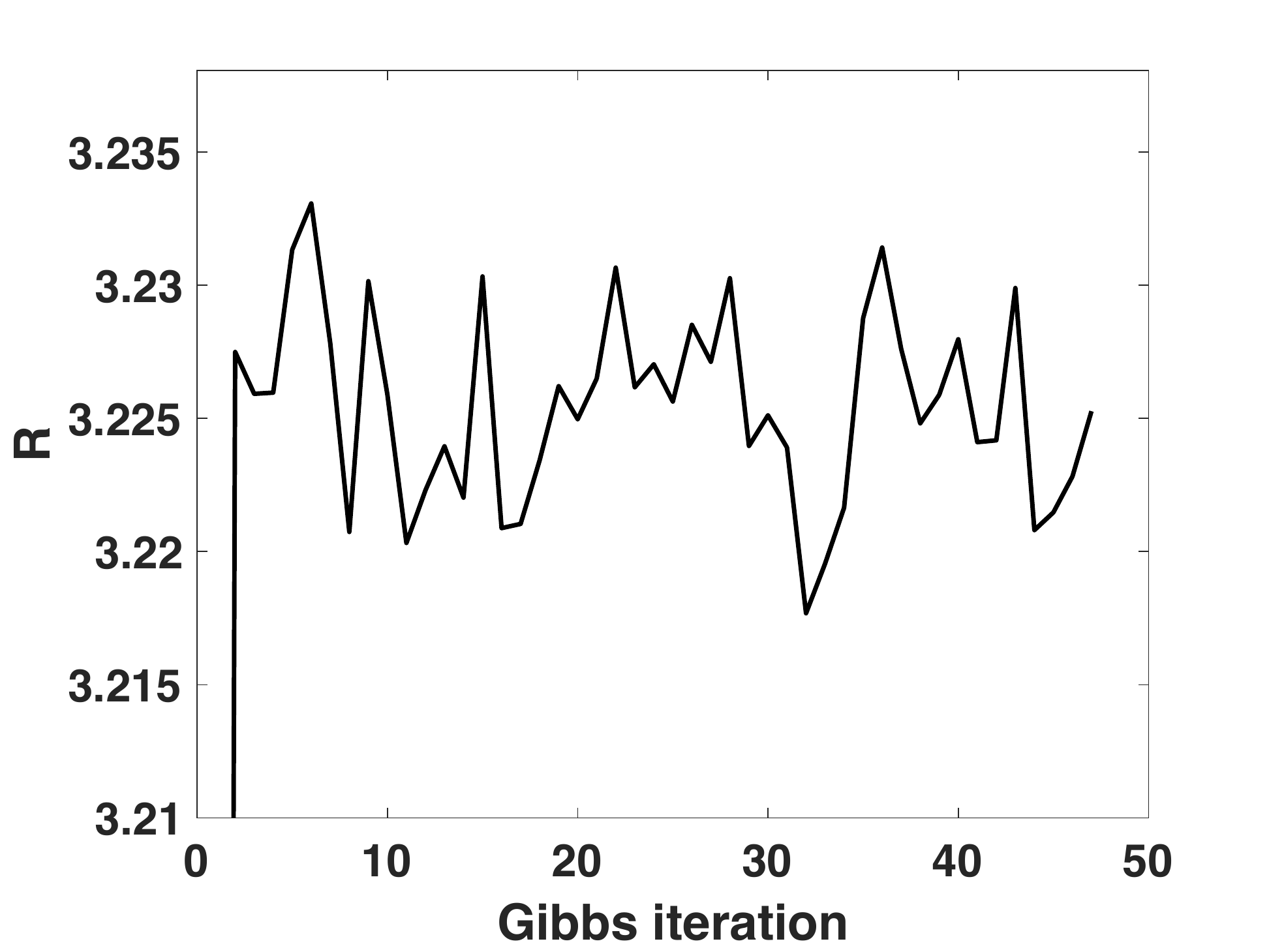}} 
	 \subfigure[]{\includegraphics[width=5.4cm]{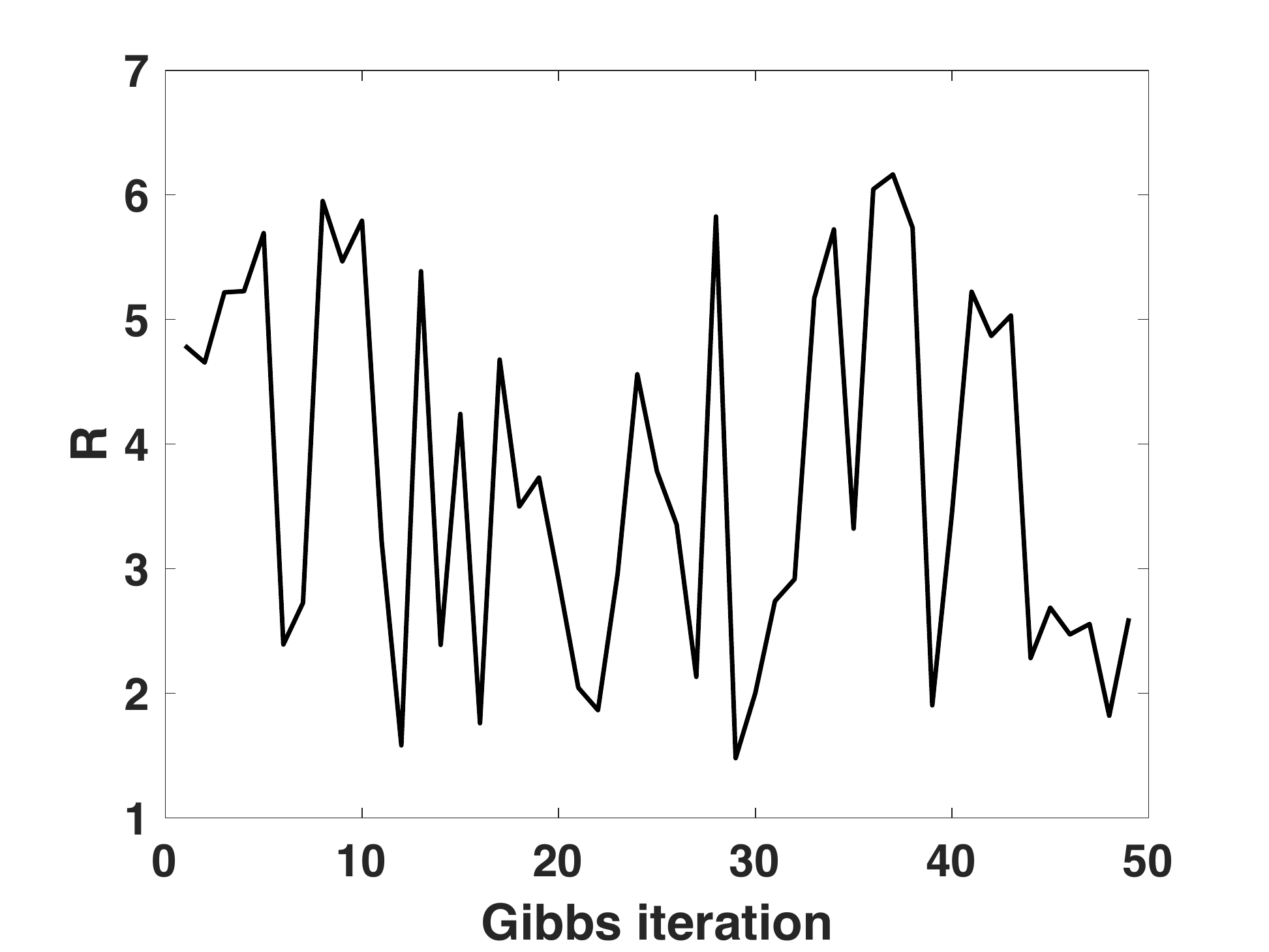}} 
	 }
	 \centerline{
 \subfigure[]{\includegraphics[width=5.4cm]{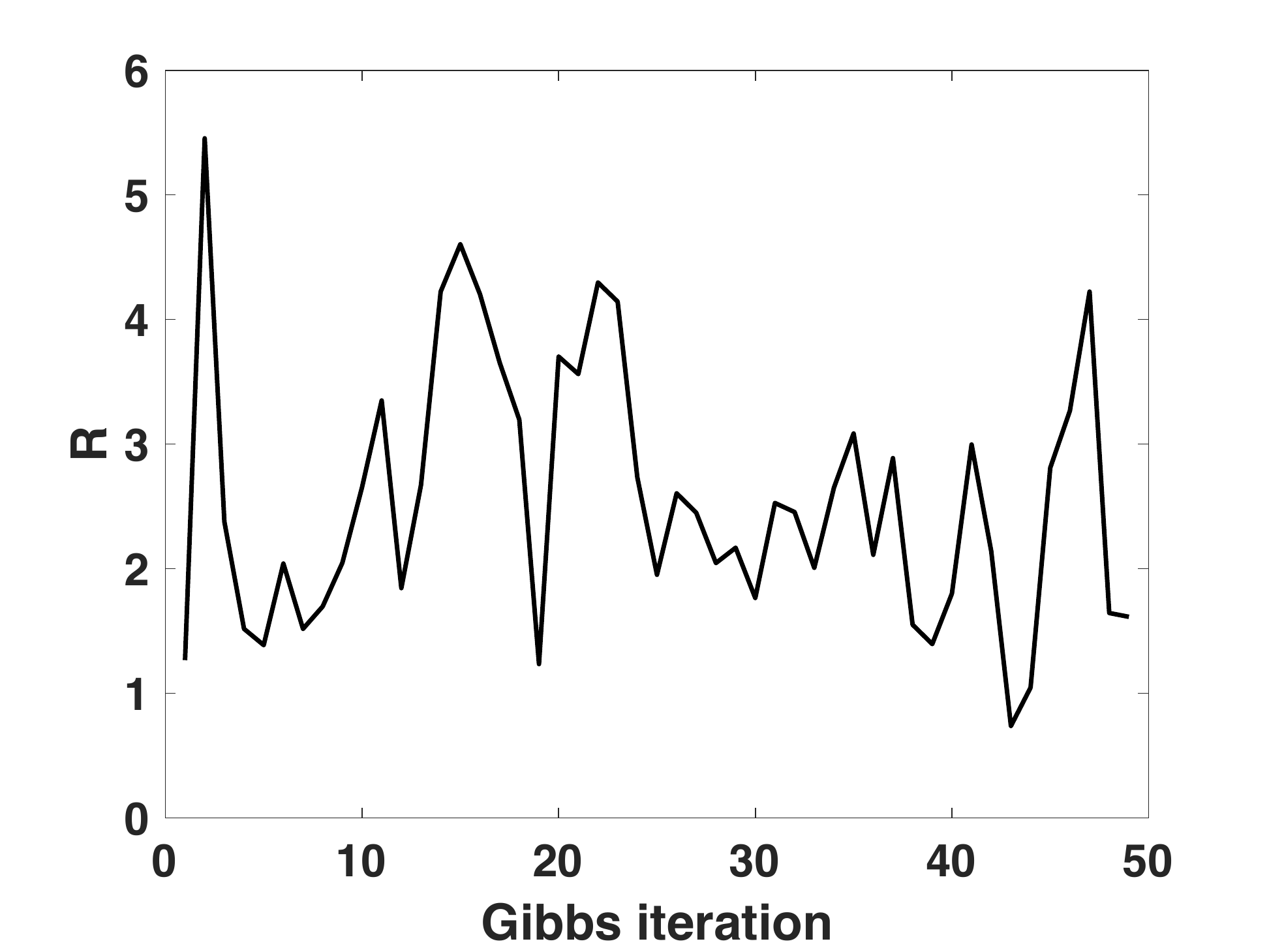}} 
  \subfigure[]{\includegraphics[width=5.4cm]{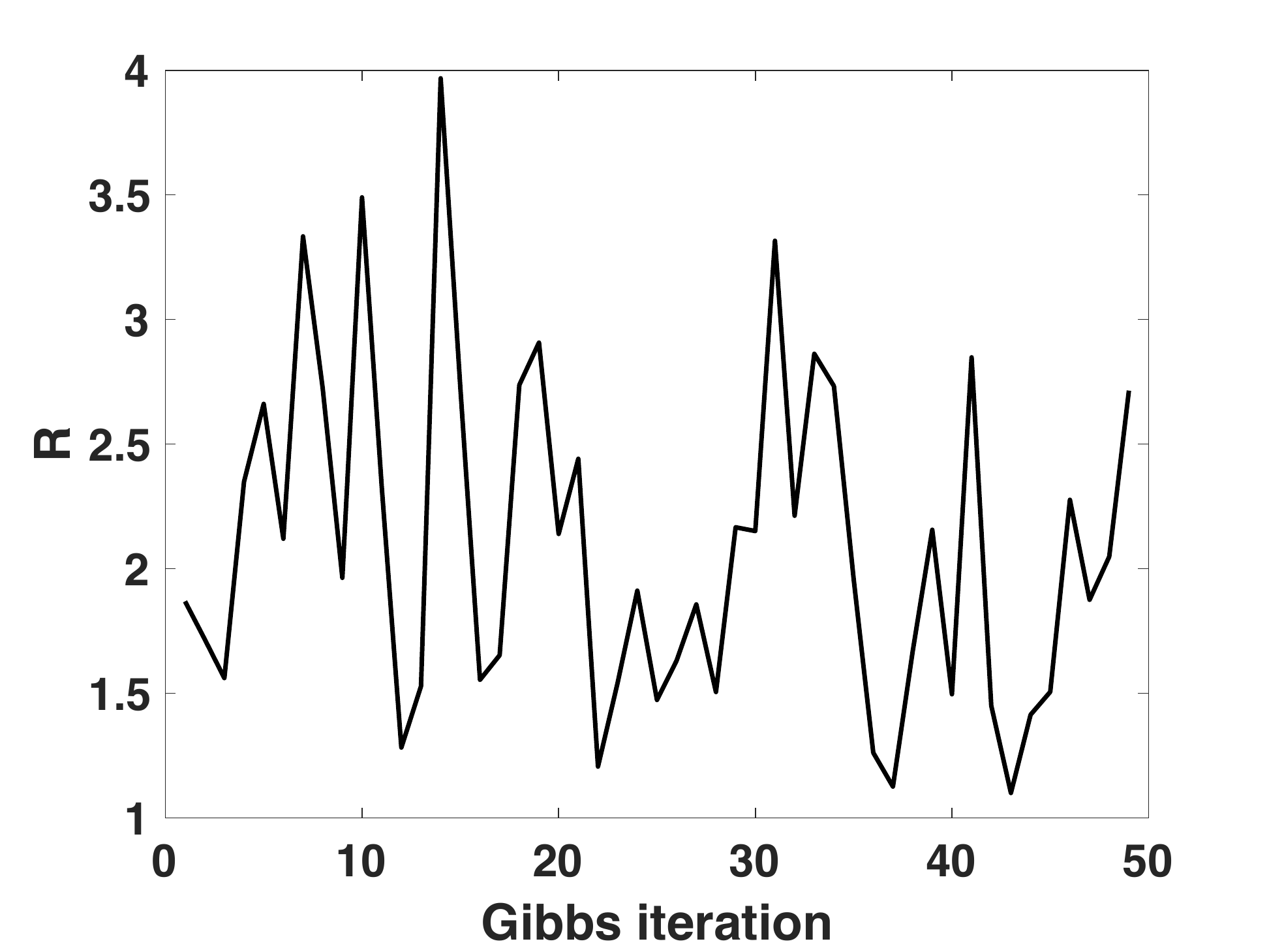}}
  }
\caption{Trace plots for $R$ in the example of Section \ref{sec:chaos}. (a) FUSS-within-Gibbs. (b)--(d) MH-within-Gibbs for $\sigma=0.2$ (b), $\sigma=1$ (c) and $\sigma=2$ (d).}
\label{fig:traceEx3}
\end{figure}

Finally, Figure \ref{fig:traceEx3} shows the trace plots for this example using the FUSS and MH algorithms, both within the Gibbs sampler, for the parameter $R$.\footnote{The behaviour of the trace plots for the other parameters (not shown) is similar.}
On the one hand, note the small variance of the chain's state around the true value of the target $R$ in Figure \ref{fig:traceEx3}(a) when using the FUSS algorithm.
Let us remark that the conditional distribution of $R$ in this example is univariate and with a very narrow peak, so having all the samples concentrated around the true value of $R$ is the desired behaviour.
On the other hand, the variance of the chain's state is much larger when using MH-within-Gibbs, Figures \ref{fig:traceEx3}(b)--(d), and the mean value is not equal to the true value of $R$ (especially when $\sigma$ increases).
This explains the poor performance shown in Table \ref{TablaResPeretti}.

\subsection{Localization in WSN and tuning of the network}
\label{sec:localization}

\vspace*{12pt}

\begin{figure}[!htb]
\centering
\includegraphics[width=6.5cm]{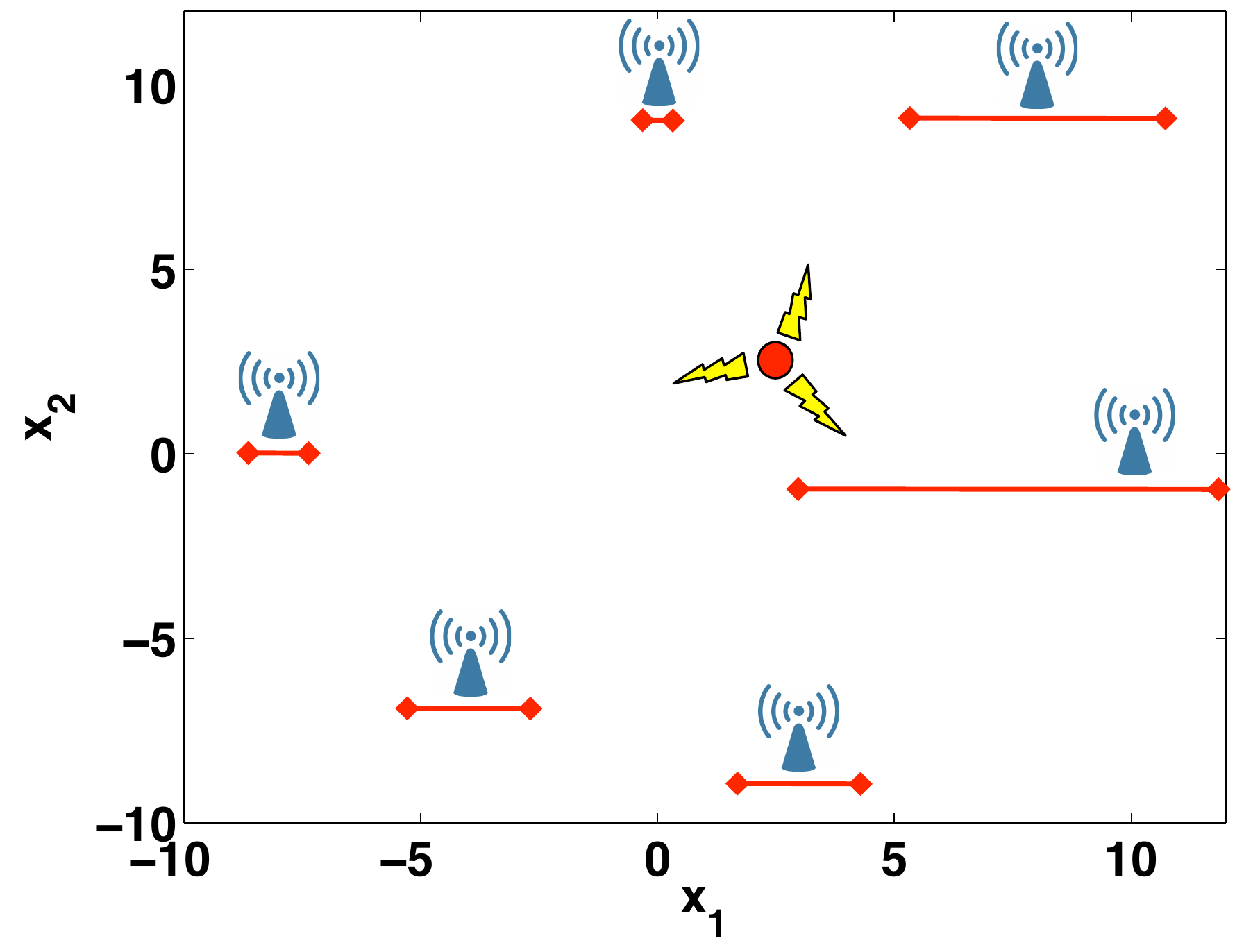}
\caption{Location of the sensors and the target for the example in Section \ref{sec:localization}.}
\label{FigSIMU_Ex2}
\end{figure}

In this second practical example, we consider the problem of localizing a target in $\mathbb{R}^{2}$ using range-only measurements in a wireless sensor network (WSN) \cite{ihler2005nonparametric,ali2009empirical}.
We assume that the measurements are contaminated by noise with an unknown power, which can be different for each sensor.
This situation is common in several practical scenarios.
The noise perturbation of each of the sensors can vary with the time and depends on the location of the sensor (due to manufacturing defects, obstacles in the reception, different physical environmental conditions, etc.).
More specifically, let us denote the target position using the random vector $\textbf{Z}=[Z_1,Z_2]^{\top}$.
The position of the target is then a specific realization $\z$. 
The range measurements are obtained from $N_S=6$ sensors located at $\textbf{h}_1=[3, -8]^{\top}$, $\textbf{h}_2=[8,10]^{\top}$, $\textbf{h}_3=[-4,-6]^{\top}$, $\textbf{h}_4=[-8,1]^{\top}$, $\textbf{h}_5=[10,0]^{\top}$ and $\textbf{h}_6=[0,10]^{\top}$, as shown in Figure \ref{FigSIMU_Ex2}.

\noindent
The observation model is
\begin{gather}
\label{IStemaejemplo}
\begin{split}
Y_{j}=20\log\left(||{\bf z}-{\bf h}_j ||\right)+B_{j}, \quad j=1,\ldots, N_S, \\
\end{split}   
\end{gather}   
where the $B_{j}$ are independent Gaussian random variables with PDFs $\gauss(b_j;0,\lambda_j^2)$ for $j=1,\ldots, N_S$.
We use ${\bm \lambda}=[\lambda_1,\ldots,\lambda_{N_S}]$ to denote the vector of standard deviations.
Given the position of the target, ${\bf z}^*=[z_1^*=2.5,z_2^*=2.5]^{\top}$, and setting ${\bm \lambda}^*=[\lambda_1^*=1,\lambda_2^*=2,\lambda_3^*=1,\lambda_4^*=0.5,\lambda_5^*=3,\lambda_{6}^*=0.2]$, we generate $N_O=20$ observations from each sensor according to the model in Eq. \eqref{IStemaejemplo}.
Then, we finally obtain a measurement matrix ${\bf Y}=[y_{k,1},\ldots, y_{k,N_S}] \in \mathbb{R}^{\ddata}$, where $\ddata=N_ON_S=120$ for $k=1,\ldots,N_O$.  
We consider a uniform prior $\mathcal{U}(\mathcal{R}_z)$ over the position $[z_1,z_2]^{\top}$ with $\mathcal{R}_z=[-30\times 30]^2$, and a uniform prior over $\lambda_j$, so that ${\bm \lambda}$ has prior $\mathcal{U}(\mathcal{R}_\lambda)$ with $\mathcal{R}_\lambda=[0,20]^{N_S}$.
Thus, the posterior PDF is 
\begin{align}
& {\bar \pi}(\parvec|\textbf{Y}) = {\bar \pi}({\bf z},{\bm \lambda}|\textbf{Y}) \nonumber \\
	\quad & = \left[\prod_{k=1}^{N_O}\prod_{j=1}^{N_S} \frac{1}{\sqrt{2\pi \lambda_j^2}}\exp\left(-\frac{1}{2\lambda_j^2}(y_{k,j}+10\log\left(||{\bf z}-{\bf h}_j ||\right)^2\right) \right]
		\mathbb{I}_{z}(\mathcal{R}_z) \mathbb{I}_{\lambda}(\mathcal{R}_\lambda), 
\end{align}
where $\parvec=[{\bf z},{\bm \lambda}]^{\top}$ is the parameter vector to be inferred, of dimension $\dpar=N_S+2=8$, and $\mathbb{I}_{c}(\mathcal{R})$ is an indicator function: $\mathbb{I}_{c}(\mathcal{R})=1$ if $c\in\mathcal{R}$, $\mathbb{I}_{c}(\mathcal{R})=0$ otherwise.

Our goal is computing the Minimum Mean Square Error (MMSE) estimator, i.e., the expected value of the posterior ${\bar \pi}(\parvec|\textbf{Y})={\bar \pi}({\bf z},{\bm \lambda}|\textbf{Y})$. 
Since the MMSE estimator cannot be computed analytically, we apply Monte Carlo methods to approximate it.
We compare the GMS algorithm, the corresponding MTM scheme, the AMIS technique, and $N$ parallel MH chains with a random walk proposal PDF.
For all of them we consider Gaussian proposal densities.
For GMS and MTM, we set $q_t(\parvec|{\bm \mu}_{n,t},\sigma^2 {\bf I})=\mathcal{N}(\parvec|{\bm \mu}_{t},\sigma^2 {\bf I})$ where ${\bm \mu}_{t}$ is adapted by considering the empirical mean of the generated samples after a training period, $t\geq 0.2 T$ \cite{Luengo+Martino:2013}, ${\bm \mu}_{0}\sim\mathcal{U}([1,5]^{\dpar})$ and $\sigma=1$.
For AMIS, we have $q_{t}(\parvec|{\bm \mu}_{t},{\bf C}_{t})=\mathcal{N}(\parvec|{\bm \mu}_{t},{\bf C}_{t})$, where ${\bm \mu}_{t}$ is as previously described (with ${\bm \mu}_{0}\sim\mathcal{U}([1,5]^{\dpar})$) and  ${\bf C}_{t}$ is also adapted using the empirical covariance matrix, starting with ${\bf C}_0= 4  {\bf I}$.
We also test the use of $N$ parallel MH chains (including the case $N=1$, which corresponds to a single chain), with a Gaussian random-walk proposal PDF, $q_n({\bm \mu}_{n,t}|{\bm \mu}_{n,t-1},\sigma^2 {\bf I})=\mathcal{N}({\bm \mu}_{n,t}|{\bm \mu}_{n,t-1},\sigma^2 {\bf I})$, and ${\bm \mu}_{n,0}\sim\mathcal{U}([1,5]^{D})$ for all $n$ and $\sigma=1$. 
 
\begin{figure}[!htb]
\centering
\includegraphics[width=6.5cm]{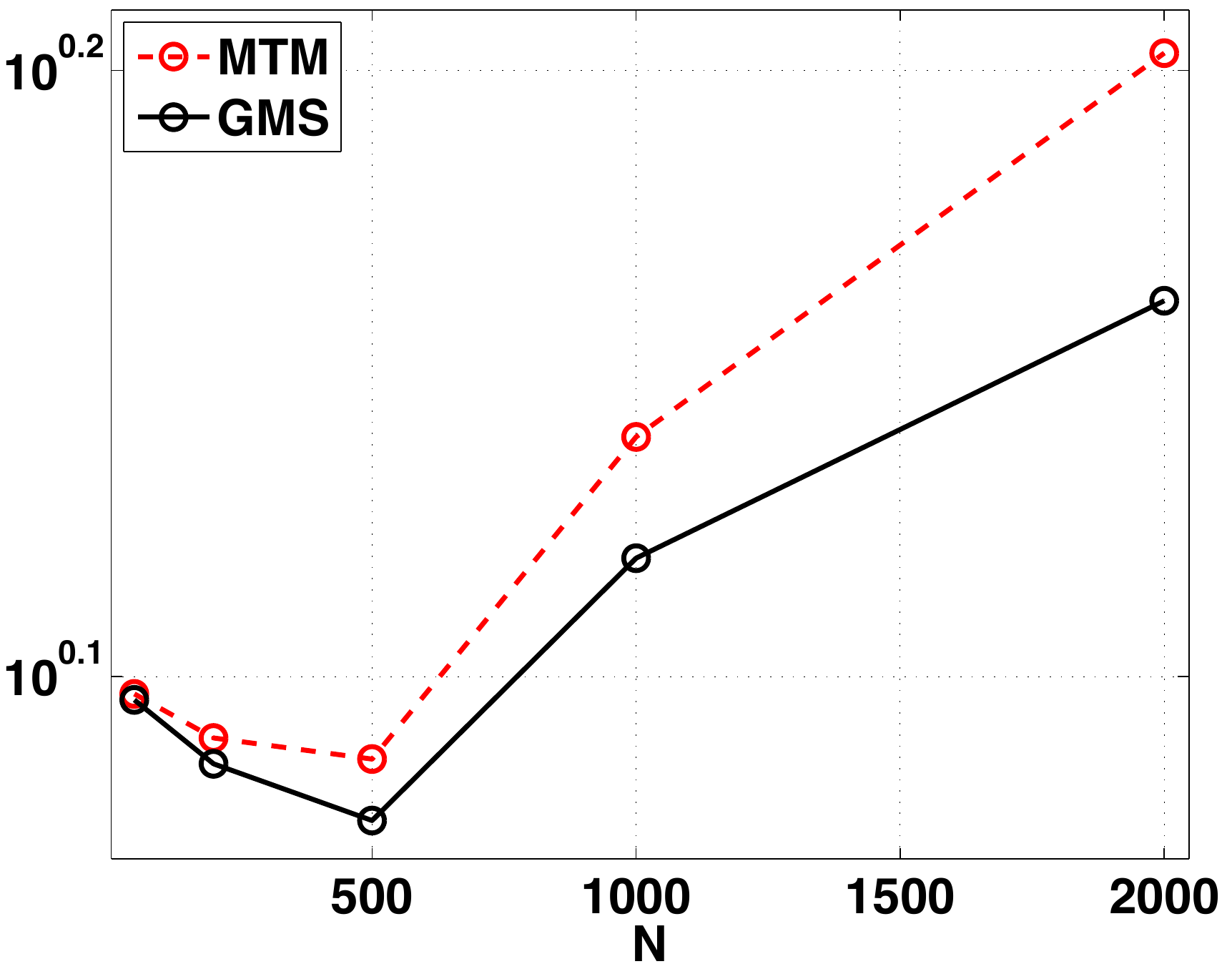}
\caption{MSE (log-scale) versus the number of candidates, $N\in\{50,200, 500, 1000, 2000\}$, obtained by GMS and the corresponding MTM algorithm, for the example in Section \ref{sec:localization}. The total number of evaluations of the posterior PDF is fixed to $E=\npart\niter=10^4$, so that $T\in\{200,50,20,10,5\}$.}
\label{FigSIMU_Ex22}
\end{figure}

We fix the total number of evaluations of the posterior density as $E=\npart\niter=10^4$.
Note that the evaluation of the posterior is usually the most costly step in MC algorithms (AMIS has the additional cost of re-weighting all the samples at each iteration according to the deterministic mixture procedure \cite{CORNUET12}).
Let us recall that $\niter$ denotes the total number of iterations and $\npart$ the number of samples drawn from each proposal at each iteration.
We consider  $\parvec^*=[{\bf z}^*,{\bm \lambda}^*]^{\top}$ as the ground-truth and compute the MSE in the estimation obtained with the different algorithms.
The results, averaged over $500$ independent runs, are provided in Tables \ref{GMS_localization}, \ref{AMIS_localization}, and \ref{MH_localization}, as well as Figure \ref{FigSIMU_Ex22}.
Note that GMS outperforms AMIS for each a pair $\{\npart,\niter\}$ (keeping $E=\npart\niter=10^4$ fixed), and also provides smaller MSE values than $N$ parallel MH chains (the case $N=1$ corresponds to a single longer chain).
Figure \ref{FigSIMU_Ex2}(b) shows the MSE versus $N$ for GMS and the corresponding MTM method.
This figure confirms again the advantage of recycling the samples in an MTM scheme.

\begin{table}[!htb]
\begin{center}
\caption{{\bf Results of the GMS algorithm for the example in Section \ref{sec:localization}.}}
\begin{tabular}{|c|c|c|c|c|c|c|c|c|}
\hline 
{\bf MSE}  &  1.30 &  1.24 &    1.22   &  1.21 &    1.22 &    {\bf 1.19} &  1.31  &    {\bf 1.44}   \\ 
\hline
\hline
$N$   &  10 & 20  & 50  & 100 & 200  & 500   & 1000&  2000  \\ 
$T$   &  1000  & 500  & 200  & 100  & 50 & 20  & 10 & 5 \\ 
\hline
$E$  &  \multicolumn{8}{c|}{$\npart\niter=10^4$}\\
\hline
{\bf MSE range} &\multicolumn{8}{c|}{ {\bf Min MSE= 1.19}  \quad   --------- \quad    {\bf Max MSE= 1.44} }     \\
\hline
\end{tabular}
\label{GMS_localization}
\end{center}
\end{table}

\begin{table}[!htb]
\begin{center}
\caption{{\bf Results of the AMIS method for the example in Section \ref{sec:localization}.}}
\begin{tabular}{|c|c|c|c|c|c|c|c|c|}
\hline 
{\bf MSE}   & 1.58 & 1.57 &  1.53   & 1.48 & 1.42  & {\bf 1.29}   &  1.48 &  {\bf 1.71}  \\  
\hline
$N$   &10 & 20 &50 & 100 & 200 & 500 & 1000 & 2000 \\
$T$   & 1000 & 500 & 200 &  100 & 50 & 20 & 10 & 5   \\ 
\hline
$E$  &  \multicolumn{8}{c|}{$NT=10^4$}\\
\hline
{\bf MSE range} &\multicolumn{8}{c|}{ {\bf Min MSE= 1.29}  \quad   --------- \quad    {\bf Max MSE= 1.71} }     \\
\hline
\end{tabular}
\label{AMIS_localization}
\end{center}
\end{table}

\begin{table}[!htb]
\begin{center}
\caption{{\bf Results of $N$ parallel MH chains with a random-walk proposal PDF for the example in Section \ref{sec:localization}.}}
\begin{tabular}{|c|c|c|c|c|c|c|c|c|}
\hline 
{\bf MSE} & 1.42 &  {\bf 1.31} &    1.44 &  2.32 &   2.73 &    {\bf 3.21} &    3.18 &    3.15  \\ 
\hline
\hline
$N$  &1 & 5 & 10 &  50 & 100 & 500 & 1000 &2000 \\
$T$   & $10^4$  & 2000    & 1000 & 200  & 100     &  20   & 10    & 5 \\ 
\hline
$E$  &  \multicolumn{8}{c|}{$NT=10^4$}\\
\hline
{\bf MSE range} &\multicolumn{8}{c|}{ {\bf Min MSE= 1.31}  \quad   --------- \quad    {\bf Max MSE=3.21 } }     \\
\hline
\end{tabular}
\label{MH_localization}
\end{center}
\end{table}

\subsection{Spectral analysis}
\label{Simu_SA}

\begin{figure*}[!tb]
\centering 
\centerline{
  \subfigure[]{\includegraphics[width=5.7cm]{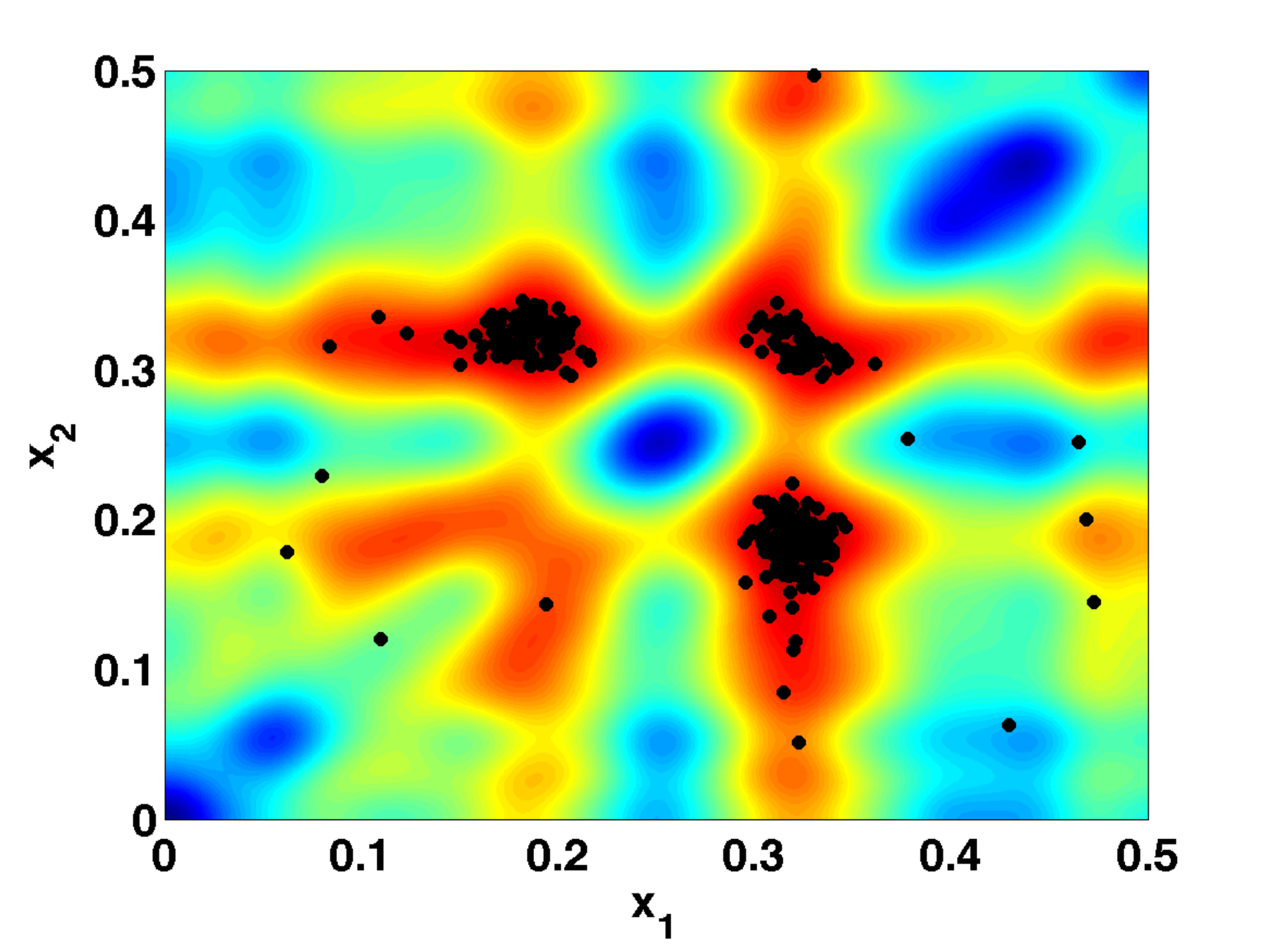}}
  \subfigure[]{\includegraphics[width=5.7cm]{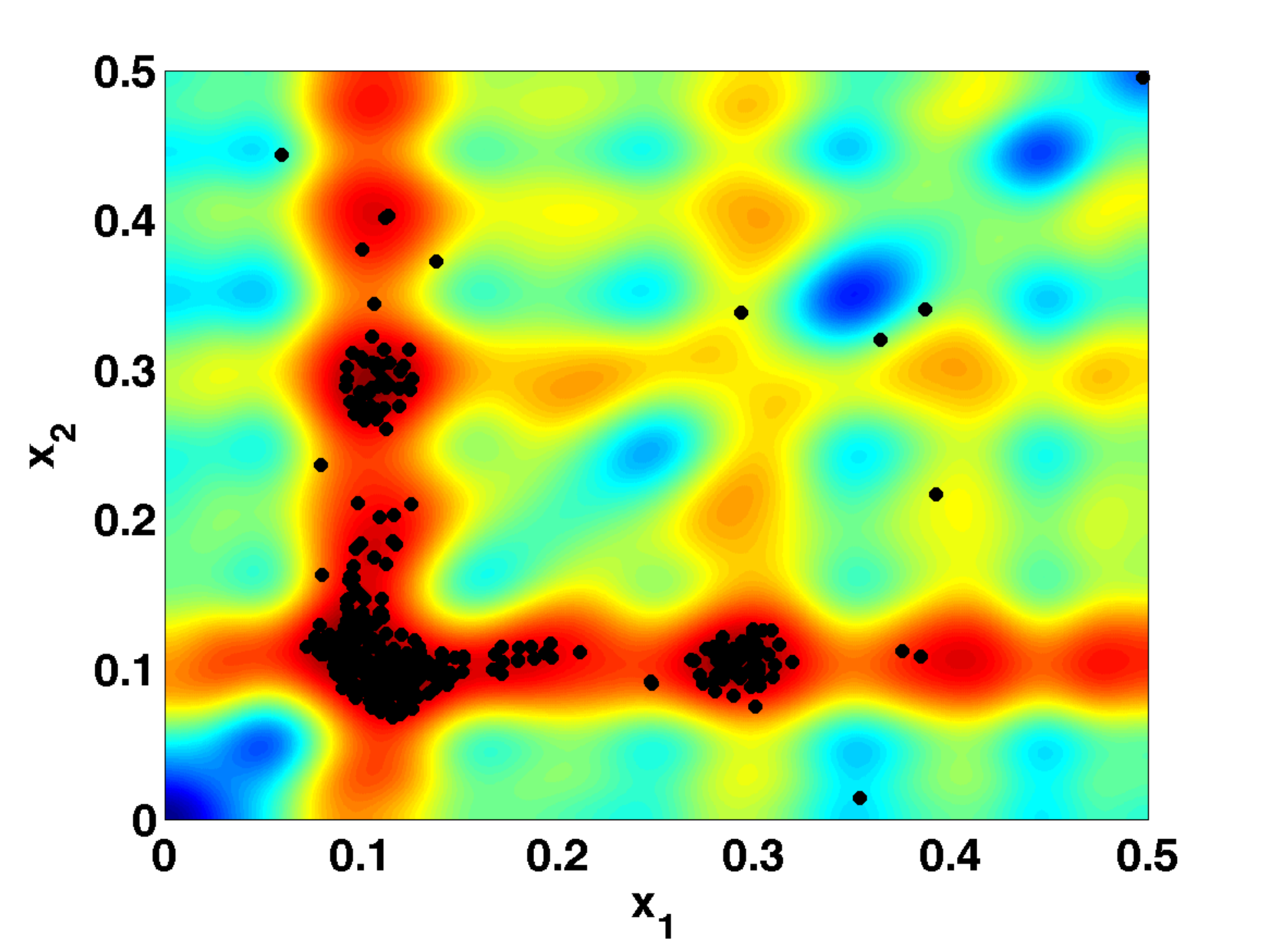}}
   \subfigure[]{\includegraphics[width=5.7cm]{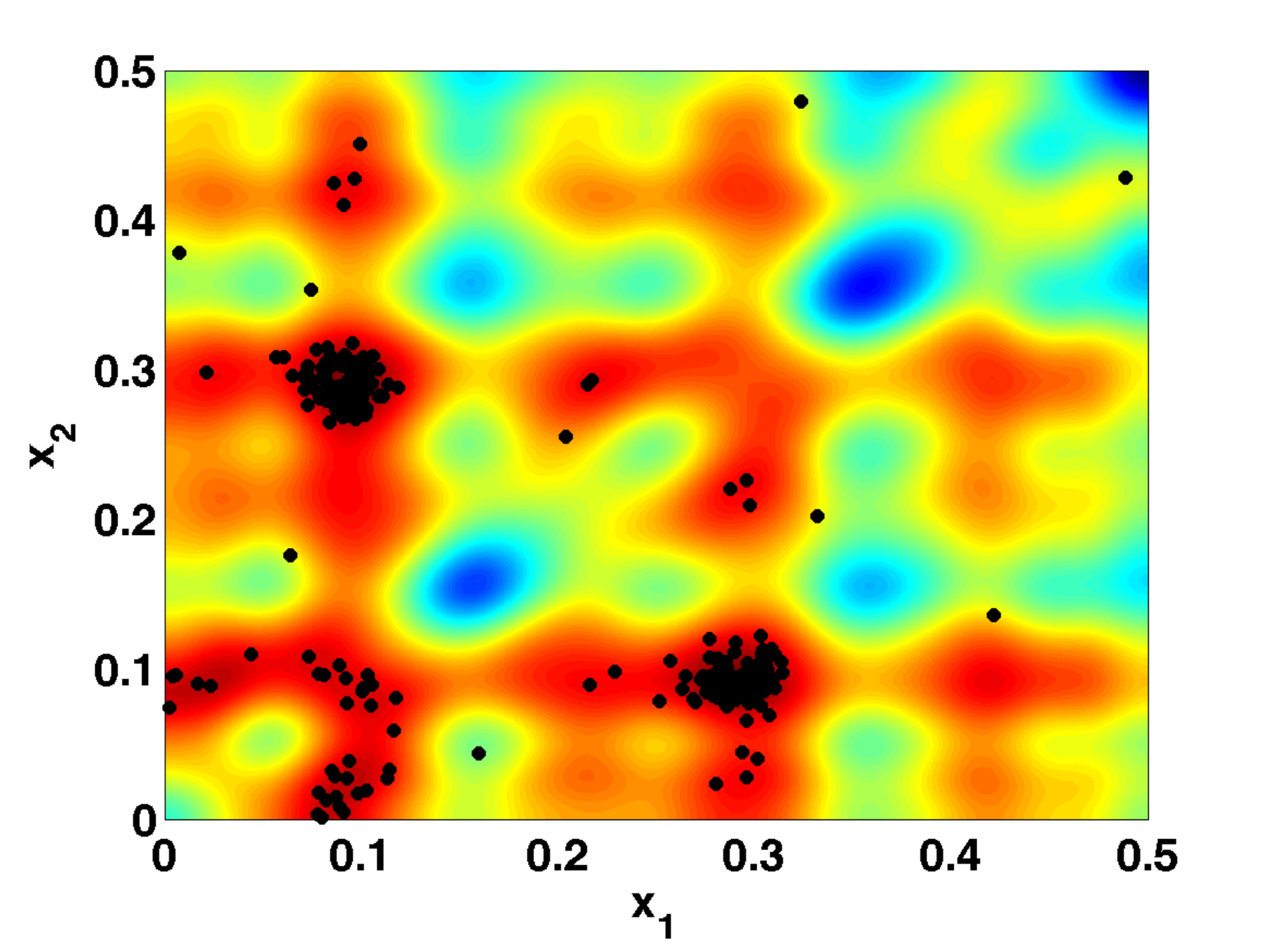}}
 }
  \caption{Several examples of the function $V({\bm \theta})=\log \pi({\bm \theta})$ with $D_\theta=2$, $L=10$, given different realizations of the measurements $y[1],\ldots,y[K]$. Black dotted points shows all the states generated throughout an O-MCMC run ($N=10$ and $T=500$)}
\label{figExFreq0}
\end{figure*}

Many problems in science and engineering require dealing with a noisy multi-sinusoidal signal, whose general form is given by
\begin{equation}
\nonumber
	y_c(\tau) = A_0 + \sum_{i=1}^{D_\theta}{A_i \cos(2\pi f_i \tau + \phi_i)} + r(\tau), \quad \tau \in \mathbb{R},
\end{equation}
where $A_0$ is a constant term, $D_\theta$ is the number of sinusoids, $\{A_i\}_{i=1}^{D_\theta}$ is the set of amplitudes, $\{2 \pi f_i\}_{i=1}^{D_\theta}$ are the frequencies, $\{\phi_i\}_{i=1}^{D_\theta}$ their phases, and $r(\tau)$ is an additive white Gaussian noise (AWGN) term.
The estimation of the parameters of this signal is required by many applications in signal processing \cite{stoica1993list,so2005linear}, in control (where a multi-harmonic disturbance is often encountered in industrial plants) \cite{bodson1997adaptive,bobtsov2012cancelation} or in digital communications (where multiple narrowband interferers can be roughly modeled as sinusoidal signals) \cite{carlemalm2000suppression,rao2005nonlinear}.
Let us assume that we have $L$ equispaced samples from $y_c(\tau)$, obtained discretizing $y_c(\tau)$ with a period $T_s < \frac{\pi}{\max_{1 \le i \le D_\theta} 2 \pi f_i}$ (in order to fulfill the sampling theorem \cite{Proakis95}):
\begin{equation}
\nonumber
	y[k] = A_0 + \sum_{i=1}^{D_\theta}{A_i \cos(\Omega_i k + \phi_i)} + r[k], \quad k=1,\ldots,L,
\end{equation}
where $y[k] = y_c(kT_s)$ for $k=0, 1, \ldots, L-1$, $\Omega_i = 2 \pi f_i T_s$ for $i=1, \ldots, D_\theta$, and $r[k] \sim \mathcal{N}(0,\sigma_w^2)$.
We apply parallel MH algorithms to provide an accurate estimate of the set of unknown frequencies, $\{\Omega_i\}_{i=1}^{D_\theta}$ or merely $\{f_i\}_{i=1}^{D_\theta}$.
In order to maintain the notation used throughout the paper, we denote the vector of frequencies to be inferred as ${\bm \theta}\in \mathbb{R}^{D_\theta}$.
Thus, considering the hyper-rectangular domain ${\bm \Theta}=\left[0,\frac{1}{2}\right]^{D_\theta}$ (it is straightforward to note the periodicity outside ${\bm \Theta}$), and  a uniform prior on ${\bm \Theta}$, the posterior distribution given $K$ data is  $\bar{\pi}({\bm \theta}) \propto \exp\left(-V({\bm \theta})\right)$, where 
\begin{eqnarray}
V(\theta_1,\ldots,\theta_{D_\theta})=\frac{1}{2\sigma_w^2} \sum_{k=1}^{L}\left(y[k]- A_0 - \sum_{i=1}^{D_\theta}{A_i \cos(\theta_i k + \phi_i)} \right)^2 \mathbb{I}_{{\bm \Theta}}({\bm \theta}), \nonumber
\end{eqnarray}
and we have used $\mathbb{I}_{{\bm \Theta}}({\bm \theta})$ to denote the indicator function such that  $\mathbb{I}_{{\bm \Theta}}({\bm \theta})=1$ if ${\bm \theta} \in {\bm \Theta}$ and $\mathbb{I}_{{\bm \Theta}}({\bm \theta})=0$ if ${\bm \theta} \notin {\bm \Theta}$.
Moreover, for the sake of simplicity we have also assumed that $S$ and $\sigma_w^2$ are known.
Furthermore, we set $A_0=0$, $A_i=A=1$ and $\phi_i=0$.\footnote{Let us remark that the estimation of all these parameters would make the inference harder, but can be easily incorporated into our algorithm.} Note that the problem is symmetric with respect to the hyperplane $\theta_1=\theta_2=\ldots=\theta_{D_\theta}$ (and, in general, multimodal).
Bidimensional examples of $V({\bm \theta})=\log \pi({\bm \theta})$ are depicted in Figure \ref{figExFreq0}. 
We apply the OMCMC method \cite{OMCMC}, where $N$ parallel interacting MH chains are used, comparing it with $N$ independent parallel MH chains (IPCs).
The proposal densities are all Gaussian random-walks proposal PDFs with diagonal covariance matrices $\vec{C} = \sigma^2 \vec{I}$.

\begin{figure*}[!tb]
\centering 
\centerline{
  \subfigure[]{\includegraphics[width=0.3\textwidth]{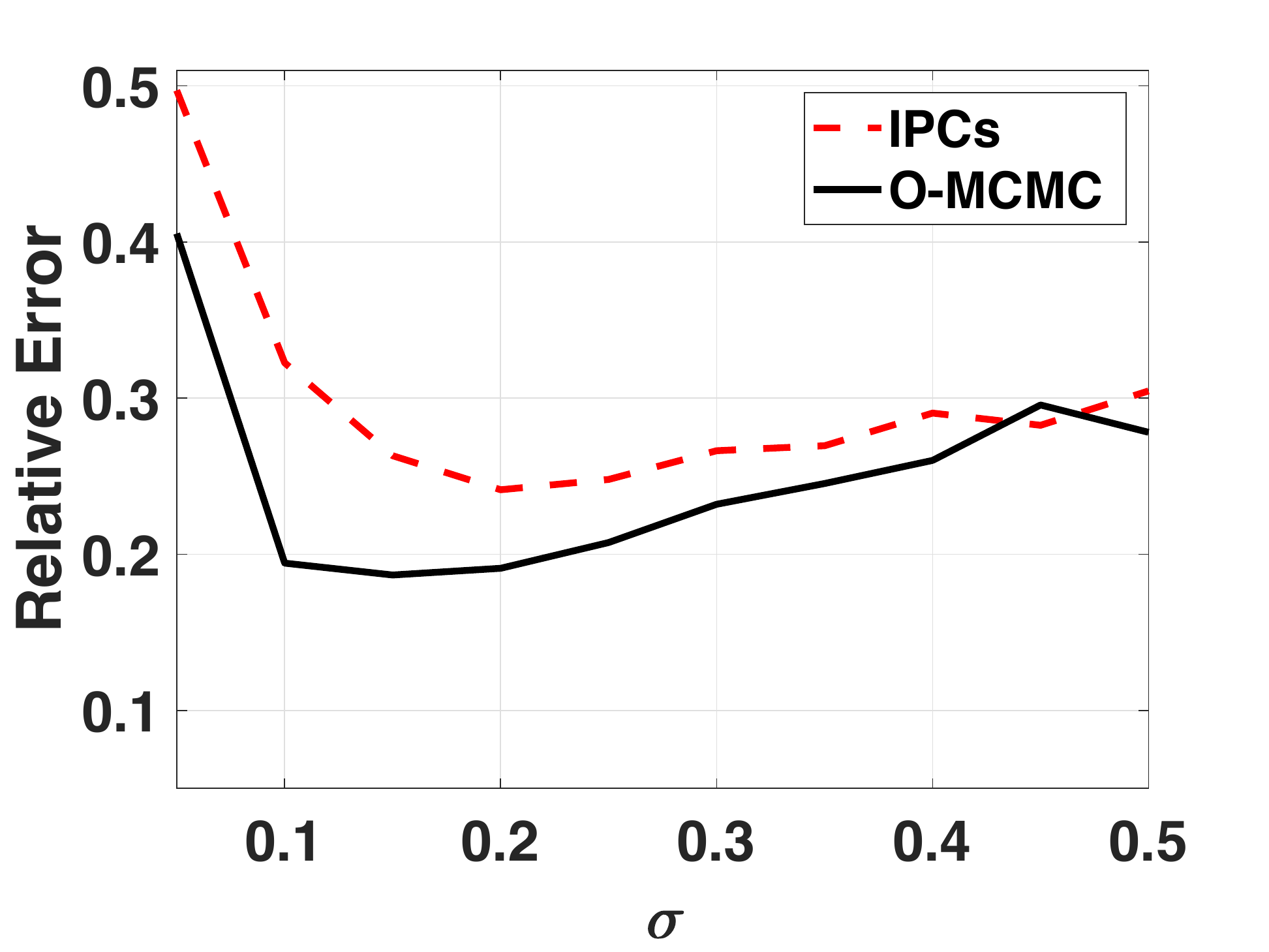}} 
  \subfigure[]{\includegraphics[width=0.3\textwidth]{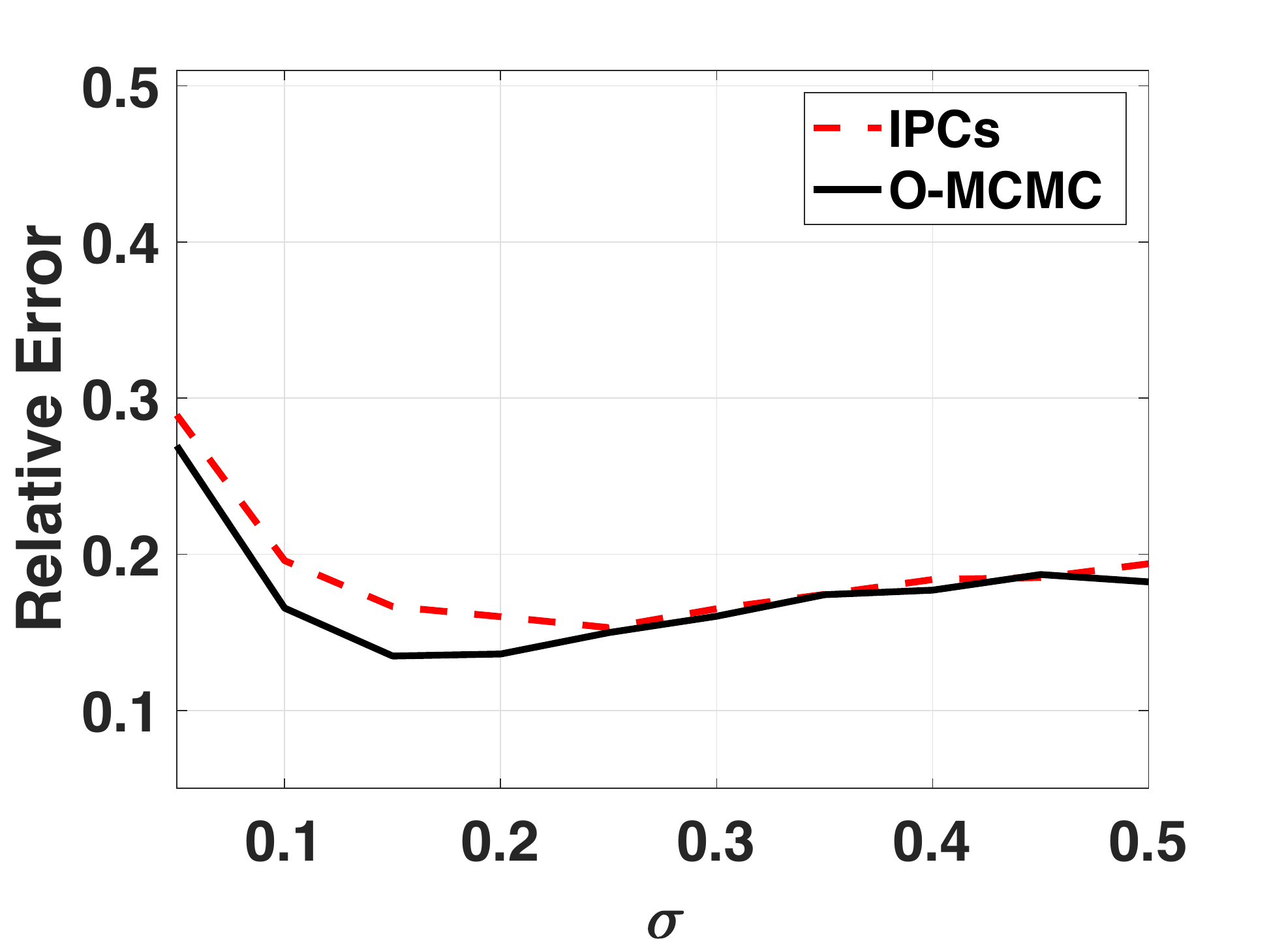}} 
  \subfigure[]{\includegraphics[width=0.3\textwidth]{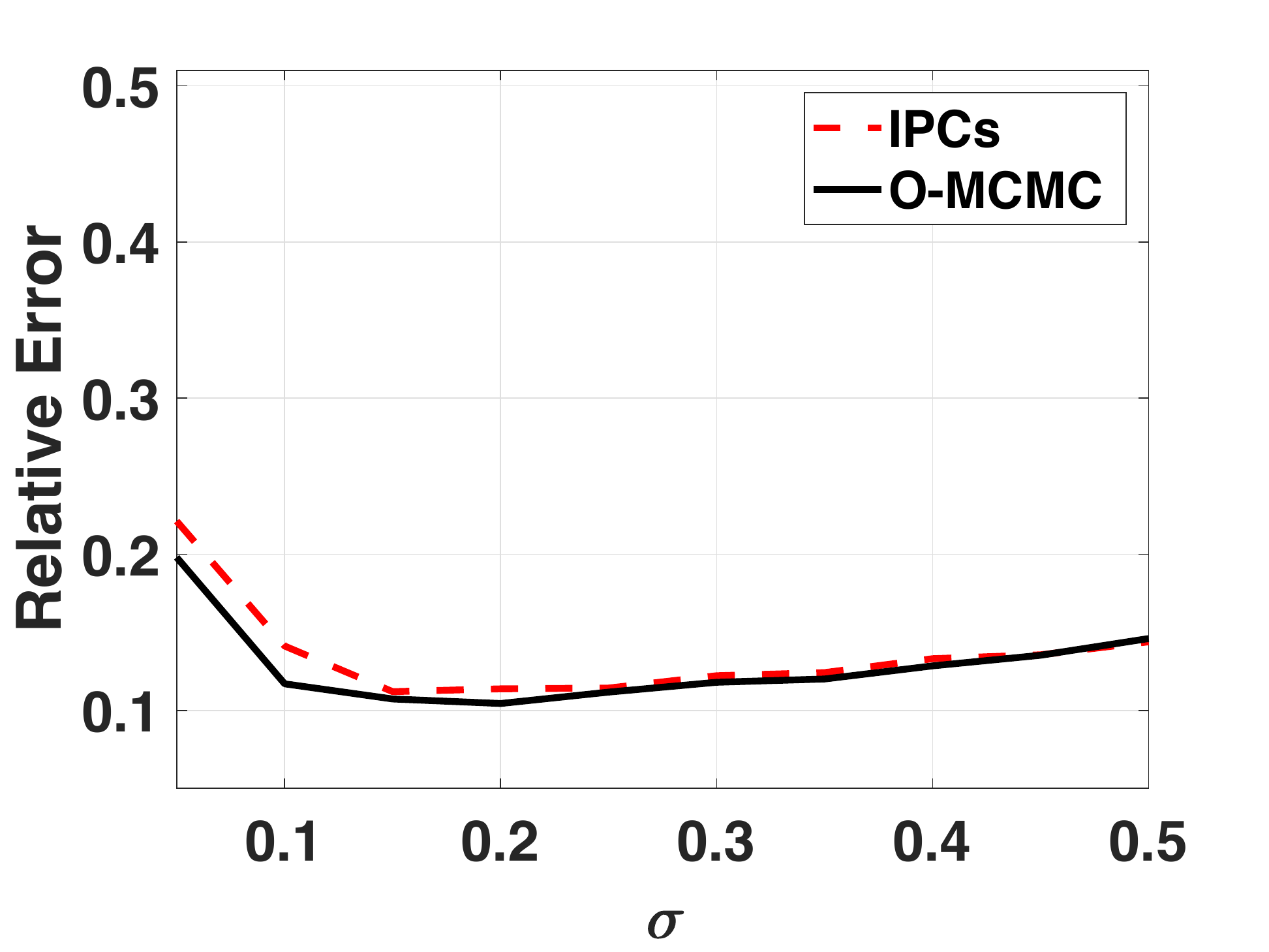}}
 }
  \caption{Relative Error (averaged over 500 runs)  for OMCMC (solid line) and IPCs (dashed line) with different computational effort $E_T$.}
\label{figExFreq1}
\end{figure*}

We set ${\bf f}=[f_1=0.1,f_2=0.3]^{\top}$ and generate $L=10$ synthetic data from the model.
Moreover, we set the total number of target evaluations for OMCMC to $E_T=M(N+1)\in \{2730,5450,10.9\cdot 10^3\}$.
For a fair comparison, we consider $N$ independent parallel chains (IPCs) choosing $T$ such that $E_T'=NT$ is equal to $E_T$, i.e., $E_T'=E_T$.
We test different values of $\sigma\in [0.05,0.5]$ and $N\in \{2,5,10\}$.
We test several combinations of the number of chains ($N$) and epochs ($M$ for OMCMC and $T$ for IPCs), always keeping $E_T$ fixed.
The Relative Error (RE) in the estimation, averaged over 500 independent runs, is shown in Figure \ref{figExFreq1}.
We can observe that  O-MCMC (solid line) outperforms IPCs (dashed line), attaining lower REs.
The performance becomes similar as the computational effort $E_T$ grows, since the state space in the first experiment, ${\bm \Theta}=\left[0,\frac{1}{2}\right]^2$, is small enough to allow for an exhaustive exploration of ${\bm \Theta}$ by independent chains. 

\begin{figure}[!htb]
  \centering
  \centerline{
	\includegraphics[width=6cm]{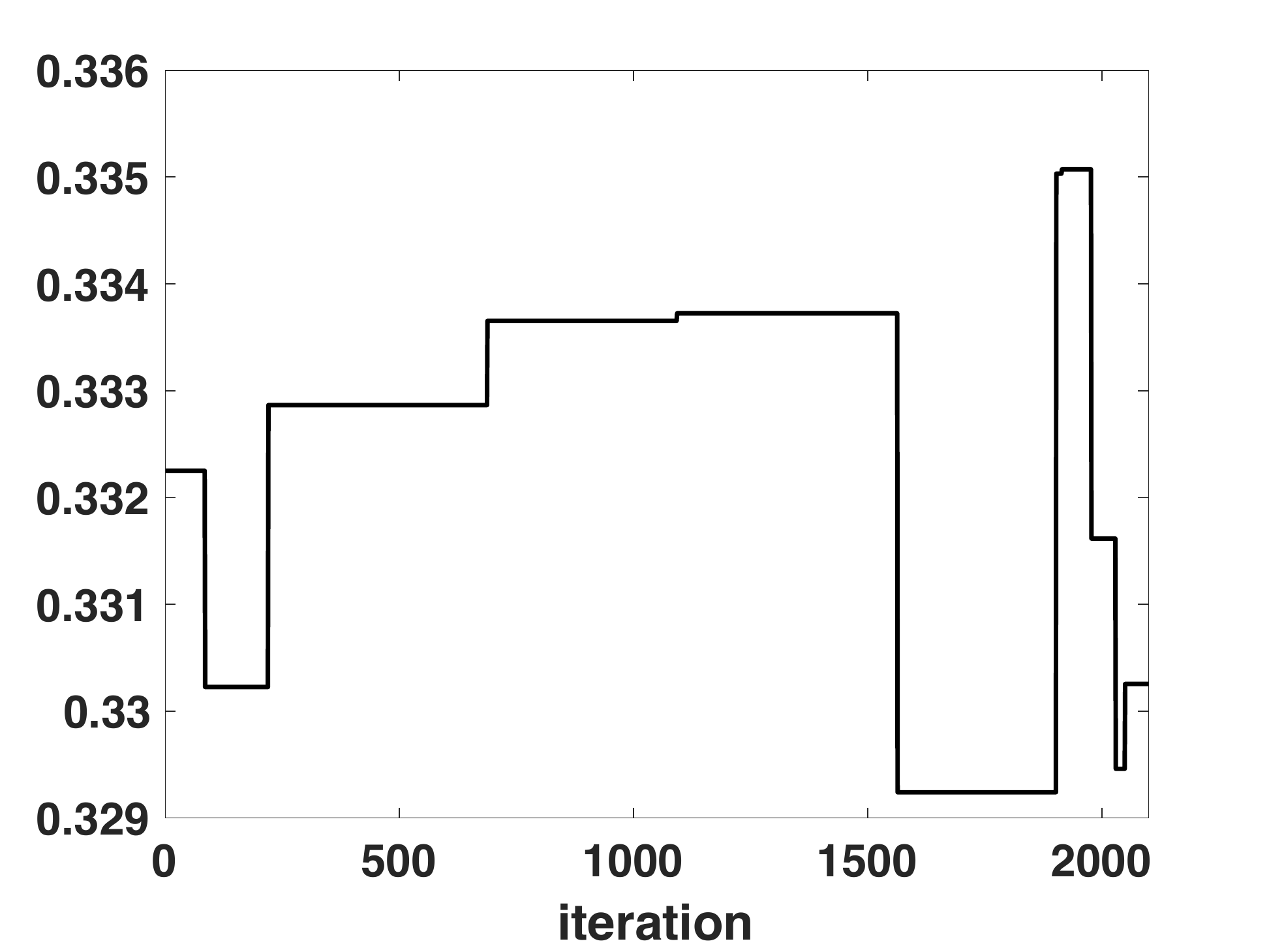}
	\hspace{2cm}
	\includegraphics[width=6cm]{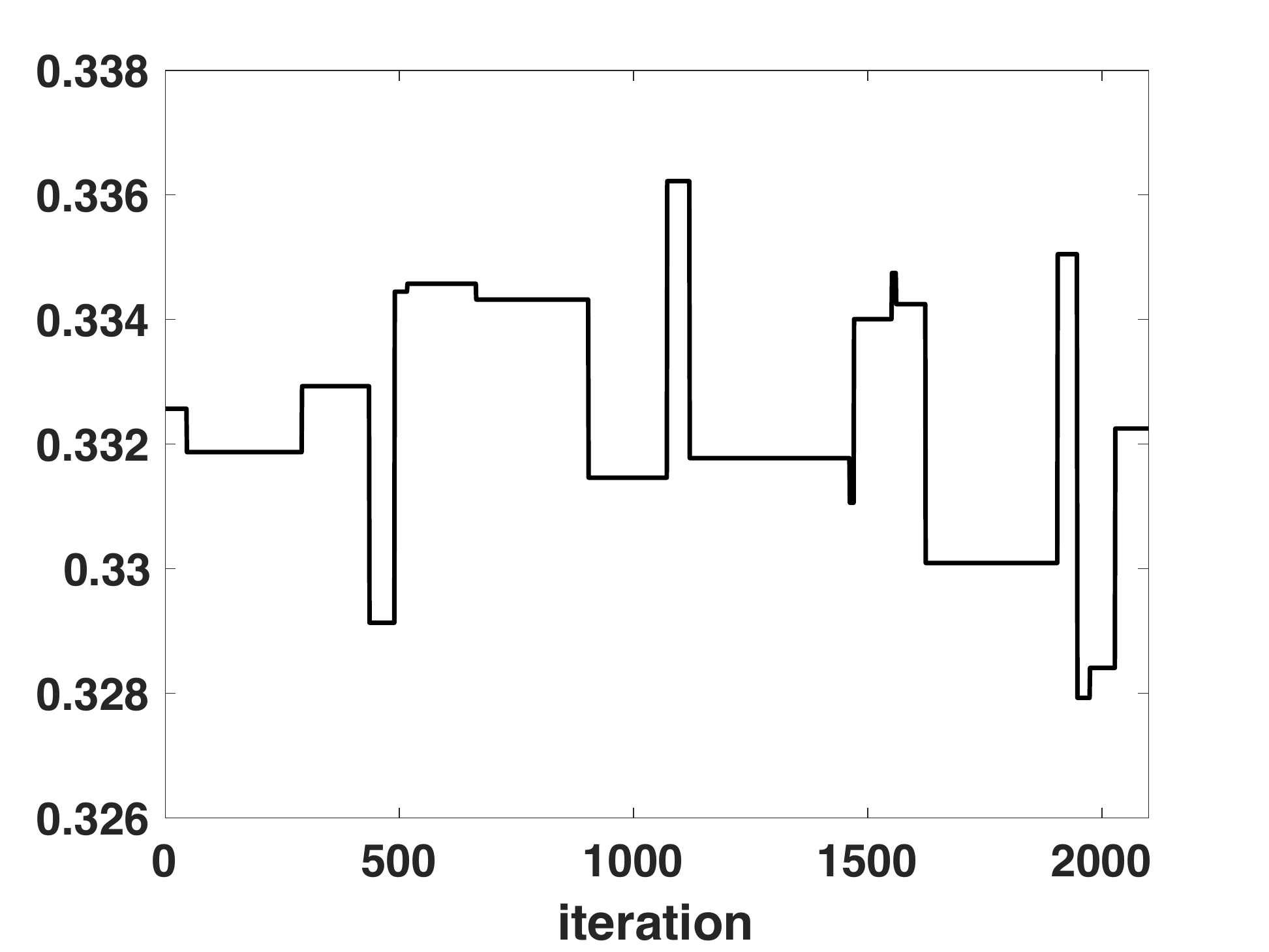}
	}
\caption{Two typical trace plots for $f_2=\frac{1}{3}$ in the example of Section \ref{Simu_SA} using the OMCMC-MTM algorithm with $T_v = 2$ vertical steps of an MH algorithm, $T_h = 1$ horizontal steps of the MTM algorithm, and $E_T = 700$ target evaluations.}
\label{fig:traceEx5}
\end{figure}

Finally, Figure \ref{fig:traceEx5} shows two typical examples of trace plots for the estimation of frequency $f_2=\frac{1}{3}$, as in Figure \ref{figExFreq0}(a).
In both cases, we use the OMCMC-MTM algorithm with $T_v = 2$ vertical steps of an MH algorithm, $T_h = 1$ horizontal steps of the MTM algorithm, and $E_T = 700$ target evaluations.
Note the fast convergence of the algorithm to frequencies close to the true one.
This is a particularly good result, since the peaks of the target PDF in this case were very narrow.

\newpage

\section{Conclusion}
\label{sec:discussion}


In this paper, we have performed a review of Monte Carlo (MC) methods for the estimation of static parameters in statistical signal processing problems.
MC methods are simulation-based techniques that are extensively used nowadays to perform approximate inference when analytical estimators cannot be computed, as it happens in many real-world signal processing applications.
We have concentrated on the description of some of the most relevant methods available in the literature, rather than focusing on specific applications.
Many different algorithms are provided throughout the text in a clear and unified format, so that signal processing practitioners can directly apply them in their specific problems.

In order to make the paper as self-contained as possible, we have started from scratch, describing first the MC method altogether with its convergence properties.
Markov chain Monte Carlo (MCMC) techniques are considered first, starting with three classical MC methods (the Metropolis-Hastings (MH) algorithm, the Gibbs sampler and MH-within-Gibbs) that are widely used by signal processing practitioners and can be considered as the basic building blocks of more recent approaches.
Then, we detail several advanced MCMC algorithms, focusing on adaptive MCMC schemes (both using parametric and non-parametric proposals) and MCMC methods with multiple candidates.
Although the focus of the paper is on MCMC methods, a brief description of importance sampling (IS) and adaptive importance sampling (AIS) methods is also included for the sake of completeness.

Two simple problems (where the analytical estimators can be computed and used to evaluate the performance of several MC methods), a challenging example that appears in several scientific fields (the estimation of the parameters of a chaotic system), and two classical signal processing applications (localization in a wireless sensor network and the spectral analysis of multiple sinusoids) are used to test many of the algorithms described.

Finally, let us remark that Monte Carlo methods can result infinite variance estimators if not properly applied.
As a cautionary note, let us mention Newton and Raftery's weighted likelihood bootstrap \cite{newton1994approximate}.
Although their approach leads to asymptotically unbiased estimators, it is also well-known that the variance of these estimators is infinite and thus practitioners may end up with estimated values which are very far away from the correct ones.

%
%

\section*{Acknowledgements}

The authors gratefully acknowledge the support for their research: Ministerio de Econom\'{\i}a y Competitividad (MINECO) of Spain under the TEC2015-64835-C3-3-R MIMOD-PLC project, Ministerio de Educaci\'on, Cultura y Deporte of Spain under CAS15/00350 grant, Universidad Polit\'ecnica de Madrid through a mobility grant for a short visit to Stony Brook University (D. Luengo); MINECO of Spain through Red de Excelencia KERMES TEC2016-81900-REDT (D. Luengo and L. Martino); the National Science Foundation under Award CCF-1617986 (M. F. Bugallo); the European Research Council (ERC) through the ERC Consolidator Grant SEDAL ERC-2014-CoG 647423 (L. Martino); the Academy of Finland project 266940 (S. S\"arkk\"a); the French National Research Agency through PISCES project ANR-17-CE40-0031-01 (V. Elvira); and from the French-American Fulbright Commission through the Fulbright scholar fellowship (V. Elvira).
They also want to thank the reviewers for their many helpful comments that have contributed to substantially improve the quality of the paper.


\bibliographystyle{bmc-mathphys} 
\bibliography{bmc_article,bib_david}      


\end{document}